\documentclass[12pt]{article}
\usepackage{ametsoc}
\usepackage{amsmath} 
\usepackage{amssymb} 
\usepackage{amsfonts}
\usepackage{graphicx,color}                 
\usepackage{lineno}
\usepackage{subfigure}
\usepackage{IEEEtrantools}

\usepackage{array}
\newcommand{\myabstract}{
In this work we propose an ensemble 4D seismic history matching framework for reservoir characterization. Compared to similar existing frameworks in reservoir engineering community, the proposed one consists of some relatively new ingredients, in terms of the type of seismic data in choice, wavelet multiresolution analysis for the chosen seismic data and related data noise estimation, and the use of recently developed iterative ensemble history matching algorithms.

Typical seismic data used for history matching, such as acoustic impedance, are inverted quantities, whereas extra uncertainties may arise during the inversion processes. In the proposed framework we avoid such intermediate inversion processes. In addition, we also adopt wavelet-based sparse representation to reduce data size. Concretely, we use intercept and gradient attributes derived from amplitude versus angle (AVA) data, apply multilevel discrete wavelet transforms (DWT) to attribute data, and estimate noise level of resulting wavelet coefficients. We then select the wavelet coefficients above a certain threshold value, and history-match these leading wavelet coefficients using an iterative ensemble smoother.        

As a proof-of-concept study, we apply the proposed framework to a 2D synthetic case originated from a 3D Norne field model. The reservoir model variables to be estimated are permeability (PERMX) and porosity (PORO) at each active gridblock. A rock physics model is used to calculate seismic parameters (velocity and density) from reservoir properties (porosity, fluid saturation and pressure), then reflection coefficients are generated using a linearized AVA equation that involves velocity and density. AVA data are obtained by computing the convolution between reflection coefficients and a Ricker wavelet function. The multiresolution analysis applied to the AVA attributes helps to obtain a good estimation of noise level and substantially reduce the data size. We compare history matching performance in three scenarios: (S1) with production data only, (S2) with seismic data only, and (S3) with both production and seismic data. In either scenario S2 or S3, we also inspect two sets of experiments, one using the original seismic data (full-data experiment) and the other adopting sparse representations (sparse-data experiment). Our numerical results suggest that, in this particular case study, using production data largely improves the estimation of permeability, but has little effect on the estimation of porosity. Using seismic data only improves the estimation of porosity, but not that of permeability. In contrast, using both production and 4D seismic data improves the estimation accuracies of both porosity and permeability. Moreover, in either scenario S2 or S3, provided that a certain stopping criterion is equipped in the iterative ensemble smoother, adopting sparse representations results in better history matching performance than using the original data set.     
}
\begin{document}
%
%
\title{\textbf{\large{An Ensemble 4D Seismic History Matching Framework with Sparse Representation Based on Wavelet Multiresolution Analysis}}}
%
%
\author{\textsc{Xiaodong Luo}
				\thanks{\textit{Corresponding author address:}
				International Research Institute Of Stavanger (IRIS), Thorm{\o}hlens Gate 55, 5008 Bergen, Norway
				\newline{E-mail: xiaodong.luo@iris.no}} \\
				\textit{\footnotesize{International Research Institute Of Stavanger (IRIS), Bergen, Norway}}				
\and
\centerline{\textsc{Tuhin Bhakta}}\\
\centerline{\textit{\footnotesize{International Research Institute Of Stavanger (IRIS), Bergen, Norway}}}
\and
\centerline{\textsc{Morten Jakobsen}}\\
\centerline{\textit{\footnotesize{University of Bergen, and International Research Institute Of Stavanger (IRIS), Bergen, Norway}}}
\and
\centerline{\textsc{Geir N\ae{}vdal}}\\
\centerline{\textit{\footnotesize{International Research Institute Of Stavanger (IRIS), Bergen, Norway}}}
}
%
\ifthenelse{\boolean{dc}}
{
\twocolumn[
\begin{@twocolumnfalse}
\amstitle

\begin{center}
\begin{minipage}{13.0cm}
\begin{abstract}
	\myabstract
	\newline
	\begin{center}
		\rule{38mm}{0.2mm}   
	\end{center}
\end{abstract}
\end{minipage}
\end{center}
\end{@twocolumnfalse}
]
}
{
\amstitle
\begin{abstract}
\myabstract 
\end{abstract}
}

\section{Introduction}\label{sec:introduction}
\noindent History matching is an inverse problem that aims to find reservoir model variables (e.g., permeabilities and porosities at gridblocks) that match observations \citep{Oliver-recent-2010}. With the advance of computer power, automatic history matching gains popularity in the community. 
At the early stage, a lot of efforts were dedicated to gradient-based deterministic methods (see, for example, \citealp{chen1974new,dadashpour2008nonlinear}). In a gradient-based deterministic method, one solves a certain deterministic optimization problem and obtains a single set of estimated reservoir model variables. Although gradient-based deterministic methods often work well in both research and practice, there are also a few noticed limitations. 

One practical limitation is the need to evaluate the Jacobian matrix of observations with respect to model variables. Such an evaluation can be done through either a numerical scheme (e.g., finite difference approximation, see, for example, \citealp{dadashpour2008nonlinear}) or an adjoint-based system (see, for example, \citealp{chen1974new}). Using a numerical scheme requires to store the Jacobian matrix in memory, therefore it is efficient when the numbers of both model variables and observations are relatively small. Adopting an adjoint-based approach can mitigate the intensive demand of storage memory from large models, however, it is typically time-consuming to develop an adjoint system. Flexibility of adjoint-based approaches is another hurdle in practice. For instance, one cannot build an adjoint system for a black-box reservoir simulator. In addition, should either observations or model variables be changed, the corresponding adjoint system may need to be re-built. 

Another limitation is the capacity of quantifying the uncertainty associated with inversion results. Such a quantification is essential for better decision-making in reservoir operations and management. In terms of uncertainty quantification, gradient-based deterministic methods appear less competent by nature, since they are mostly developed in the context of deterministic inverse problems (see, for example, \citealp{Engl2000-regularization}). Although it is possible to relate a gradient-based deterministic method to a stochastic approach, e.g., in the context of Bayesian inverse problems (see, for example, \citealp{Tarantola-inverse}), this connection is established by confining the distributions of model variables and observations to some specific choices (e.g., Gaussian distribution), which may not always be valid in practice.          

In recent years, ensemble-based history matching methods, for instance, the ensemble Kalman filter (EnKF, see, for example, \citealp{naevdal2005reservoir,Aanonsen-ensemble-2009}), the ensemble smoother (ES, see, for example, \citealp{van1eeuwen996data}), the ensemble Kalman smoother (EnKS, see, for example, \citealp{Evensen2000}) and their iterative versions (see, for example, \citealp{chen2013-levenberg,emerick2012ensemble,Gu2007-iterative,Li2009-iterative,luo2015Iterative}), have received a lot of attention in the community. In contrast to gradient-based deterministic methods, advantages of ensemble-based approaches include simplicity and flexibility in implementation, and better capacity of uncertainty quantification. 
Ensemble-based methods are derivative-free in that they do not evaluate Jacobian matrices in their implementations. They are also non-intrusive approaches and can work with generic forward simulators (including black-box systems). Having these two features (derivative-free and non-intrusive), ensemble-based methods emerge as a convenient toolset for history matching and can be adopted in various situations that may involve reservoir models with different sizes, and different types of model parameterization and/or observation data (see, for example, \citealp{Aanonsen-ensemble-2009}). In addition, ensemble-based methods can be interpreted as stochastic versions of gradient-based deterministic approaches, and under suitable conditions, they are Bayesian approaches, regardless of the statistical distributions of model variables and observations (see, for example, the analysis in \citealp{luo2015Iterative}). This forms the ground of uncertainty quantification in the context of Bayesian inversion.

In history matching problems it is customary to assimilate production data (e.g., bottom hole pressures, oil production rates, water cuts etc.) into reservoir models. Production data are often frequent in time, but have relatively sparse coverage in space. Complementary to production data, 4D (or time-lapse) seismic data are usually denser in space, and thus provide valuable additional information for history matching. Seismic history matching using ensemble-based methods has been carried out in some recent works. For instance, \cite{abadpour20134d,fahimuddin2010ensemble,katterbauer2015history,leeuwenburgh2014distance,skjervheim2007incorporating,trani2012seismic} use the EnKF or a combination of the EnKF and EnKS, whereas \cite{emerick2012history,emerick2013history} employ the ensemble smoother with multiple data assimilation (ES-MDA). To reduce the computational cost in forward simulations, most of seismic history matching studies adopt inverted seismic properties that are obtained through seismic inversions. Such inverted properties can be, for instance, acoustic impedance (see, for example, \citealp{emerick2012history,emerick2013history,fahimuddin2010ensemble,skjervheim2007incorporating}) or saturation front (see, for example, \citealp{abadpour20134d,leeuwenburgh2014distance,trani2012seismic}). One issue in using inverted seismic properties is that, there may not be associated uncertainty quantification, as inverted seismic properties are often obtained using deterministic inversion algorithms. Another issue, which is present in either direct or inverted seismic data, is the ``big data'' problem, due to spatially dense distributions of seismic data. As discussed in \cite{Aanonsen-ensemble-2009,emerick2012history}, in ensemble-based methods large data size may lead to some numerical problems, e.g., ensemble collapse and high computational cost in computing Kalman gain matrices. Apart from these, many history matching algorithms are developed for under-determined inverse problems, whereas a large data size may make the inverse problem become over-determined instead. This may thus affect the performance of history matching algorithms, as will be shown in the numerical example later. As a result, certain procedures (e.g., local analysis, sparse representation) may need to be adopted to tackle the aforementioned problems. 

In this work we propose an ensemble-based 4D seismic history matching framework with sparse representation based on wavelet multiresolution analysis. Compared to similar existing frameworks (see, for example, \citealp{emerick2012history,emerick2013history,gosselin2003history,skjervheim2007incorporating}), our proposed one has some relatively new ingredients, in terms of the type of seismic data, wavelet multiresolution analysis for sparse representation and related noise estimation, and the use of a recently developed iterative ensemble smoother based on a regularized inversion algorithm \citep{luo2015Iterative}. The seismic data we used are intercept and gradient attributes of amplitude versus angle (AVA). These two AVA attributes are processed data that are transformed from raw seismic data but without involving any inversion, therefore they avoid uncertainty associated with the inversion process. Wavelet multiresolution analysis is proposed for the following two purposes: one is to reduce the size of seismic data by exploiting the sparse representation nature of wavelet basis functions, and the other is to exploit its capacity of noise estimation in the wavelet domain \citep{donoho1995adapting}.    

This work is organized as follows. First we introduce the ingredients that comprise the proposed framework, including forward simulation of AVA attributes, sparse representation and noise estimation in the wavelet domain, and the iterative ensemble history matching algorithm. As a proof-of-concept study, we adopt a 2D synthetic case, which is originated from a 3D Norne field model, to demonstrate the performance of the proposed framework in various situations. Finally, we conclude the work and discuss some possible future work on top of current investigations.   
  
\section{The proposed ensemble 4D seismic history matching framework}\label{sec:framework}

\subsection{Forward modeling of AVA attributes}
\begin{figure*}[!htb]
	\centering
	\includegraphics[scale=0.5]{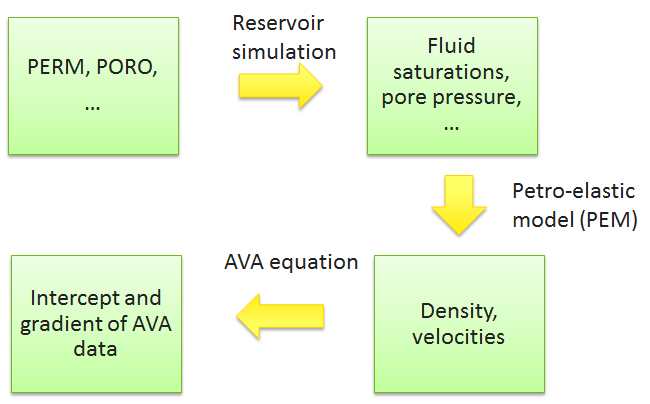}
	\caption{\label{fig:flow_charts} Flowchart of forward modeling of AVA attributes.}
\end{figure*}

As indicated in Figure \ref{fig:flow_charts}, the forward modeling of AVA attributes involves a few steps. First we use petro-physical parameters (e.g., permeability and porosity) to generate pore pressure and fluid saturations through reservoir flow simulation. Then we adopt a petro-elastic model (PEM) to link pore pressure and fluid saturations to P- and S-wave velocities and formation density. Finally we employ a certain AVA equation to compute intercept and gradient attributes.  

The changes in pressure and fluid saturations within a reservoir may have significant effects on seismic parameters (such as P-wave velocity, S-wave velocity and density) \citep{Domenico:1974,Nur:1989}. The rock-physics analysis provides the link between seismic and reservoir parameters. Building a proper PEM is therefore very crucial, whereas it is challenging at the same time. To interpret the changes observed in seismic response over the time, a thorough knowledge of rock and fluid properties is required \citep{jack2001coming}. For example, the P-wave velocity can be linked with the fluid saturation change using the Gassmann model \citep{Gassmann:1951}, whereas the Hertz-Mindlin model \citep{Mindlin:1949} can be used as a link between the P-wave velocity and the effective stress change. By combining the Gassmann model and the Hertz-Mindlin model, we can then express the changes in seismic parameters in terms of the changes in fluid-saturation and pore-pressure. 

Here, in this study the above mentioned rock-physics models are used. The Hertz-Mindlin model \citep{Mindlin:1949} is used to estimate effective dry moduli of the reservoir rock as a function of effective stress. The effective bulk modulus $(K_{d})$ and shear modulus $(\mu_{d})$ of a dry random packing of identical spheres is given by \citep{Mindlin:1949}
\begin{equation}
K_{d}= \sqrt[n]{\frac{C_{p}^2(1- \phi)^2 \mu_{s}^2}{18\pi^2(1-\nu_{s})^2}P_{eff}} \, ,
\end{equation}
and  
\begin{equation}
\mu_{d}= \frac{5-4\nu_{s}}{5(2-\nu_{s})}\sqrt[n]{\frac{3C_{p}^2(1- \phi)^2 \mu_{s}^2}{2\pi^2(1-\nu_{s})^2}P_{eff}} \, ,
\end{equation}
where $\mu_{s}$, $\nu_{s}$, $P_{eff}$ are grain shear modulus, Poisson's ratio, and effective stress, respectively. The coordination number $C_{p}$ denotes the average number of contacts per sphere and $n$ is the degree of root. For perfect, identical-sphere packing $n$ is equal to 3. For a better fit of real data we need to adjust between $C_{p}$ and $n$. In this study, $n$ equals to 5, and this is the same as that in the previous work \citep{dadashpour2009reservoir}. 

The above equations can be further simplified as a function of effective stress $P_{eff}$, namely, 
\begin{equation}
K_{d} \propto \mu_{d} \propto P_{eff}^{\frac{1}{n}} \, ,
\end{equation}
if other parameters remain unchanged over time. Therefore, we can calculate new dry effective bulk and shear moduli at new effective stress if we know the initial effective stress and initial dry frame moduli ($K_{d}$ and $\mu_{d}$). Here, we use the same values of initial $K_{d}$ and $\mu_{d}$ as in  \citet{dadashpour2009reservoir}. Once the pressure effect is modeled for each reservoir grid block, the saturation effect can be estimated by using the Gassmann model \citep{Gassmann:1951}. Saturated bulk $(K_{sat})$ and shear moduli $(\mu_{sat})$ can be expressed as 
\begin{equation}
K_{sat}= K_{d} + \frac{(1-\frac{K_{d}}{K_{s}})^2}{\frac{\phi}{K_{f}} + \frac{1 - \phi}{K_{s}} - \frac{K_{d}}{K_{s}^2}  }  \, , 
\end{equation}
and 
\begin{equation}
\mu_{sat}= \mu_{d} \, ,
\end{equation}
where $K_{d}$, $\mu_{d}$, $K_{s}$, $K_{f}$, $\phi$ are dry/frame bulk modulus, dry/frame shear modulus, solid/mineral bulk modulus, effective fluid bulk modulus, and porosity, respectively. Here, the effective fluid bulk modulus ($K_{f}$) is estimated by using the Reuss average \citep{reuss1929berechnung}. $K_{f}$ for three phase fluid mixture is given by 
\begin{equation}
K_{f}= (\frac{S_{w}}{K_{w}} + \frac{S_{o}}{K_{o}} + \frac{S_{g}}{K_{g}})^{-1} \, , 
\end{equation}
where $K_{w}$, $K_{o}$, $K_{g}$, $S_{w}$, $S_{o}$ and $S_{g}$ are bulk modulus of water/brine, bulk modulus of oil, bulk modulus of gas, saturation of water/brine, saturation of oil and saturation of gas, respectively.

Further, the saturated density \citep{mavko2009rock} can be written as (for three phase fluid)
\begin{equation}\label{eqn:density}
\rho_{sat} = (1-\phi)\rho_{m} + \phi S_{w}\rho_{w} + \phi S_{o}\rho_{o} + \phi S_{g}\rho_{g} \, ,
\end{equation}
where $\rho_{sat}$, $\rho_{m}$, $\rho_{w}$, $\rho_{o}$ and $\rho_{g}$ are saturated density of rock, mineral density, water/brine density, oil density and gas density, respectively. In this study $S_{w}$, $S_{o}$ and $S_{g}$ are obtained from reservoir flow simulation. 

Using the above equations we can obtain P- and S-wave velocities, which can be expressed as \citep{mavko2009rock} 
\begin{equation}
V_{P} = \sqrt{\frac{K_{sat} + \frac{4}{3}\mu_{sat}}{\rho_{sat}}} \, ,
\end{equation}
and
\begin{equation}
V_{S} =  \sqrt{\frac{\mu_{sat}}{\rho_{sat}}} \, ,
\end{equation}
where $V_{P}$ and $V_{S}$ are P- and S-wave velocities, respectively.

Once the seismic parameters are estimated as a function of reservoir parameters by using the PEM, we can then generate synthetic seismogram based on these seismic parameters. Here, we exploit the amplitude versus angle (AVA) relation by using the linearized AVA equation \citep{Smith:1987}.  For a two-layer model where the reservoir layer is below an overburden layer, the P-wave reflectivity $(R_{pp})$ can be expressed as \citep{Smith:1987}
\begin{equation}\label{eqn:Smith-Gidlow}
R_{pp}(\theta) = \frac{1}{2}(\frac{\Delta\rho}{\rho}+ \frac{\Delta\alpha}{\alpha} ) - \frac{2\beta^{2}}{\alpha^{2}}(\frac{\Delta\rho}{\rho}+ \frac{2\Delta\beta}{\beta} )\sin^2 \theta + \frac{\Delta\alpha}{2\alpha}\tan^2 \theta 
\end{equation}
where, $\alpha = \frac{1}{2}(\alpha_{1} + \alpha_{2})$, $\beta = \frac{1}{2}(\beta_{1} + \beta_{2})$ and $\rho = \frac{1}{2}(\rho_{1} + \rho_{2})$ represent average P-wave velocity, S-wave velocity and density, between overburden and reservoir layers; and $\Delta\alpha = \alpha_{2} - \alpha_{1} $, $\Delta\beta = \beta_{2} - \beta_{1}$ and $\Delta\rho = \rho_{2} - \rho_{1}$ are lithological contrasts in P-wave velocity, S-wave velocity and density, respectively, between overburden and reservoir layers. $\alpha_{1}$, $\beta_{1}$ and $\rho_{1}$ are the P-wave velocity, S-wave velocity and density of the overburden, respectively. Similarly, $\alpha_{2}$, $\beta_{2}$ and $\rho_{2}$ are the P-wave velocity, S-wave velocity and density of the reservoir, respectively. $\theta$ is the incidence angle. Eq. (\ref{eqn:Smith-Gidlow}) is the linearized version of the exact reflectivity equation, i.e., the Zoeppritz's equation. The first term in Eq. (\ref{eqn:Smith-Gidlow}) expresses the zero-offset reflectivity, whereas the other two terms include the offset dependency in reflectivity. This equation works reasonably well for small incident angles (up to $30^{\circ}$) and small contrast between the layers. Now, considering the conventional AVO formula, $ R = R_{0} + G\sin^2 \theta $ and assuming $ \tan^2 \theta \approx \sin^2 \theta $, then Eq. \ref{eqn:Smith-Gidlow} can be further simplified and split into two terms ($R_{0}$ and G) as \citep{landro2001discrimination}   
\begin{equation}
R_{0} = \frac{1}{2}(\frac{\Delta\rho}{\rho}+ \frac{\Delta\alpha}{\alpha} ) 
\end{equation}
and
\begin{equation}
G =  - \frac{2\beta^{2}}{\alpha^{2}}(\frac{\Delta\rho}{\rho}+ \frac{2\Delta\beta}{\beta} ) + \frac{\Delta\alpha}{2\alpha}
\end{equation}
Here, $R_{0}$ and G are intercept and gradient AVA attributes. For multi-layer cases, we need to calculate intercept and gradient reflectivity series as a function of two-way travel time (see, for example, \citealp{buland2003bayesian}). Here, travel time is computed from the P-wave velocity and vertical thickness of each grid block. Further, we convolve the reflectivity series with a Ricker wavelet of 30 Hz dominant frequency to obtain the seismic AVA attributes (here $R_{0}$ and $G)$. 

\subsection{Sparse representation of seismic data and noise estimation through wavelet multiresolution analysis}

Let $\mathbf{m}^{ref} \in \mathcal{R}^{m}$ denote the reference reservoir model. To simplify the discussions below, it is assumed that the AVA attributes are 2D data and saved in the form of $p_1 \times p_2$ matrices, such that the forward simulator of AVA attributes $\mathbf{g}: \mathcal{R}^{m} \rightarrow \mathcal{R}^{p_1 \times p_2}$. The observed AVA attributes $\mathbf{d}^o$ are supposed to be contaminated by some additive noise $\boldsymbol{\epsilon}$, such that  
\begin{linenomath*} 
\begin{equation} \label{eq:obs_system}
\mathbf{d}^o = \mathbf{g}(\mathbf{m}^{ref}) + \boldsymbol{\epsilon} \, .
\end{equation}     
\end{linenomath*}  
It is further assumed that, for a given AVA attribute (intercept or gradient), the elements of $\boldsymbol{\epsilon}$ are independently and identically distributed (i.i.d) Gaussian white noise, with zero mean but unknown variance $\sigma^2$, where $\sigma$ will be estimated through wavelet multiresolution analysis below.    

For the purpose of reservoir characterization, we aim to find an estimate $\hat{\mathbf{m}}$ of $\mathbf{m}^{ref}$ based on the observation $\mathbf{d}^o$. The observation system Eq. (\ref{eq:obs_system}) suffices for this purpose from the algorithmic perspective. However, as  aforementioned, certain numerical issues may arise due to the large data size. To mitigate this problem, one feasible strategy is to employ sparse representation by transforming the data into a different domain, such that the main features of the original data are kept in relatively few, but dominant components in the transform domain. An illustration example in this aspect is the low-rank representation of a matrix through a truncated singular value decomposition (TSVD), in which the original matrix is approximated by the summation of the products of left and right singular vectors associated with the singular values above a certain threshold (see, for example, \citealp{chen2013-levenberg,luo2015Iterative,Luo-ensemble}). In the TSVD example, the singular values can be interpreted as a representation of the matrix, and the set of the products of left and right singular vectors comprises the basis functions of the transform domain. Sparse representation is achieved by keeping leading singular values above a certain threshold, whereas setting those below the threshold value to zero. 
 
One feature of TSVD is that its singular vectors are data-dependent. This is not desirable for our purpose, since history matching involves a comparison between observed and simulated data. Given different singular vectors (hence basis functions), it is not sensible to compare representations (truncated singular values) of observed and simulated data from different transform domains. To avoid this problem, one of the alternatives is to adopt a discrete wavelet transform (DWT), in which wavelet basis functions are independent of data. Wavelet-based multiresolution analysis is already used in many works in the community. For instance, it is adopted in \cite{awotunde2014multiresolution,chen2012multiscale,gentilhomme2015ensemble,sahni2005multiresolution} for sparse (or multiscale) representations of reservoir models. It is also used in \cite{awotunde2014multiresolution} for sparse representation of time-lapse saturation maps. However, it seems that \cite{awotunde2014multiresolution} does not consider how to estimate the variance of noise of the transformed data (i.e., wavelet coefficients), which is an issue to be addressed later.   

\begin{figure*}[!htb]
	\centering
	\includegraphics[scale=0.5]{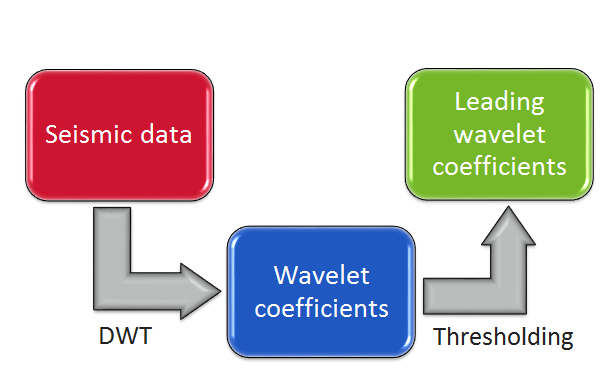}
	\caption{\label{fig:wavelet_ssr} Workflow of sparse representation of seismic data through discrete wavelet transform (DWT).}
\end{figure*}

Figure \ref{fig:wavelet_ssr} indicates a similar workflow of sparse representation of seismic data through DWT. Given a set of seismic data, one first applies a multilevel DWT to the data. For a chosen family of wavelet basis functions, the data are represented by the corresponding wavelet coefficients. For the purpose of sparse representation, one nice feature of wavelet transform is that, in many cases (e.g., when dealing with noisy signals with structures), small wavelet coefficients are dominated by noise, while large coefficients mainly carry signal information \citep{jansen2012noise}. Similar to TSVD, sparse representation can be achieved by only keeping the wavelet coefficients above a certain threshold, while setting those below the threshold value to zero. Let $\mathcal{W}$ and $\mathcal{T}$ denote wavelet transform and thresholding operators, respectively, then after DWT and thresholding, the effective observation system becomes          
\begin{linenomath*} 
\begin{equation} \label{eq:tw_obs_system}
\mathcal{T} \circ \mathcal{W} (\mathbf{d}^o) = \mathcal{T} \circ \mathcal{W} \circ \mathbf{g}(\mathbf{m}^{ref}) + \mathcal{T} \circ \mathcal{W} (\boldsymbol{\epsilon}) \, .
\end{equation}     
\end{linenomath*} 

For succinctness, an introduction to wavelet theory is omitted here. Interested readers are referred to, for example, \cite{mallat1999wavelet}. For our purpose, we need to adopt a 2D orthogonal wavelet transform for sparse representation and noise estimation of AVA attributes. The orthogonality requirement is based on the following observation: if $\mathcal{W}$ is orthogonal, then the wavelet transform preserves the energy of Gaussian white noise. In addition, similar to the power spectral distribution of white noise in frequency domain (e.g., as in the situation of Fourier transform), the noise energy in the wavelet domain is uniformly distributed among all wavelet coefficients \citep{jansen2012noise}. This implies that, if one can estimate the noise level of small wavelet coefficients, then this estimation can also be used to infer the noise level of leading coefficients used in history matching.

\begin{figure*}[!htb]
	\centering
	\includegraphics[scale=0.4]{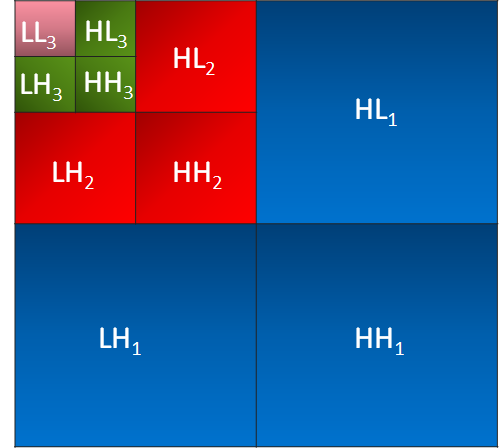}
	\caption{\label{fig:wavelet_decomposition} Three-level wavelet decomposition.}
\end{figure*}   

Figure \ref{fig:wavelet_decomposition} illustrates wavelet multiresolution analysis through a three-level 2D wavelet decomposition. At each level (say, level $j$ for $j = 1,2,3$) of decomposition, one obtains four sub-bands of wavelet coefficients, namely, $LL_j$ (upper left), $HL_j$ (upper right), $LH_j$ (lower left) and $HH_j$ (lower right). Here $LL_j$ ($HH_j$, respectively) is obtained by applying low-pass (high-pass, respectively) filters to the data corresponding to the sub-band $LL_{j-1}$, along both horizontal and vertical directions \citep{jansen2012noise} ($LL_{0}$ corresponding to the original noisy data itself), whereas $HL_j$ ($LH_j$, respectively) is attained by first applying a high-pass filter along the vertical (horizontal, respectively) direction, then a low-pass filter along the horizontal (vertical, respectively) direction. As a result, to estimate the standard deviation (std) $\sigma$ of white noise, it is customary to use the sub-band $HH_1$, since it is dominated by high-frequency components that typically correspond to noise. Concretely, let $\tilde{\mathbf{d}}^o = \mathcal{W} (\mathbf{d}^o)$ be the whole set of wavelet coefficients corresponding to $\mathbf{d}^o$, and $\tilde{\mathbf{d}}^o_{HH_1}$ those of the sub-band $HH_1$. In case of Gaussian white noise, the std $\sigma$ of noise can be estimated using the median absolute deviation (MAD) estimator \citep{donoho1995adapting}:              
\begin{linenomath*} 
\begin{equation} \label{eq:noise_std_mad}
\sigma = \dfrac{\operatorname{median}(\operatorname{abs}(\tilde{\mathbf{d}}^o_{HH_1}))}{0.6745} \, ,
\end{equation}     
\end{linenomath*} 
where $\operatorname{abs}(\bullet)$ is an element-wise operator, and takes the absolute value of an input vector. For colored noise, one may apply Eq. (\ref{eq:noise_std_mad}) on a level-dependent basis, see, for example, the discussion in \cite{johnstone1997wavelet}. After estimating $\sigma$ in an $n$-level wavelet decomposition, we apply hard thresholding and select leading wavelet coefficients in $HH_j$, $HL_j$ and $LH_j$ ($j = 1,2,\dotsb,n$) sub-bands in a way such that 
\begin{linenomath*} 
\begin{equation} \label{eq:hard_thresholding}
\mathcal{T}(\tilde{\mathbf{d}}^o) = 
 \begin{cases}
    0       & \quad \text{if } \tilde{\mathbf{d}}^o \in \{HH_j, HL_j, LH_j \vert 1 \leq j \leq n \} \text{ and } \tilde{\mathbf{d}}^o < \lambda \, ,\\
    \tilde{\mathbf{d}}^o  & \quad \text{otherwise} \, ,\\
  \end{cases}
\end{equation}     
\end{linenomath*} 
whereas in the current study the coefficients in the $LL_n$ sub-band are not modified, in light of the assumption that the $LL_n$ sub-band is dominated by low-frequency components which correspond to signals. The threshold value $\lambda$ is selected using the universal rule \citep{donoho1994ideal}
\begin{linenomath*} 
\begin{equation} \label{eq:universal_rule}
\lambda = \sqrt{2 \, \operatorname{ln}(\# \mathbf{d}^o)} \, \sigma \, ,
\end{equation}     
\end{linenomath*}  
with $\# \mathbf{d}^o$ being the number of elements in $\mathbf{d}^o$. Since we use leading wavelet coefficients in $HH_j$, $HL_j$ and $LH_j$  sub-bands and those in $LL_n$ for history matching, the corresponding observation system becomes
\begin{linenomath*} 
\begin{equation} \label{eq:reduced_obs_system}
\tilde{\mathbf{d}}^o = \mathcal{W} \circ \mathbf{g}(\mathbf{m}^{ref}) + \mathcal{W} (\boldsymbol{\epsilon}) \, , \text{ for } \tilde{\mathbf{d}}^o \geq \lambda \text{ or } \mathbf{d}^o \in LL_n \, ,
\end{equation}     
\end{linenomath*}  
where $\tilde{\mathbf{d}}^o$ is a vector containing all selected wavelet coefficients, and $\mathcal{W} (\boldsymbol{\epsilon})$ the corresponding noise component in the wavelet domain, with zero mean and covariance $\mathbf{C}_{\tilde{\mathbf{d}}^o} = \sigma^2 \mathbf{I}$ (here $\mathbf{I}$ is the identity matrix with a suitable dimension).

\newcommand{\nScale}{0.3}
\begin{figure*}
		\centering
		\subfigure[Reference intercept data]{ \label{subfig:ref_data}
			\includegraphics[scale=0.3]{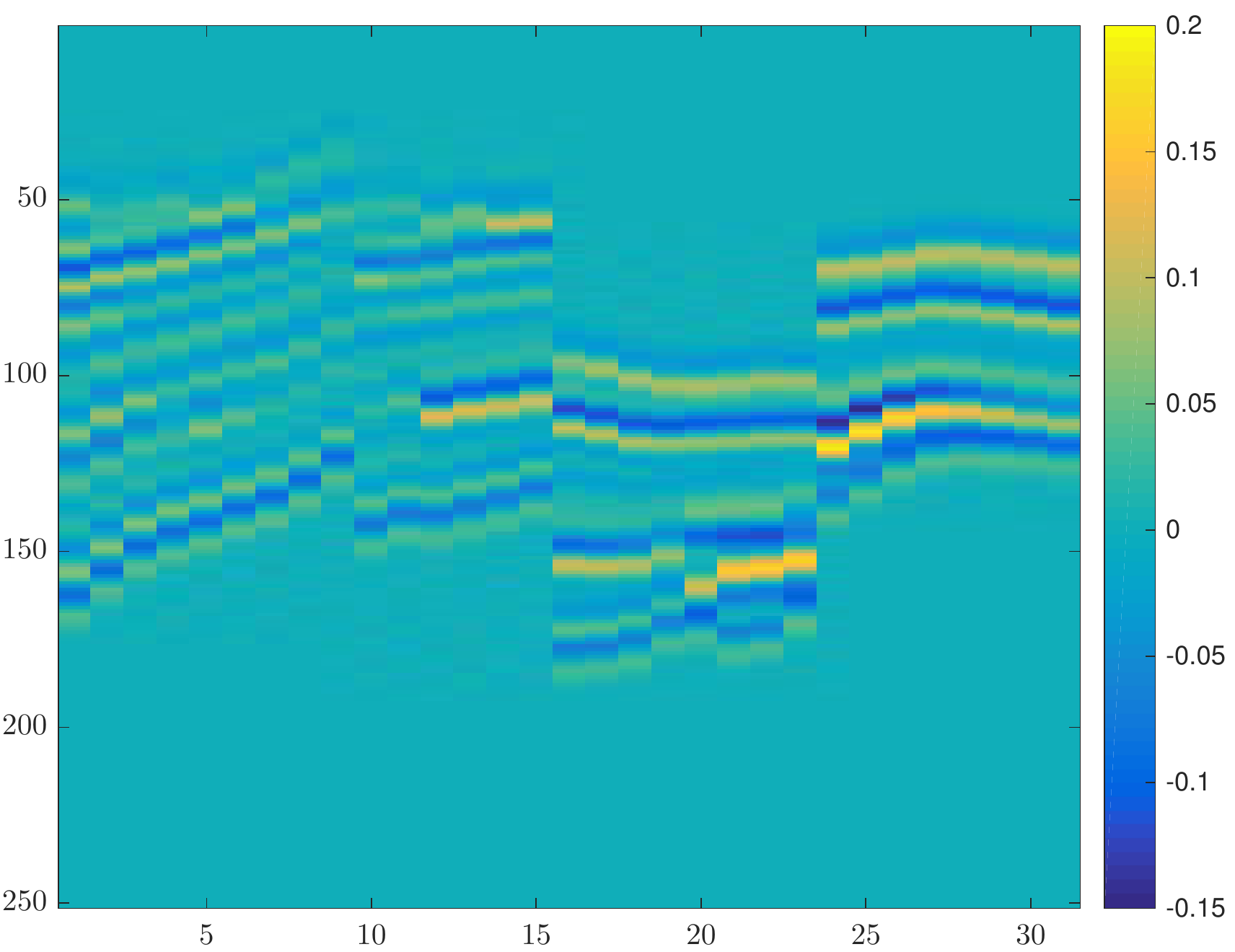}
		}%
		\subfigure[Noisy intercept data]{ \label{subfig:noisy_data}
			\includegraphics[scale=0.3]{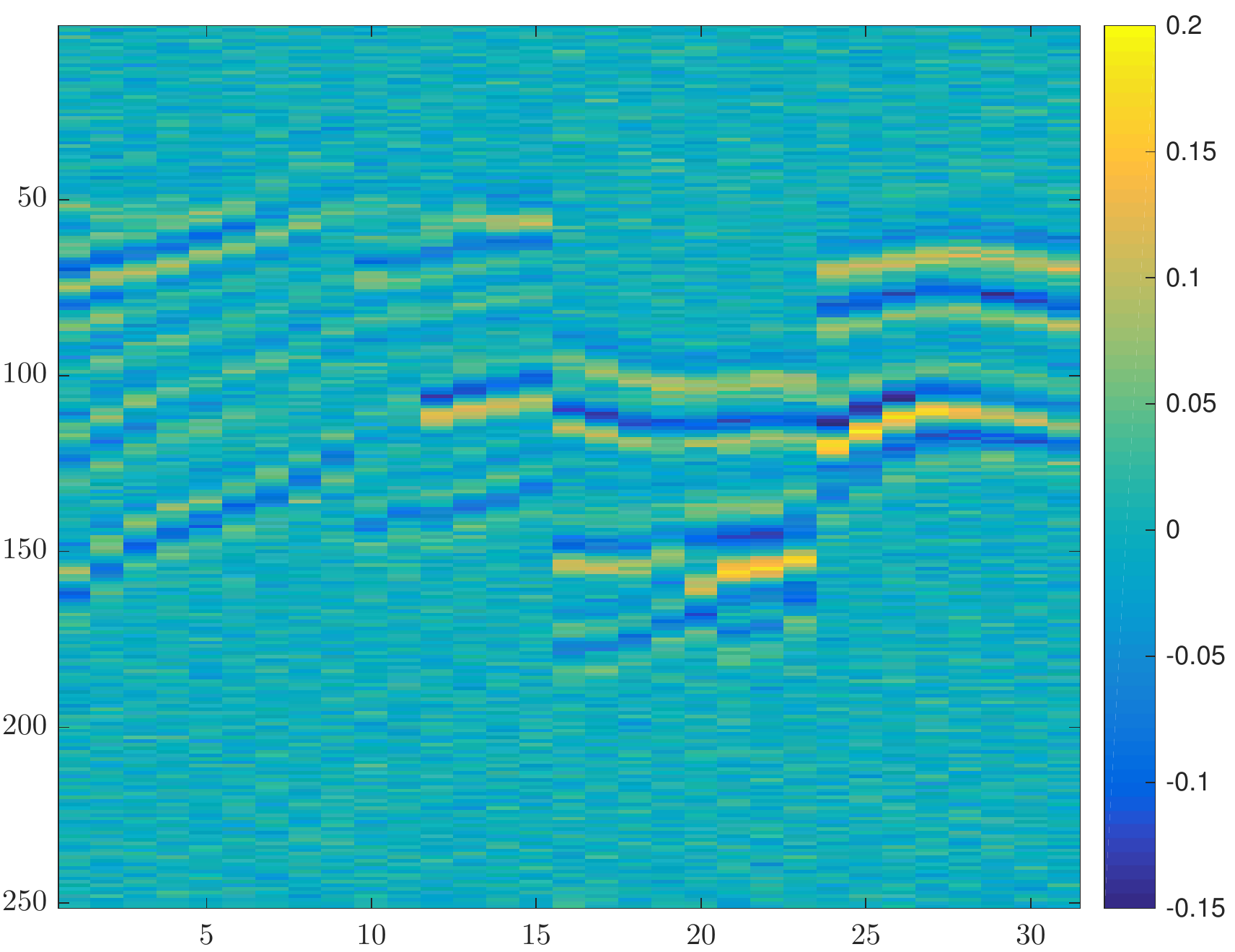}
		}%
		\subfigure[Reconstructed intercept data]{ \label{subfig:denoised_data}
			\includegraphics[scale=0.3]{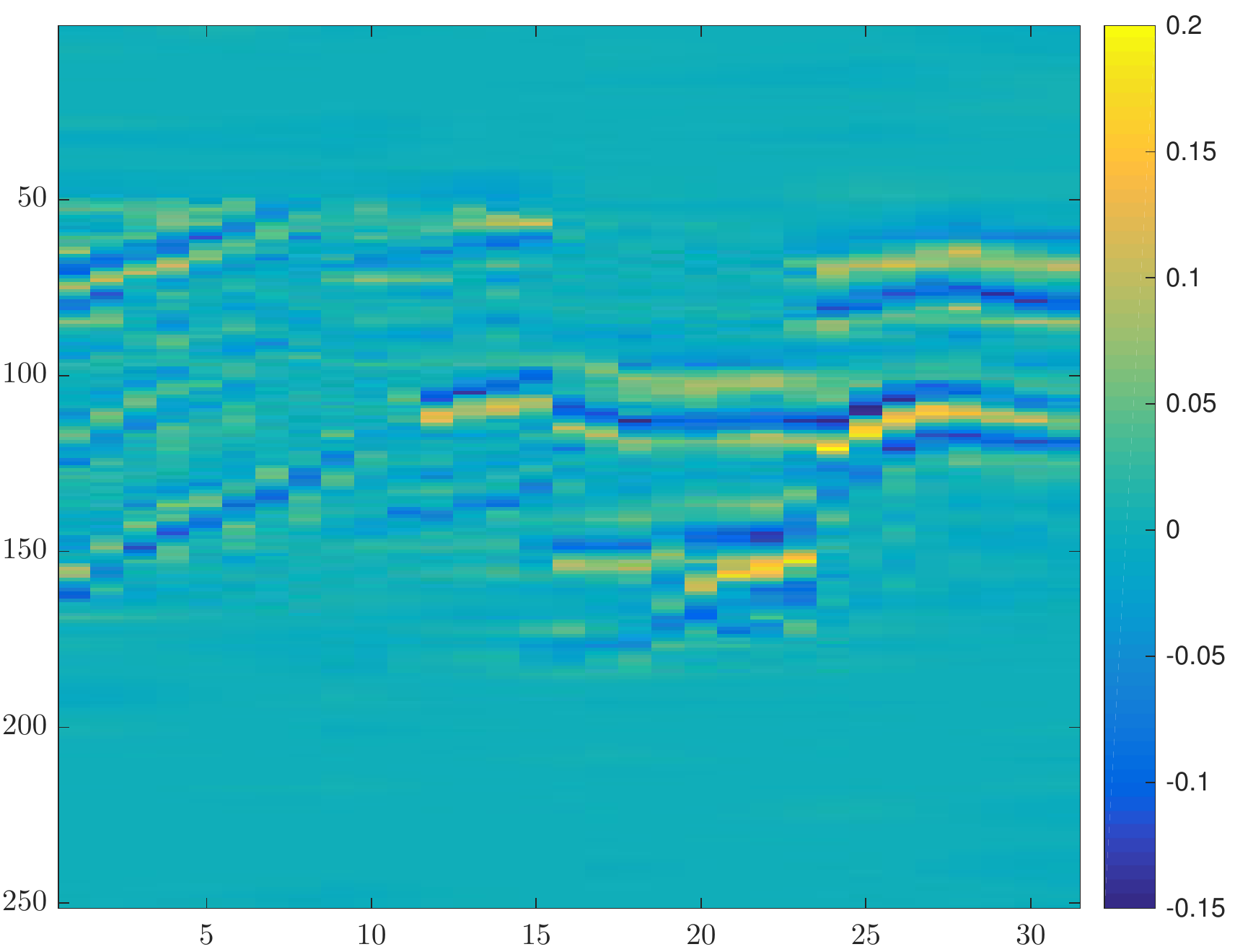}
		}
	\caption{\label{fig:illustration_data} An illustration of sparse representation of intercept data generated by a forward AVA attribute simulation. (a) Reference intercept data; (b) Noisy intercept data obtained by adding Gaussian white noise (noise level = 30\%) to the reference data; (c) Reconstructed intercept data obtained by first conducting a 2D DWT on the noisy data, then applying hard thresholding (using the universal threshold value) to wavelet coefficients, and finally reconstructing the data using an inverse 2D DWT based on the modified wavelet coefficients.} 
\end{figure*} 

\renewcommand{\nScale}{0.33}
\begin{figure*}
		\centering
		\subfigure[Coefficients of reference data]{ \label{subfig:coeff_ref_data}
			\includegraphics[scale=0.3]{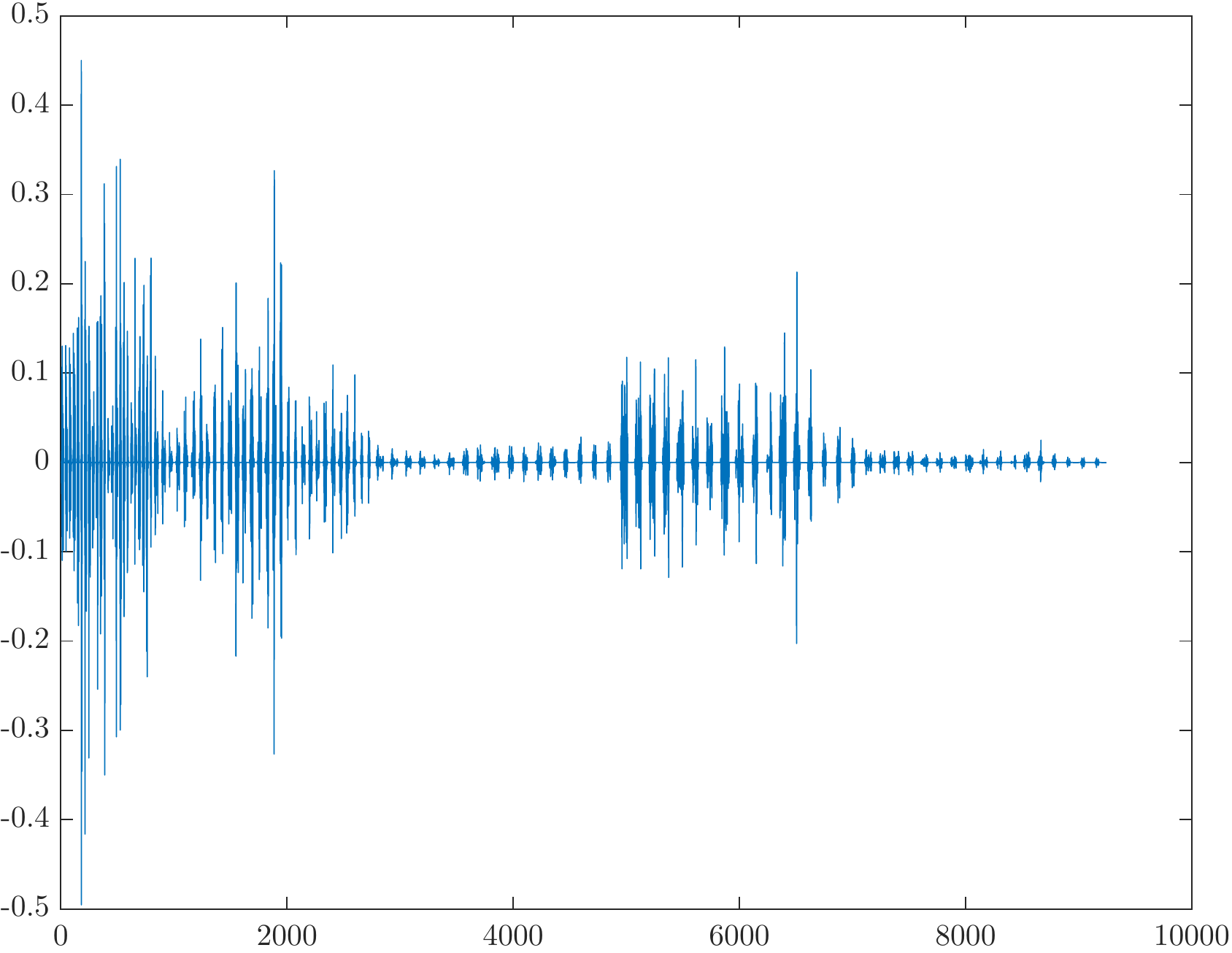}
		}%
		\subfigure[Coefficients of noisy data]{ \label{subfig:coeff_noisy_data}
			\includegraphics[scale=0.3]{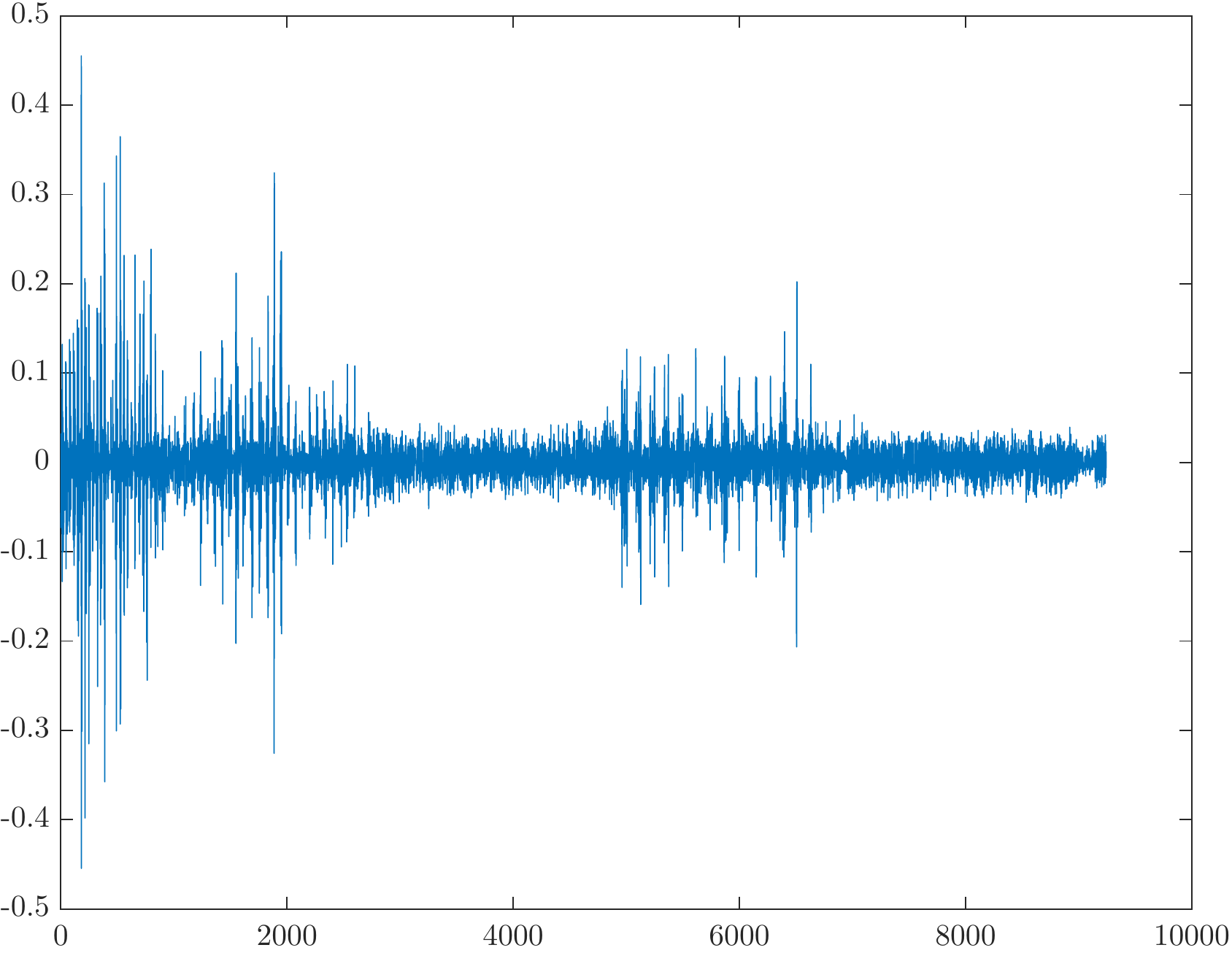}
		}%
		\subfigure[Coefficients of reconstructed data]{ \label{subfig:coeff_denoised_data}
			\includegraphics[scale=0.3]{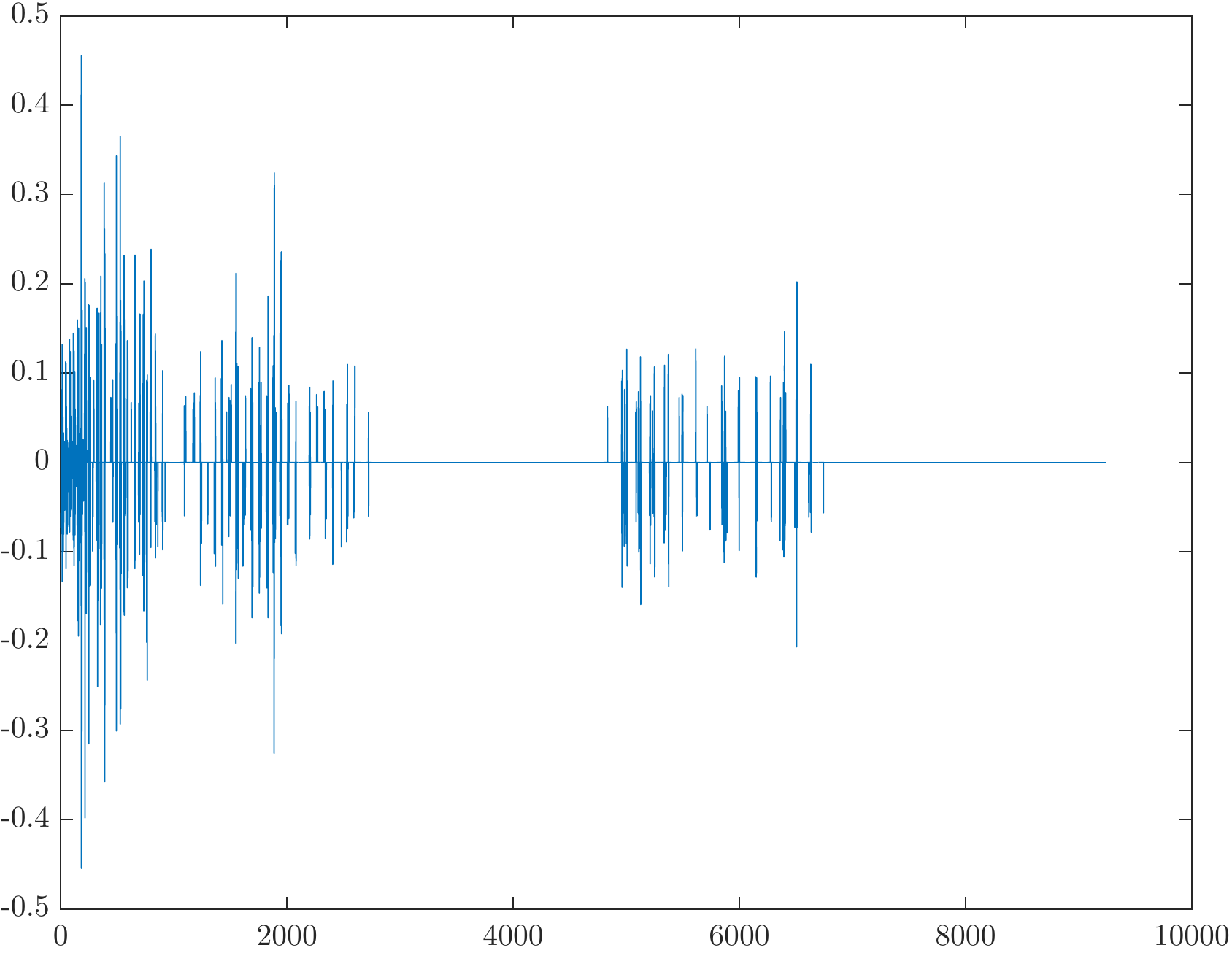}
		}
	\caption{\label{fig:illustration_coeff} As in Figure \ref{fig:illustration_data}, but with (a) Wavelet coefficients of reference data; (b) Wavelet coefficients of noisy data; (c) Wavelet coefficients after thresholding. These are used to reconstruct the data in Figure \ref{subfig:denoised_data} through an inverse 2D DWT.} 
\end{figure*} 

\renewcommand{\nScale}{0.33}
\begin{figure*}
		\centering
		\subfigure[Reference noise]{ \label{subfig:ref_noise}
			\includegraphics[scale=0.3]{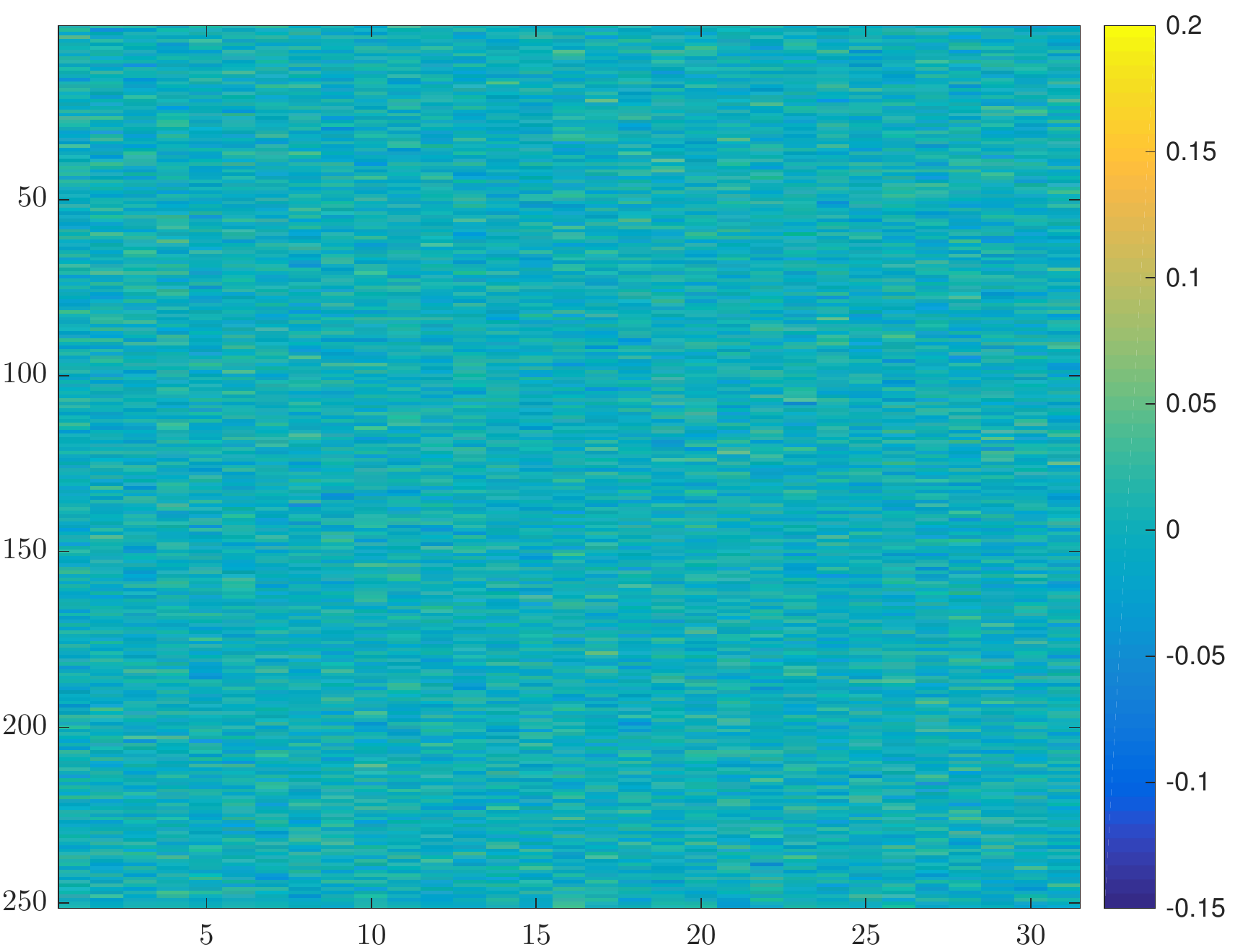}
		}%
		\subfigure[Estimated noise]{ \label{subfig:est_noise}
			\includegraphics[scale=0.3]{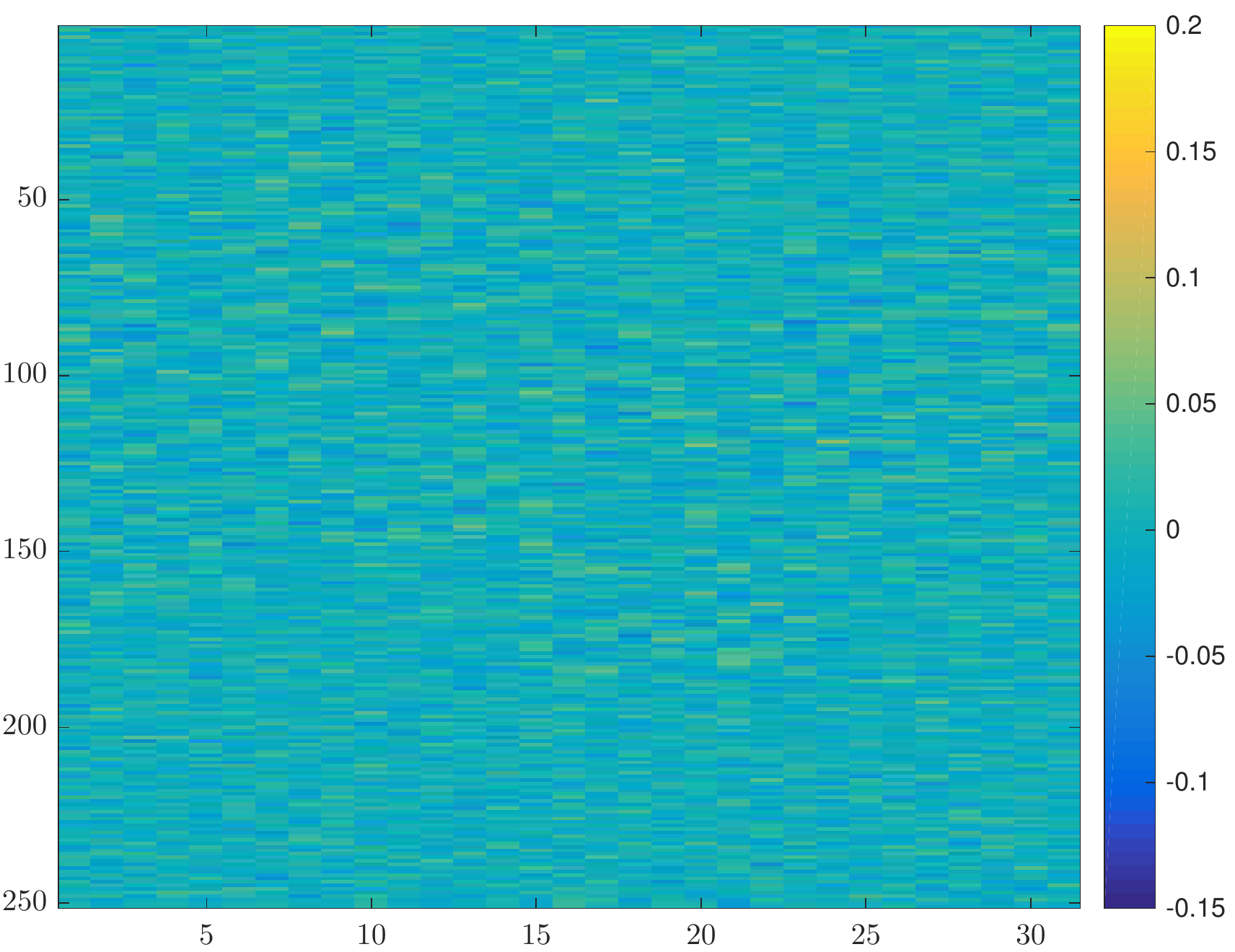}
		}%
		\subfigure[Noise difference]{ \label{subfig:diff_data}
			\includegraphics[scale=0.3]{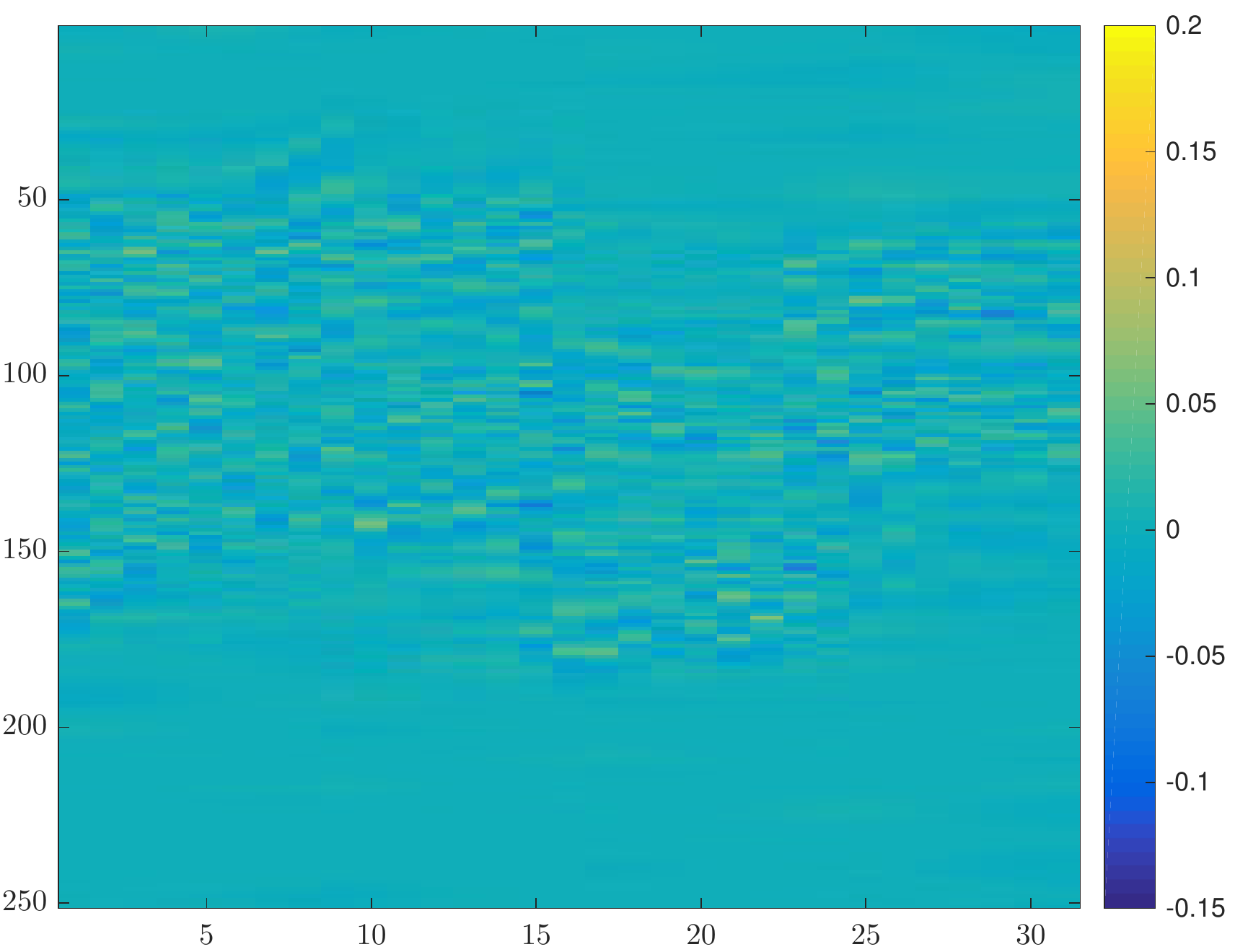}
		}
	\caption{\label{fig:illustration_noise} As in Figure \ref{fig:illustration_data}, but with (a) reference noise, defined as the difference between noisy data and reference data; (b) Estimated noise, defined as the difference between noisy data and reconstructed data; (c) Noise difference, defined as the difference between estimated noise in (b) and reference noise in (a).} 
\end{figure*} 

We use an example to illustrate the performance of sparse representation and noise estimation through wavelet multiresolution analysis. In the example, we first generate reference intercept data (see Figure \ref{subfig:ref_data}) using the forward AVA attribute simulator, and then add some Gaussian white noise to obtain the noisy intercept data in Figure \ref{subfig:noisy_data}. In this case, the size of intercept data is $251 \times 31 = 7781$, and the noise level is $30\%$. Here, noise level is defined as: 
\begin{linenomath*} 
\begin{equation} \label{eq:def_noise_level}
\text{Noise level } = \dfrac{\text{variance of noise}}{\text{variance of pure signal}} \, .  
\end{equation}     
\end{linenomath*} 
We apply a three-level DWT to the noisy data using Daubechies wavelets with two vanishing moments \citep{mallat1999wavelet}, and use hard thresholding combined with the universal rule (Eqs. (\ref{eq:noise_std_mad}) -- (\ref{eq:universal_rule})) to select leading wavelet coefficients. After thresholding, the number of leading wavelet coefficients becomes 581, only $7.5\%$ of the original data size. In addition, by applying Eq. (\ref{eq:noise_std_mad}), we have the estimated std of noise to be $0.0141$, which is very close to the true noise std $0.0148$ (we have also applied the same wavelet multiresolution analysis procedure to data with different attribute (gradient) and noise levels ($0\%$ and $10\%$), and attained similar results). By applying an inverse DWT to the thresholded wavelet coefficients, we obtain the reconstructed intercept data as shown in Figure \ref{subfig:denoised_data}. Comparing the figures in Figure \ref{fig:illustration_data}, we see that the reconstructed data capture the main features of the reference data, whereas substantially remove the noise components (in fact, thresholding is a standard procedure used in wavelet-based denoising algorithms, for example, see \citealp{donoho1995adapting,donoho1994ideal,jansen2012noise}).                  

For further illustrations, Figures \ref{fig:illustration_coeff} and \ref{fig:illustration_noise} show wavelet coefficients associated with the reference, noisy and reconstructed attributes, and reference and estimated noise, respectively. As one can see, after applying thresholding to wavelet coefficients of noisy data (Figure \ref{subfig:coeff_noisy_data}), the modified coefficients (Figure \ref{subfig:coeff_denoised_data}) preserve those with large amplitudes. In general, the modified coefficients appear similar to those of reference data, but lose some details in some places with small coefficients (e.g., for indices larger than $8000$). Accordingly, in Figure \ref{fig:illustration_noise} we see that, to some extent, the estimated noise (Figure \ref{subfig:est_noise}) looks similar to the reference noise (Figure \ref{subfig:ref_noise}), although there is some ``signal leakage'' into the estimated noise (Figure \ref{subfig:diff_data}) due to the fact that some small wavelet coefficients of the reference data are smeared out in the course of thresholding.

\subsection{The ensemble history matching algorithm}

A popular ensemble-based method for reservoir history matching is the EnKF (see, for example,  \citealp{naevdal2005reservoir,Aanonsen-ensemble-2009}). Recently, the ES (see, for example, \citealp{skjervheim2011ensemble}) and iterative ES (iES) (see, for example, \citealp{chen2013-levenberg,emerick2012history,luo2015Iterative}) have also attracted attention in the community. Compared to the EnKF, the ES and iES have certain technical advantages in terms of algorithm implementation and execution, see, for example, the discussions in \cite{luo2015Iterative,skjervheim2011ensemble}.

In this work we use the iES in \cite{luo2015Iterative} for seismic history matching, whereas other iES variants may serve the same purpose. We provide a short introduction to the chosen iES, and pay special attention to one of the stopping criteria associated with this iES. The stopping criterion to be discussed below has a substantial impact on history matching results of the numerical example in this work.  

Similar to Eq. (\ref{eq:obs_system}), but without loss of generality, let $\mathbf{d}^o$ denote $p$-dimensional observations in history matching, which may represent values in the ordinary data space (e.g., 2D AVA attributes by reshaping matrices into vectors), or their sparse representation in the transform domain (e.g., leading wavelet coefficients). Here $\mathbf{d}^o$ is assumed to be contaminated by certain observation errors with zero mean and covariance $\mathbf{C}_d$. Also suppose that $\mathbf{g}$ is the forward simulator that takes an $m$-dimensional reservoir model $\mathbf{m}$ as the input and outputs simulated observations $\mathbf{d} \equiv \mathbf{g}(\mathbf{m})$. In the context of iES, let $\mathbf{M}^i \equiv \{ \mathbf{m}_j^i \}_{j=1}^{{N_e}}$ be an ensemble of ${N_e}$ reservoir models obtained at the $i$th iteration step, then the iES aims to update $\mathbf{M}^i$ to a new ensemble $\mathbf{M}^{i+1} \equiv \{ \mathbf{m}_j^{i+1} \}_{j=1}^{{N_e}}$ by solving an optimization problem, which is discussed below. For convenience, we call $\mathbf{S}_m^i$, which is defined as
\begin{linenomath*} 
\begin{IEEEeqnarray}{clc} \label{eq:model_sqrt}
  & \mathbf{S}_m^i = \frac{1}{\sqrt{{N_e}-1}}\left[\mathbf{m}_1^i - \bar{\mathbf{m}}^i,\dotsb, \mathbf{m}_{N_e}^i - \bar{\mathbf{m}}^i \right] \, ; & \quad \bar{\mathbf{m}}^i = \frac{1}{{N_e}} \sum_{j=1}^{{N_e}} \mathbf{m}_j^i \, ,
\end{IEEEeqnarray}
\end{linenomath*}   
a \textit{model square root matrix} in the sense that $\mathbf{C}_{m}^{i} \equiv \mathbf{S}_m^i \left( \mathbf{S}_m^i \right)^T$ equals the sample covariance matrix of the ensemble $\mathbf{M}^i$. Similarly, we call $\mathbf{S}_d^i$, which is defined as  
\begin{linenomath*} 
\begin{IEEEeqnarray}{clc} \label{eq:data_sqrt}
  & \mathbf{S}_d^i = \frac{1}{\sqrt{{N_e}-1}}\left[\mathbf{g}(\mathbf{m}_1^i) - \mathbf{g}(\bar{\mathbf{m}}^i),\dotsb, \mathbf{g}(\mathbf{m}_{N_e}^i) - \mathbf{g}(\bar{\mathbf{m}}^i) \right] \, ,
\end{IEEEeqnarray}
\end{linenomath*}   
a \textit{data square root matrix} with respect to the ensemble $\mathbf{M}^i$. 

Given $\mathbf{M}^i$, a new ensemble $\mathbf{M}^{i+1}$ is obtained by solving the following minimum-average-cost (MAC) problem
\begin{linenomath*}    
\begin{IEEEeqnarray}{lll} \label{eq:wls_rlm_mac}
\underset{\{\mathbf{m}^{i+1}_j\}_{j=1}^{N_e}}{\operatorname{argmin}} & \dfrac{1}{N_e} \sum\limits_{j=1}^{N_e} & \, \left[ \left( \mathbf{d}^o_j - \mathbf{g} \left( \mathbf{m}^{i+1}_j \right) \right)^T \mathbf{C}_{d}^{-1} \left( \mathbf{d}^o_j - \mathbf{g} \left( \mathbf{m}^{i+1}_j \right) \right) \right. \\
& & \quad \left. +  \gamma^{i} \left( \mathbf{m}^{i+1}_j - \mathbf{m}^{i}_j \right)^T \left(  \mathbf{C}_{m}^{i} \right)^{-1} \left( \mathbf{m}^{i+1}_j - \mathbf{m}^{i}_j \right) \right] \, , \nonumber
\end{IEEEeqnarray}
\end{linenomath*}   
where $\mathbf{d}^o_j$ ($j = 1,2,\dotsb, N_e$) are perturbations of $\mathbf{d}^o$ generated by drawing $N_e$ samples from the normal distribution $N(\mathbf{d}^o,\mathbf{C}_{d})$, and $\gamma^i$ a positive scalar adaptive with iteration step (which is automatically chosen using a procedure similar to back-tracking line search in \citealp{luo2015Iterative}). Through linearization, one obtains an approximate solution to the MAC problem as follows: 
\begin{linenomath*}  
\begin{IEEEeqnarray}{crlc} \label{eq:rlm_mac}
& \mathbf{m}^{i+1}_j = \mathbf{m}^{i}_j + \mathbf{S}_m^i (\mathbf{S}_d^i)^T \left( \mathbf{S}_d^i (\mathbf{S}_d^i)^T + \gamma^i \, \mathbf{C}_{d}  \right)^{-1} \left( \mathbf{d}^o_j - \mathbf{g} \left( \mathbf{m}^{i}_j \right) \right) \, , \text{ for } j = 1, 2, \dotsb, N_e \, .& &
\end{IEEEeqnarray}
\end{linenomath*}  
As discussed in \cite{luo2015Iterative}, the iES using Eq. (\ref{eq:rlm_mac}) corresponds to an ensemble implementation of
the regularized Levenburg-Marquardt method \citep{jin2010regularized}) for inverse problems. For this reason, this iES is called regularized Levenburg-Marquardt method for the minimum-average-cost problem (\textit{RLM-MAC} in short).
 
Numerous studies (see, for example, \citealp{Engl2000-regularization}) indicate that the stopping criterion is crucial for the performance of an iterative inversion algorithm. In practice, an iteration process should stop after reaching moderate data mismatch that is neither too large nor too small, whereas the latter requirement is imposed to avoid over-fitting observations in the presence of noise \citep{Engl2000-regularization}. Let 
\begin{linenomath*}    
\begin{IEEEeqnarray}{lll} \label{eq:avg_data_mismatch}
\boldsymbol{\Xi}^i \equiv \dfrac{1}{N_e} \sum\limits_{j=1}^{N_e} & \, \left[ \left( \mathbf{d}^o_j - \mathbf{g} \left( \mathbf{m}^{i}_j \right) \right)^T \mathbf{C}_{d}^{-1} \left( \mathbf{d}^o_j - \mathbf{g} \left( \mathbf{m}^{i}_j \right) \right) \right] & 
\end{IEEEeqnarray}
\end{linenomath*}  
be the average (normalized) data mismatch with respect to the ensemble $\mathbf{M}^i$, then following Proposition 6.3 of \cite{Engl2000-regularization}, we stop the iteration in Eq. (\ref{eq:rlm_mac}) when 
\begin{linenomath*}    
\begin{IEEEeqnarray}{lll} \label{eq:stopping_criterion_ndm}
\boldsymbol{\Xi}^i < 4 p
\end{IEEEeqnarray}
\end{linenomath*} 
for the first time, where $p$ is the number of observations, and the factor $4$ is a critical value below which the iteration process starts to transit from convergence to divergence \citep[p.158]{Engl2000-regularization}. Note that this critical value is proven on top of the assumption that the noise level is precisely known, therefore to apply the stopping criterion (\ref{eq:stopping_criterion_ndm}) in practice, it requires a good estimation of noise level. In addition, using the critical value $4$ is only a choice for robustness, and it does not guarantee the optimality in terms of estimation accuracy. As a result, in some cases it is possible that one may obtain better inversion results in terms of, for example, root mean square error (RMSE), by using a factor less than 4. Finally, many inversion algorithms (including RLM) are developed for under-determined inverse problems, in which the numbers $p$ of observations are less than the model sizes $m$. Given a large amount of seismic data, however, the history matching problem may become over-determined instead, and this could affect the performance of history matching algorithms to some extent.

\section{A numerical example}\label{sec:example}

\subsection{Experimental settings}
\begin{figure*}[!htb]
	\centering
	\includegraphics[scale=0.297]{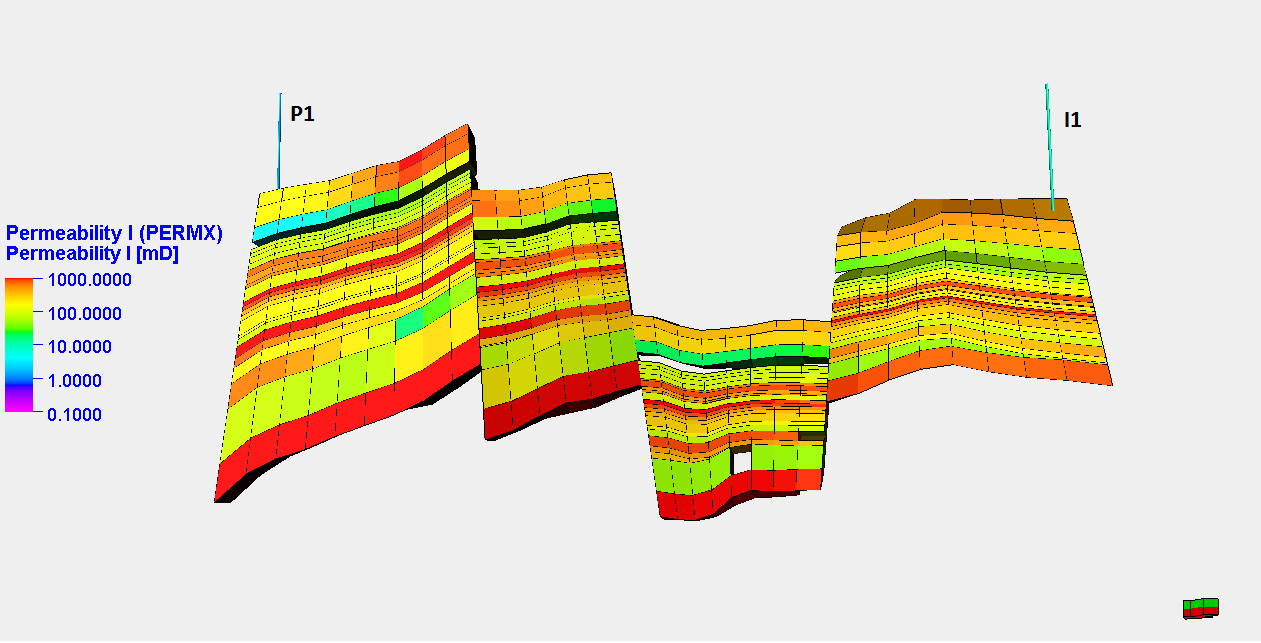}
	\includegraphics[scale=0.3]{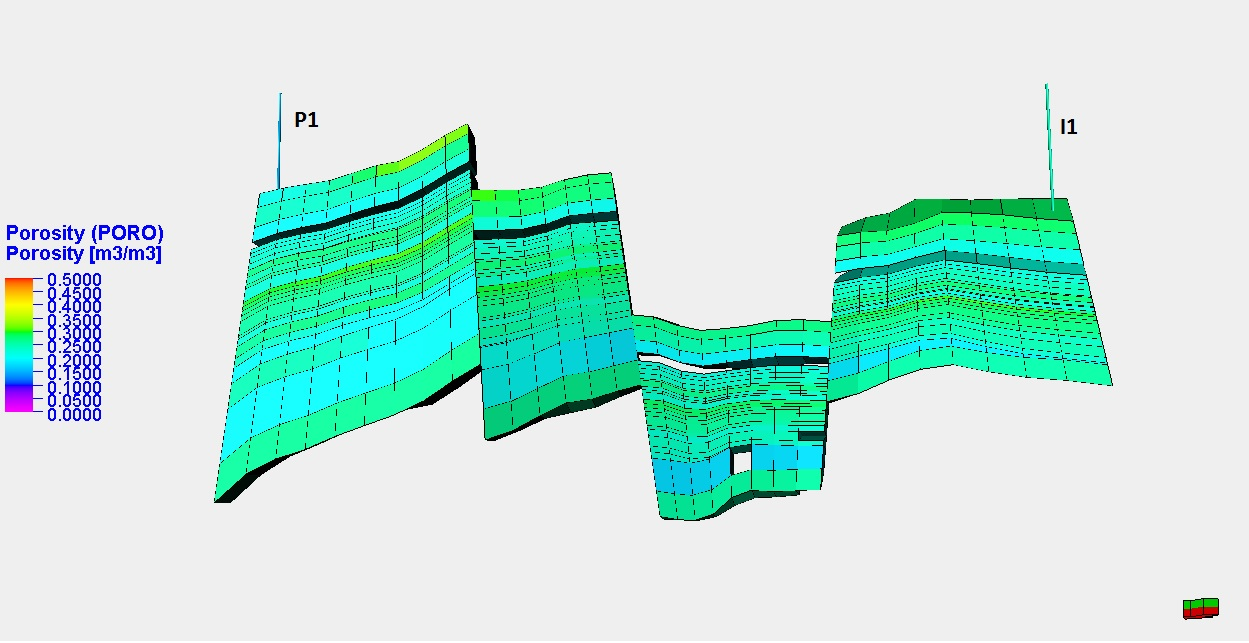}
	\caption{\label{fig:norne2D_ref_petrel} PERMX and PORO fields of the reference reservoir model used in the case study. Figures are generated using PETREL$^\copyright$.}
\end{figure*}

We use a 2D synthetic model for a proof-of-concept study. This reservoir model consists of three phases (water, oil and gas), and is a vertical section of a 3D Norne field model. This 2D model has 26 layers, with 39 gridblocks at each layer. Thus there are 1014 gridblocks overall, whereas 739 of them are active cells. This reservoir model was previously used in \cite{dadashpour2008nonlinear}. The petro-physical properties to be estimated include the isotropic permeability field and the porosity field, whereas other parameters, e.g., z-direction transmissibility multiplier (MULTZ) and relative permeability curves, are assumed to be known to us. Nevertheless, we do not have the reference permeability and porosity fields used in \cite{dadashpour2008nonlinear}. Therefore we create another reference model based on a data file associated with the 3D Norne filed. Figure \ref{fig:norne2D_ref_petrel} indicates the x-direction permeability (PERMX) and porosity (PORO) of our reference model, and Table \ref{tab:norne_2d} summarizes some critical information of the case study.   
\begin{table*}
\centering
\caption{\label{tab:norne_2d} Information summary of the 2D model case study}
\setlength\tabcolsep{2pt}
\begin{tabular}{|l|l|}
\hline  
Model dimension & $39 \times 26$ (1014 gridblocks), with 739 out of 1014 being active cells  \\ 
\hline   
Gridblock size & Irregular. Average $\Delta X \approx 80 m$, average $\Delta Z \approx 5 m$   \\ 
\hline 
Reservoir simulator  & ECLIPSE 100 (control mode RESV)   \\ 
\hline 
Well completion  & Injector I1: layer 1 -- 25; Producer P1: layer 1 -- 12  \\ 
\hline 
Production period & 3750 days (with 27 report time) \\ 
\hline 
Production data  & I1: BHP; P1: BHP, GOR, OPT, WCT. Total number: $5 \times 27 = 135$  \\ 
\hline 
Seismic survey time  & Base: day 0; Monitor (1st): day 2040; Monitor (2nd): day 3750   \\ 
\hline 
Seismic data  & AVA intercepts and gradients at survey time. Total number: 46686   \\ 
\hline 
DWT (seismic)  & 2D Daubechies wavelets with two vanishing moments   \\ 
\hline 
Thresholding  & Hard thresholding based on universal rule \\ 
\hline 
History matching method  & iES (RLM-MAC) with an ensemble of 100 reservoir models \\ 
\hline 
\end{tabular}
\end{table*}
 
On average, the gridblock size of the reservoir is around $80$ meters horizontally, and $5$ meters vertically. There are one injection well (labelled as I1 in Figure \ref{fig:norne2D_ref_petrel}) and one production well (labelled as P1 in Figure \ref{fig:norne2D_ref_petrel}). The completion of I1 is from layer 1 to layer 25, and that of P1 is from layer 1 to layer 12. The production period is 3750 days, and there are 27 report time instances from the ECLIPSE$^\copyright$ 100 black-oil simulator. The bottom hole pressures (BHP) from I1 and P1, and gas-oil ratio (GOR), oil production total (OPT) and water cut (WCT) from P1 at each report time are used for history matching. So overall the size of production data is $5 \times 27 = 135$. Gaussian white noise (with zero mean) is added to each production data as the synthetic observation error. For BHP data, the standard deviations of noise are 1 bar, and for GOR, OPT and WCT, the standard deviations are $10\%$ of the magnitudes of production data (and if the standard deviations are less than $10^{-6}$, they are reset to $10^{-6}$ to avoid the numerical issue of division by zero). It is assumed that observations errors between different types of production data are uncorrelated, so that the corresponding observation error covariance matrix is diagonal.

Seismic surveys are conducted at three time instances: day 0 (base survey), day 2040 (1st monitor survey) and day 3750 (2nd monitor survey). The intercepts and gradients of AVA data at all survey time are used as 4D seismic data. At each survey, the data size of each AVA attribute is 7781, therefore the total number of seismic data is $3 \times 2 \times 7781 = 46686$. In history matching study, Gaussian white noise (with zero mean) is also added to each attribute data, with the noise level (as defined in Eq. (\ref{eq:def_noise_level})) being $30\%$. For the purpose of comparison, we consider two sets of experiments. In one of them, the data to be history-matched are the original attributes without any transform, therefore the observation size is 46686. In addition, we assume that the standard deviations of observation noise are already known, without involving any estimation method (such as the one in Eq. (\ref{eq:noise_std_mad})). For distinction, we call this \textit{full-data experiment}. In the other set of experiment, we apply three-level 2D DWT to all AVA attributes using Daubechies wavelets with two vanishing moments. We then select leading wavelet coefficients by adopting hard thresholding in combination with the universal rule. These leading wavelet coefficients are used as the data for history matching. In the current case study, the total number of leading wavelet coefficients is 2746, roughly $5.9\%$ of the original data size. For each attribute, Eq. (\ref{eq:noise_std_mad}) is applied to estimate the standard deviation of observation noise. For distinction, we call this \textit{sparse-data experiment}.            

We consider three history matching scenarios, which involve: (S1) production data only; (S2) 4D seismic data only; and (S3) both production and 4D seismic data, whereas in both (S2) and (S3) we compare the results of full- and sparse-data experiments. The petro-physical properties to be estimated are PERMX (in the scale of natural logarithm) and PORO at active gridblocks. The initial ensemble of log PERMX and PORO are generated using Gaussian random function simulation in PETREL$^\copyright$. Log PERMX and PORO are generated separately, using the same spherical variogram model whose horizontal range is $1901$ meters and vertical one is $12.8$ meters. In the course of generating log PERMX and PORO realizations, the simulation processes are conditioned on synthetic well log data of I1 and P1 generated from the reference reservoir model. The total number of realizations (ensemble size) is 100.     

RLM-MAC introduced in the proceeding section is used for all history matching scenarios. Except for the aforementioned stopping criterion (\ref{eq:stopping_criterion_ndm}), we introduce two additional stopping conditions for run-time control:
 \begin{linenomath*}    
 \begin{IEEEeqnarray}{lll} 
 \text{(C1) RLM-MAC stops if it reaches a maximum of 10 outer iteration steps;} \nonumber \\
 \text{(C2) RLM-MAC stops if the relative change of average data mismatch over two consecutive iteration steps is less than } 0.01\% \, . \nonumber
 \end{IEEEeqnarray}
 \end{linenomath*}     
For more information of the implementation of RLM-MAC, readers are referred to \cite{luo2015Iterative}. To see the effect of the stopping criterion (\ref{eq:stopping_criterion_ndm}), we also compare the estimation results obtained by only invoking the stopping conditions (C1) and (C2) and those obtained by invoking (C1), (C2) and the criterion (\ref{eq:stopping_criterion_ndm}).    

\subsection{Results of scenario S1 (using production data only)}     
\begin{figure*}
	\centering
	\includegraphics[scale=0.4]{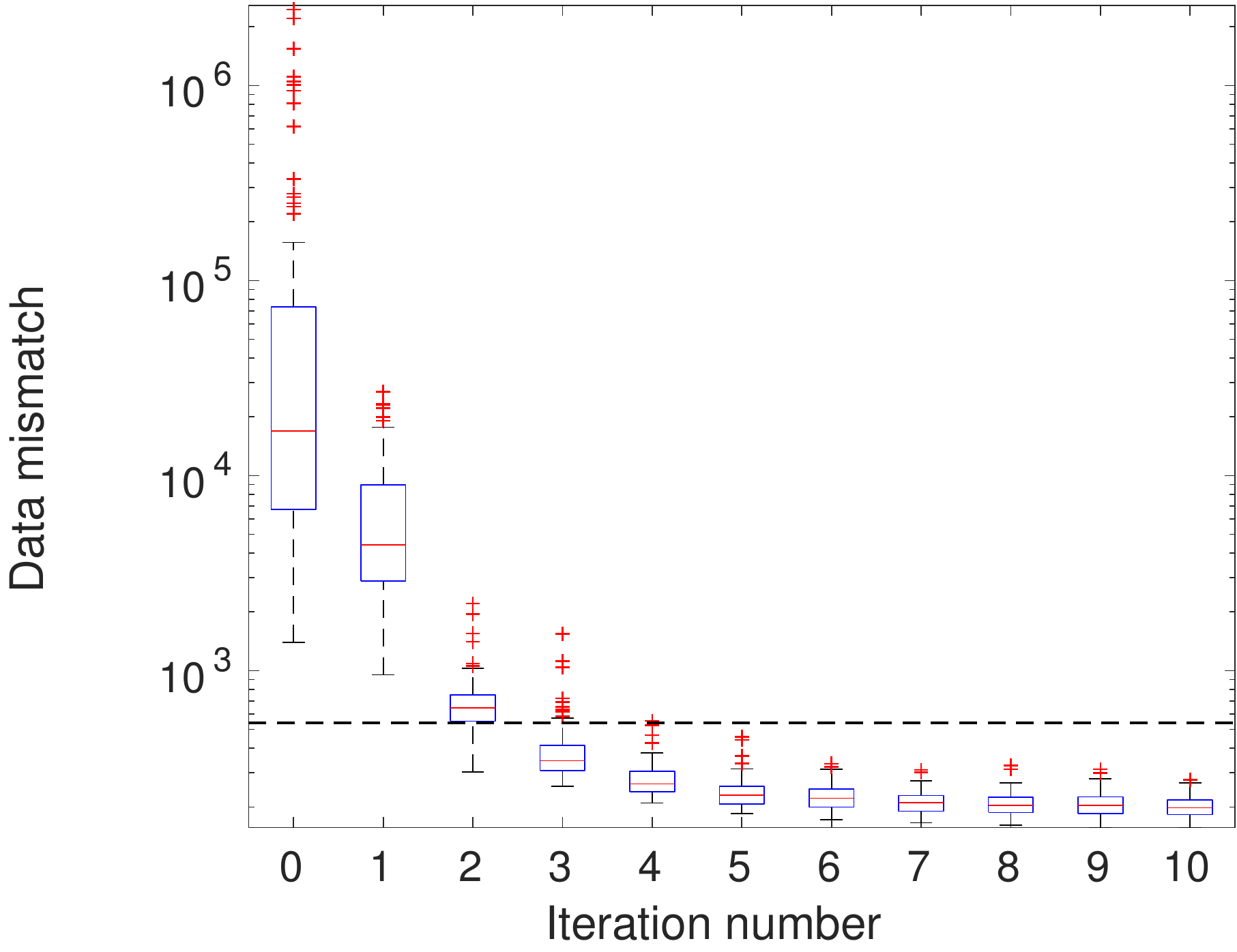}
	\caption{\label{fig:Norne2D_RLM-MAC_boxplot_objRealIter_S1} Boxplots of data mismatch as a function of iteration step (scenario S1). The horizontal dashed line indicates the threshold value ($4 \times 135 = 540$) for the stopping criterion (\ref{eq:stopping_criterion_ndm}). For visualization, the vertical axis is in the logarithmic scale. In each box plot, the horizontal line (in red) inside the box denotes the median; the top and bottom of the box represent the 75th and 25th percentiles, respectively; the whiskers indicate the ranges beyond which the data are considered outliers, and the whiskers’ positions are determined using the default setting of MATLAB$^\copyright$ R2015b, while the outliers themselves are plotted individually as plus signs (in red).}
\end{figure*}   

In this case we only use production data in history matching. Figure \ref{fig:Norne2D_RLM-MAC_boxplot_objRealIter_S1} shows the boxplots of data mismatch as a function of iteration step. At iteration 0, the average data mismatch of the initial ensemble is $1.57 \times 10^5$. If one only adopts the stopping conditions (C1) and (C2), then RLM-MAC stops after 10 iterations, with the corresponding average data mismatch being $202.12$. If the extra stopping criterion (\ref{eq:stopping_criterion_ndm}) is used, then RLM-MAC stops after 3 iterations, at which the average data mismatch is $402.19$, lower than the critical value $4 \times 135 = 540$ (horizontal dashed line in Figure \ref{fig:Norne2D_RLM-MAC_boxplot_objRealIter_S1}) for the first time.
   
\renewcommand{\nScale}{0.4} 
\begin{figure*}
	\centering
	\subfigure[RMSEs of log PERMX]{ \label{subfig:rmse_PERMX_boxplot_ensemble_S1}
				\includegraphics[scale=\nScale]{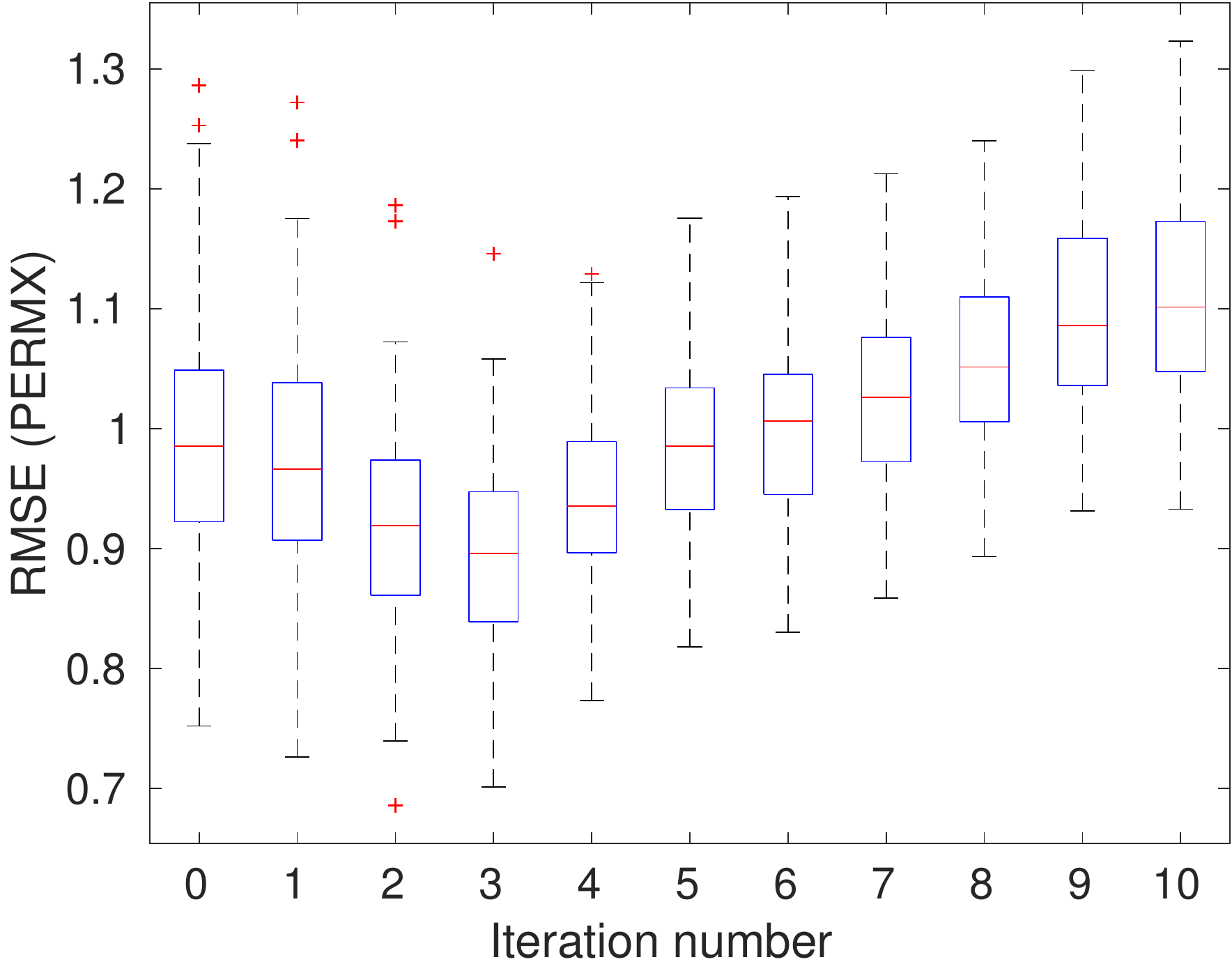}
			}
	\subfigure[RMSEs of PORO]{ \label{subfig:rmse_PORO_boxplot_ensemble_S1}
					\includegraphics[scale=\nScale]{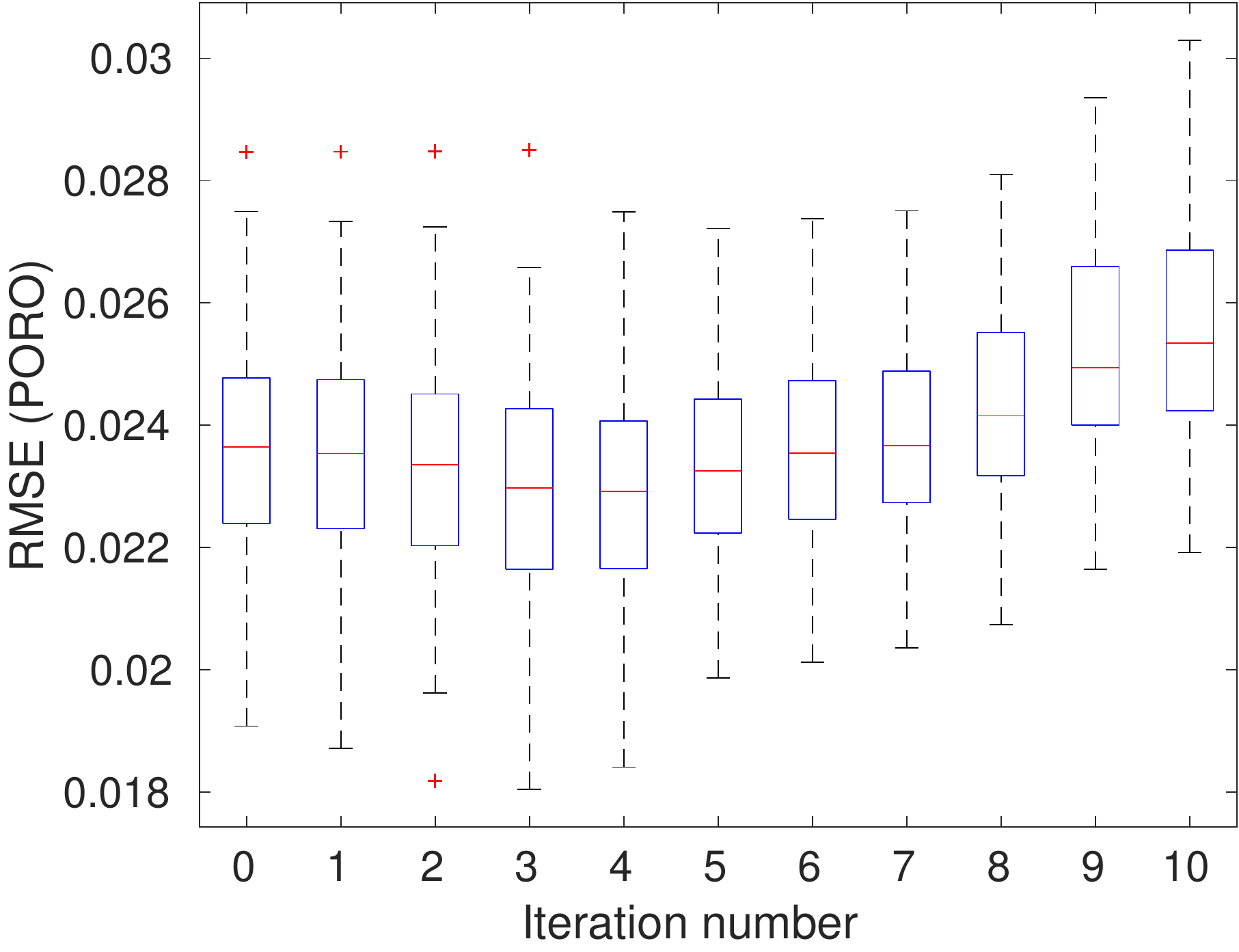}
				}
	\caption{\label{fig:Norne2D_RLM-MAC_RMSE_S1} Boxplots of RMSEs of (a) log PERMX and (b) PORO as functions of iteration step (scenario S1).}
\end{figure*}   

Figure \ref{fig:Norne2D_RLM-MAC_RMSE_S1} examines the RMSEs of log PERMX and PORO obtained at different iteration steps. Let $\mathbf{v}^{tr}$ be the $\ell$-dimensional true property, and $\hat{\mathbf{v}}$ an estimation of the truth, then in this work RMSE $e_{\mathbf{v}}$ of $\hat{\mathbf{v}}$ is defined as follows:  
\begin{linenomath*} 
\begin{IEEEeqnarray}{lll} \label{eq:RMSE_def}
e_{\mathbf{v}} = \dfrac{\Vert \hat{\mathbf{v}} - \mathbf{v}^{tr} \Vert_2}{\ell} \, ,
\end{IEEEeqnarray}
\end{linenomath*} 
where $\Vert \bullet \Vert_2$ denotes the Euclidean norm. For both log PERMX and PORO, their RMSEs decrease at the first few iterations but rebound as iteration proceeds further. As a result, the RMSEs at 10th iteration step tend to be higher than those of the initial ensemble. In contrast, the RMSEs at the 3rd iteration step tend to be the lowest for both log PERMX and PORO. This indicates the benefit of equipping the stopping criterion (\ref{eq:stopping_criterion_ndm}) in this particular case. A further remark is that, if one uses the ensemble at the 3rd iteration step as the final history-matched models, then compared to the initial ensemble, there appear to be more changes in log PERMX than PORO. In other words, in this particular case production data appear more sensitive to log PERMX than to PORO.

\begin{figure*} %
	\centering
	\subfigure[BHP (bar)]{ \label{subfig:WBHP_I1_init_forecasts_S1}
				\includegraphics[scale=0.3]{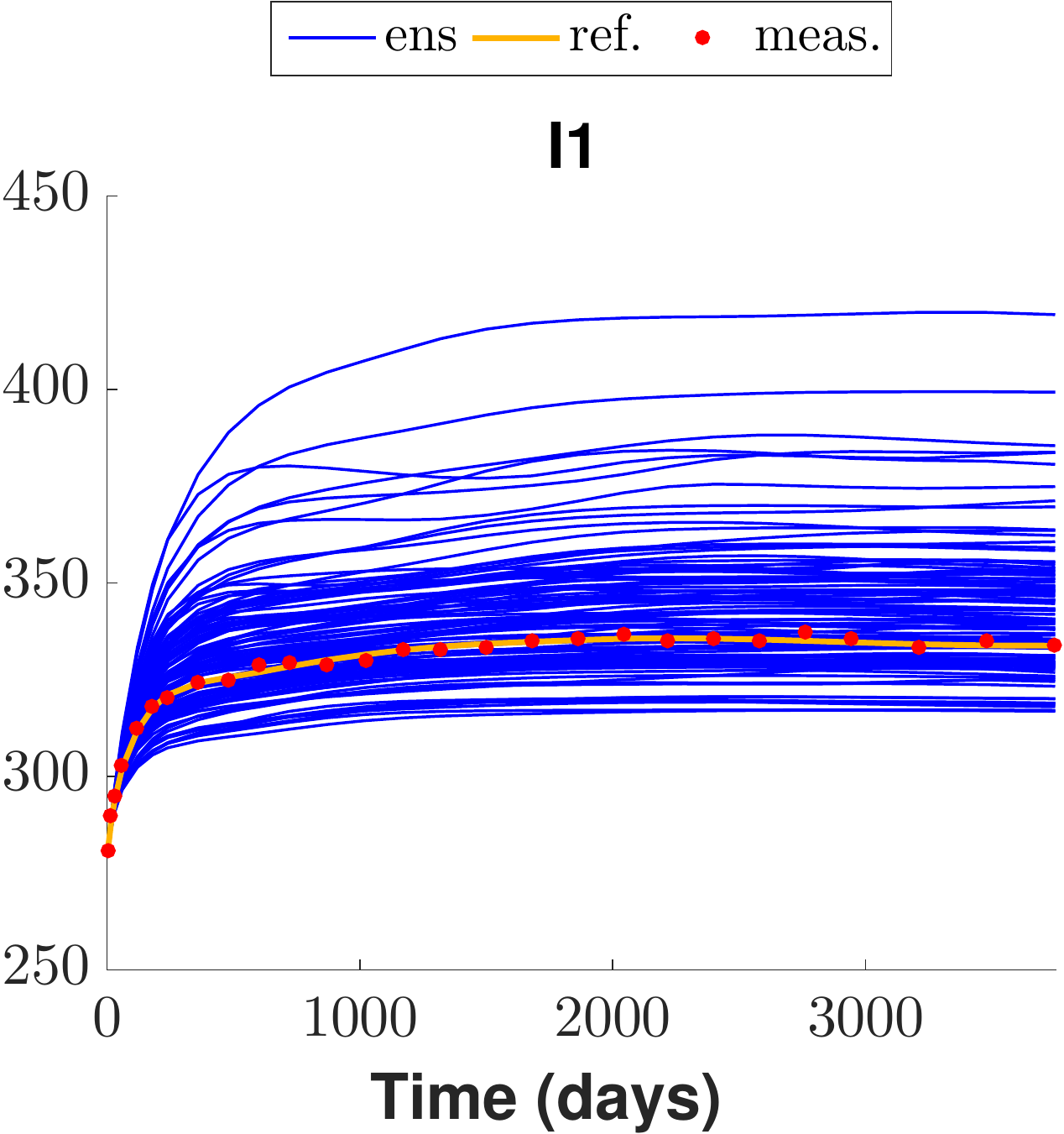}
			}%
	\subfigure[BHP (bar)]{ \label{subfig:WBHP_P1_init_forecasts_S1}
					\includegraphics[scale=0.3]{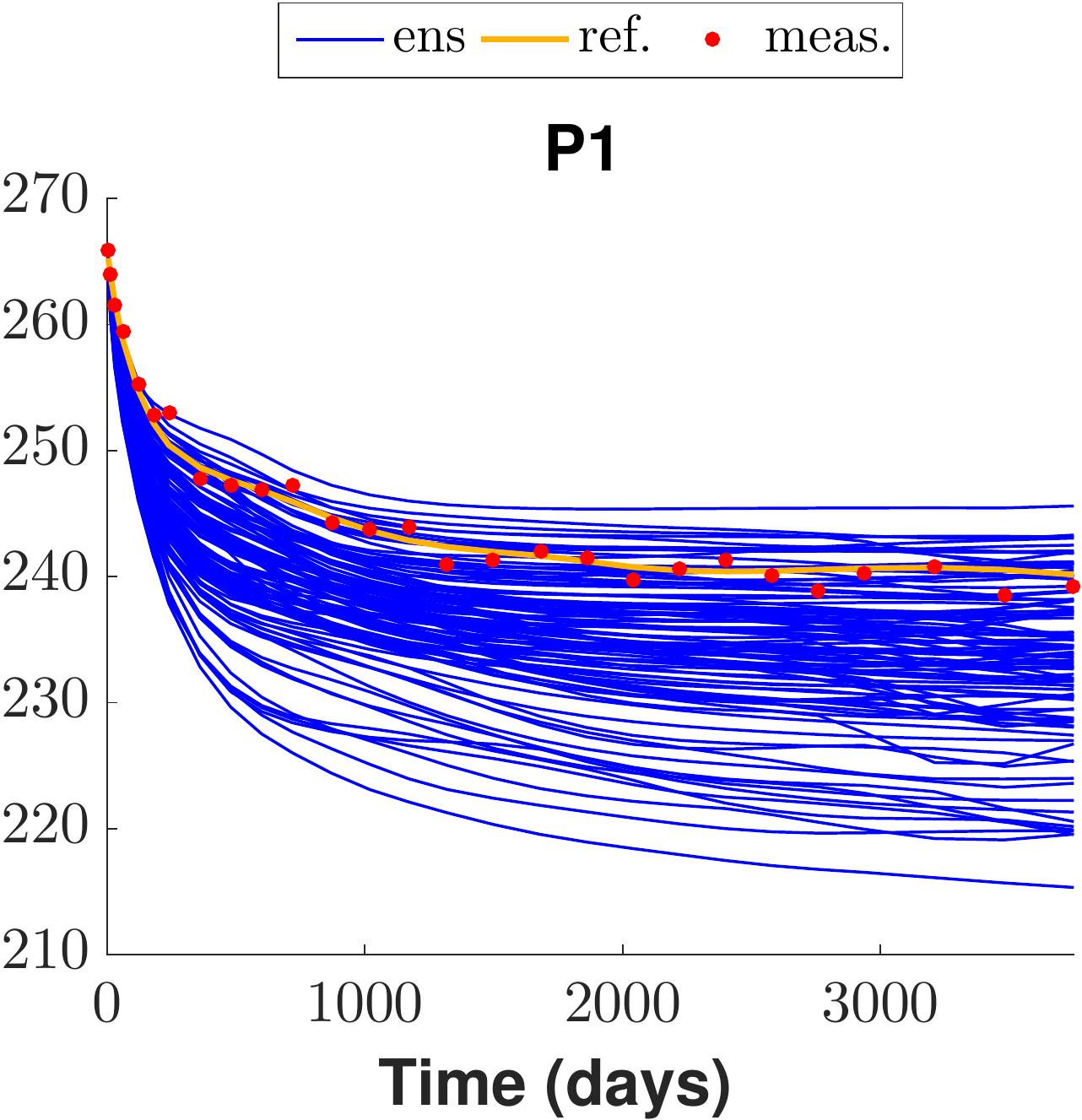}
				}
				
	\subfigure[GOR]{ \label{subfig:WGOR_P1_init_forecasts_S1}
					\includegraphics[scale=0.3]{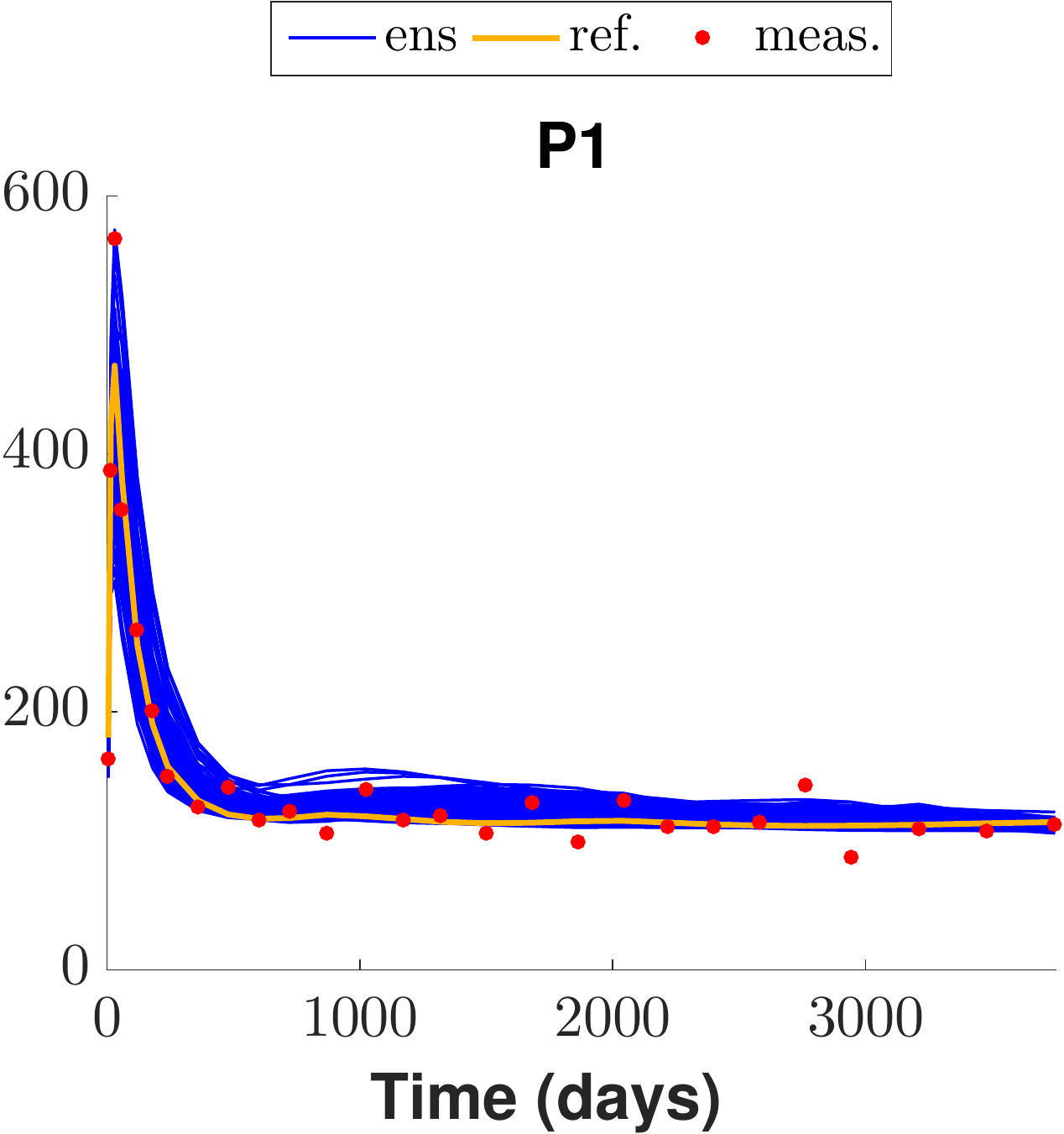}
				}%
	\subfigure[OPT (sm$^3$)]{ \label{subfig:WOPT_P1_init_forecasts_S1}
					\includegraphics[scale=0.3]{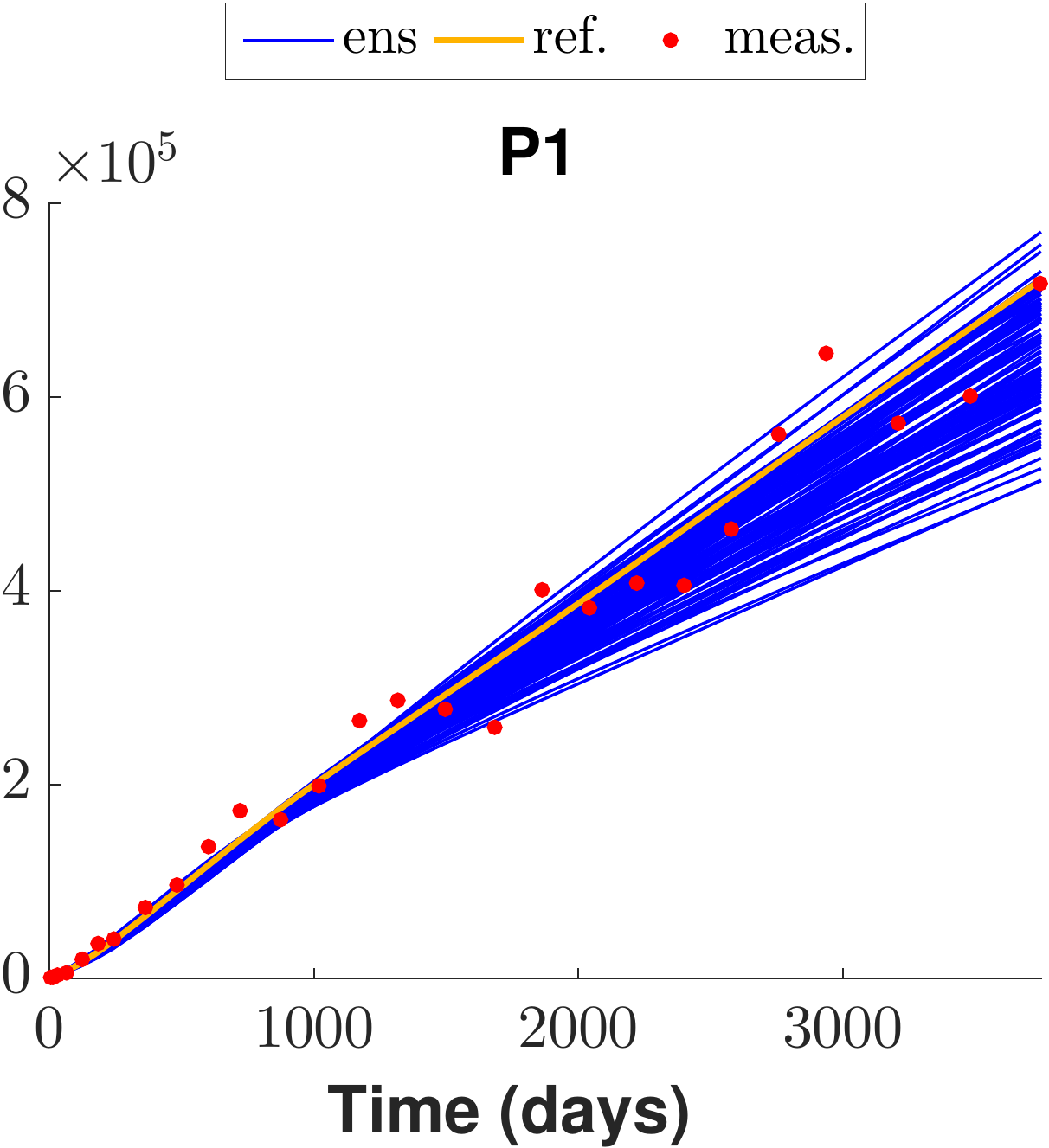}
				}%
	\subfigure[WCT]{ \label{subfig:WWCT_P1_init_forecasts_S1}
					\includegraphics[scale=0.3]{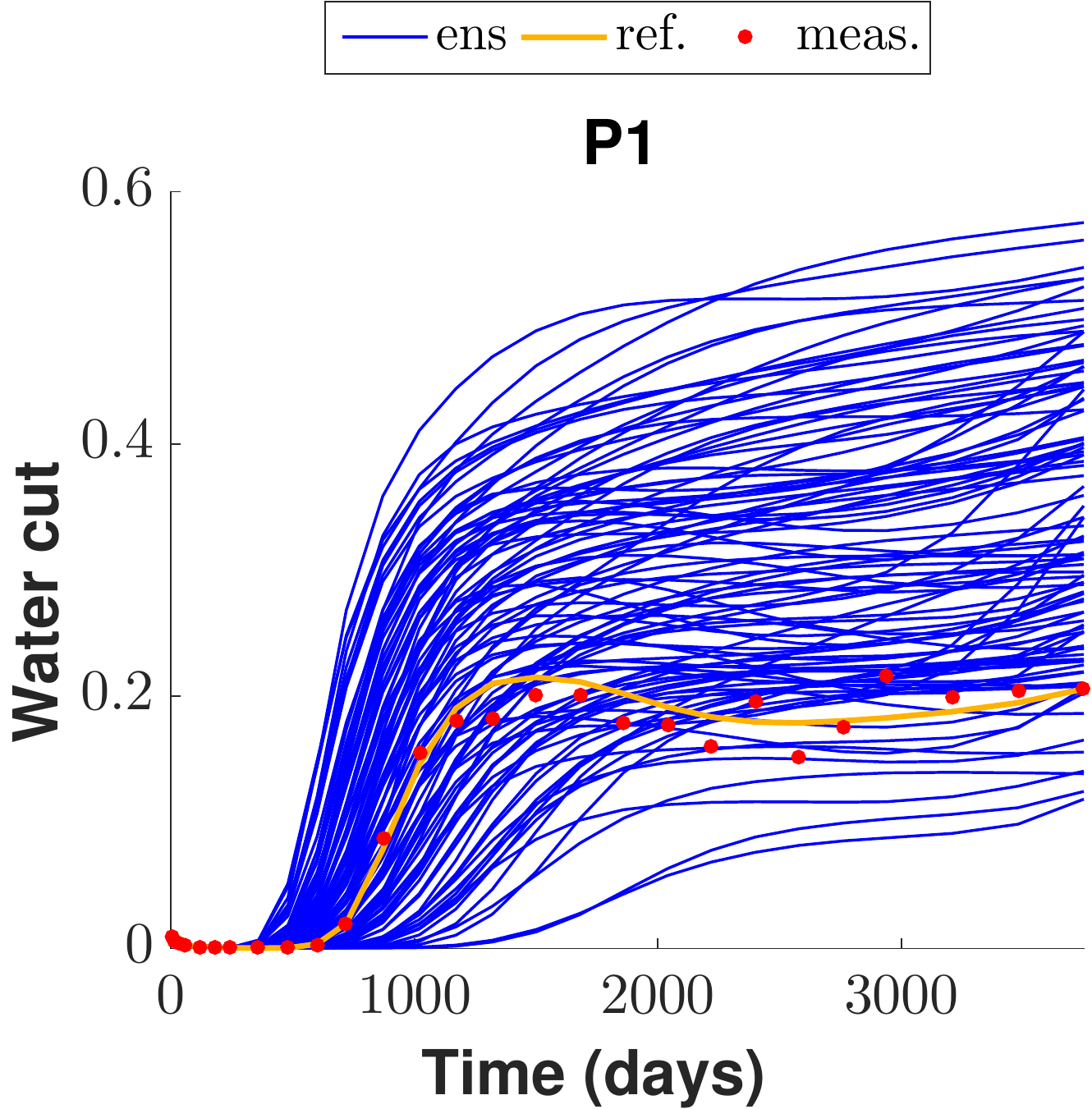}
				}											
	\caption{\label{fig:Norne2D_production_profile_init_S1} Production data profiles (scenario S1). (a) BHP at I1; (b) BHP at P1; (c) GOR; (d) OPT; (e) WCT. In each sub-figure, blue curves represent production data forecasts with respect to the initial ensemble, orange curve corresponds to true data of the reference model, and red dots are observations.}
\end{figure*}

\begin{figure*} %
	\centering
	\subfigure[BHP (bar)]{ \label{subfig:WBHP_I1_final_forecasts_S1}
				\includegraphics[scale=0.3]{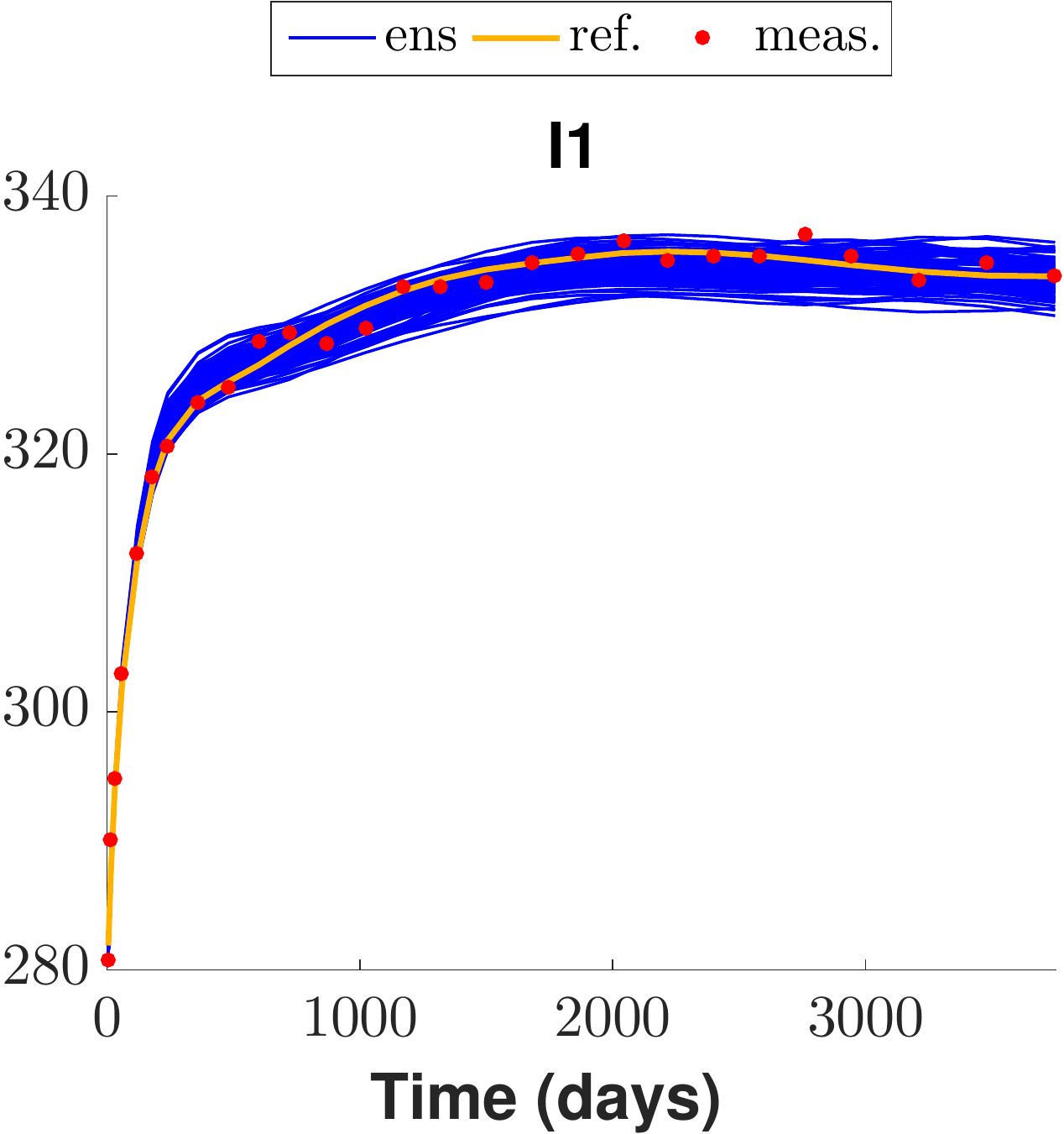}
			}%
	\subfigure[BHP (bar)]{ \label{subfig:WBHP_P1_final_forecasts_S1}
					\includegraphics[scale=0.3]{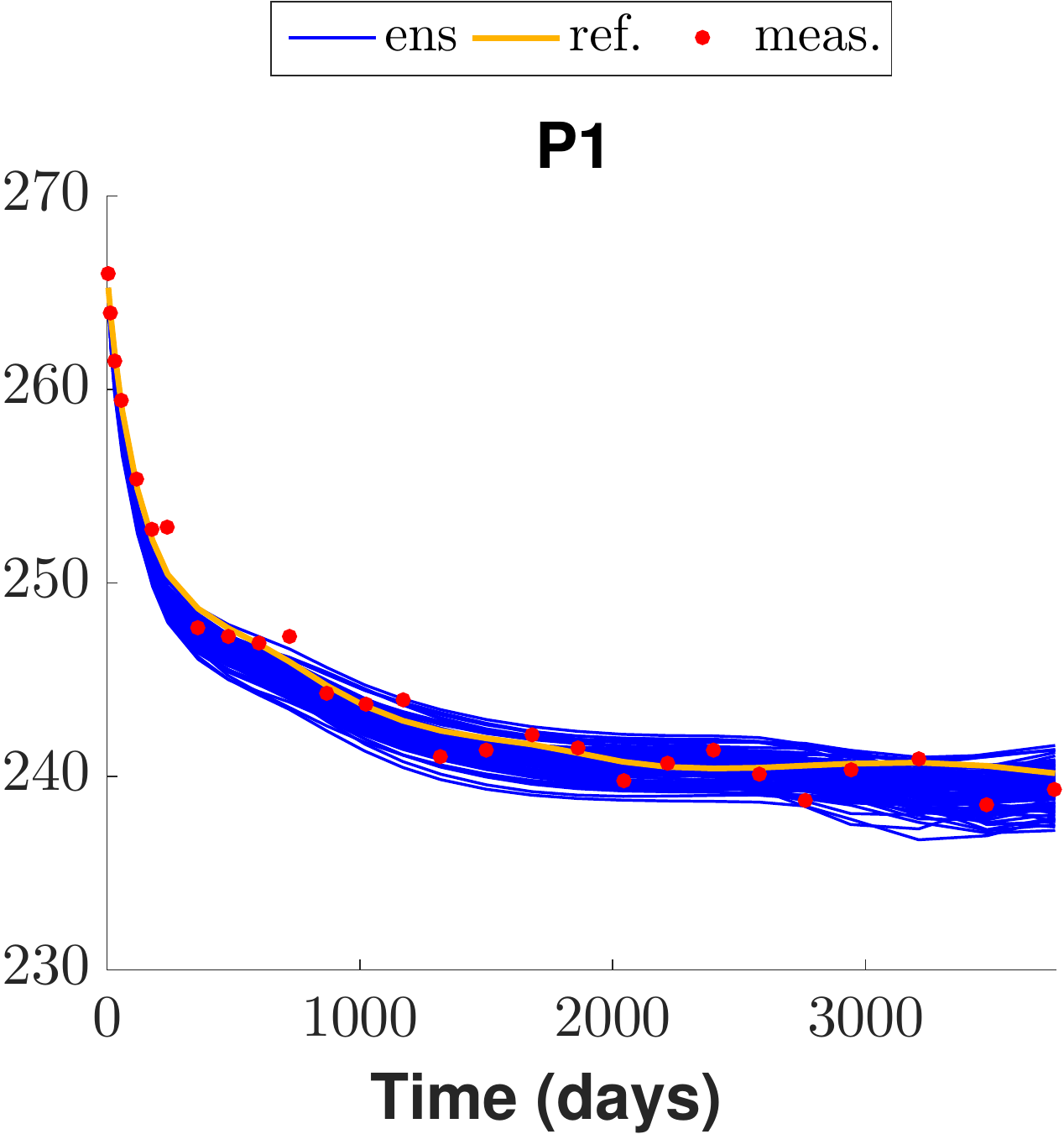}
				}
				
	\subfigure[GOR]{ \label{subfig:WGOR_P1_final_forecasts_S1}
					\includegraphics[scale=0.3]{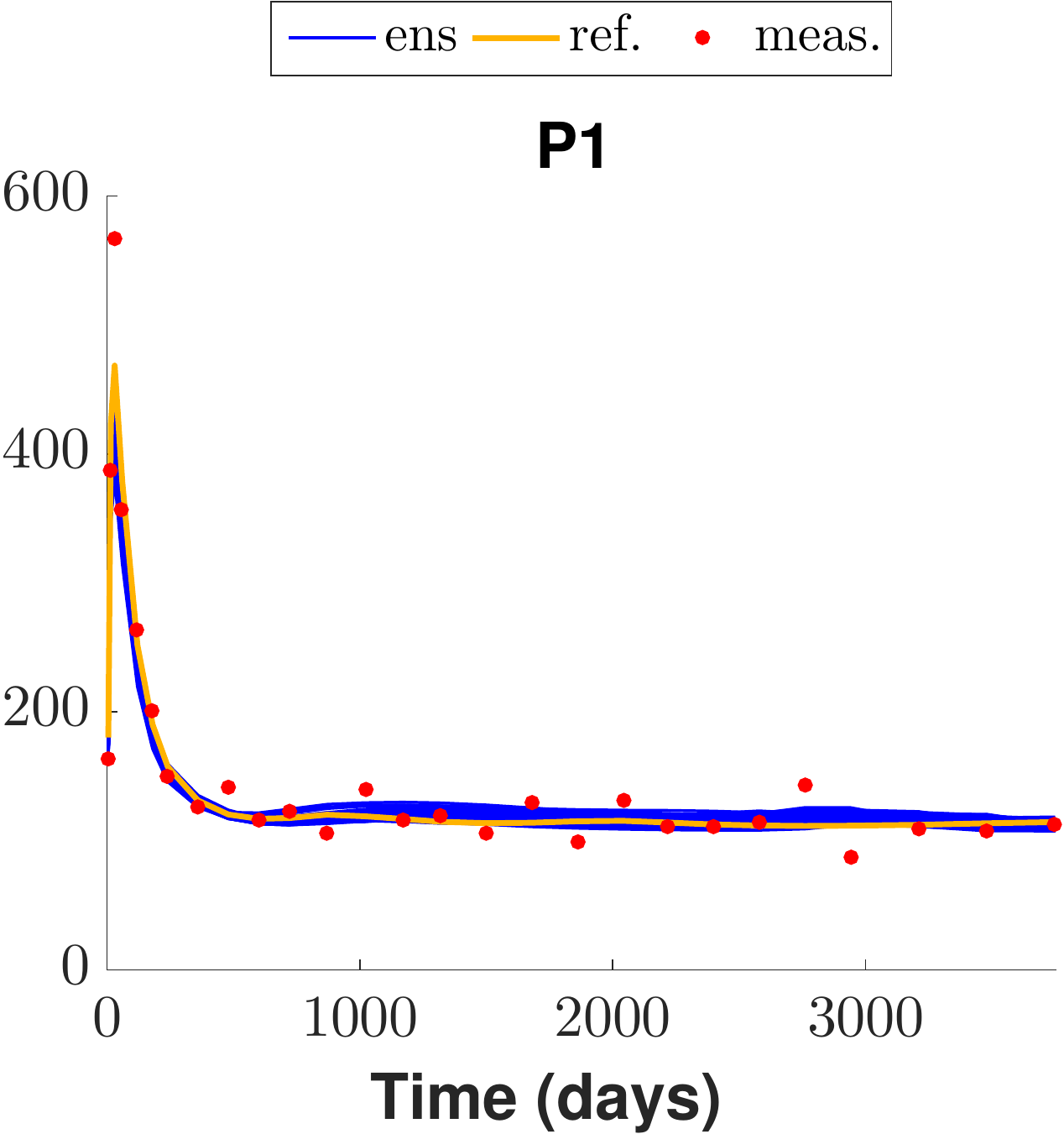}
				}%
	\subfigure[OPT (sm$^3$)]{ \label{subfig:WOPT_P1_final_forecasts_S1}
					\includegraphics[scale=0.3]{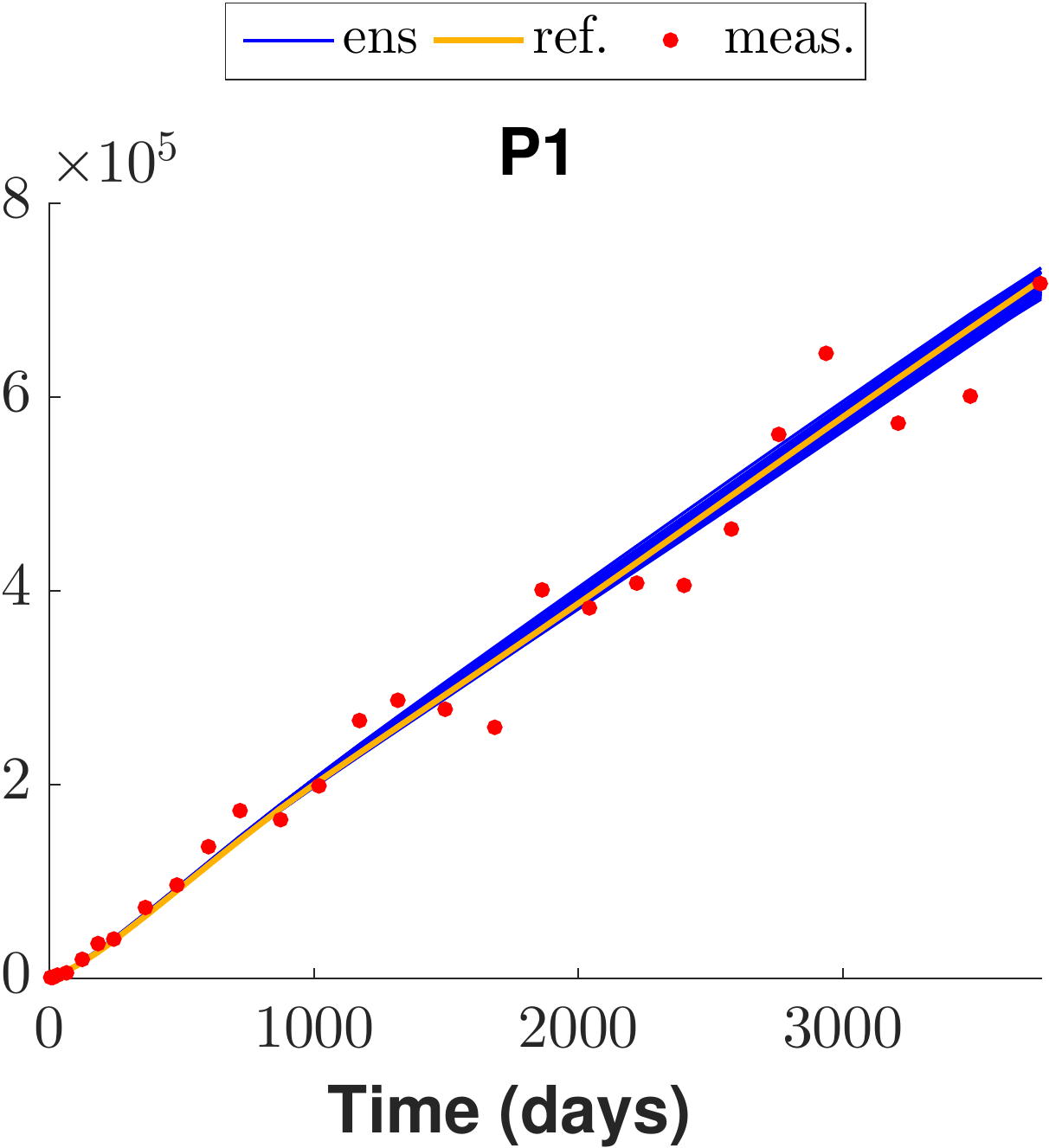}
				}%
	\subfigure[WCT]{ \label{subfig:WWCT_P1_final_forecasts_S1}
					\includegraphics[scale=0.3]{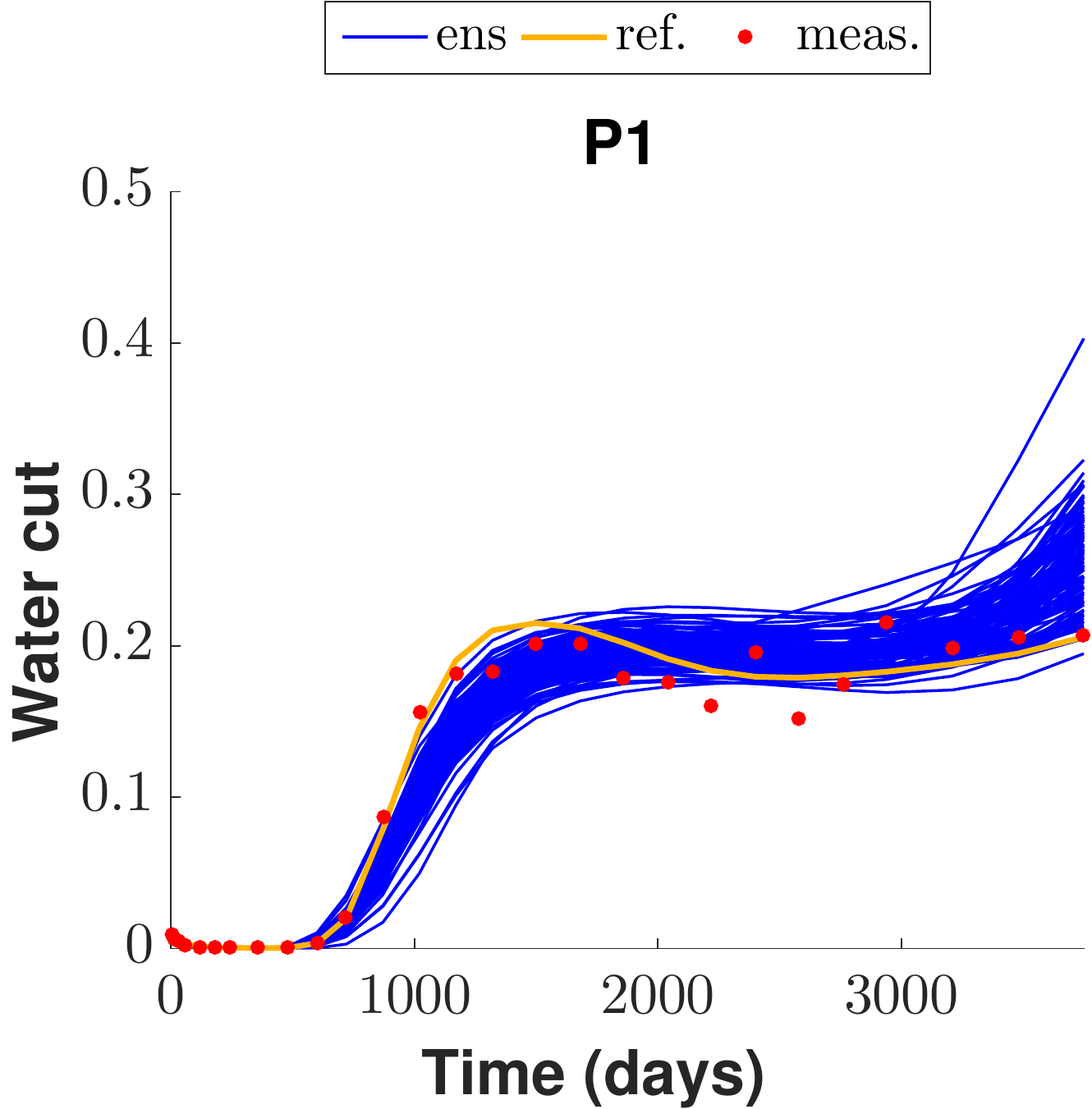}
				}											
	\caption{\label{fig:Norne2D_production_profile_final_S1} As in Figure \ref{fig:Norne2D_production_profile_init_S1}, but now blue curves correspond to production data forecasts with respect to the ensemble obtained at the 3rd iteration step.}
\end{figure*}

In the sequel we show production data profiles and distributions of log PERMX and PORO with respect to the reference reservoir model, the initial and final ensembles of reservoir models, respectively. Here, the final ensemble is taken as the one from the 3rd iteration step. Production data profiles are indicated in Figures \ref{fig:Norne2D_production_profile_init_S1} and \ref{fig:Norne2D_production_profile_final_S1}. As one can see there, after history matching, the final ensemble has better predictions of production data and reduced prediction spreads. This is consistent with the results in Figure \ref{fig:Norne2D_RLM-MAC_boxplot_objRealIter_S1}. 

Distributions of log PERMX and PORO are reported in Figures \ref{fig:Norne2D_PERMX_S1} and \ref{fig:Norne2D_PORO_S1}, respectively. In each figure, we show the reference property in the first row, then the mean and two sample properties of the initial ensemble in the second row, and finally the mean and two sample properties of the final ensemble in the 3rd one. One can see differences between the means (or samples) of the initial and final ensembles. In this particular case, the changes in log PERMX appear more substantial than those in PORO, consistent with the results in Figure \ref{fig:Norne2D_RLM-MAC_RMSE_S1}. 
  
\renewcommand{\nScale}{0.3}

\renewcommand{\nScale}{0.3}
 
\renewcommand{\nScale}{0.45}
\begin{figure*} 
	\centering
	\subfigure[Reference]{ \label{subfig:permx_ref}
				\includegraphics[scale=\nScale]{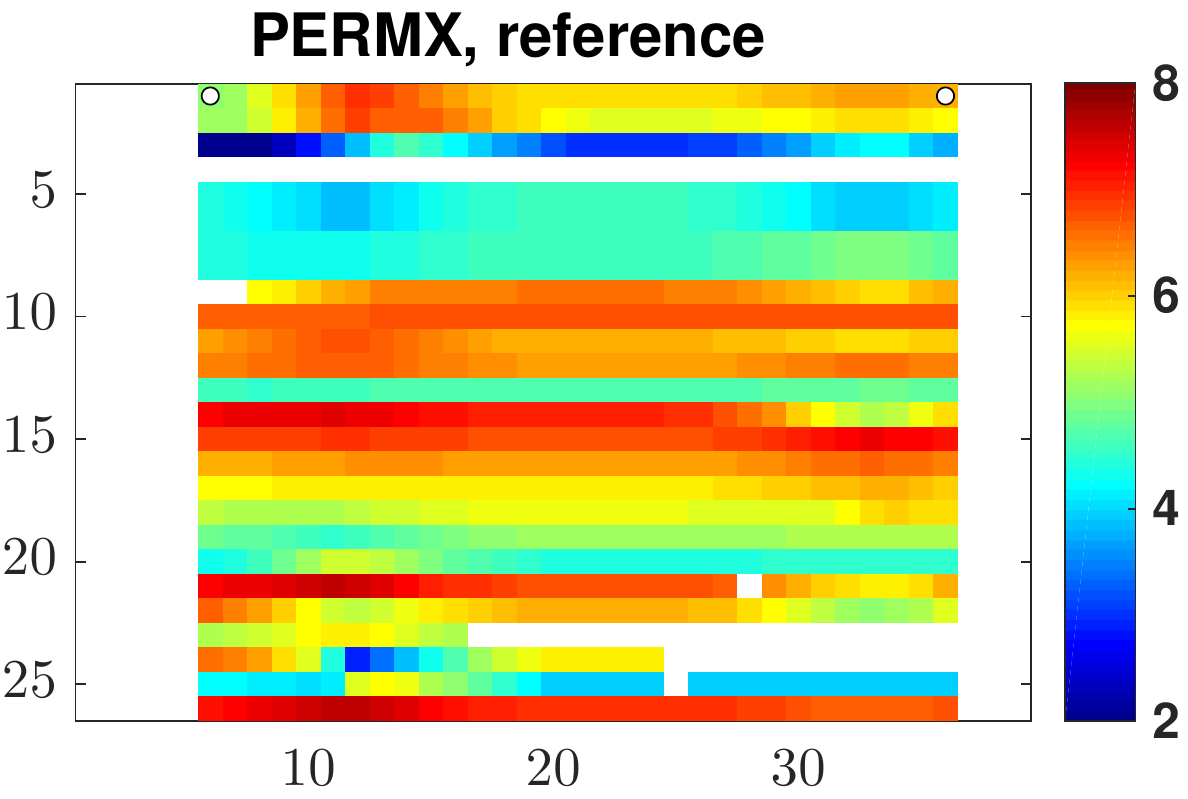}
			}
			
	\subfigure[Initial mean]{ \label{subfig:field_PERMX_mean_init_ensemble}
					\includegraphics[scale=0.28]{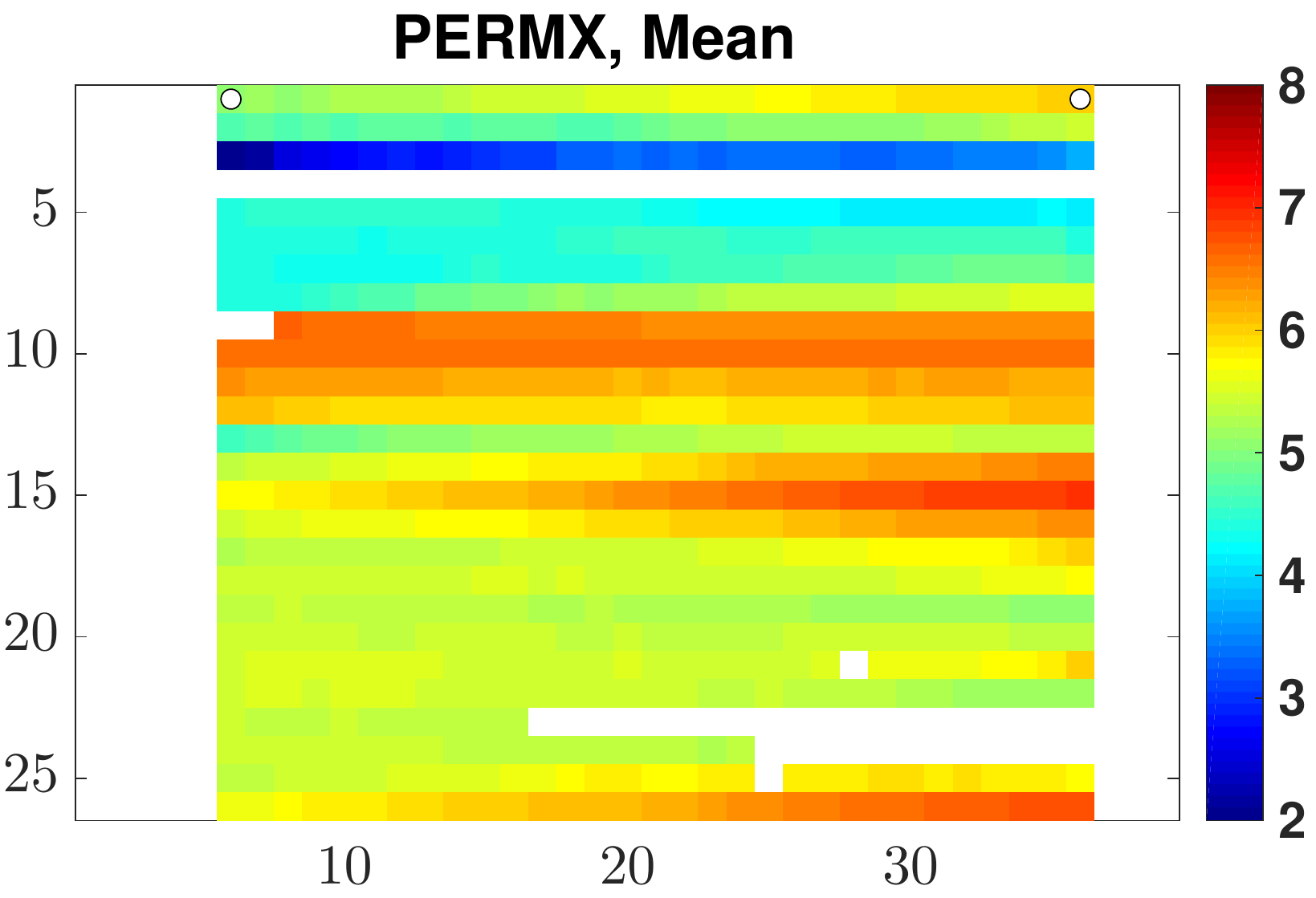}
				}			
	\subfigure[Initial member 1]{ \label{subfig:field_PERMX_1_1_init_ensemble}
					\includegraphics[scale=\nScale]{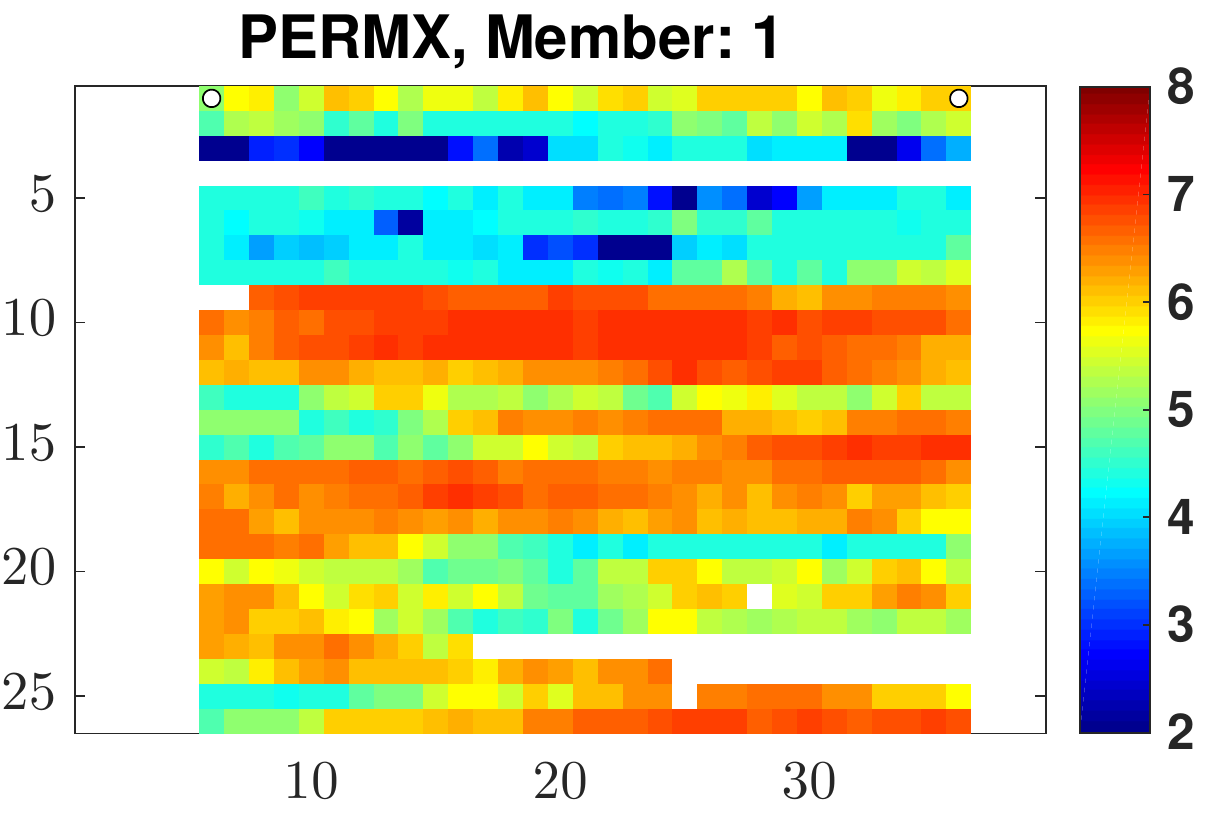}
				}
	\subfigure[Initial member 2]{ \label{subfig:field_PERMX_1_2_init_ensemble}
					\includegraphics[scale=\nScale]{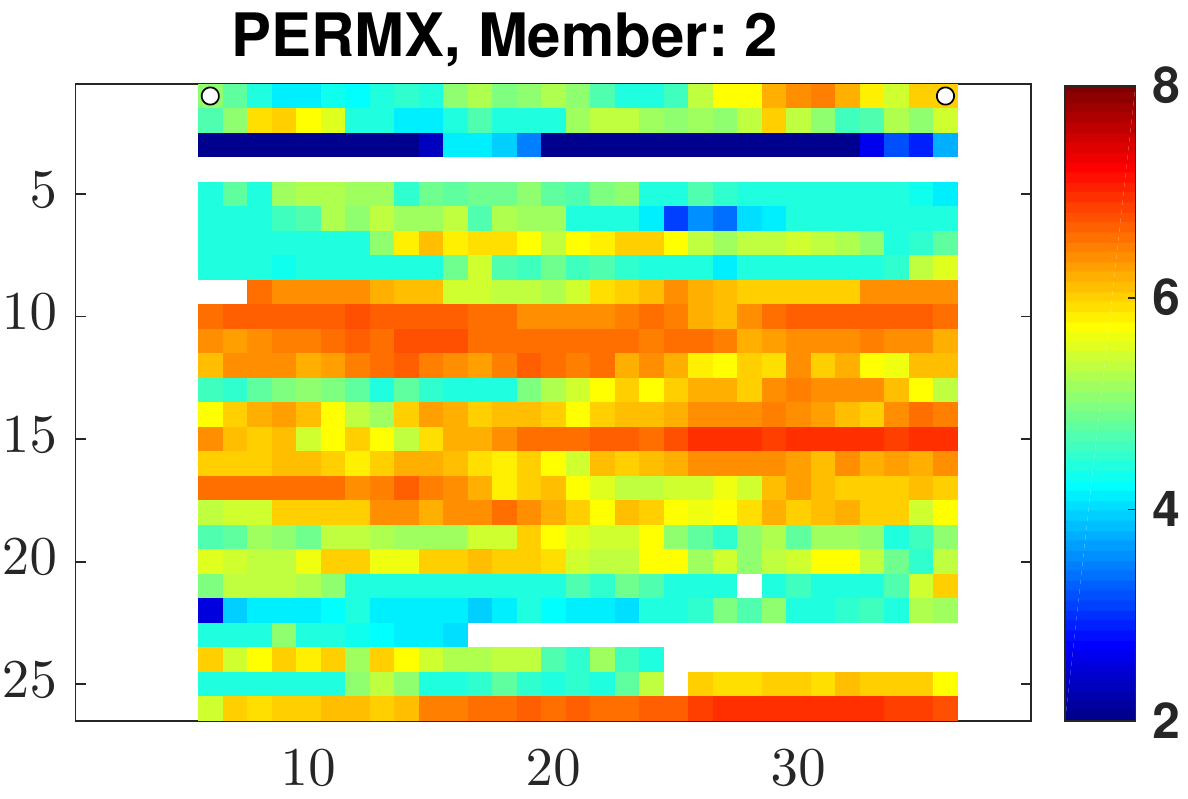}
				}		
				
	\subfigure[Final mean]{ \label{subfig:field_PERMX_mean_ensemble3}
					\includegraphics[scale=0.28]{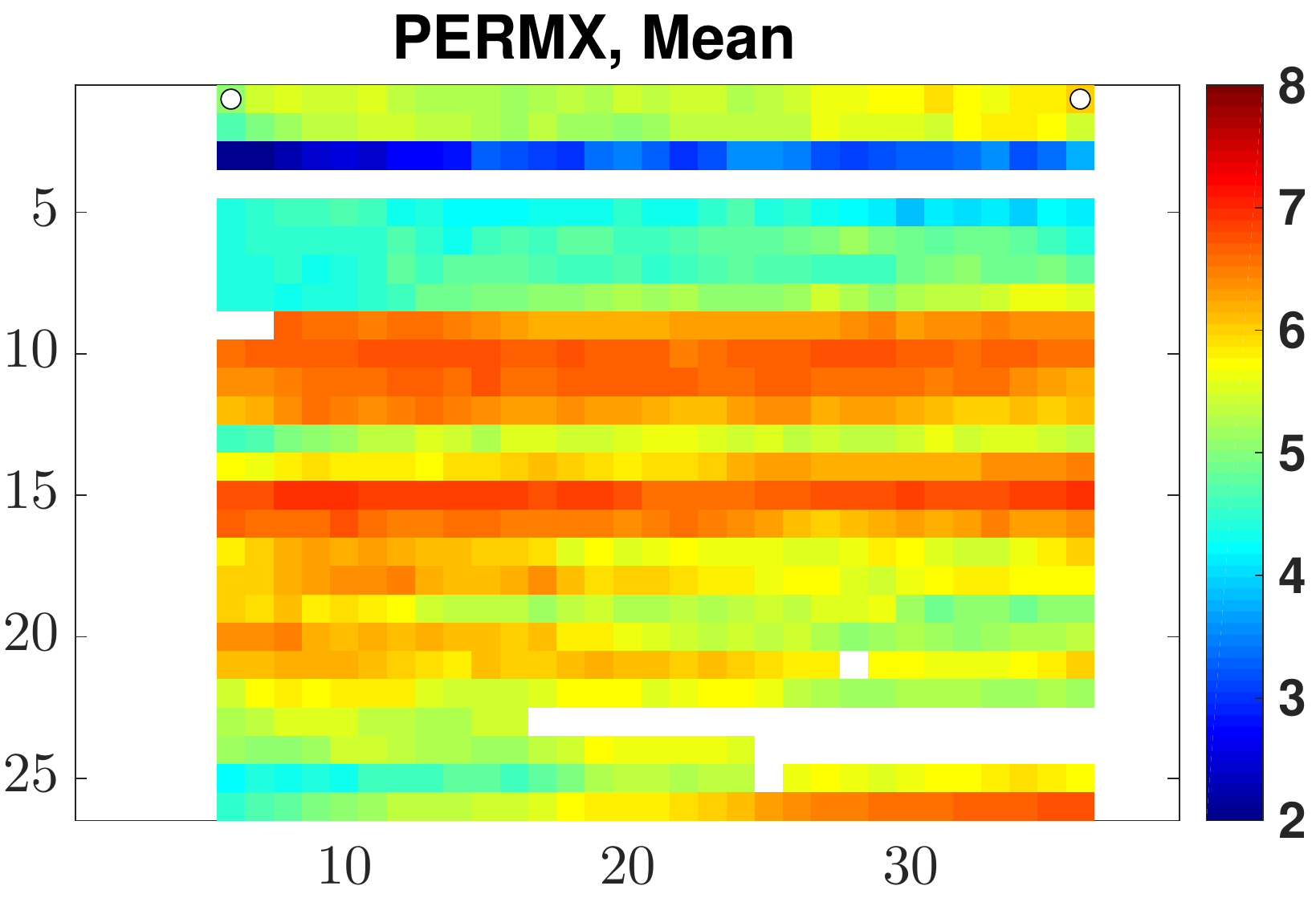}
				}			
	\subfigure[Final member 1]{ \label{subfig:field_PERMX_1_1_ensemble3}
					\includegraphics[scale=\nScale]{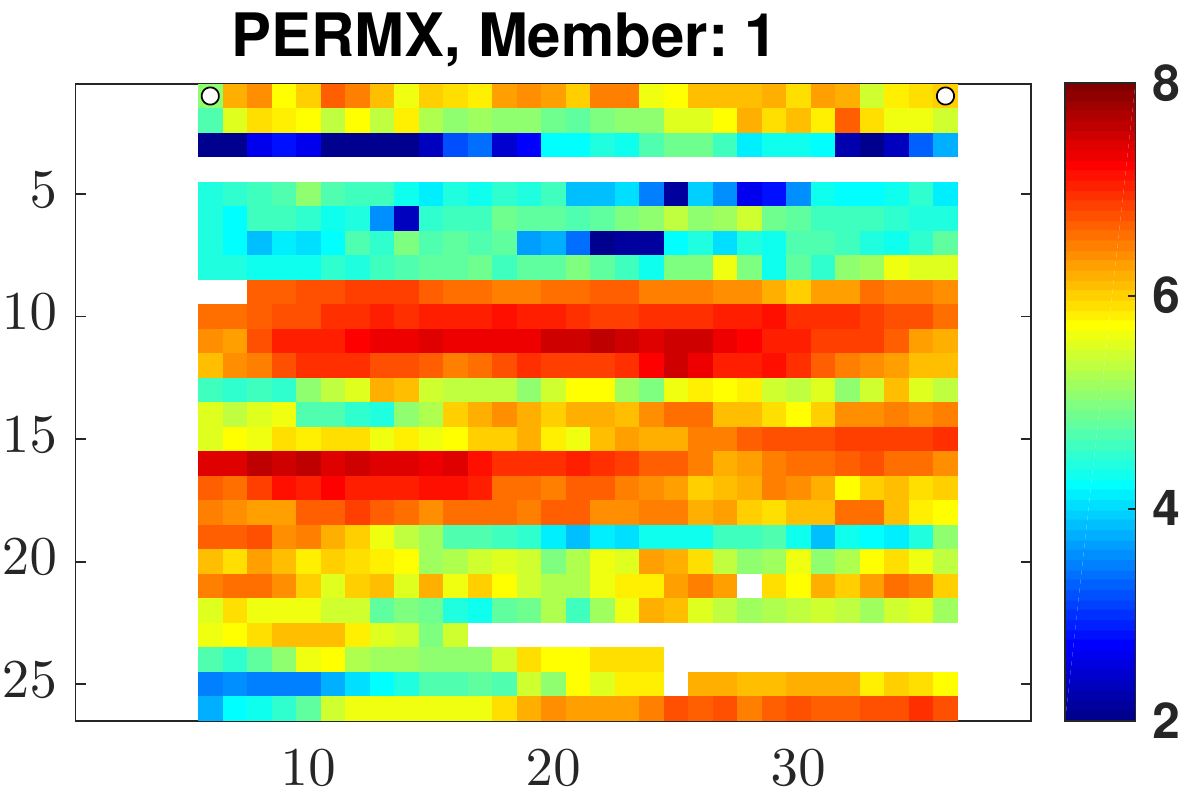}
				}
	\subfigure[Final member 2]{ \label{subfig:field_PERMX_1_2_ensemble3}
					\includegraphics[scale=\nScale]{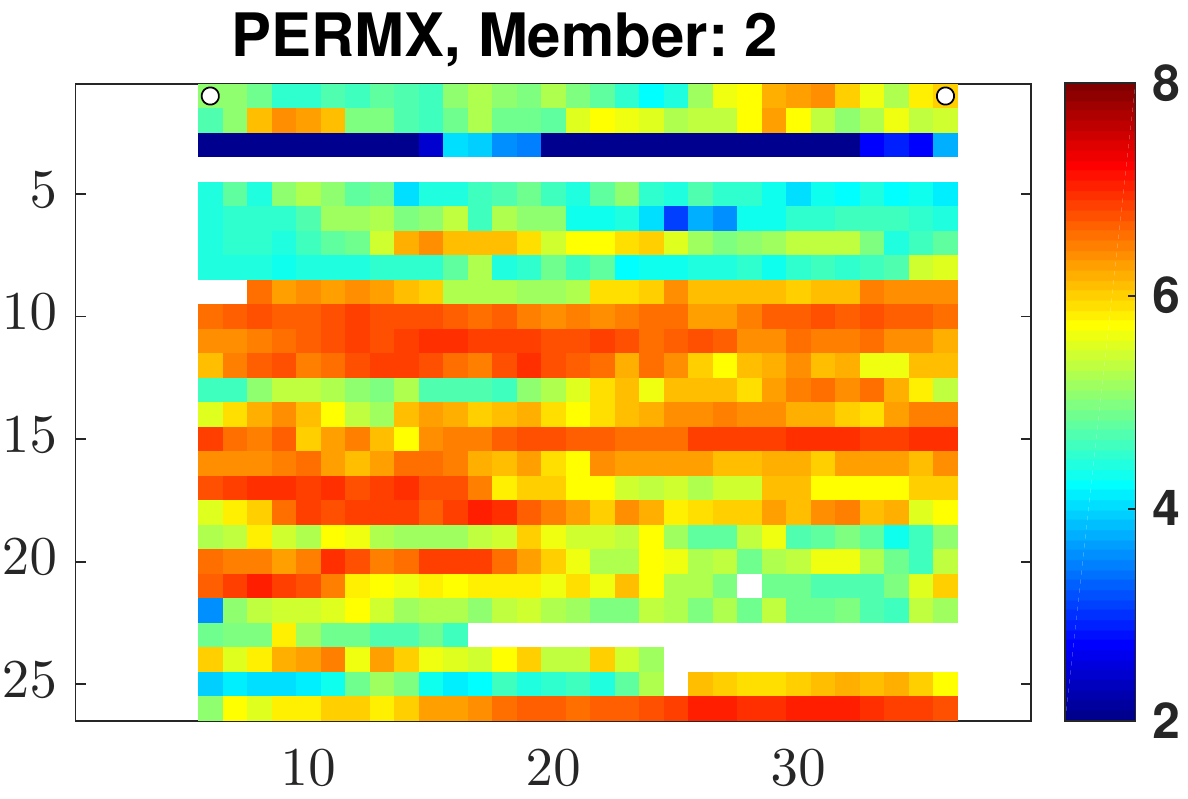}
				}															
	\caption{\label{fig:Norne2D_PERMX_S1} Distributions of log PERMX (scenario S1). (a) Reference model; (b) -- (d) Mean and 2 sample realizations of the initial ensemble of log PERMX; (e) -- (g) Corresponding mean and 2 sample realizations of the final ensemble. In each sub-figure, there are two white dots on the top layer, and they are used to indicate locations of the wells.}
\end{figure*}   

\renewcommand{\nScale}{0.45}
\begin{figure*} %
	\centering
	\subfigure[Reference]{ \label{subfig:poro_ref}
				\includegraphics[scale=\nScale]{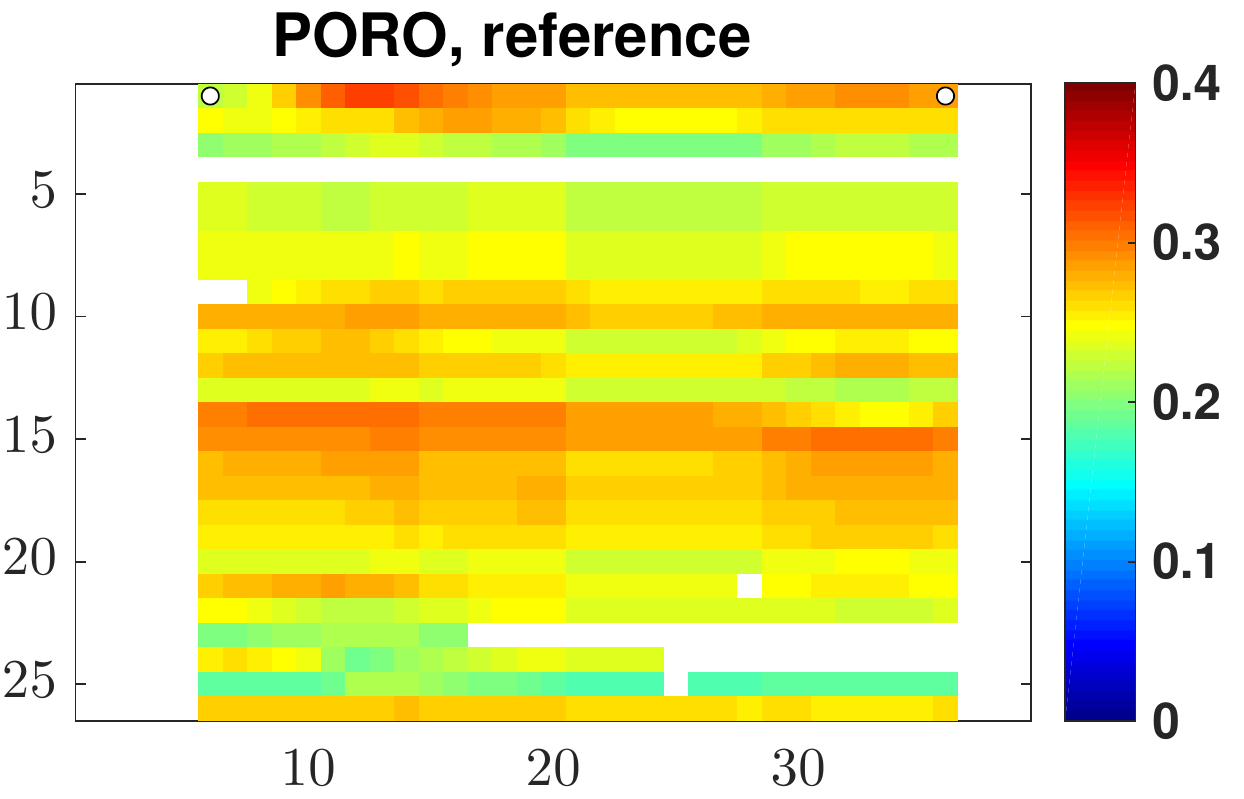}
			}
			
	\subfigure[Initial mean]{ \label{subfig:field_PORO_mean_init_ensemble}
					\includegraphics[scale=0.28]{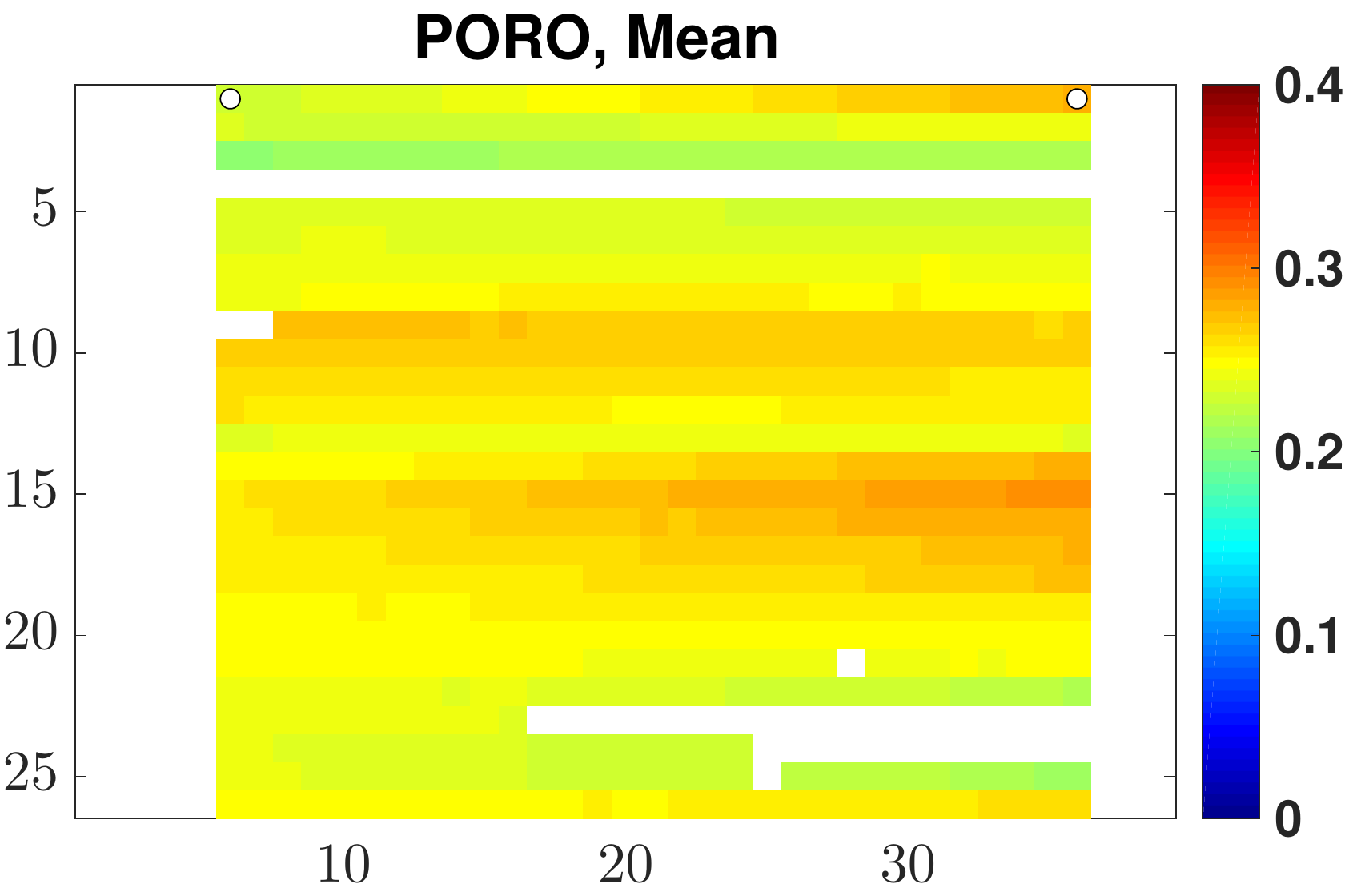}
				}			
	\subfigure[Initial member 1]{ \label{subfig:field_PORO_1_1_init_ensemble}
					\includegraphics[scale=\nScale]{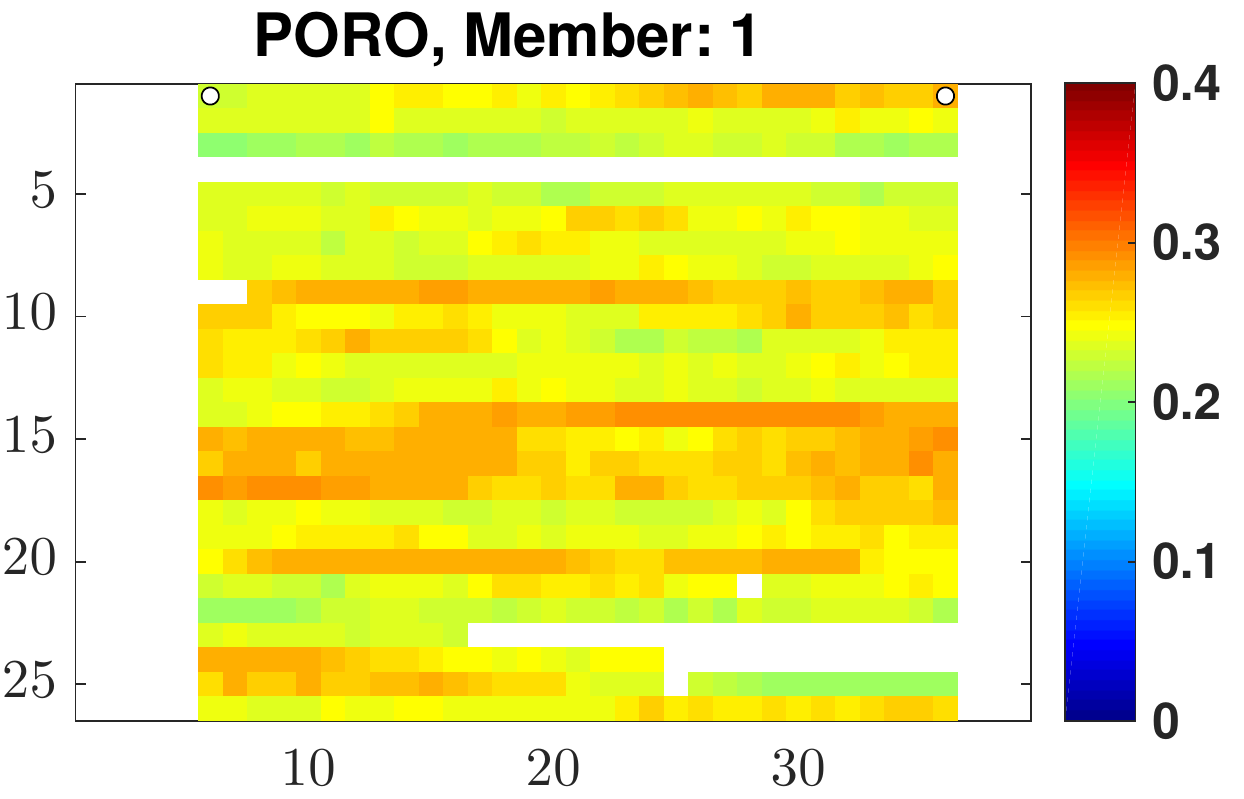}
				}
	\subfigure[Initial member 2]{ \label{subfig:field_PORO_1_2_init_ensemble}
					\includegraphics[scale=\nScale]{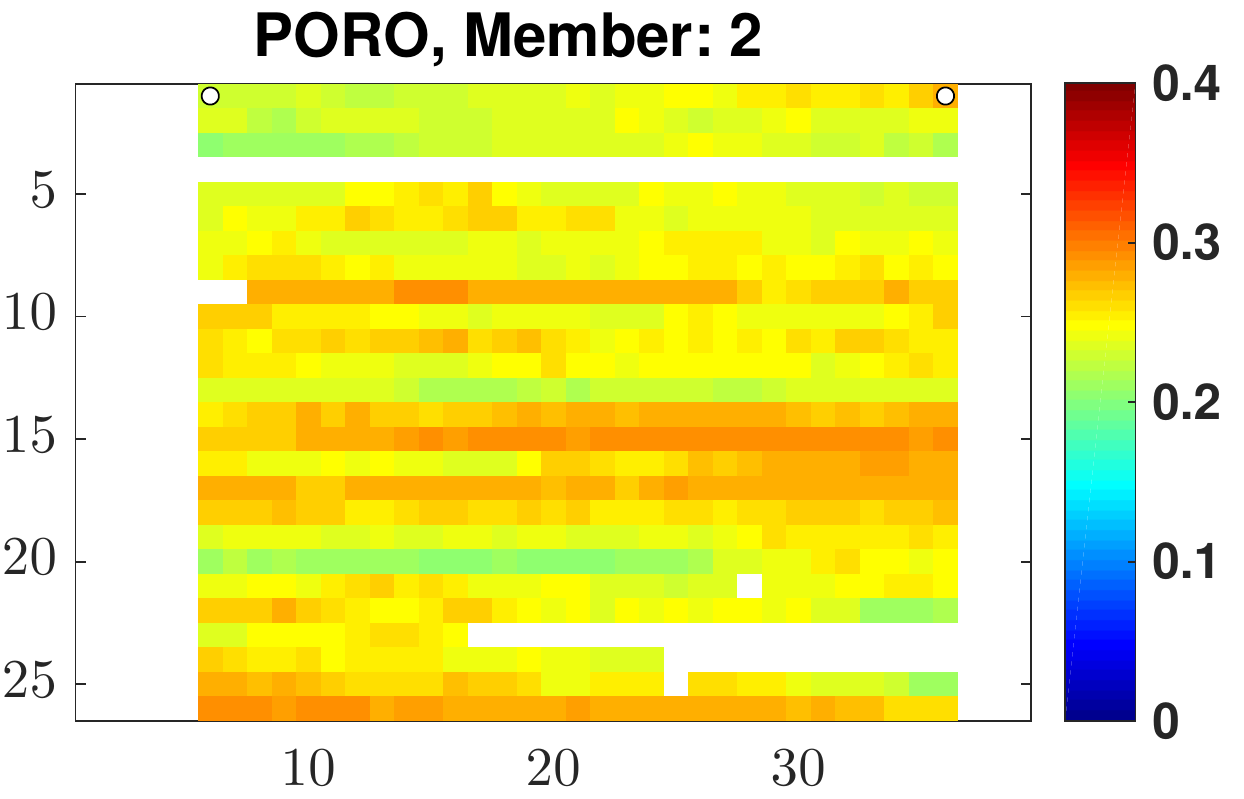}
				}		
				
	\subfigure[Final mean]{ \label{subfig:field_PORO_mean_ensemble3}
					\includegraphics[scale=0.28]{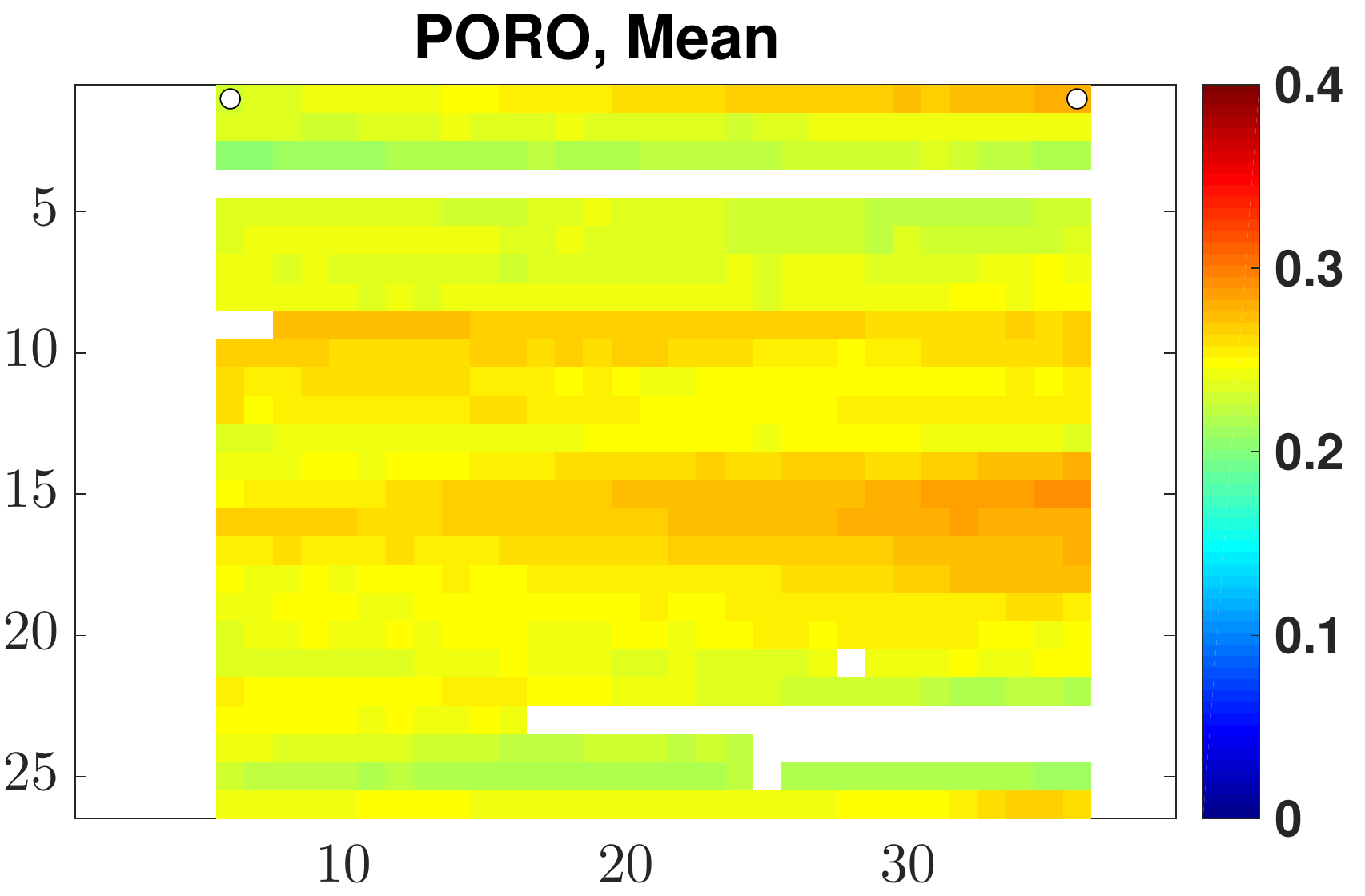}
				}			
	\subfigure[Final member 1]{ \label{subfig:field_PORO_1_1_ensemble3}
					\includegraphics[scale=\nScale]{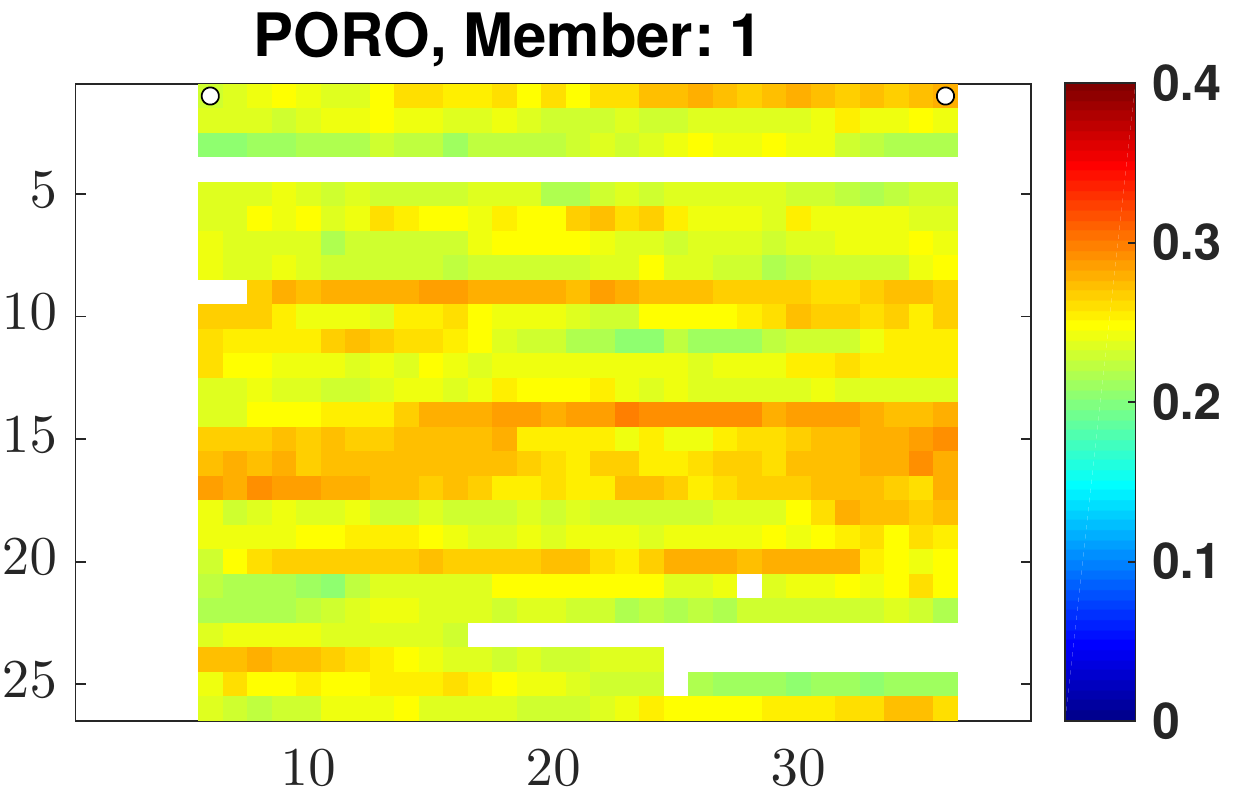}
				}
	\subfigure[Final member 2]{ \label{subfig:field_PORO_1_2_ensemble3}
					\includegraphics[scale=\nScale]{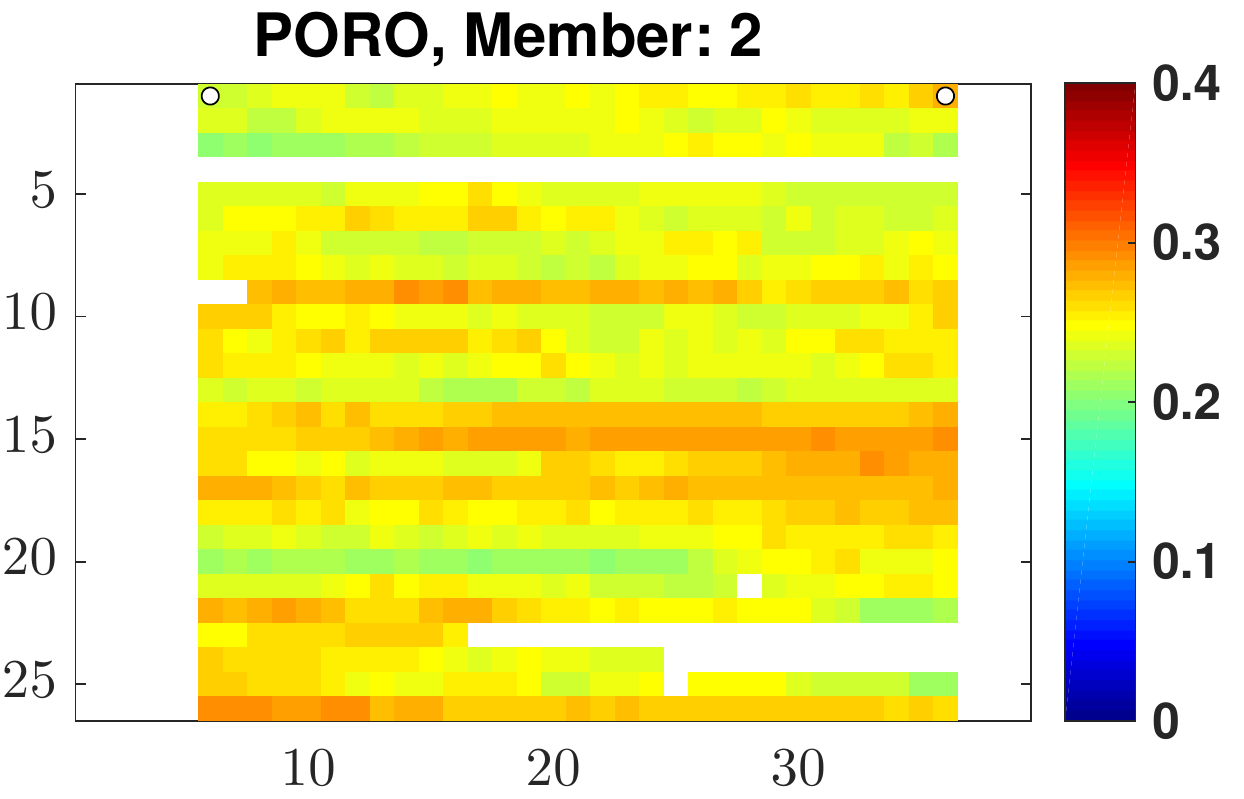}
				}															
	\caption{\label{fig:Norne2D_PORO_S1} As in Figure \ref{fig:Norne2D_PERMX_S1}, but now for the distributions of PORO in scenario S1.}
\end{figure*}   

\subsection{Results of scenario S2 (using 4D seismic data only)} 

\begin{figure*} %
	\centering	
			
	\subfigure[Intercept (day 0)]{ \label{subfig:R0_obs_seisTimeStep1_S2}
					\includegraphics[scale=0.3]{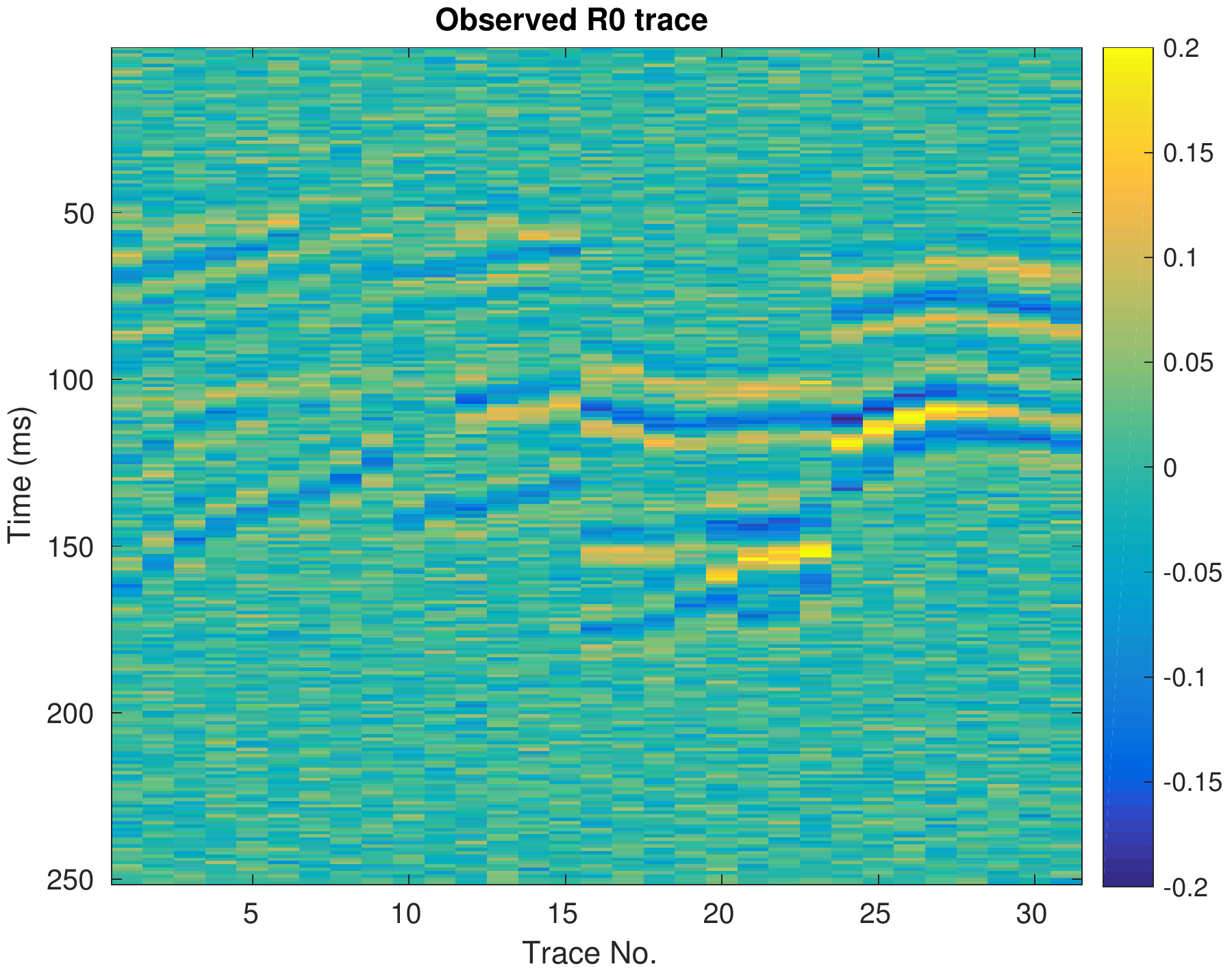}
				}%
	\subfigure[Intercept (day 2040)]{ \label{subfig:R0_obs_seisTimeStep2_S2}
					\includegraphics[scale=0.3]{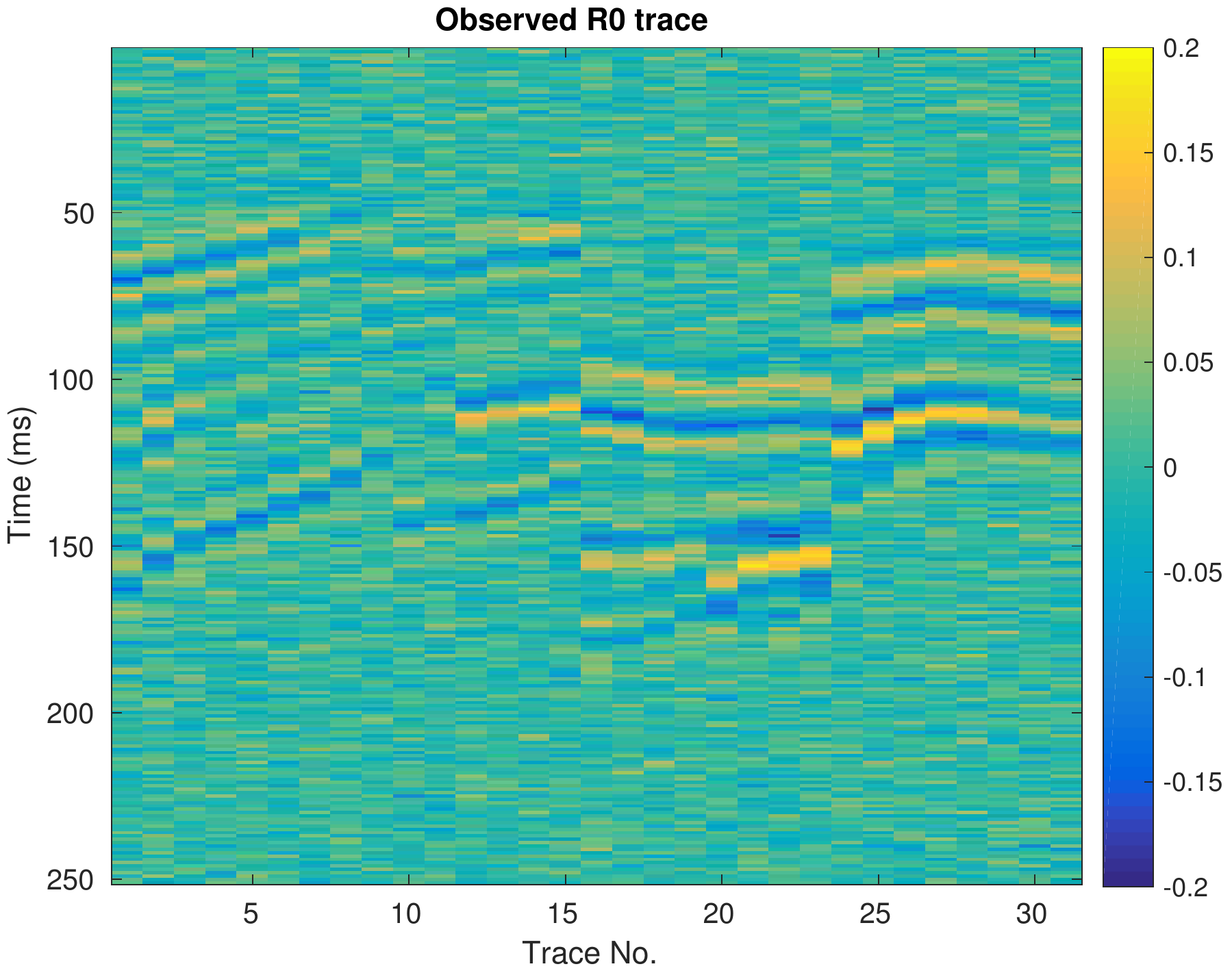}
				}%
	\subfigure[Intercept (day 3750)]{ \label{subfig:R0_obs_seisTimeStep3_S2}
					\includegraphics[scale=0.3]{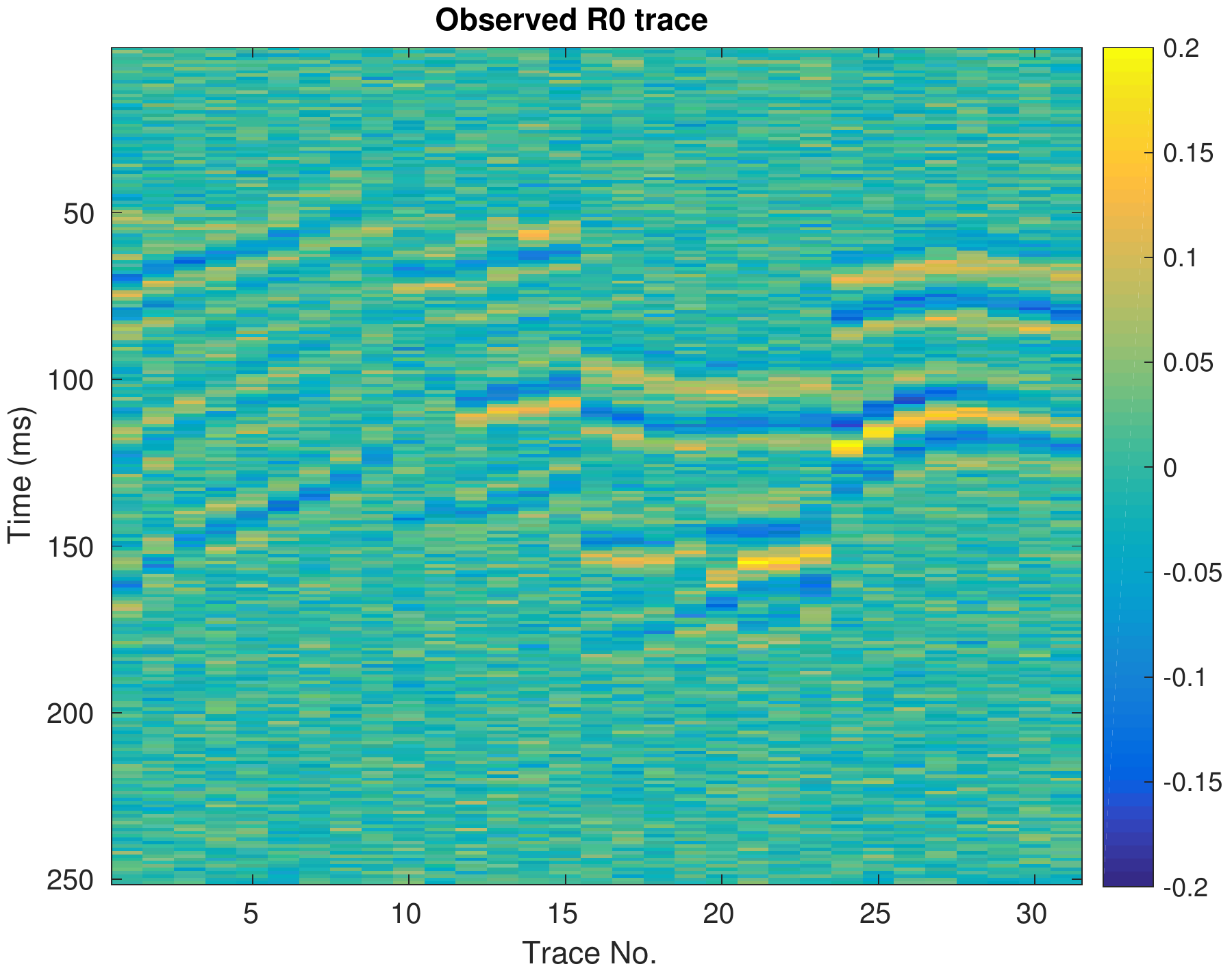}
				}		
				
	\subfigure[Gradient (day 0)]{ \label{subfig:G_obs_seisTimeStep1_S2}
					\includegraphics[scale=0.3]{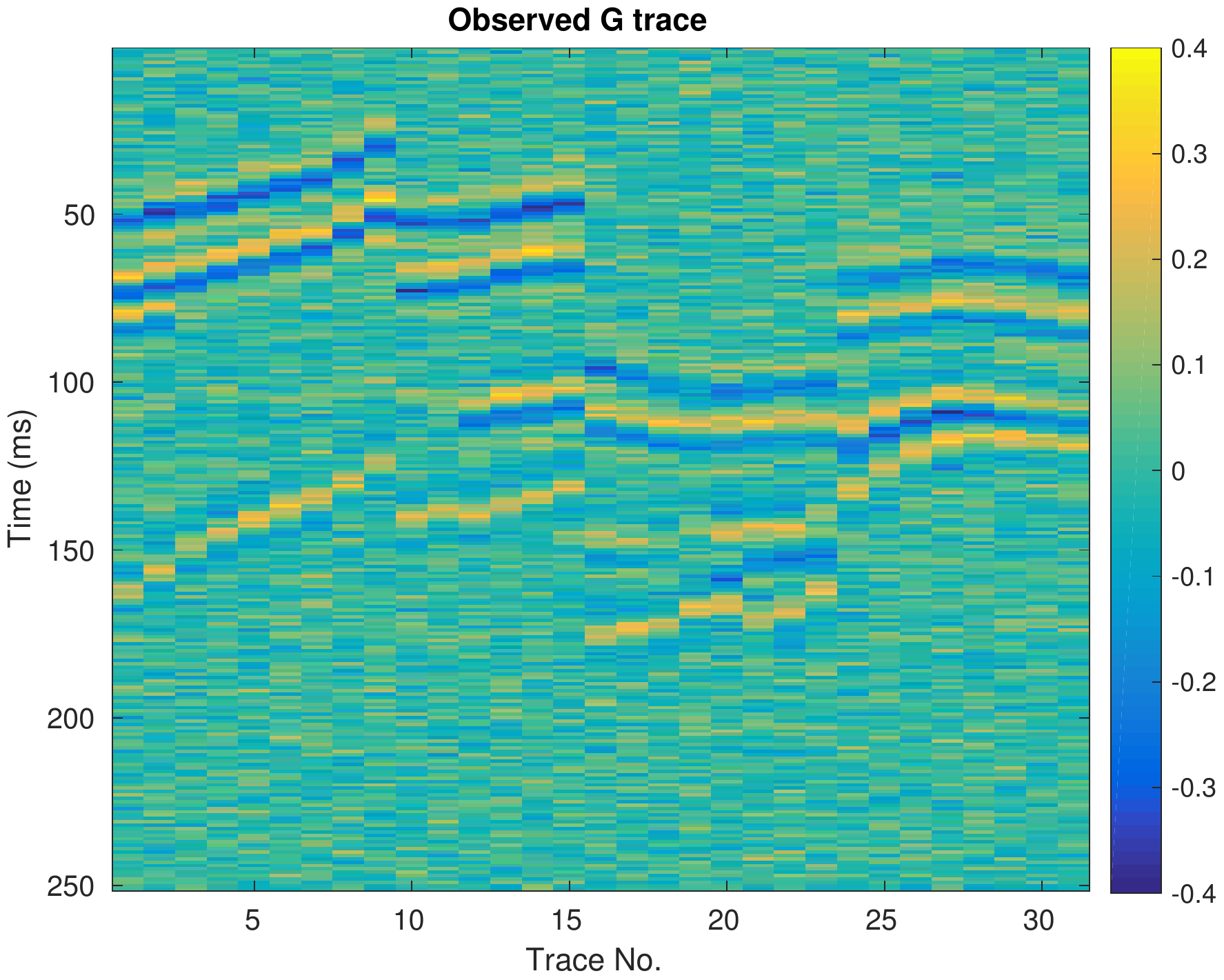}
				}%
	\subfigure[Gradient (day 2040)]{ \label{subfig:G_obs_seisTimeStep2_S2}
					\includegraphics[scale=0.3]{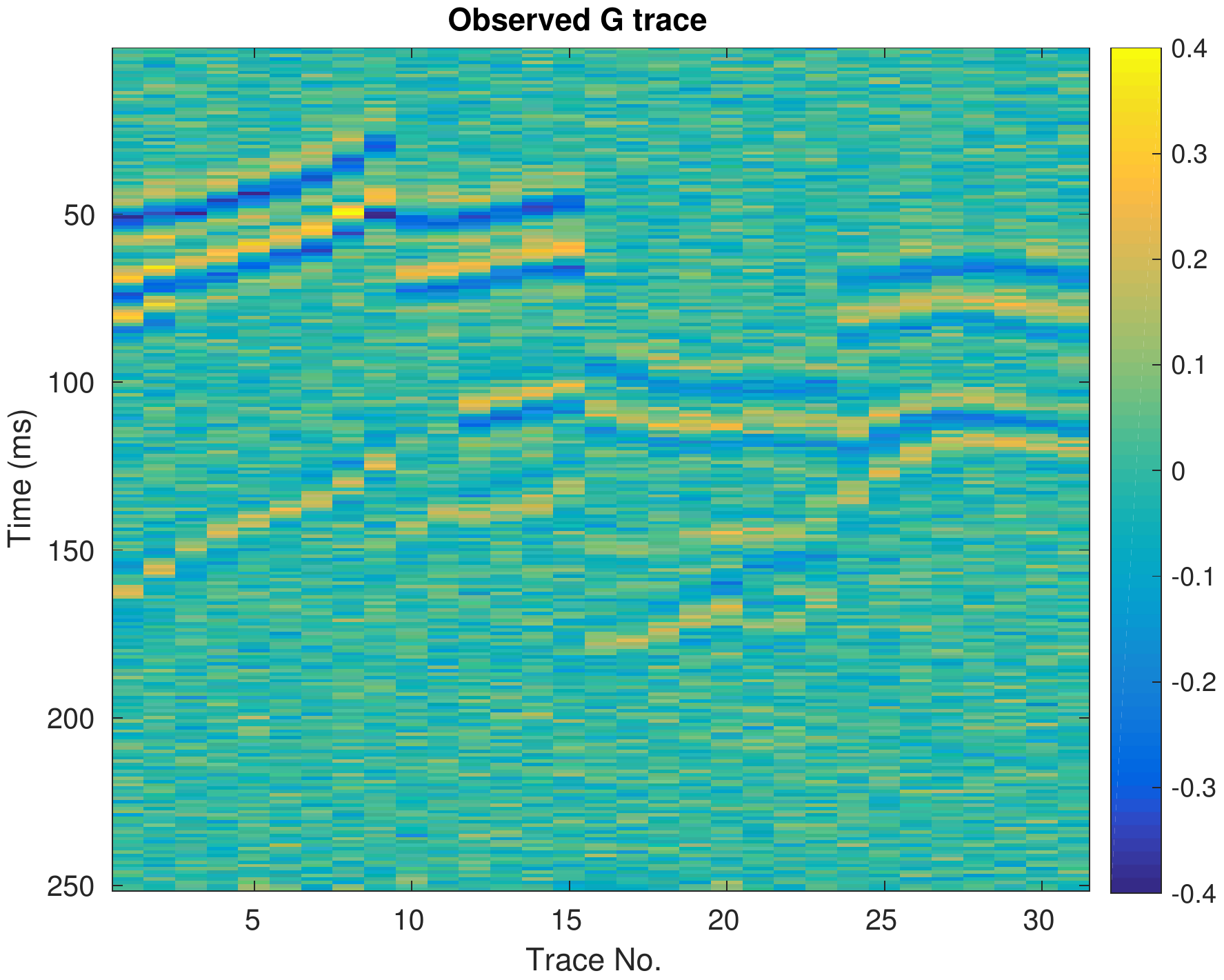}
				}%
	\subfigure[Gradient (day 3750)]{ \label{subfig:G_obs_seisTimeStep3_S2}
					\includegraphics[scale=0.3]{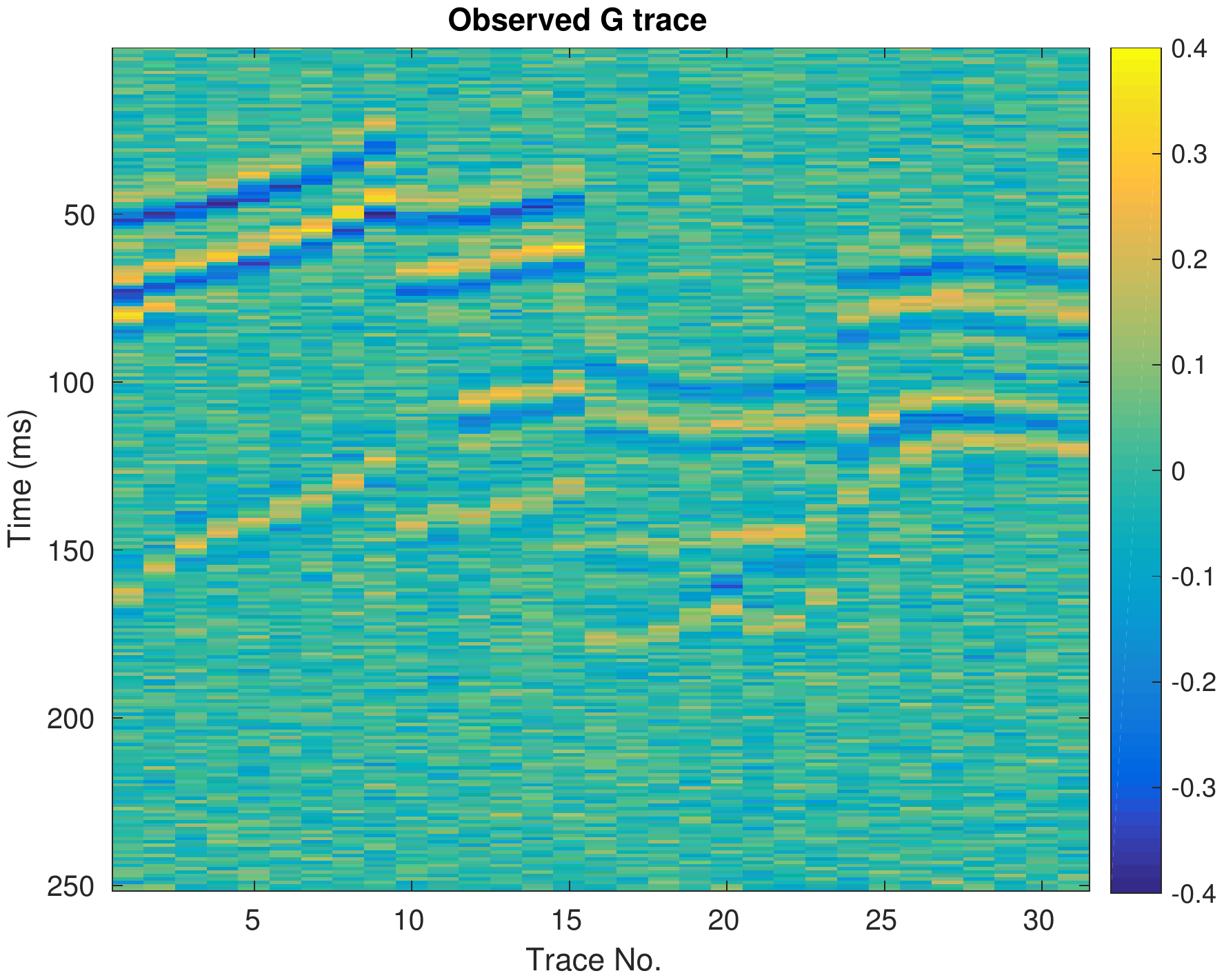}
				}
					
	\caption{\label{fig:Norne2D_attributes_S2} Noisy AVA attributes (intercept and gradient) at three survey time instances. These attributes are used as the observations of full-data experiment.}
\end{figure*}

\begin{figure*} %
	\centering
				
	\subfigure[Intercept (day 0)]{ \label{subfig:denoised_R0_trace_seisTimeStep1_S2}
					\includegraphics[scale=0.3]{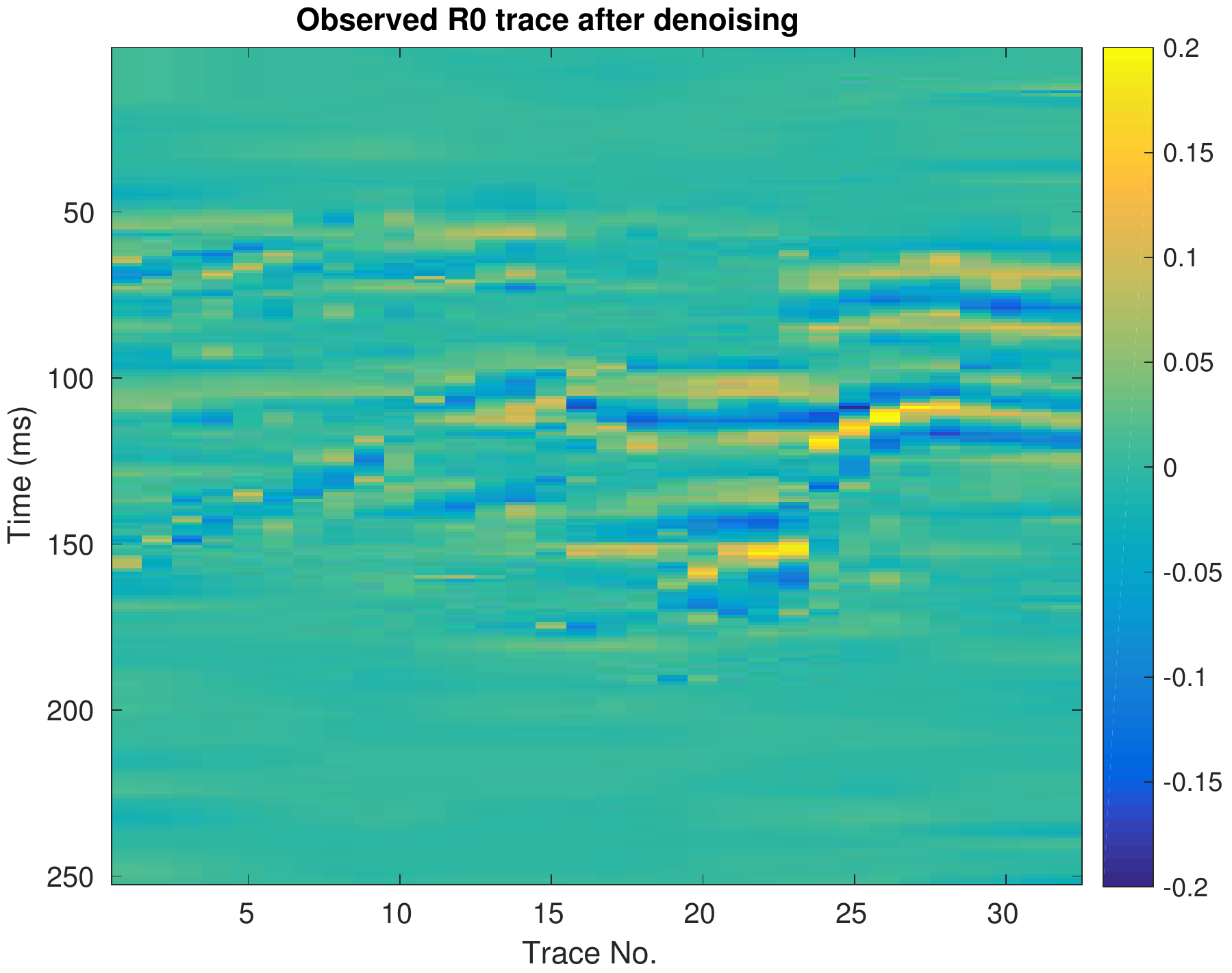}
				}
	\subfigure[Intercept (day 2040)]{ \label{subfig:denoised_R0_trace_seisTimeStep2_S2}
					\includegraphics[scale=0.3]{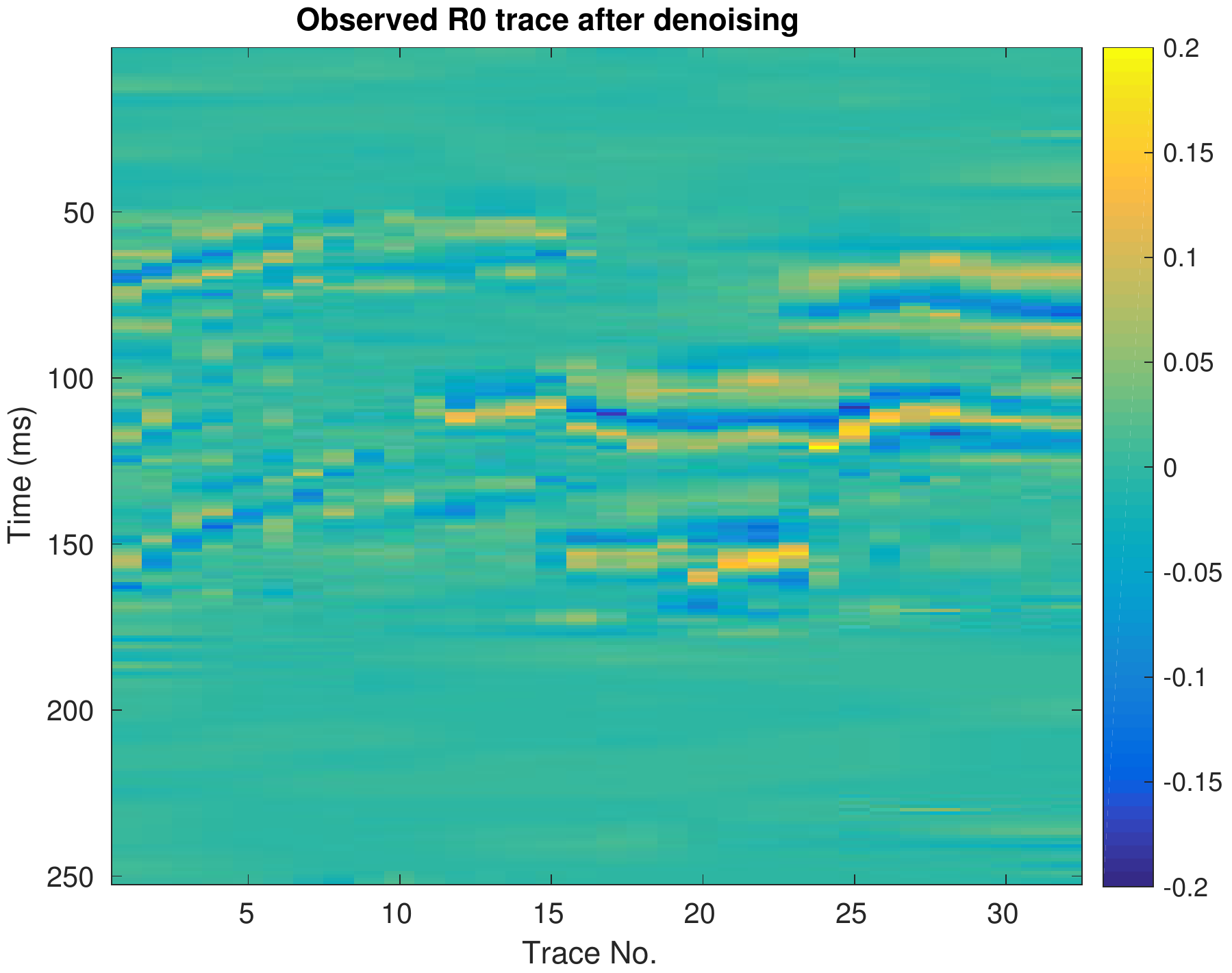}
				}
	\subfigure[Intercept (day 3750)]{ \label{subfig:denoised_R0_trace_seisTimeStep3_S2}
					\includegraphics[scale=0.3]{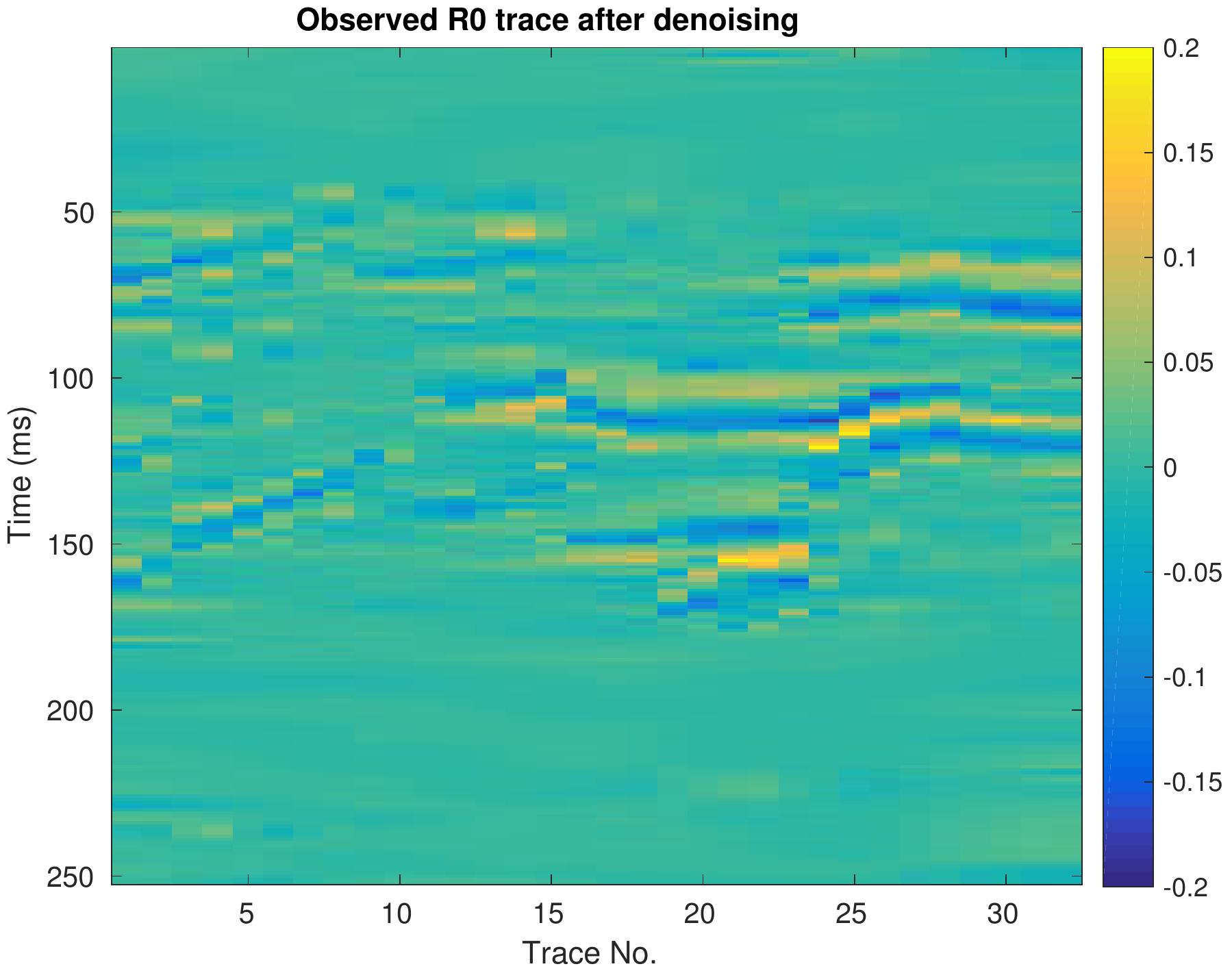}
				}		
				
	\subfigure[Gradient (day 0)]{ \label{subfig:denoised_G_trace_seisTimeStep1_S2}
					\includegraphics[scale=0.3]{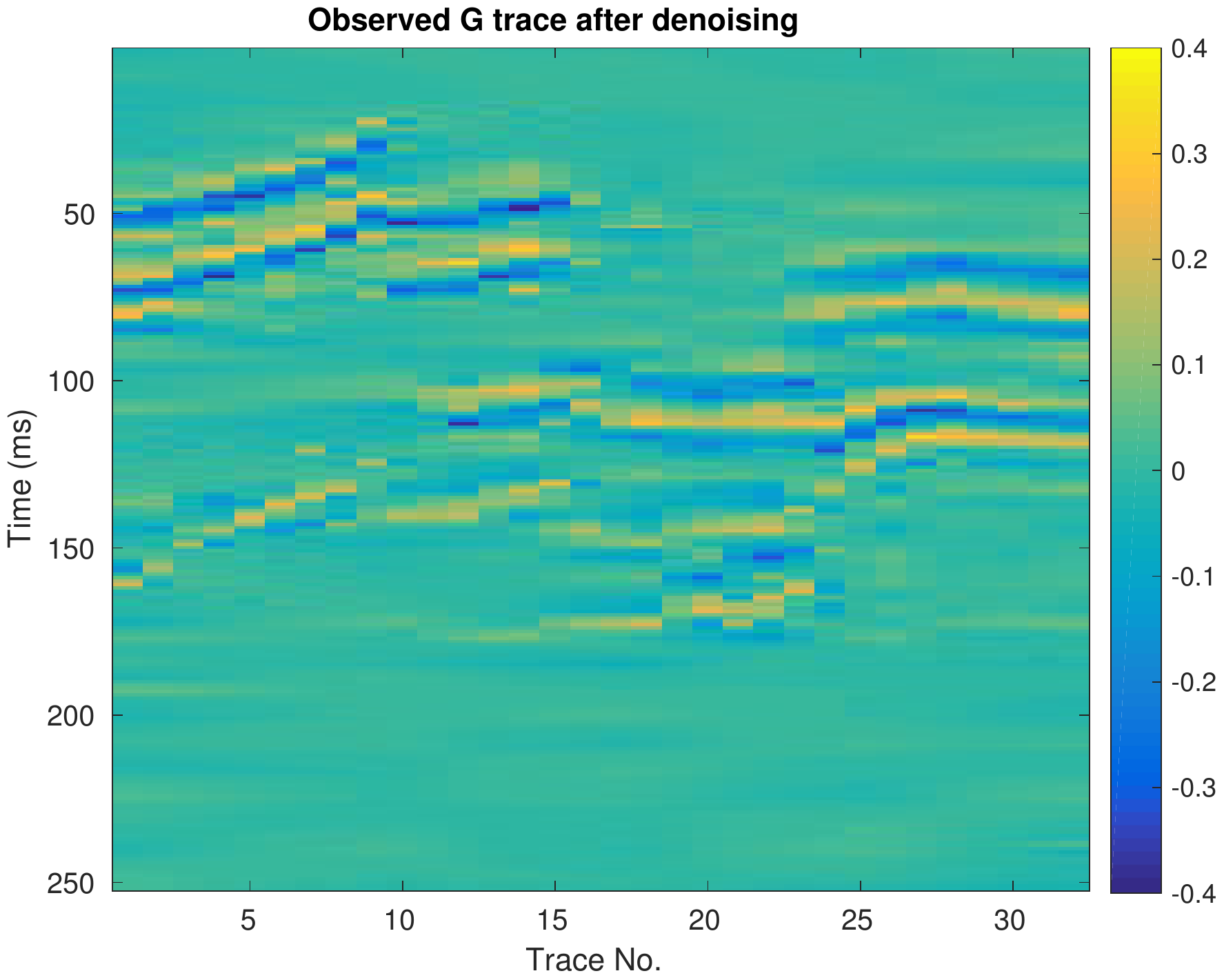}
				}
	\subfigure[Gradient (day 2040)]{ \label{subfig:denoised_G_trace_seisTimeStep2_S2}
					\includegraphics[scale=0.3]{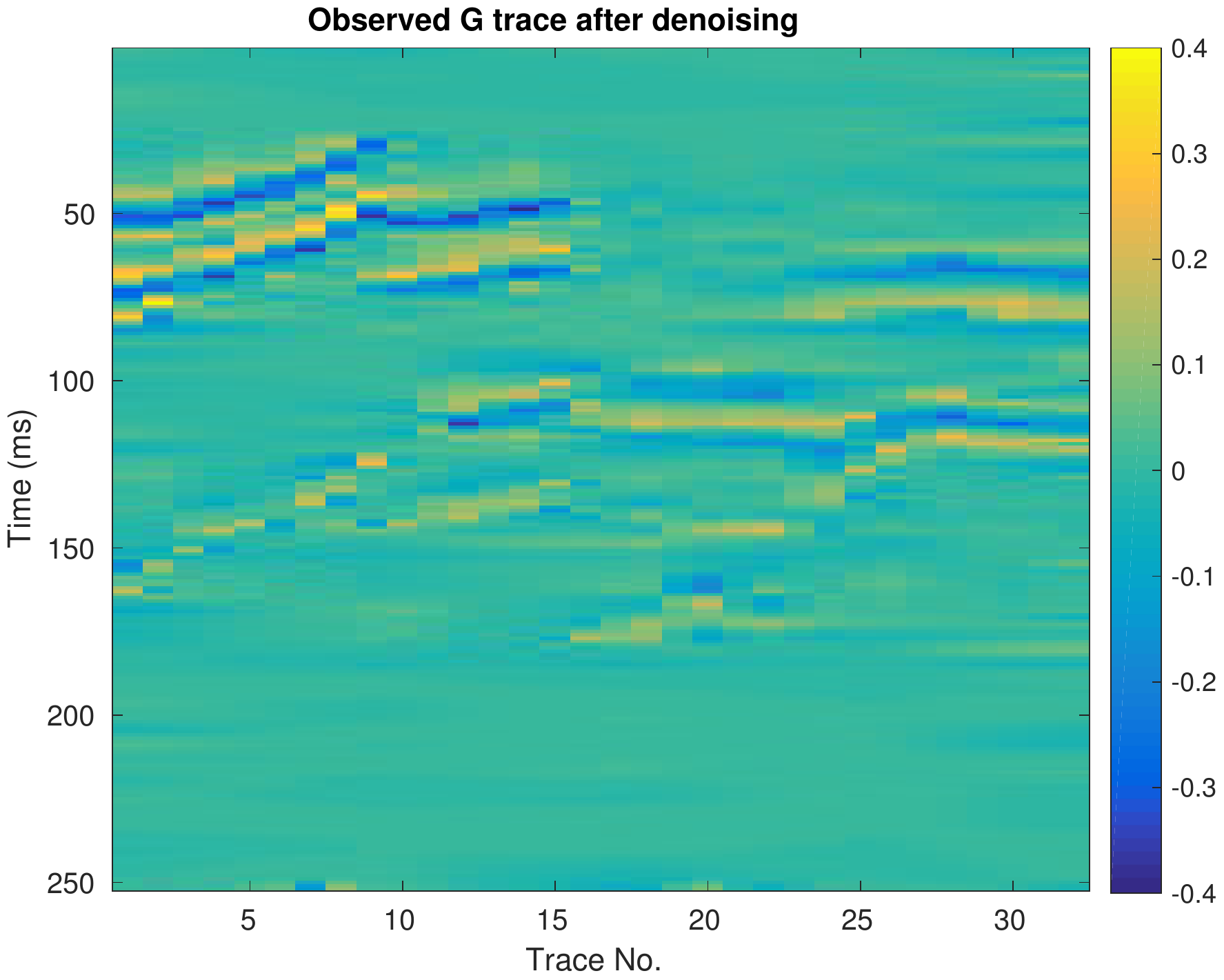}
				}
	\subfigure[Gradient (day 3750)]{ \label{subfig:denoised_G_trace_seisTimeStep3_S2}
					\includegraphics[scale=0.3]{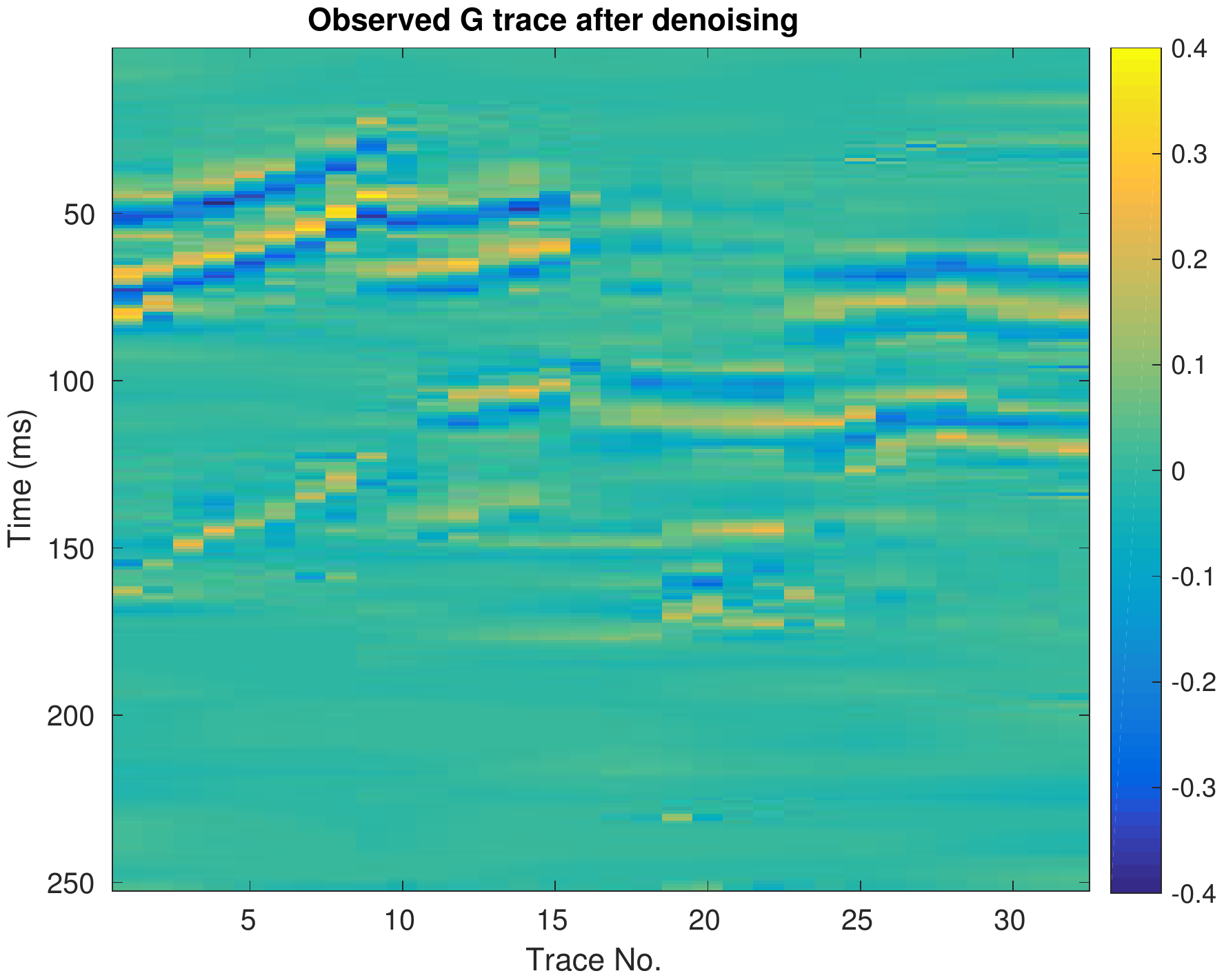}
				}					
	\caption{\label{fig:Norne2D_denoised_attributes_S2} As in Figure \ref{fig:Norne2D_attributes_S2}, but attributes are reconstructed from inverse DWT using leading wavelet coefficients after thresholding. Note that it is leading wavelet coefficients that are used as the observations of sparse-data experiment.}
\end{figure*}

Here we consider the case in which only seismic data are used in history matching. As aforementioned, we consider both full- and sparse-data experiments for the purpose of comparison. In the full-data experiment, the observations are noisy AVA intercepts (denoted by $R_0$) and gradients (denoted by $G$) at different survey time (see Figure \ref{fig:Norne2D_attributes_S2}), whereas in the sparse-data experiment, the observations are leading wavelet coefficients of noisy AVA attributes after thresholding. The data sizes of AVA attributes and the corresponding leading wavelet coefficients are $46686$ and $2746$, respectively. Figure \ref{fig:Norne2D_denoised_attributes_S2} indicates AVA attributes reconstructed from inverse DWT. In these inverse transforms, we keep leading wavelet coefficients while set the others to zero. Comparing Figures \ref{fig:Norne2D_attributes_S2} and \ref{fig:Norne2D_denoised_attributes_S2}, it is clear that the reconstructed attributes tend to capture the main features in the original data but remove the noise components, and this is the rationale behind wavelet-based denoising algorithms (see, for example, \citealp{jansen2012noise,donoho1995adapting,donoho1994ideal}). 

\renewcommand{\nScale}{0.0005}

\renewcommand{\nScale}{0.3}

\renewcommand{\nScale}{0.4}
\begin{figure*} %
	\centering
				
	\subfigure[Full-data experiment]{ \label{subfig:Norne2D_boxplot_objRealIter_S2_full}
					\includegraphics[scale=\nScale]{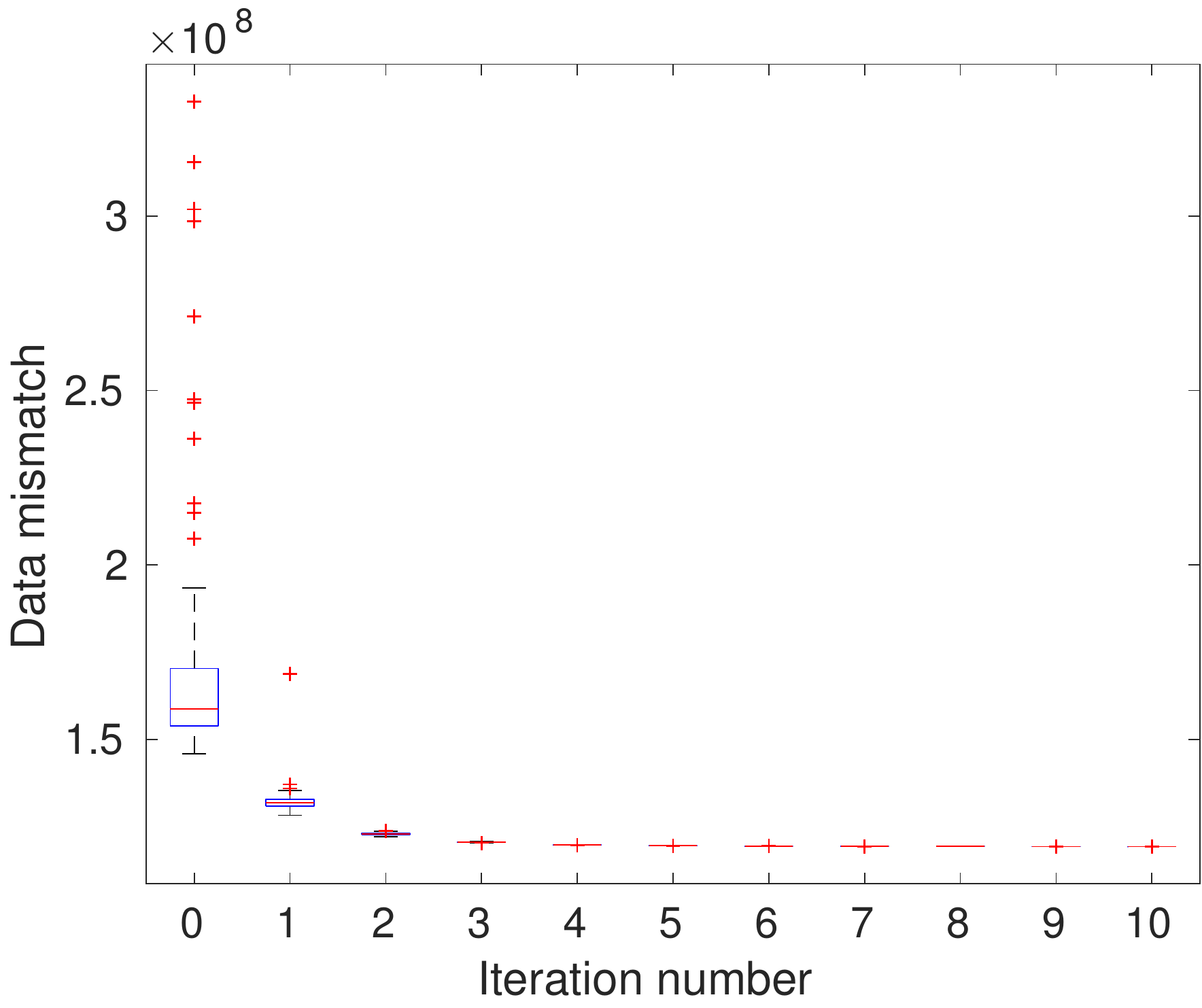}
				} 			
	\subfigure[Sparse-data experiment]{ \label{subfig:Norne2D_boxplot_objRealIter_S2_sparse}
					\includegraphics[scale=\nScale]{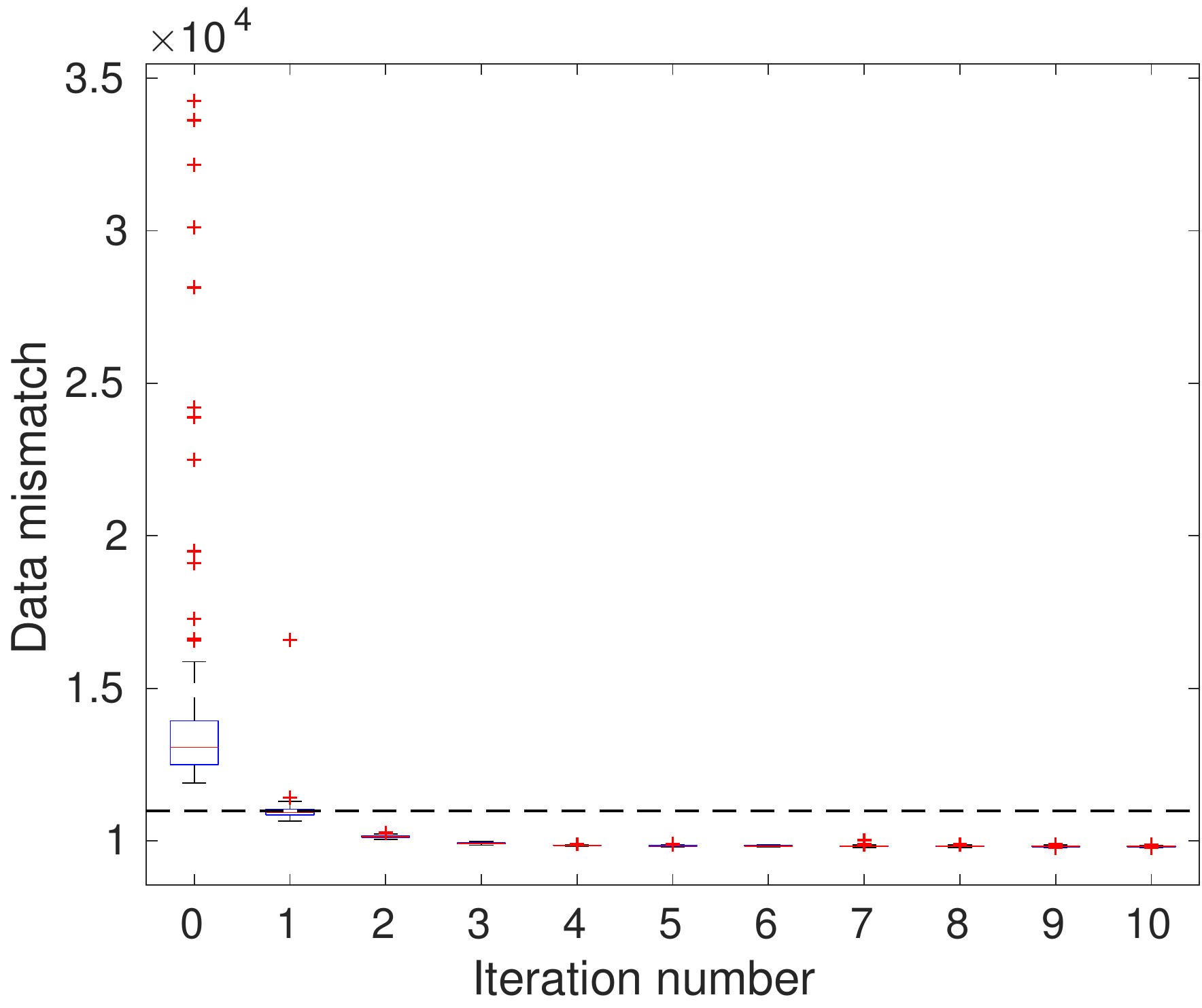}
				}				
	\caption{\label{fig:Norne2D_obj_S2} Boxplots of data mismatch as functions of iteration step (scenario S2). (a) Full-data experiment; and (b) Sparse-data experiment.  The horizontal dashed line in (b) represents the threshold of the stopping criterion (\ref{eq:stopping_criterion_ndm}), whereas data mismatch in (a) is too high to hit at the threshold value.}
\end{figure*} 

\renewcommand{\nScale}{0.4}
\begin{figure*} %
	\centering
				
	\subfigure[RMSEs of log PERMX (full-data)]{ \label{subfig:rmse_PERMX_boxplot_S2_full}
					\includegraphics[scale=\nScale]{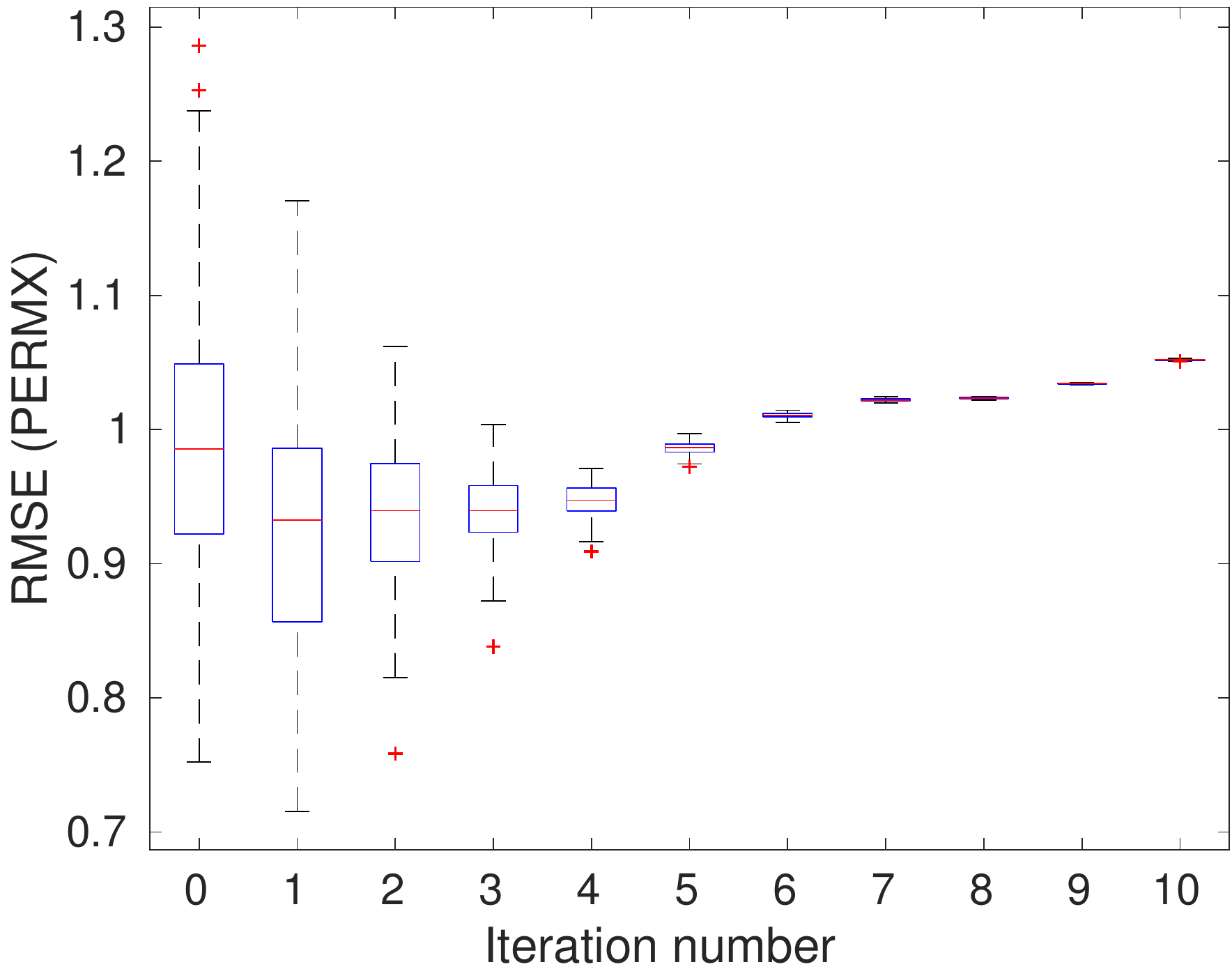}
				} 			
	\subfigure[RMSEs of PORO (full-data)]{ \label{subfig:rmse_PORO_boxplot_S2_full}
					\includegraphics[scale=\nScale]{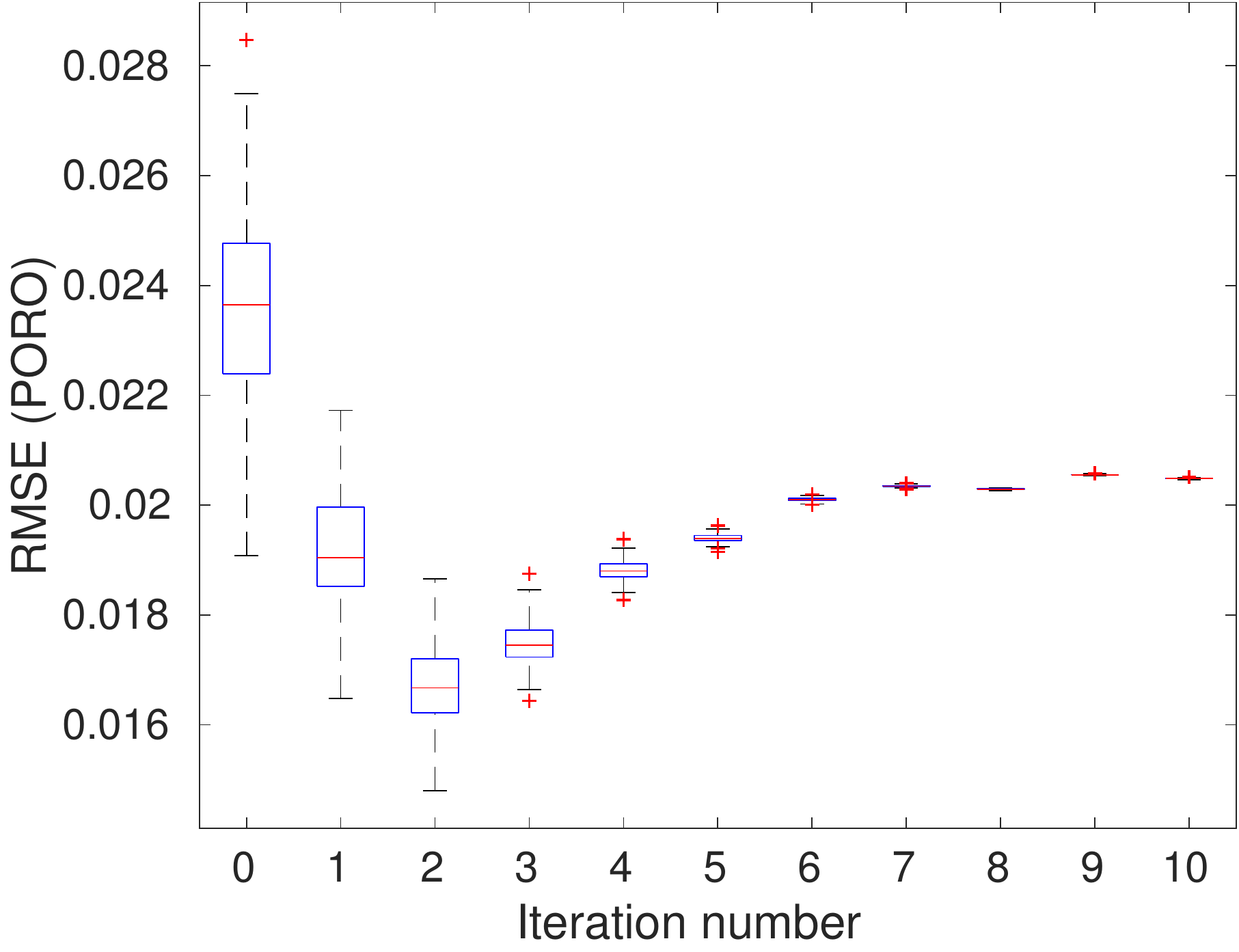}
				}		
				
	\subfigure[RMSEs of log PERMX (sparse-data)]{ \label{subfig:rmse_PERMX_boxplot_S2_sparse}
					\includegraphics[scale=\nScale]{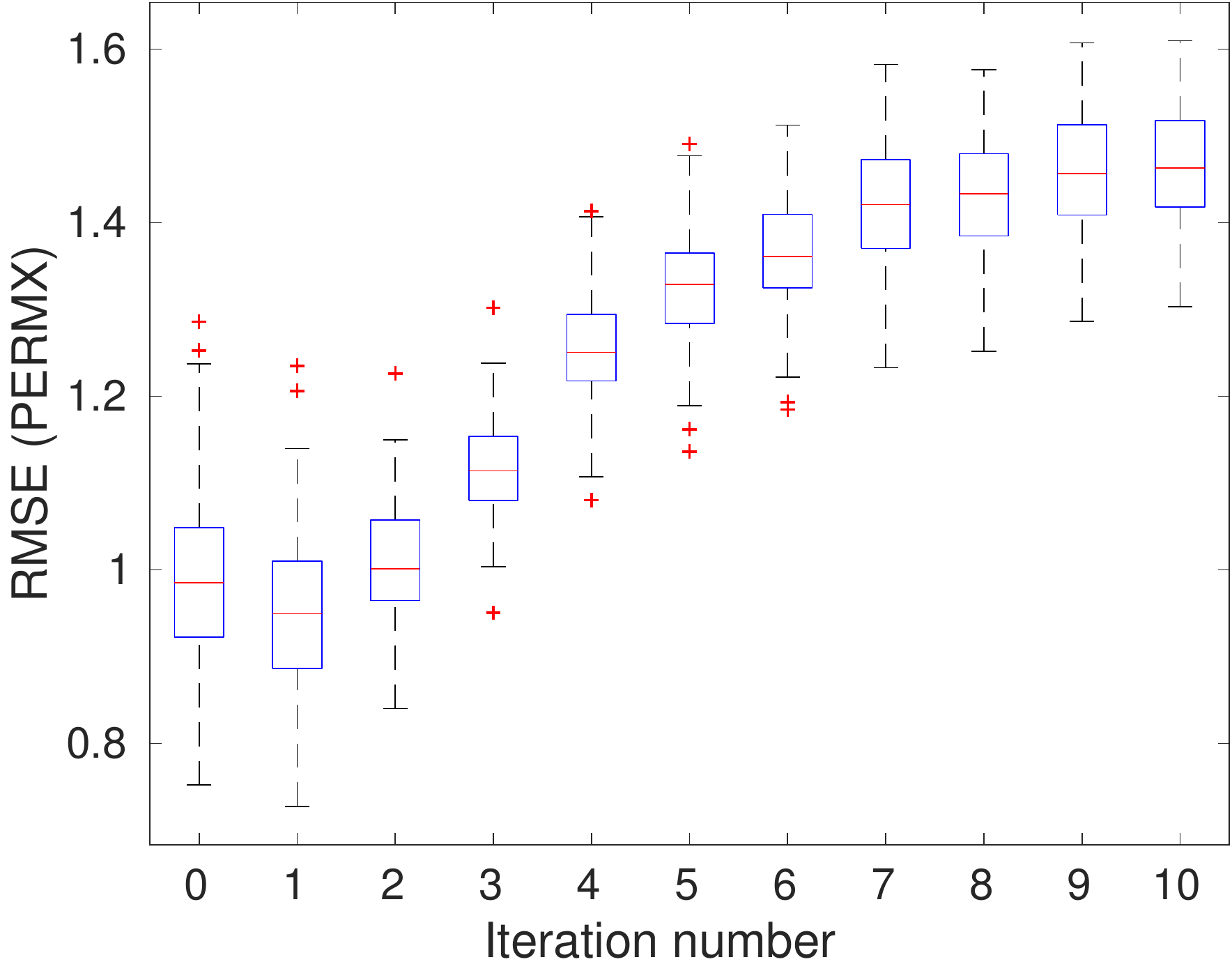}
				} 			
	\subfigure[RMSEs of PORO (sparse-data)]{ \label{subfig:rmse_PORO_boxplot_S2_sparse}
					\includegraphics[scale=\nScale]{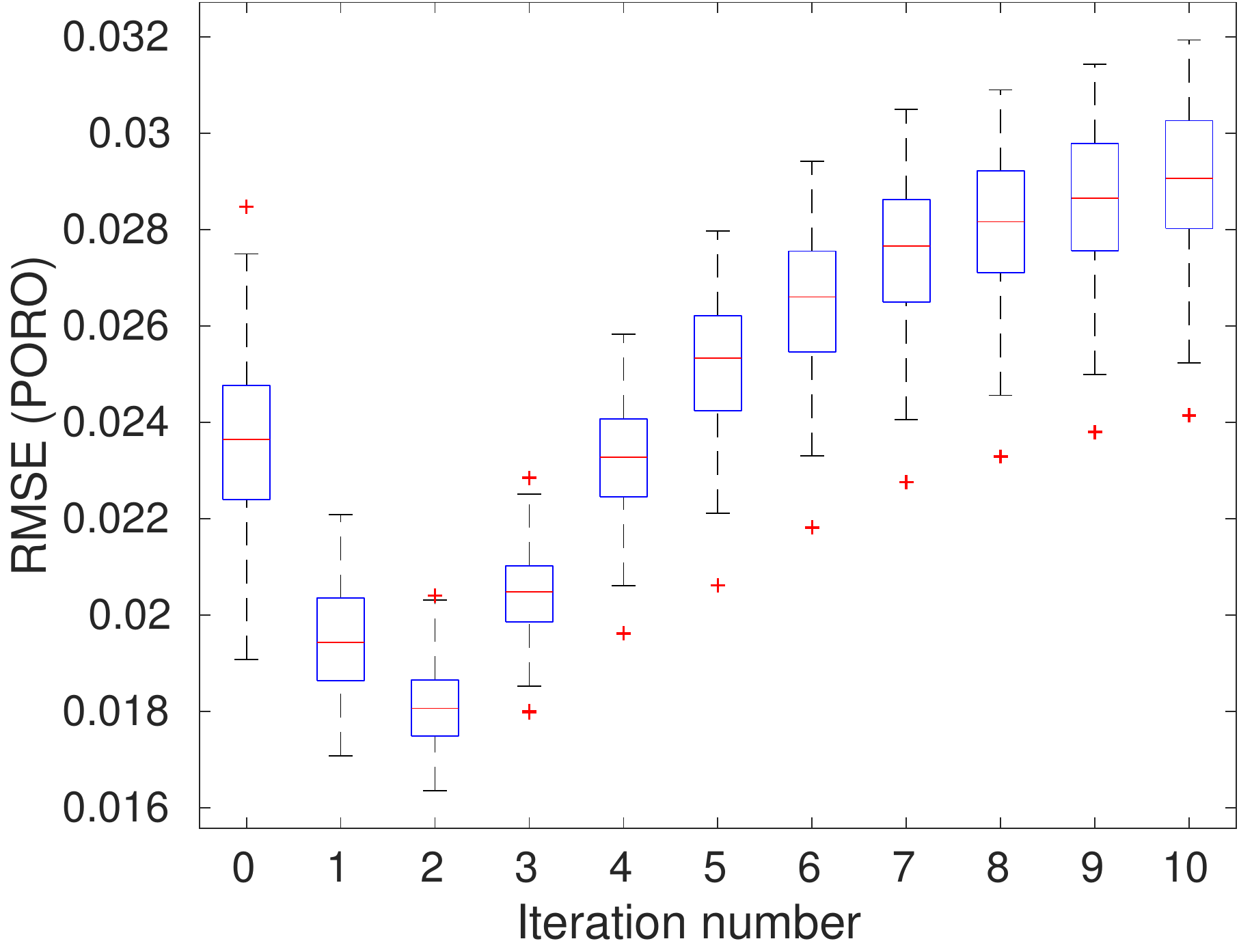}
				}								
	\caption{\label{fig:Norne2D_rmse_S2} Boxplots of RMSEs as functions of iteration step (scenario S2). Sub-figures (a) and (b) in the first row are for the results of full-data experiment, and (c) and (d) in the second row for the results of sparse-data experiment.}
\end{figure*} 

Figure \ref{fig:Norne2D_obj_S2} shows boxplots of data mismatch as functions of iteration step in full- and sparse-data experiments, respectively. In the full-data experiment (Figure \ref{subfig:Norne2D_boxplot_objRealIter_S2_full}), data mismatch is computed in the original data space. The initial average data mismatch is $1.72 \times 10^8$. Using the stopping conditions (C1) and (C2), RLM-MAC stops after 10 iterations and attains the final average data mismatch of $1.19 \times 10^8$. In this case, the final value is above the threshold $4 \times 46686 \approx 1.87 \times 10^5$, such that the extra stopping criterion (\ref{eq:stopping_criterion_ndm}) is not activated. In contrast, in the sparse-data experiment (Figure \ref{subfig:Norne2D_boxplot_objRealIter_S2_sparse}), data mismatch is calculated in the wavelet domain. The initial average data mismatch is $1.46 \times 10^4$. By only adopting the stopping conditions (C1) and (C2), RLM-MAC stops after 10 iterations, with the final average data mismatch being $9.82 \times 10^3$. Meanwhile, by applying the stopping criterion (\ref{eq:stopping_criterion_ndm}), one has the threshold value $4 \times 2746 \approx 1.10 \times 10^4$ (corresponding to the horizontal dashed line in Figure \ref{subfig:Norne2D_boxplot_objRealIter_S2_sparse}), indicating that RLM-MAC should stop at the 2nd iteration step instead.  

In the full-data experiment, the number of attribute data is much larger than the number of reservoir model variables to be estimated (which is $2 \times 739 = 1478$). This means that history matching is an over-determined inverse problem, thus some of the observations might not be matched well. As a result, the initial average data mismatch per observation, which is defined as the ratio of average data mismatch to the number of observations, appears very large (in the order of $10^3$), and RLM-MAC cannot help reduce this ratio very much because of the nature of ``over-determinedness'' in the inverse problem (In contrast, in scenario S1, the number of production data is less than the size of reservoir model, therefore history matching is an under-determined inverse problem, and RLM-MAC substantially reduces data mismatch, despite the high initial average data mismatch per observation, also in the order of $10^3$). Through sparse representation, the size of observations is significantly reduced, although it is still larger than the size of reservoir model. By only comparing leading wavelet coefficients of observed and simulated AVA attributes, the initial average data mismatch per observation becomes much lower (around 5.32), and through iteration, RLM-MAC reduces this ratio further to some values below $4$, such that the stopping criterion (\ref{eq:stopping_criterion_ndm}) can be triggered.   
          
Figure \ref{fig:Norne2D_rmse_S2} shows RMSEs of log PERMX and PORO as functions of iteration step. In all cases, the RMSEs exhibit U-turn behaviour and their values tend to decrease at the first few iteration steps and then arise. According to Figure \ref{fig:Norne2D_obj_S2}, if using the stopping conditions (C1) and (C2) only, one would take the ensembles at the 10th iteration steps as the final results. This leads to almost the worst estimation (except for PORO in Figure \ref{subfig:rmse_PORO_boxplot_S2_full}), in terms of RMSE, among all possible choices of the final ensembles. By adopting the extra stopping criterion (\ref{eq:stopping_criterion_ndm}) in sparse-data experiment, RLM-MAC stops at the 2nd iteration. This leads to the lowest RMSEs of PORO, compared to other choices of final iteration steps (Figure \ref{subfig:rmse_PORO_boxplot_S2_sparse}). On the other hand, the RMSEs of log PERMX at the 2nd iteration step are not the lowest, and even appear slightly higher than those of the initial ensemble (Figure \ref{subfig:rmse_PERMX_boxplot_S2_sparse}). A possible explanation of this result is that, from the perspective of forward simulation, porosity affects AVA attributes more directly than permeability. Indeed, in our forward simulation, apart from porosity, AVA attributes depend on two reservoir dynamical variables, namely, pressure and saturation, whereas permeability affects pressure and saturation through reservoir simulation. Therefore the relation between AVA attributes and permeability is more complicated, and perhaps also more nonlinear, than that between AVA attributes and porosity. This extra complexity thus makes it more difficult to invert permeability than porosity in the current scenario. 

A comparison of the results of full- and sparse-data experiments (Figure \ref{subfig:rmse_PERMX_boxplot_S2_full} versus Figure \ref{subfig:rmse_PERMX_boxplot_S2_sparse}, or Figure \ref{subfig:rmse_PORO_boxplot_S2_full} versus Figure \ref{subfig:rmse_PORO_boxplot_S2_sparse}) reveals pros and cons of using full seismic data, when the stopping criterion (\ref{eq:stopping_criterion_ndm}) is not adopted. On one hand, using full seismic data results in a high degree of ``over-determinedness'' in full-data experiment. This provides more constraints in history matching, and seems to prevent the estimation results from being too poor (as a contrast, one may see the last few steps of Figures \ref{subfig:rmse_PERMX_boxplot_S2_sparse} and \ref{subfig:rmse_PORO_boxplot_S2_sparse}). On the other hand, Figures \ref{subfig:rmse_PERMX_boxplot_S2_full} and \ref{subfig:rmse_PORO_boxplot_S2_full} indicate signs of ensemble collapses at the last few iteration steps. These, however, appear avoided in Figures \ref{subfig:rmse_PERMX_boxplot_S2_sparse} and \ref{subfig:rmse_PORO_boxplot_S2_sparse}.       

\renewcommand{\nScale}{0.3}
\begin{figure*} %
	\centering
	
	\subfigure[Initial ensemble, base survey]{ \label{subfig:diff_initEnsMean_R0_traces_seisTimeStep1_S2}
					\includegraphics[scale=0.3]{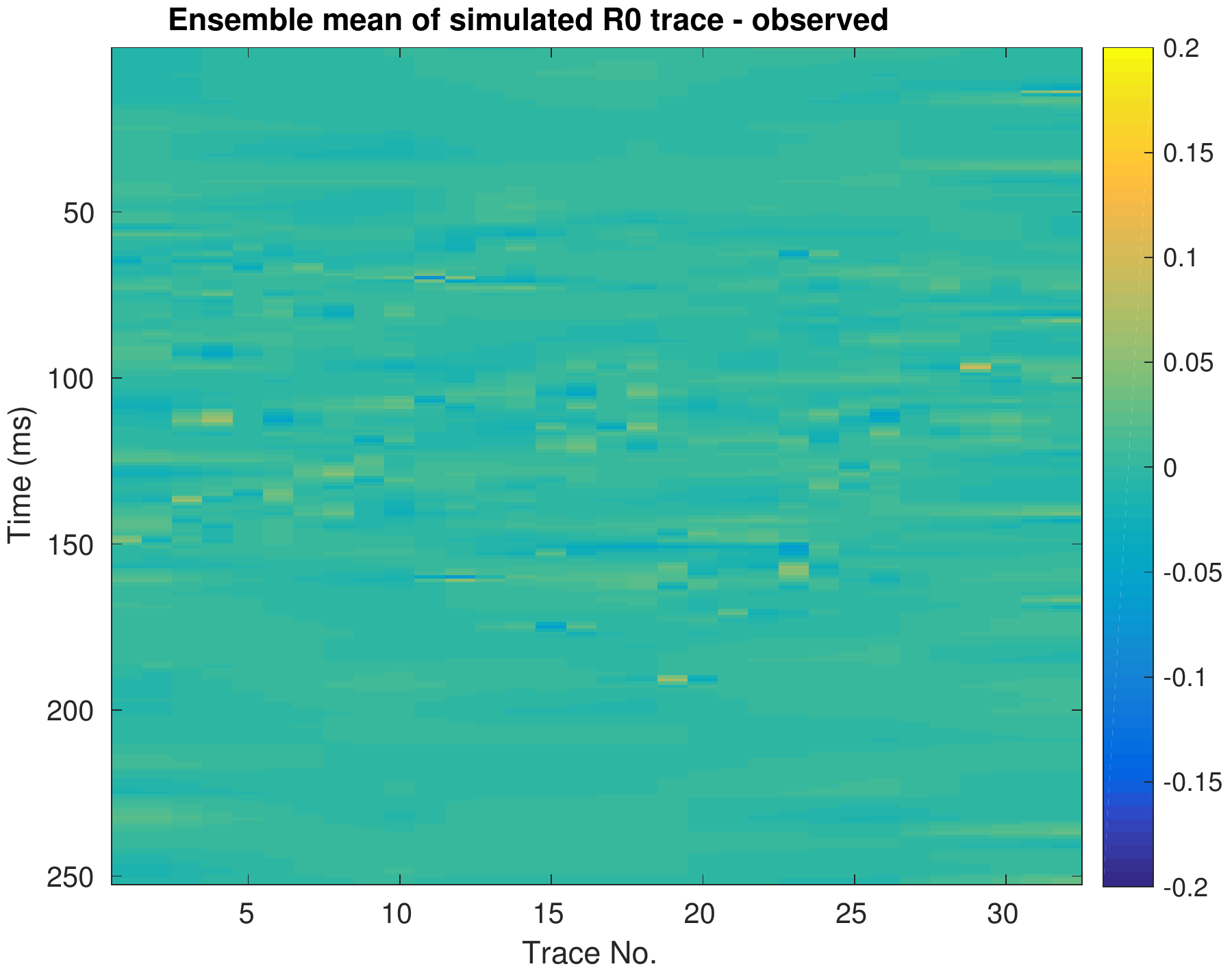}
				}%
	\subfigure[Initial ensemble, 1st monitor survey]{ \label{subfig:diff_initEnsMean_R0_traces_seisTimeStep2_S2}
					\includegraphics[scale=0.3]{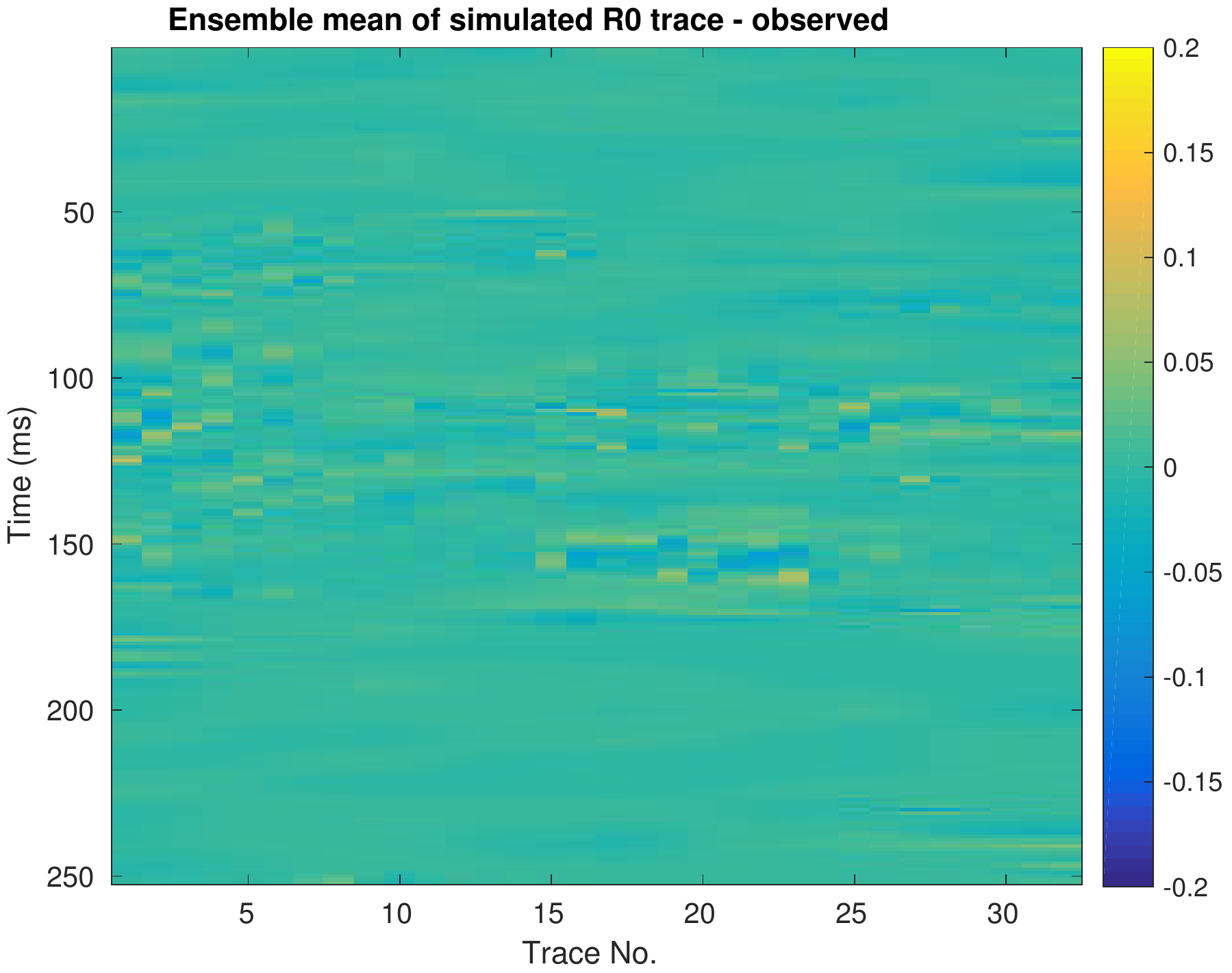}
				}%
	\subfigure[Initial ensemble, 2nd monitor survey]{ \label{subfig:diff_initEnsMean_R0_traces_seisTimeStep3_S2}
						\includegraphics[scale=0.3]{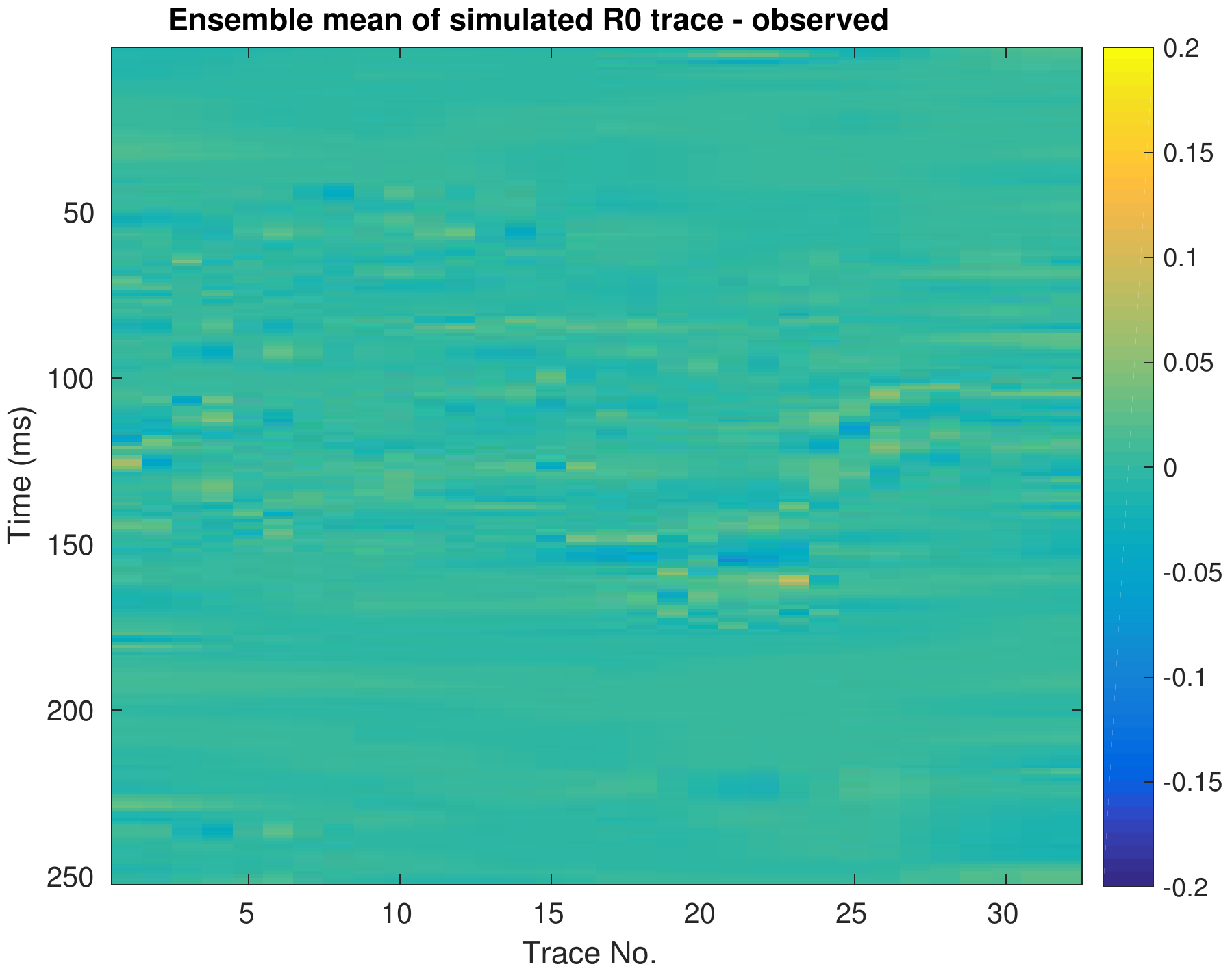}
					} 		
				
	\subfigure[Final ensemble, base survey]{ \label{subfig:diff_finalEnsMean_R0_traces_seisTimeStep1_S2}
					\includegraphics[scale=0.3]{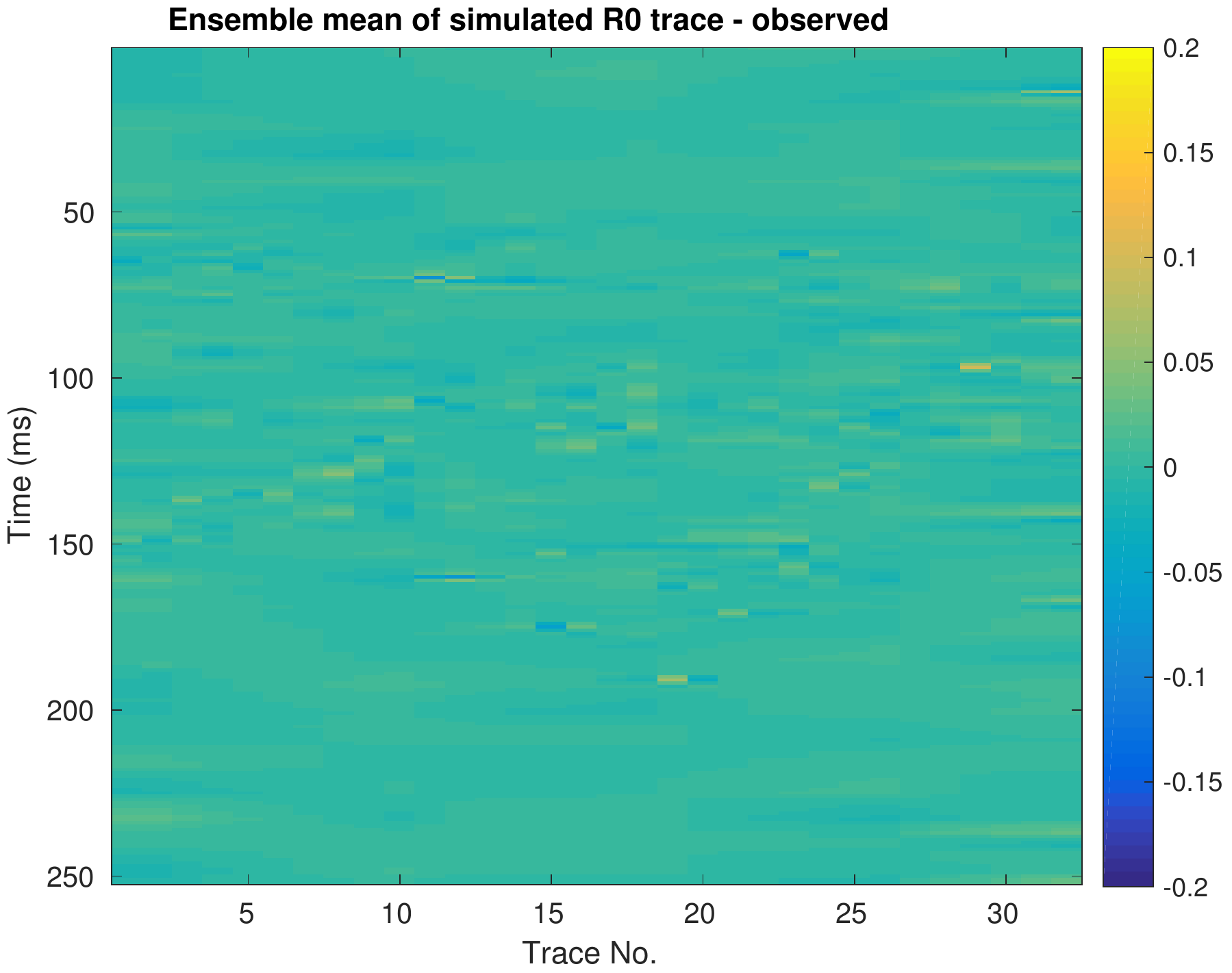}
				}%
	\subfigure[Final ensemble, 1st monitor survey]{ \label{subfig:diff_finalEnsMean_R0_traces_seisTimeStep2_S2}
					\includegraphics[scale=0.3]{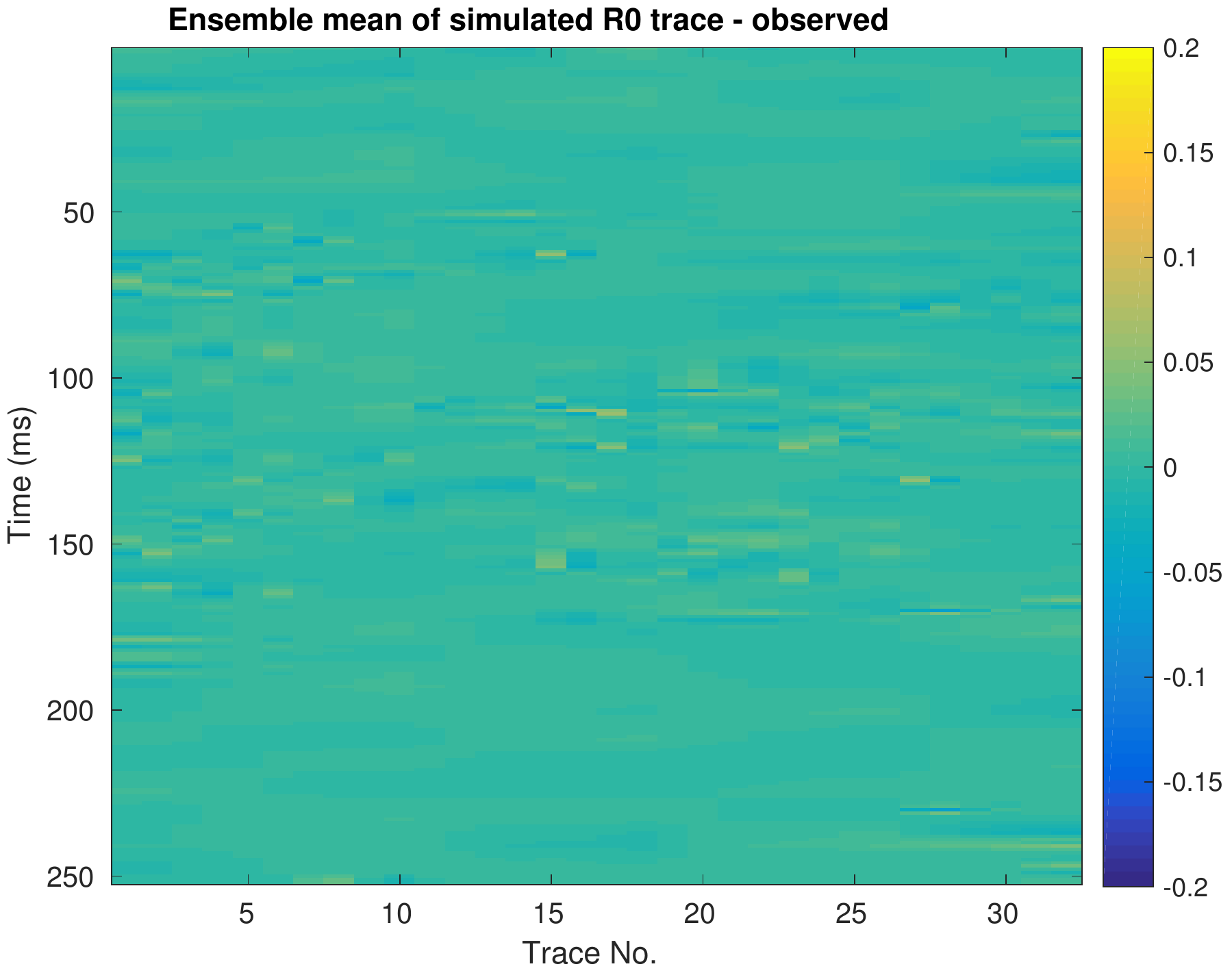}
				}%
	\subfigure[Final ensemble, 2nd monitor survey]{ \label{subfig:diff_finalEnsMean_R0_traces_seisTimeStep3_S2}
						\includegraphics[scale=0.3]{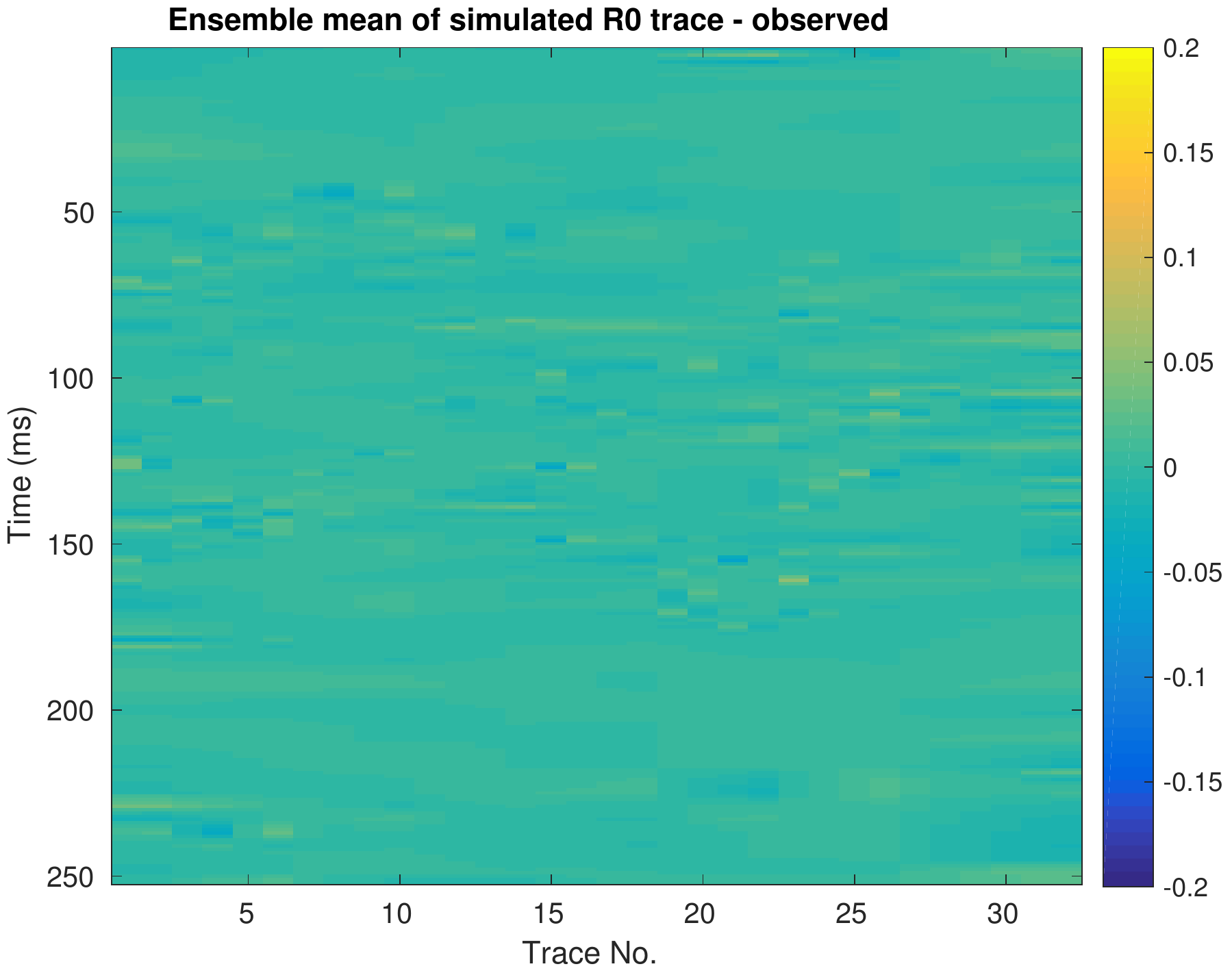}
					}						
	\caption{\label{fig:Norne2D_diff_R0_S2} Differences between ensemble means of reconstructed simulated intercepts and reconstructed observed intercepts at three survey time instances in scenario S2. Sub-figures in the first row (a--c) are for the results with respect to the initial ensemble, whereas those in the second row (d -- f) for the results with respect to the ensemble obtained at the 2nd iteration step in sparse-data experiment.}
\end{figure*} 

\renewcommand{\nScale}{0.3}
\begin{figure*} %
	\centering
	
	\subfigure[Initial ensemble, base survey]{ \label{subfig:diff_initEnsMean_G_traces_seisTimeStep1_S2}
					\includegraphics[scale=0.3]{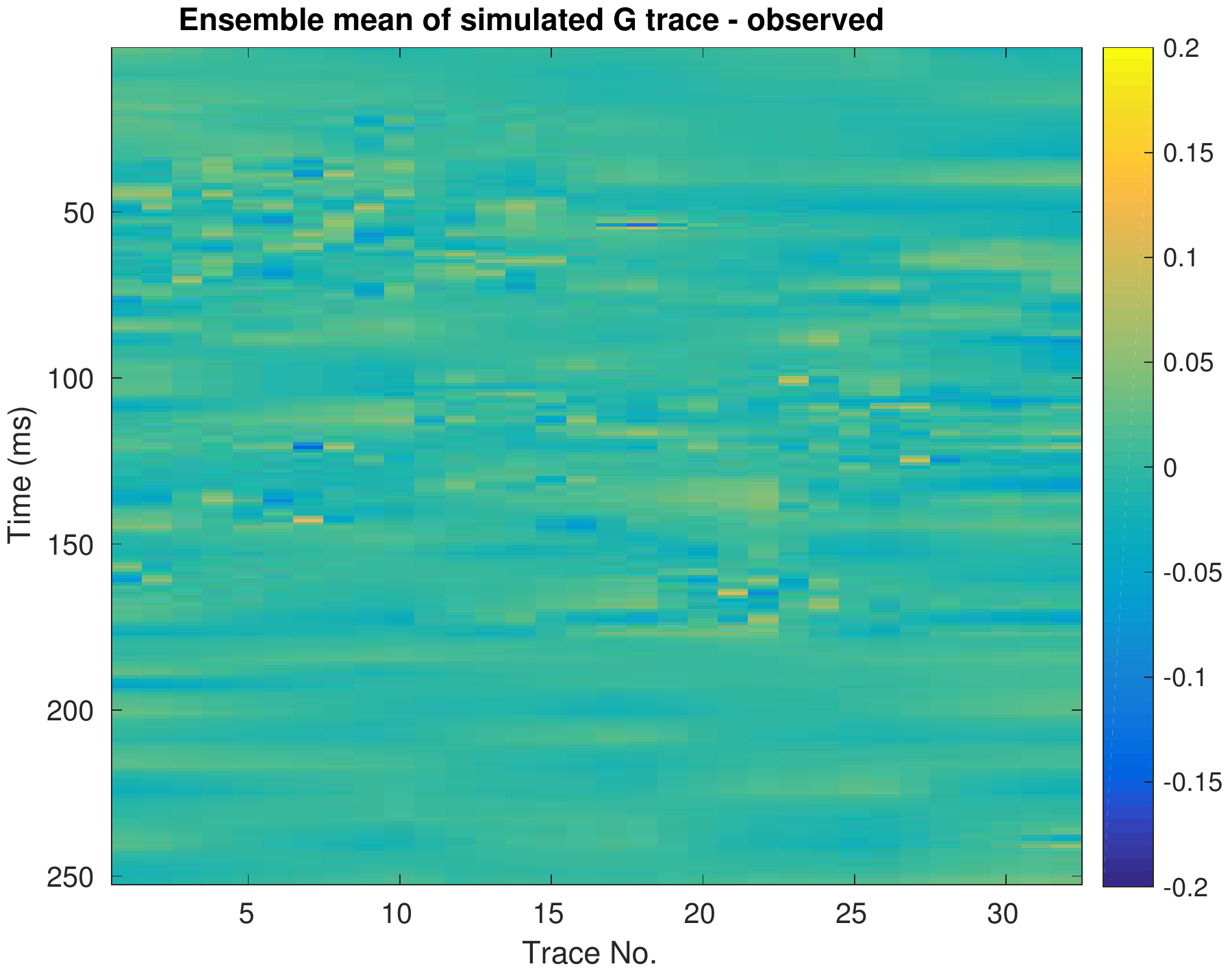}
				}%
	\subfigure[Initial ensemble, 1st monitor survey]{ \label{subfig:diff_initEnsMean_G_traces_seisTimeStep2_S2}
					\includegraphics[scale=0.3]{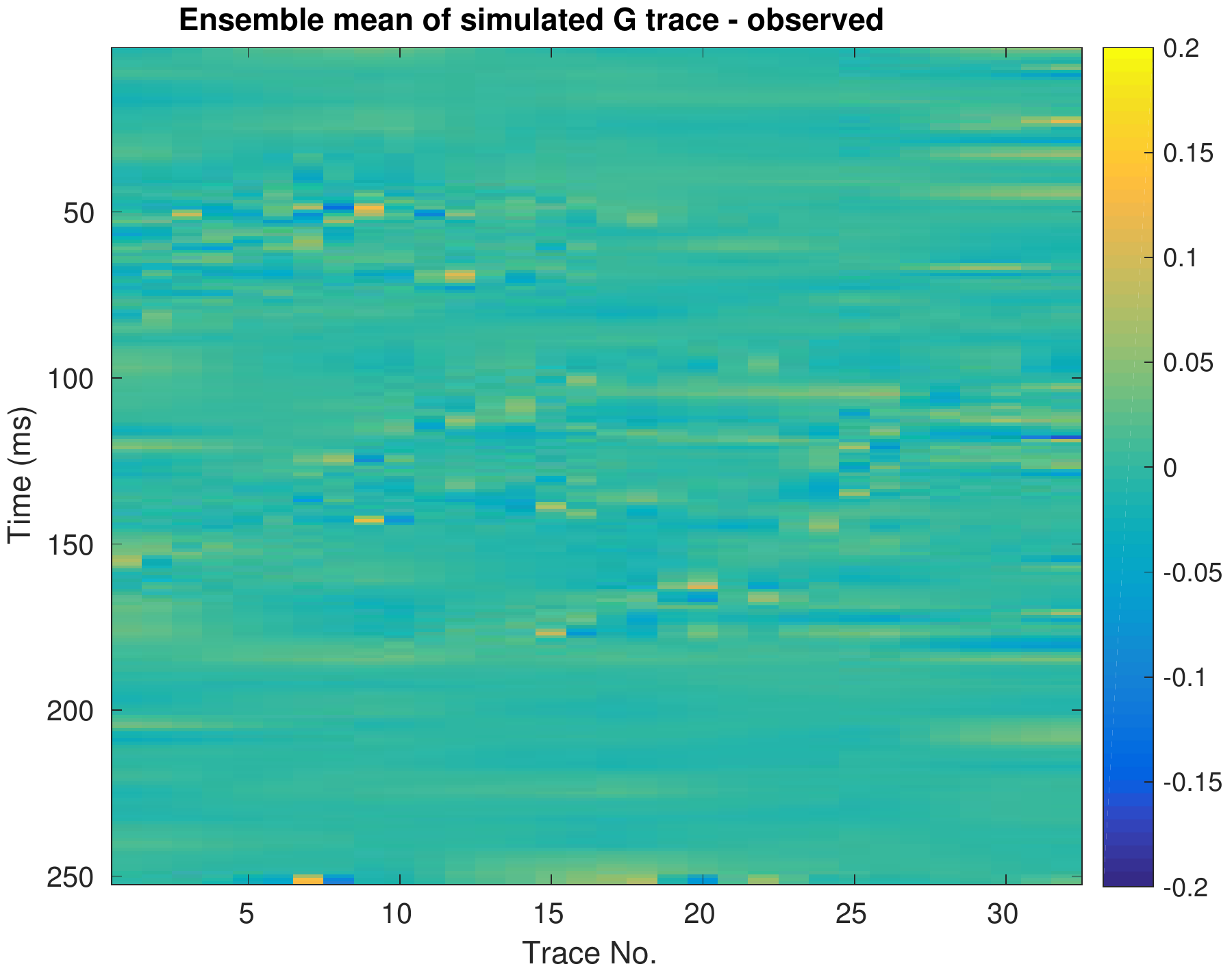}
				}%
	\subfigure[Initial ensemble, 2nd monitor survey]{ \label{subfig:diff_initEnsMean_G_traces_seisTimeStep3_S2}
						\includegraphics[scale=0.3]{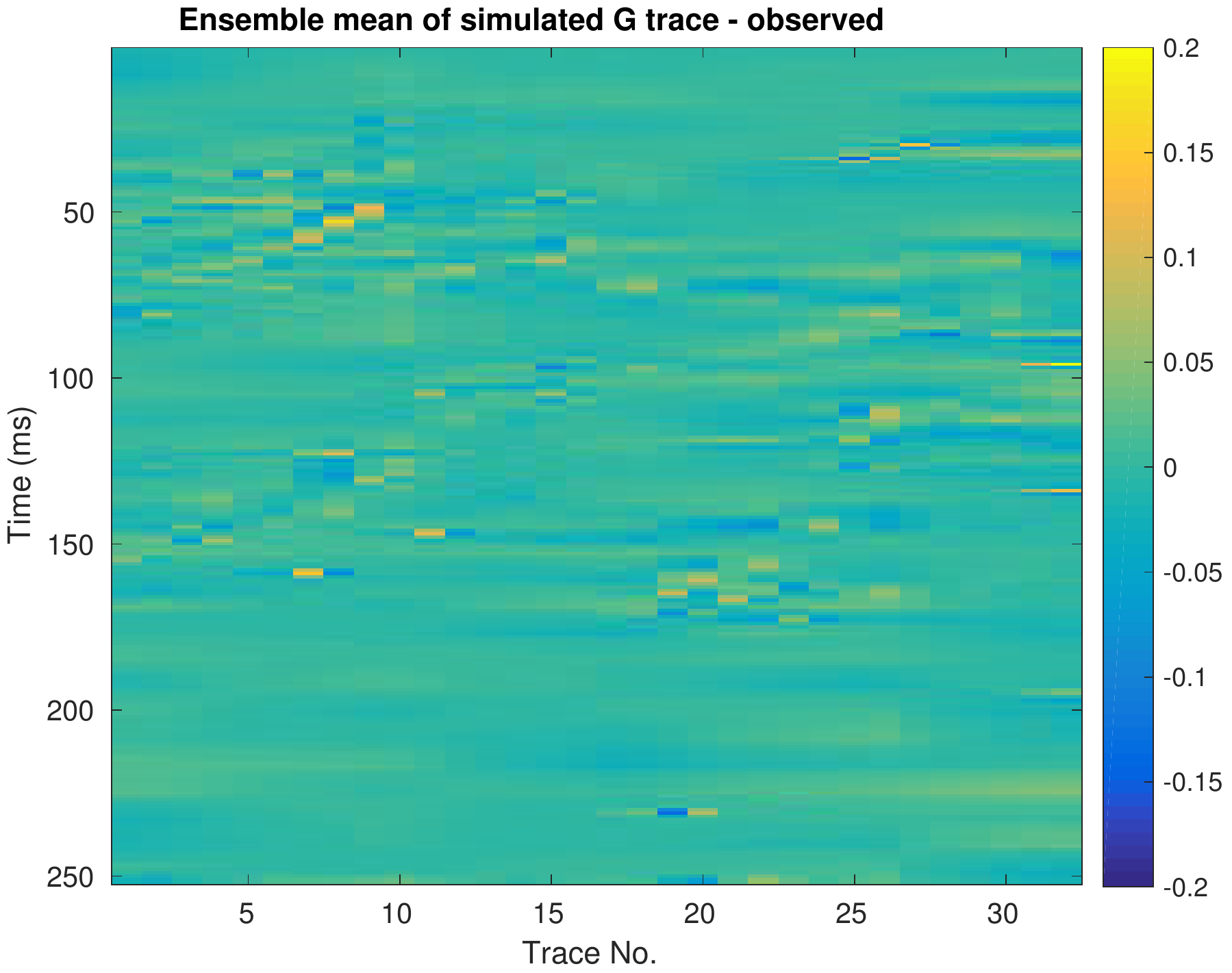}
					} 		
				
	\subfigure[Final ensemble, base survey]{ \label{subfig:diff_finalEnsMean_G_traces_seisTimeStep1_S2}
					\includegraphics[scale=0.3]{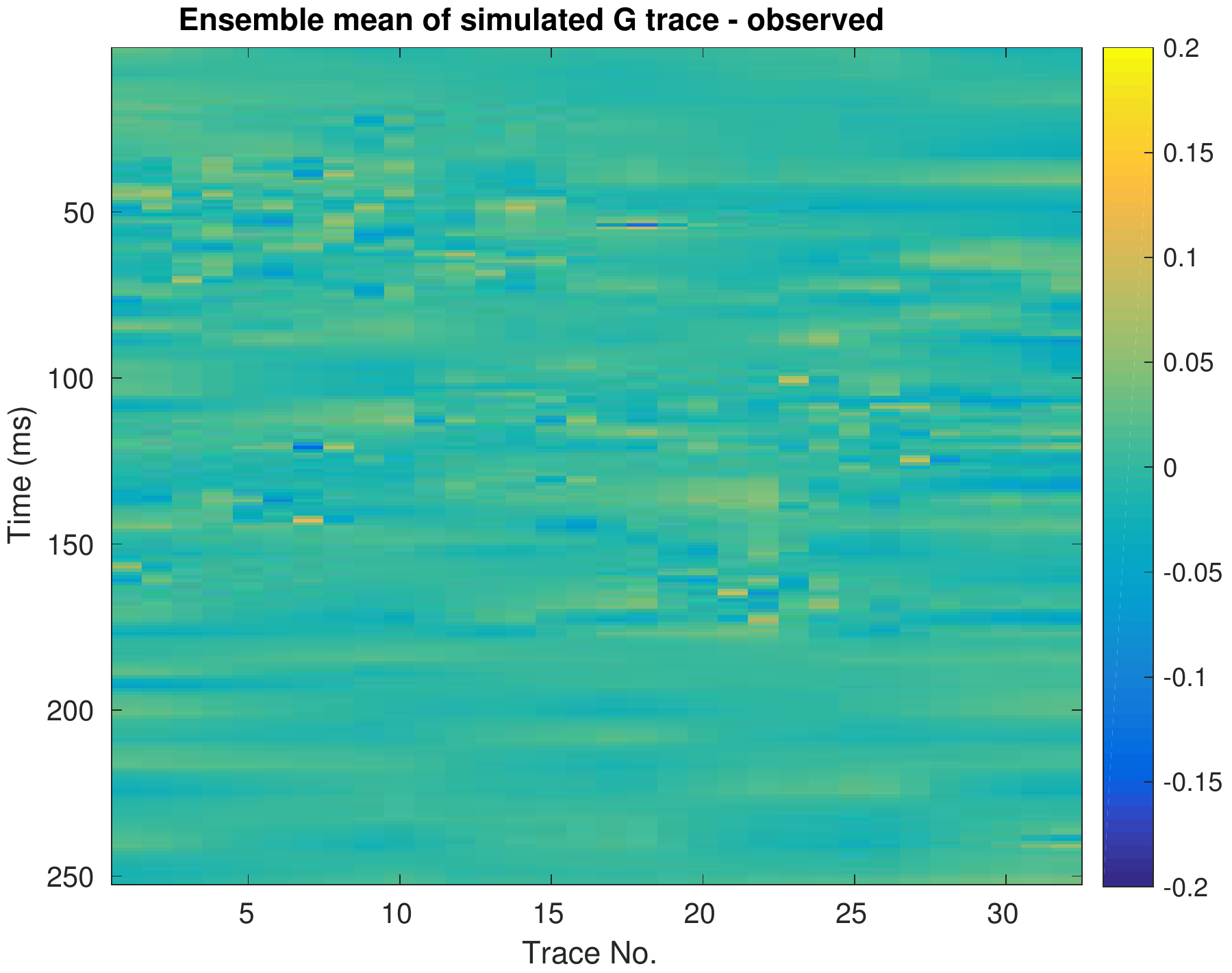}
				}%
	\subfigure[Final ensemble, 1st monitor survey]{ \label{subfig:diff_finalEnsMean_G_traces_seisTimeStep2_S2}
					\includegraphics[scale=0.3]{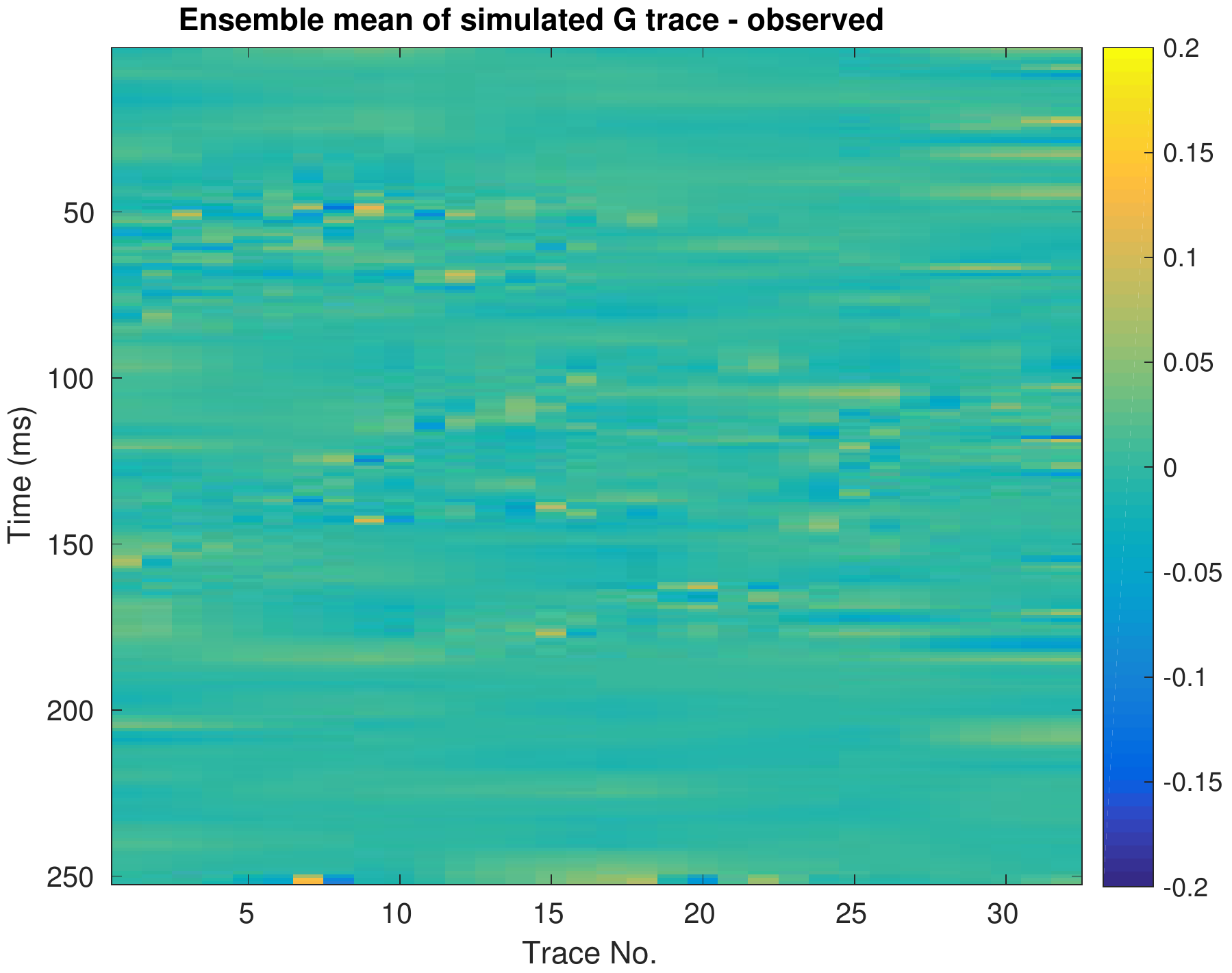}
				}%
	\subfigure[Final ensemble, 2nd monitor survey]{ \label{subfig:diff_finalEnsMean_G_traces_seisTimeStep3_S2}
						\includegraphics[scale=0.3]{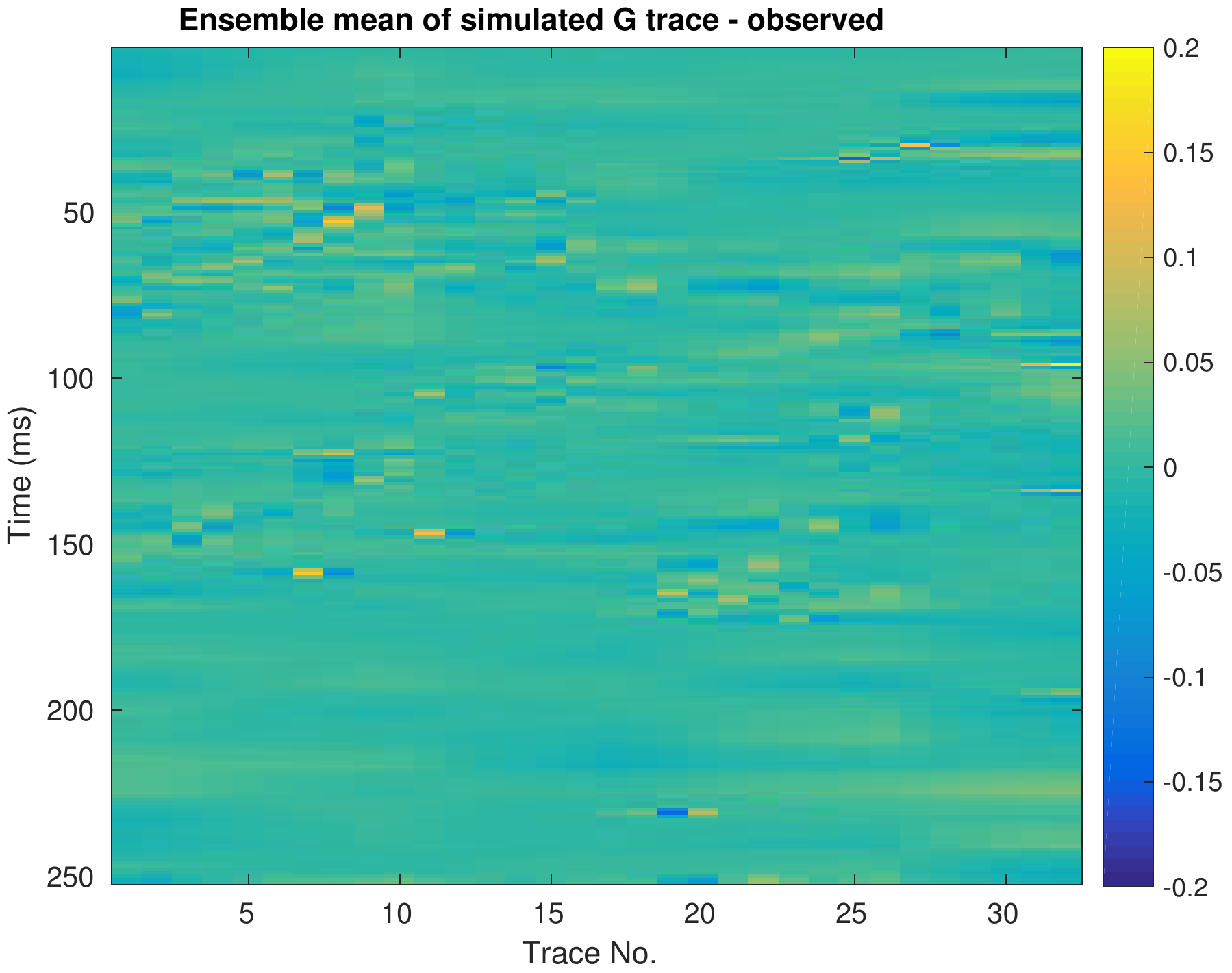}
					}						
	\caption{\label{fig:Norne2D_diff_G_S2} As in Figure \ref{fig:Norne2D_diff_R0_S2}, but now for gradient attributes in scenario S2.}
\end{figure*} 

Next we compare observed and simulated data. For succinctness, in what follows we present the results of sparse-data experiment only. To this end, we show intercept ($R_0$) and gradient ($G$) attributes that are reconstructed using leading wavelet coefficients of the original noisy attributes (see Figure \ref{fig:Norne2D_attributes_S2}), and leading wavelet coefficients of simulated AVA attributes, respectively. The former reconstructed attributes are plotted in Figure \ref{fig:Norne2D_denoised_attributes_S2}. For ease of visualization, we show differences between ensemble means of reconstructed simulated attributes, and reconstructed observed attributes. Figure \ref{fig:Norne2D_diff_R0_S2} illustrates the differences of intercept attributes at different survey time instances. There, sub-figures in the first row correspond to the results with respect to the initial ensemble, while those in the second row to the results with respect to the final ensemble (which is the ensemble obtained at the 2nd iteration step). Comparing the results of initial and final ensembles, it appears that the distributions of differences at the base survey (Figures \ref{subfig:diff_initEnsMean_R0_traces_seisTimeStep1_S2} and \ref{subfig:diff_finalEnsMean_R0_traces_seisTimeStep1_S2}) are similar, and there are more distinctions at two monitor surveys. Similarly, Figure \ref{fig:Norne2D_diff_G_S2} shows differences of gradient attributes. The results with respect to the initial and final ensembles stay close at all survey time. This is possibly because in this particular case, intercept is a more sensitive attribute than gradient. 

\begin{figure*} 
	\centering
	\subfigure[Reference]{ \label{subfig:permx_ref_S2}
				\includegraphics[scale=0.3]{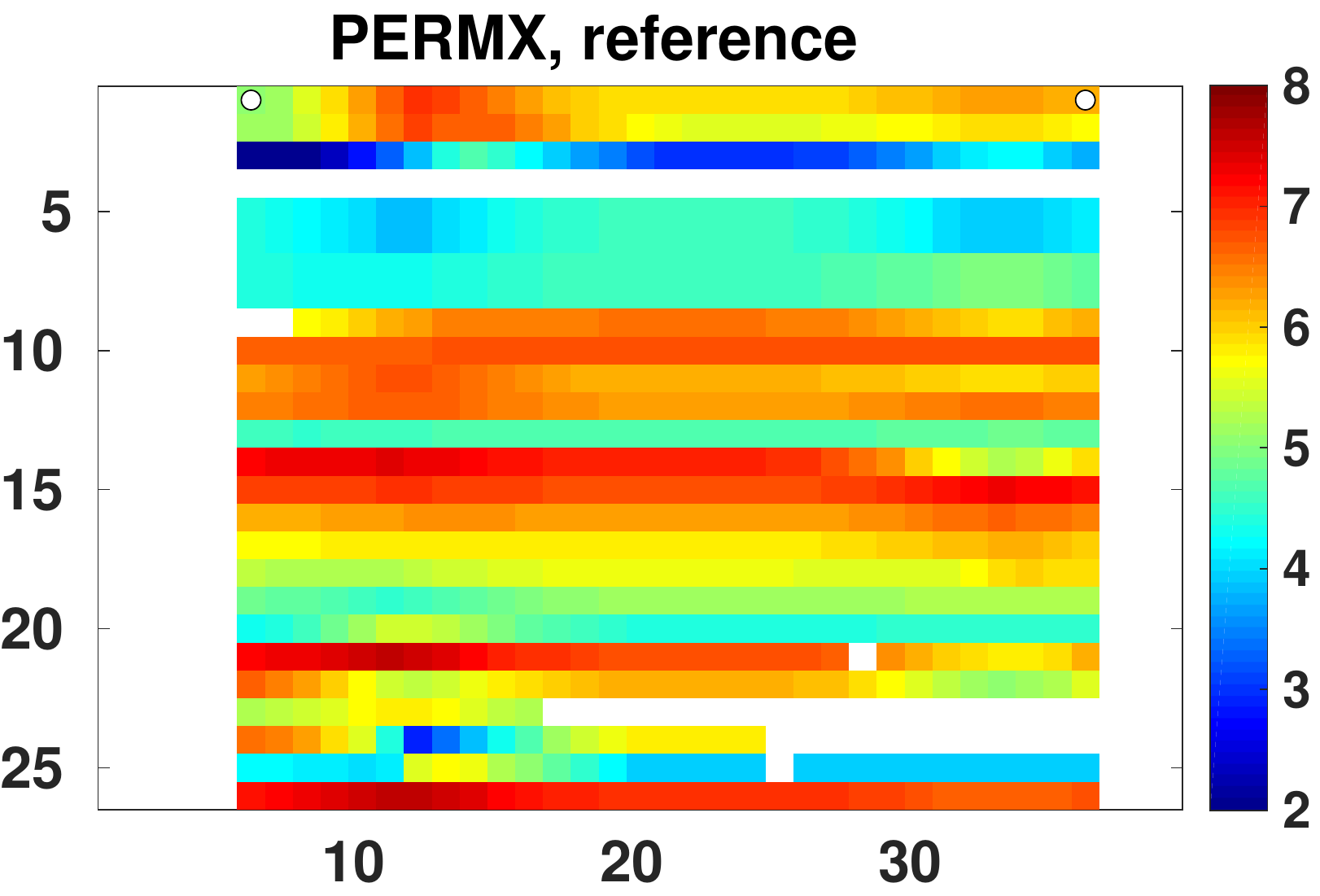}
			}
			
	\subfigure[Initial mean]{ \label{subfig:field_PERMX_mean_init_ensemble_S2}
					\includegraphics[scale=0.3]{./figures/field_PERMX_mean_init_ensemble.eps}
				}%
	\subfigure[Initial member 1]{ \label{subfig:field_PERMX_1_1_init_ensemble_S2}
					\includegraphics[scale=0.3]{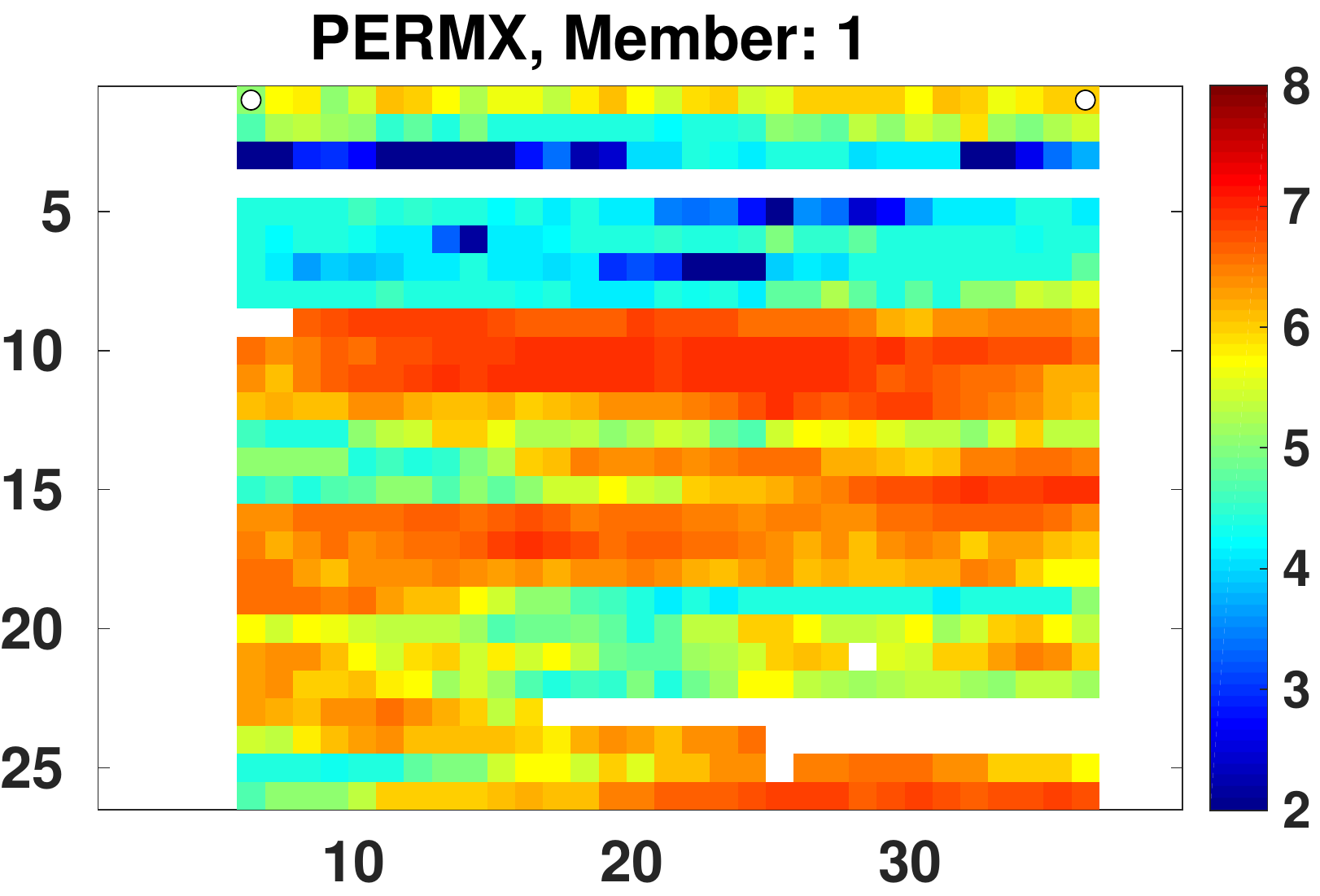}
				}%
	\subfigure[Initial member 2]{ \label{subfig:field_PERMX_1_2_init_ensemble_S2}
					\includegraphics[scale=0.3]{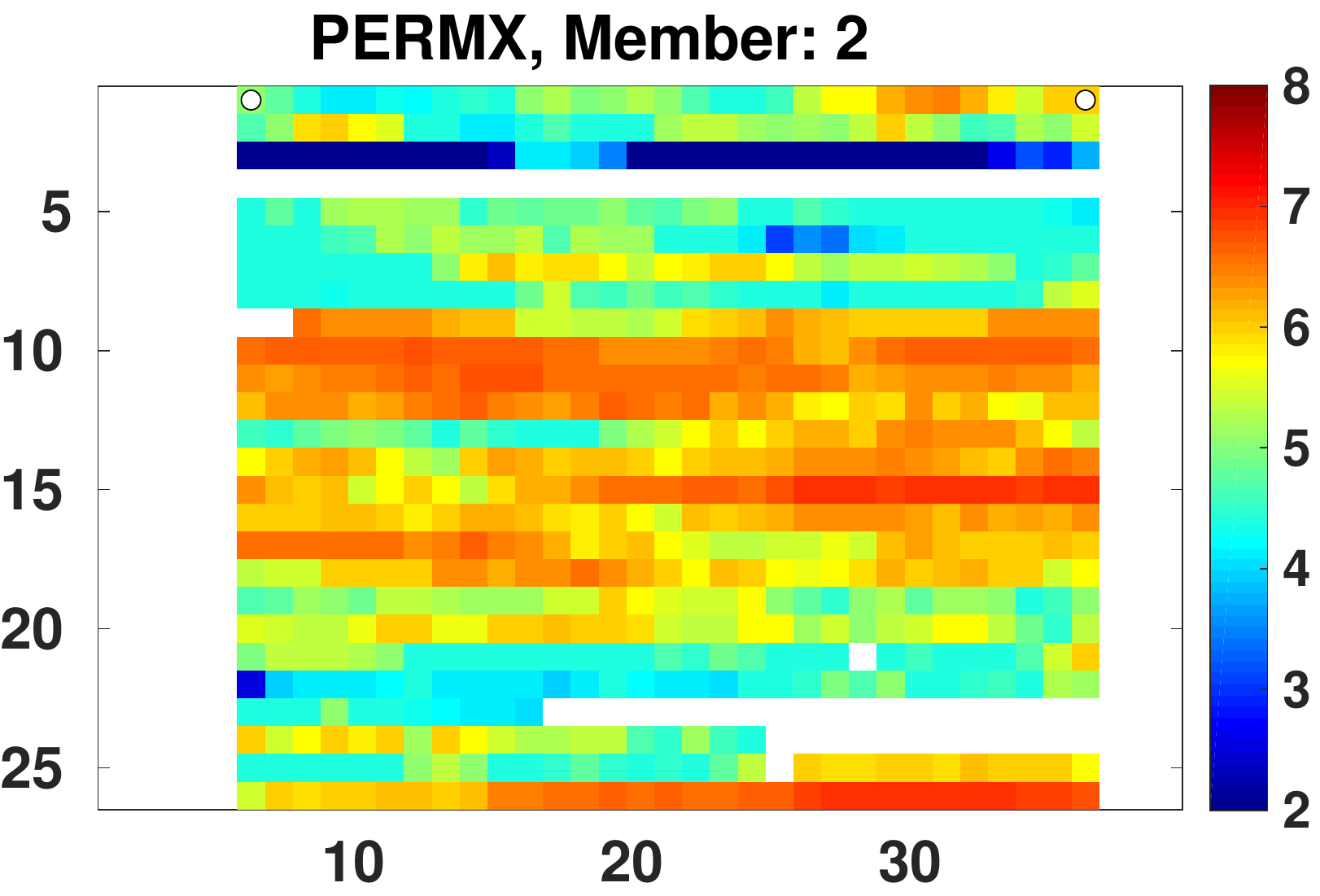}
				}		
				
	\subfigure[Final mean]{ \label{subfig:field_PERMX_mean_ensemble2_S2}
					\includegraphics[scale=0.3]{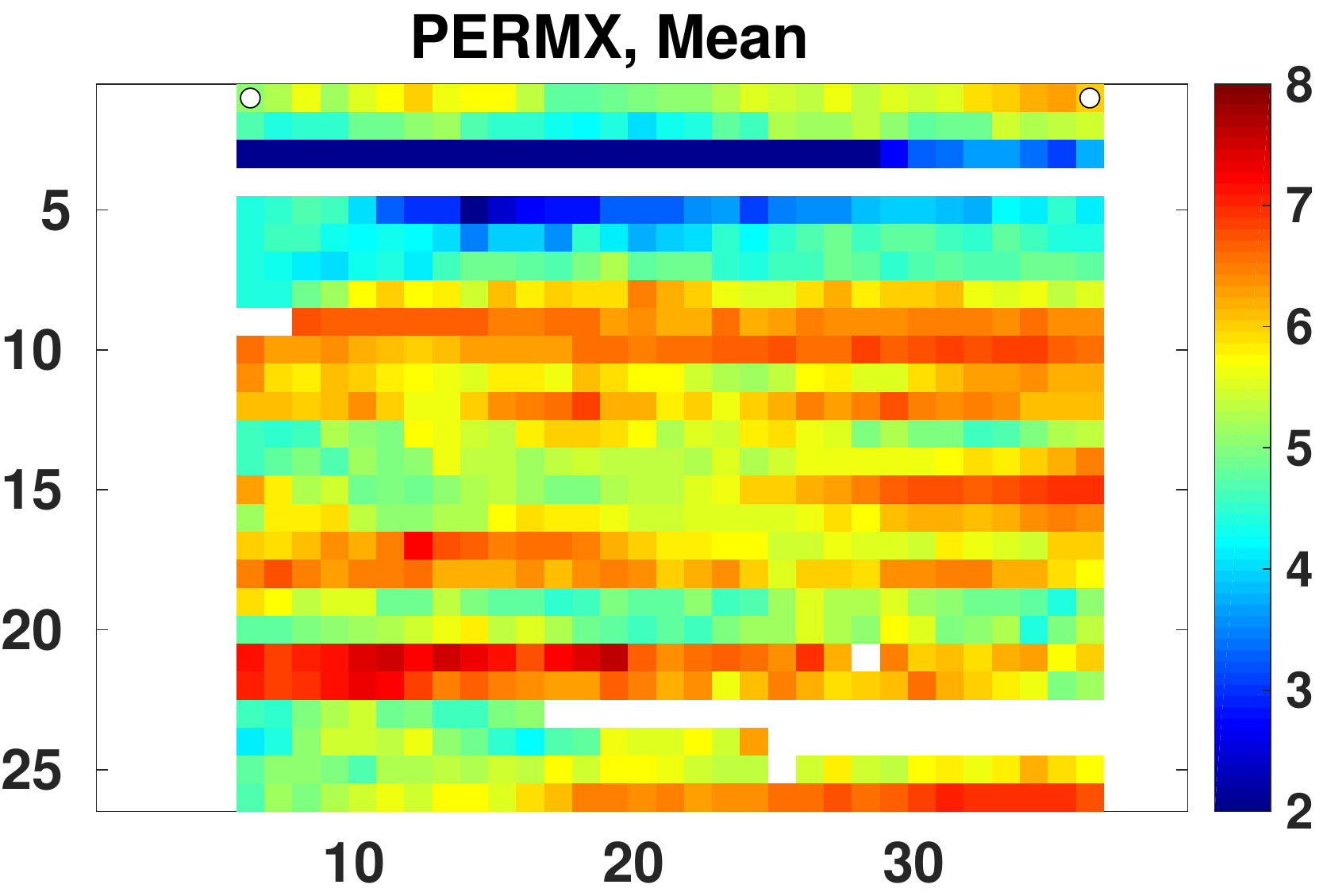}
				}%
	\subfigure[Final member 1]{ \label{subfig:field_PERMX_1_1_ensemble2_S2}
					\includegraphics[scale=0.3]{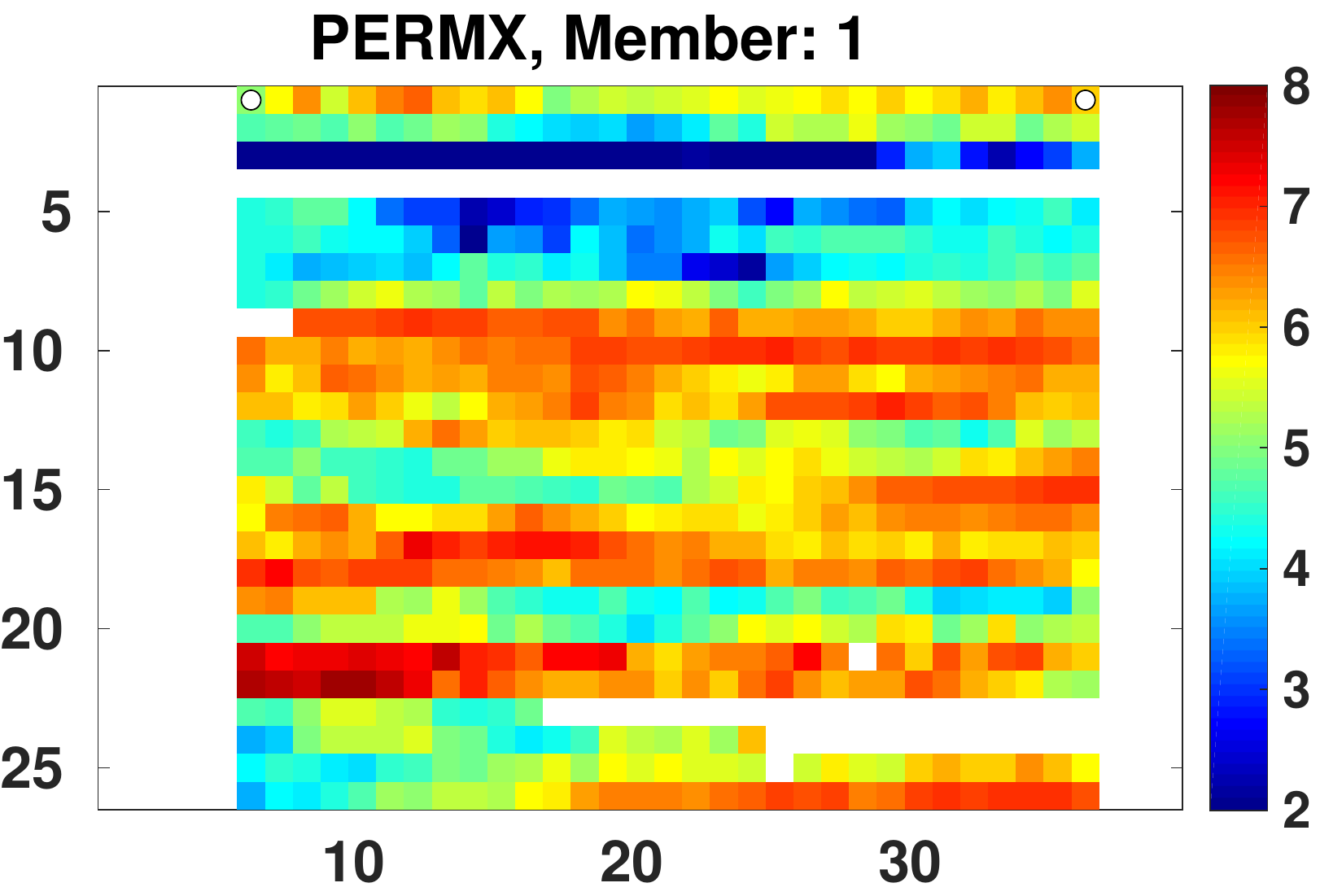}
				}%
	\subfigure[Final member 2]{ \label{subfig:field_PERMX_1_2_ensemble2_S2}
					\includegraphics[scale=0.3]{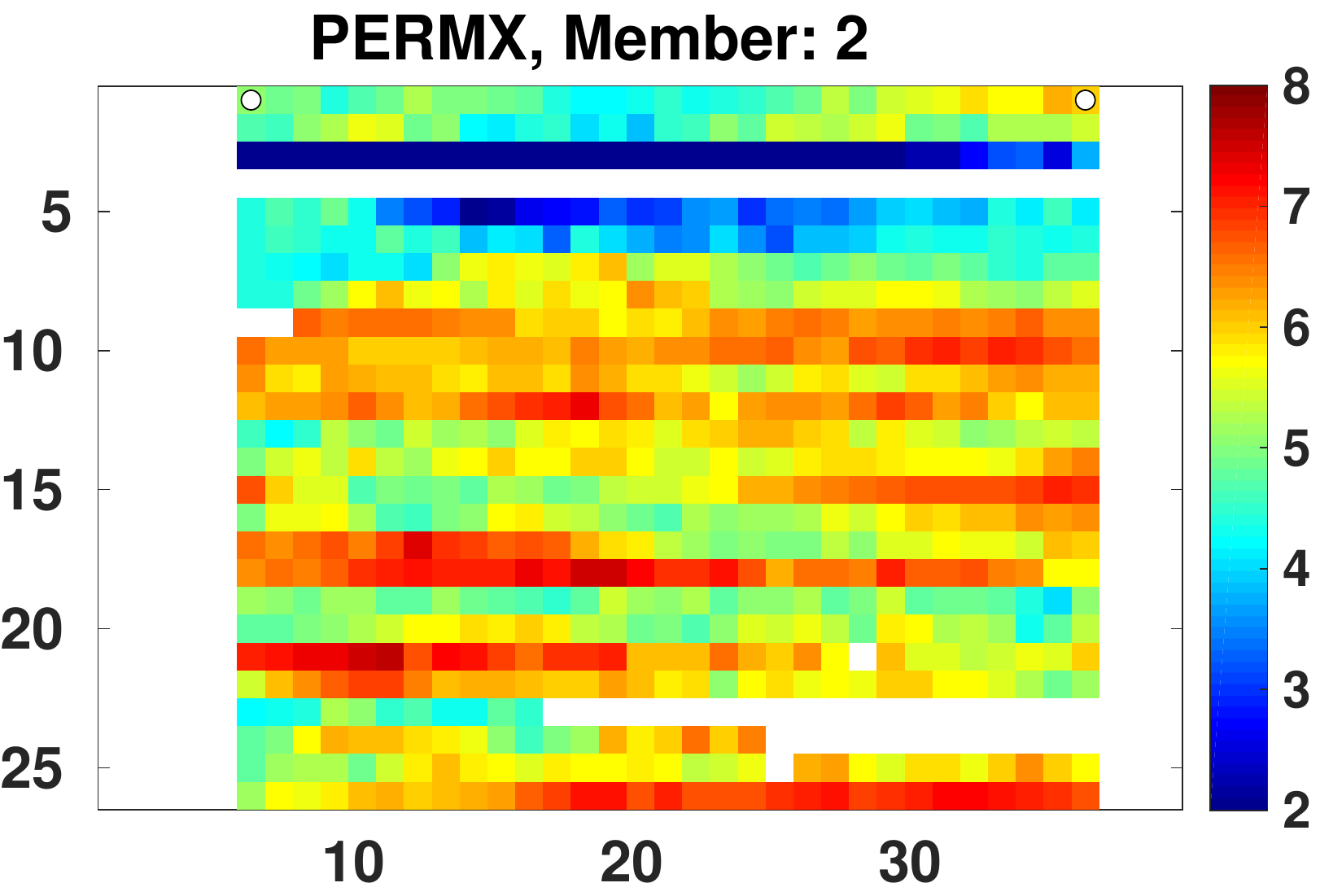}
				}															
	\caption{\label{fig:Norne2D_PERMX_S2} Distributions of log PERMX (scenario S2). (a) Reference model; (b) -- (d) Mean and 2 sample realizations of the initial ensemble of log PERMX; (e) -- (g) Corresponding mean and 2 sample realizations of the final ensemble.}
\end{figure*}

\begin{figure*} %
	\centering
	\subfigure[Reference]{ \label{subfig:poro_ref_S2}
				\includegraphics[scale=0.3]{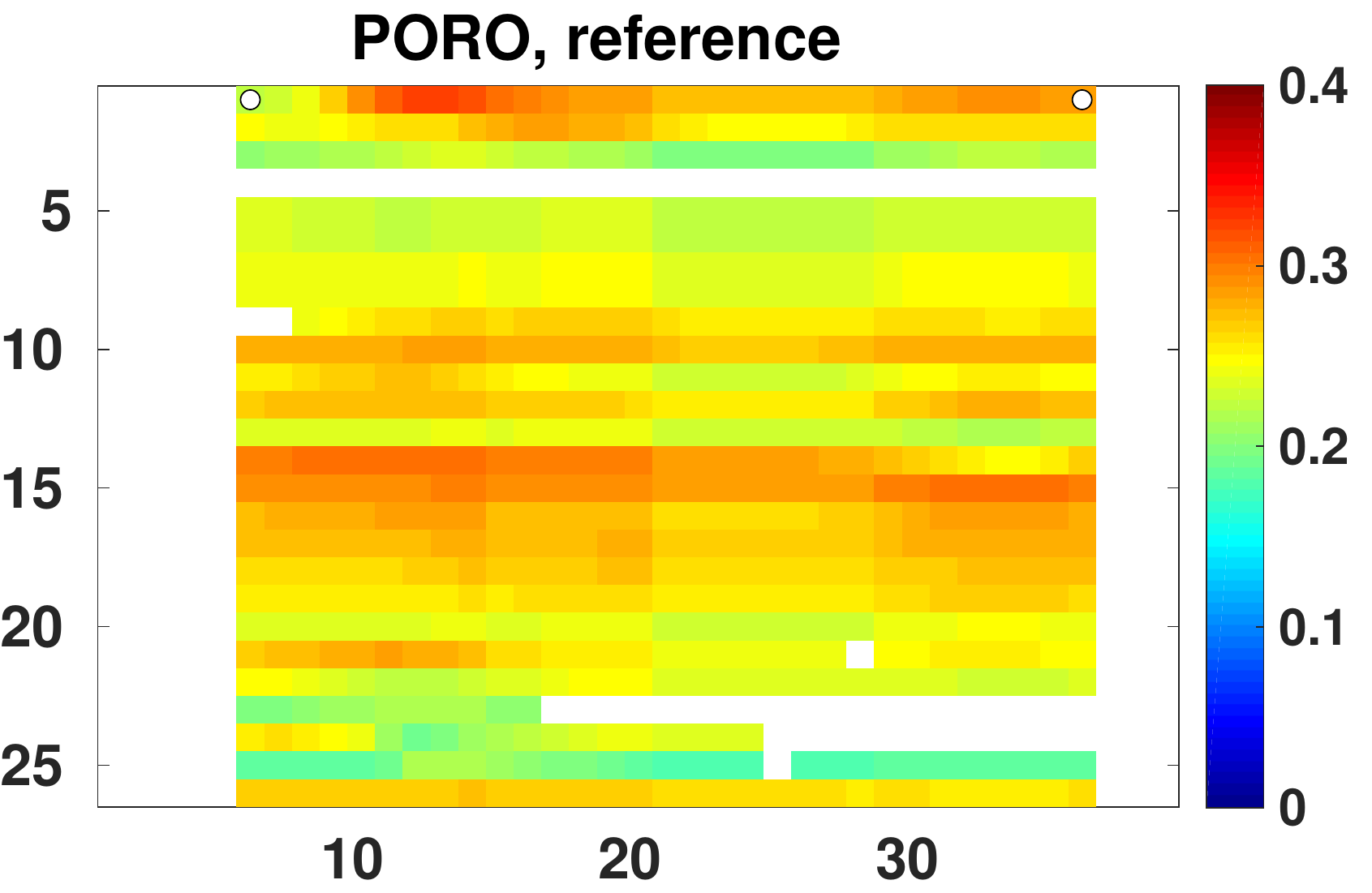}
			}
			
	\subfigure[Initial mean]{ \label{subfig:field_PORO_mean_init_ensemble_S2}
					\includegraphics[scale=0.28]{./figures/field_PORO_mean_init_ensemble.eps}
				}%
	\subfigure[Initial member 1]{ \label{subfig:field_PORO_1_1_init_ensemble_S2}
					\includegraphics[scale=0.3]{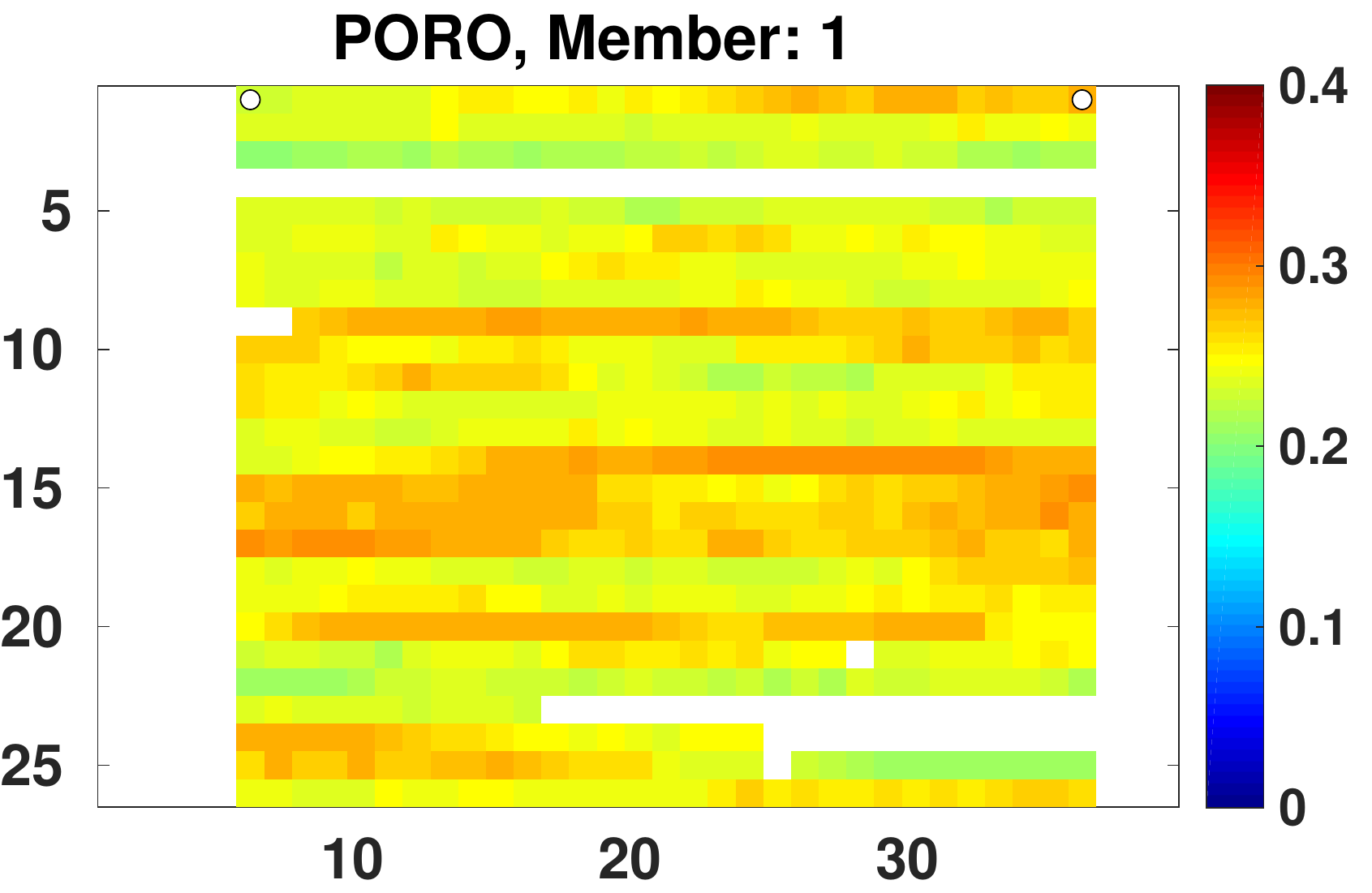}
				}%
	\subfigure[Initial member 2]{ \label{subfig:field_PORO_1_2_init_ensemble_S2}
					\includegraphics[scale=0.3]{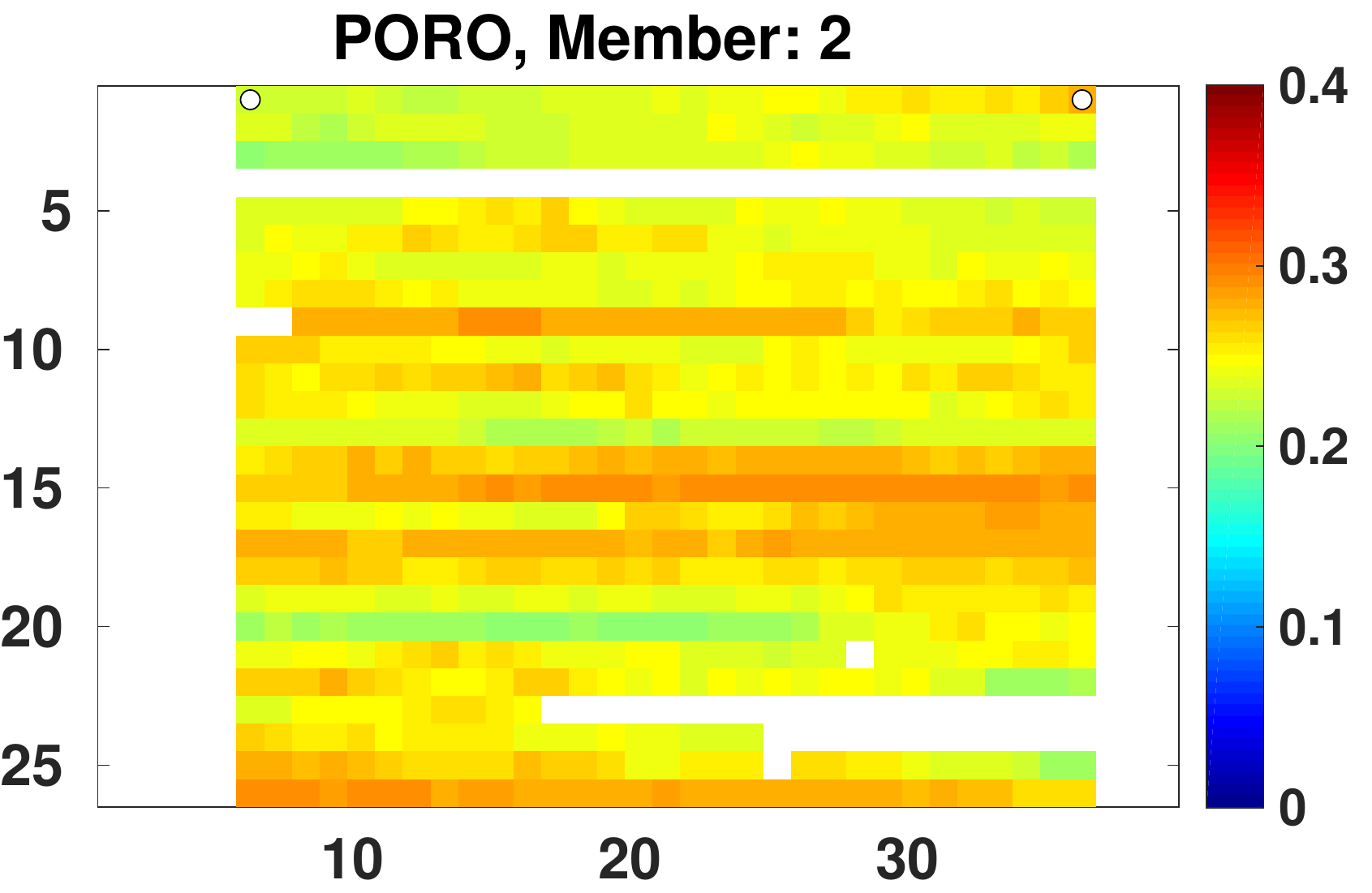}
				}		
				
	\subfigure[Final mean]{ \label{subfig:field_PORO_mean_ensemble2_S2}
					\includegraphics[scale=0.28]{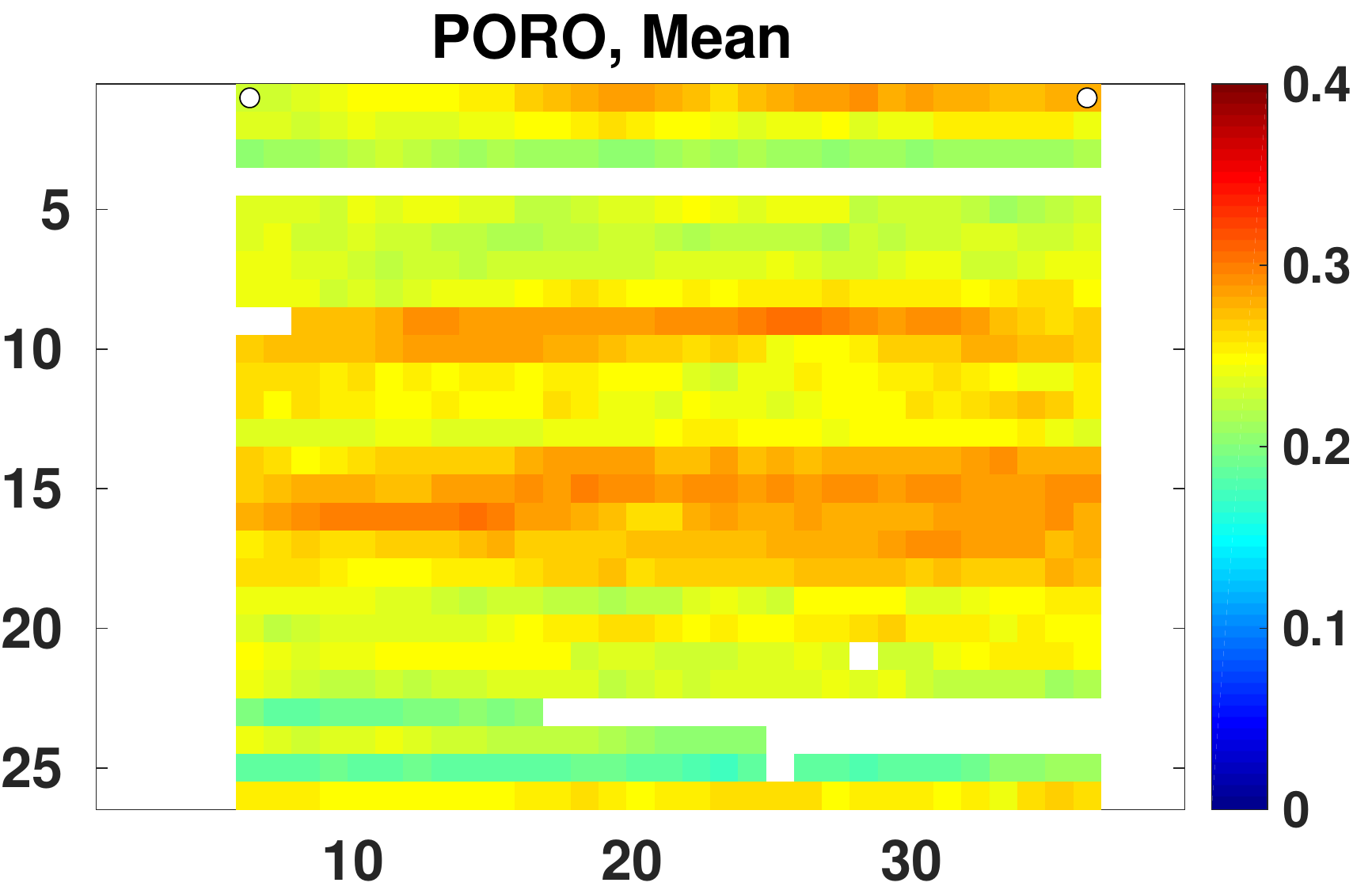}
				}%
	\subfigure[Final member 1]{ \label{subfig:field_PORO_1_1_ensemble2_S2}
					\includegraphics[scale=0.3]{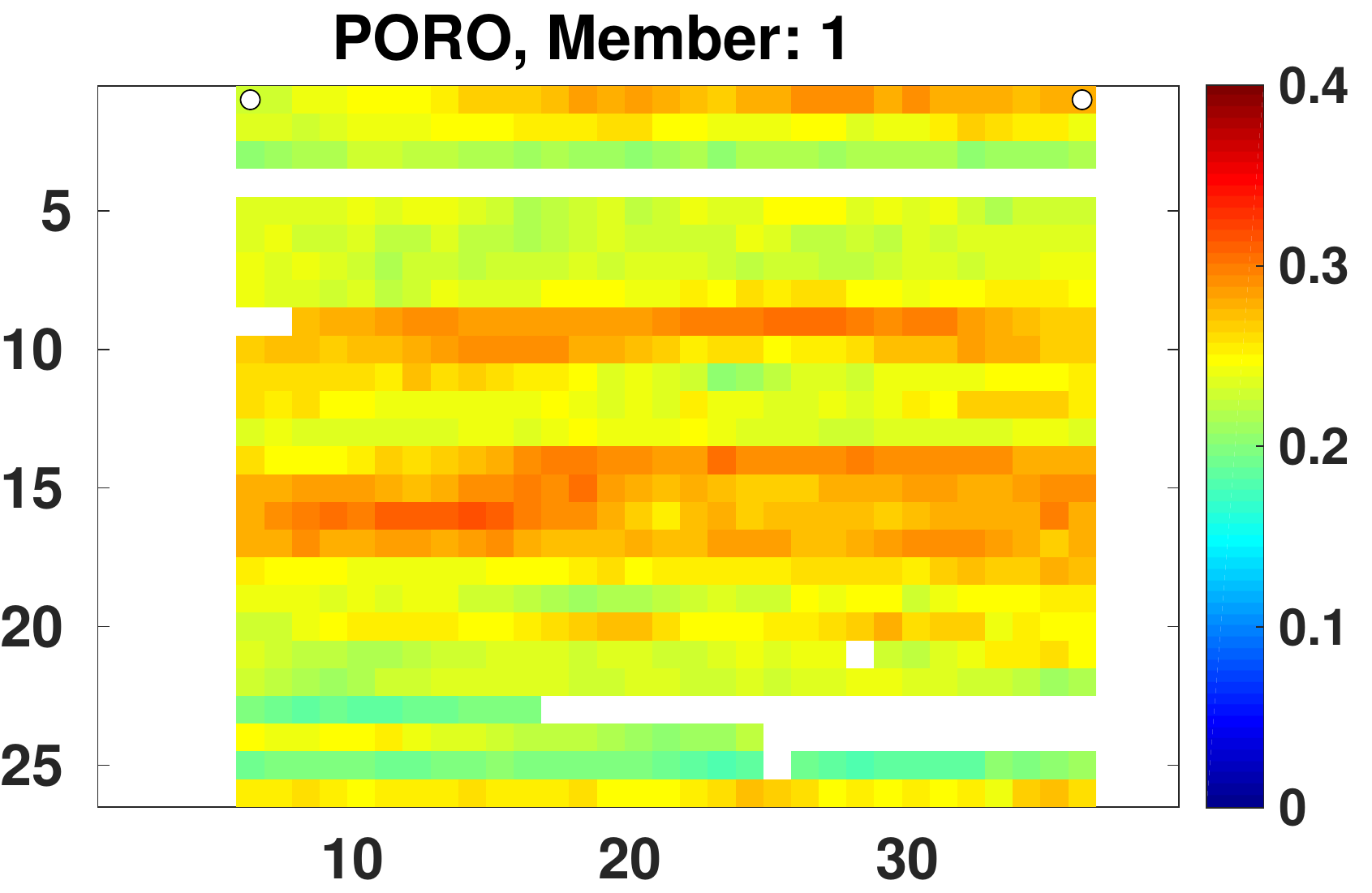}
				}%
	\subfigure[Final member 2]{ \label{subfig:field_PORO_1_2_ensemble2_S2}
					\includegraphics[scale=0.3]{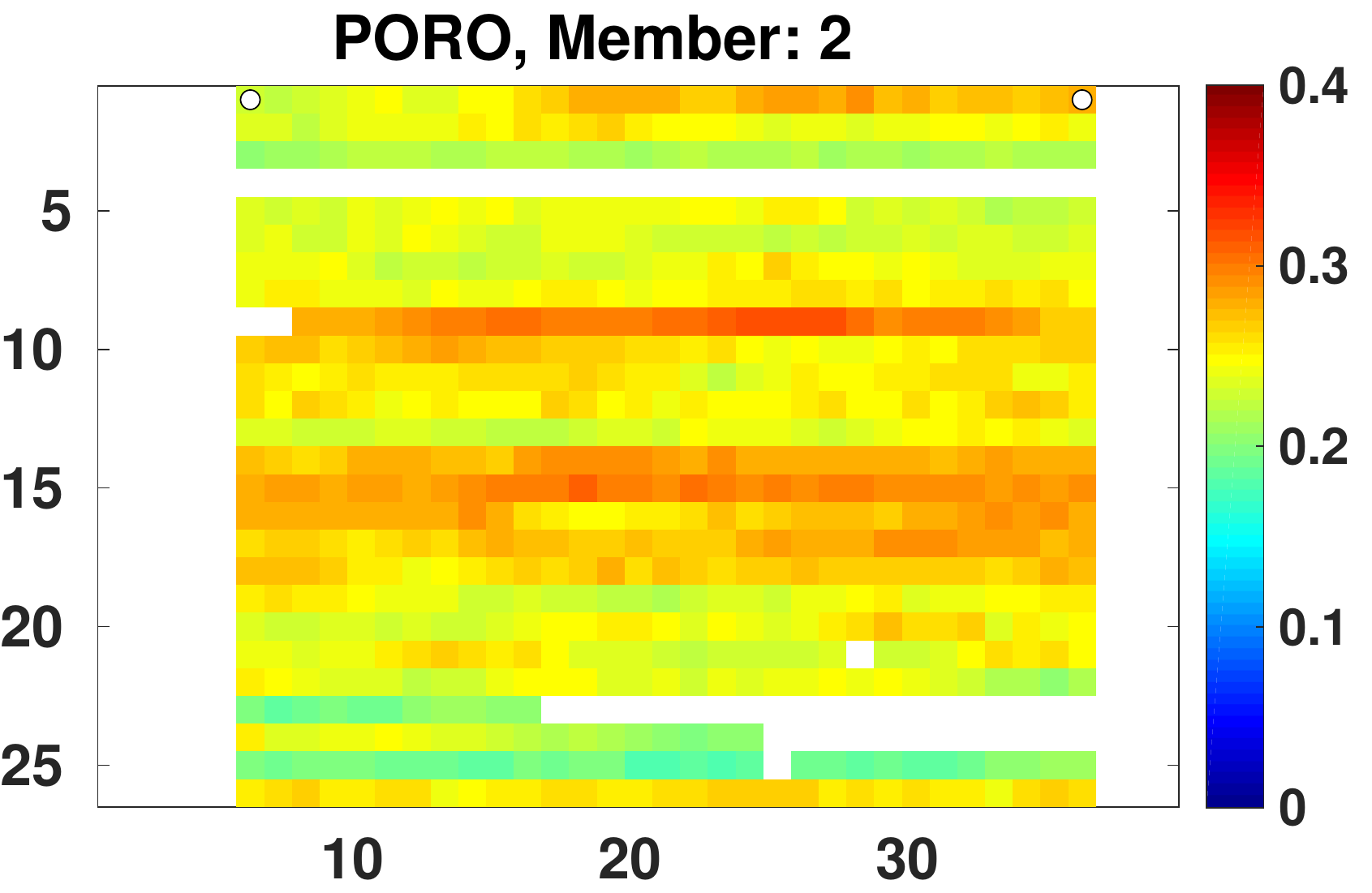}
				}															
	\caption{\label{fig:Norne2D_PORO_S2} As in Figure \ref{fig:Norne2D_PERMX_S2}, but now for the distributions of PORO in scenario S2.}
\end{figure*}

Figures \ref{fig:Norne2D_PERMX_S2} and \ref{fig:Norne2D_PORO_S2} show distributions of log PERMX and PORO. In each figure, the reference distribution is plotted in the first row, the distributions of mean and two samples of the initial ensemble in the second row, and the distributions of mean and two samples of the final ensemble in the third row. Comparing the results of initial and final ensembles (e.g., Figure \ref{subfig:field_PERMX_mean_init_ensemble_S2} versus Figure \ref{subfig:field_PERMX_mean_ensemble2_S2}), we see substantial changes in both log PERMX and PORO after history matching, consistent with the results in Figure \ref{fig:Norne2D_rmse_S2}.    
   
\renewcommand{\nScale}{0.33}

\renewcommand{\nScale}{0.33}

\subsection{Results of scenario S3 (using both production and 4D seismic data)}   

\renewcommand{\nScale}{0.4}
\begin{figure*} %
	\centering

	\subfigure[Production data, full-data experiment]{ \label{subfig:Norne2D_objReal_prod_boxplot_objRealIter_S3_full}
					\includegraphics[scale=\nScale]{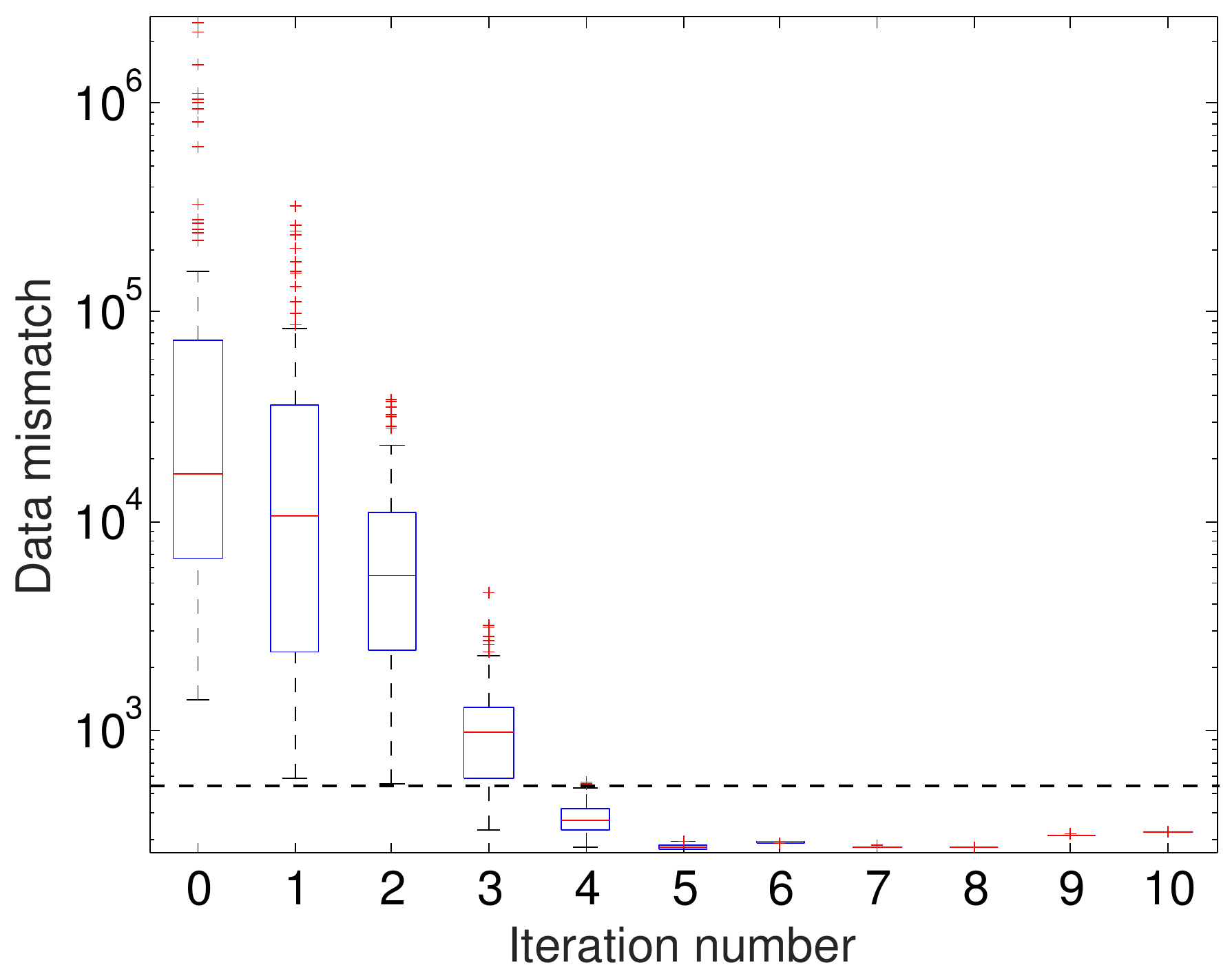}
				} 			
	\subfigure[Seismic data, full-data experiment]{ \label{subfig:Norne2D_objReal_seis_boxplot_objRealIter_S3_full}
					\includegraphics[scale=\nScale]{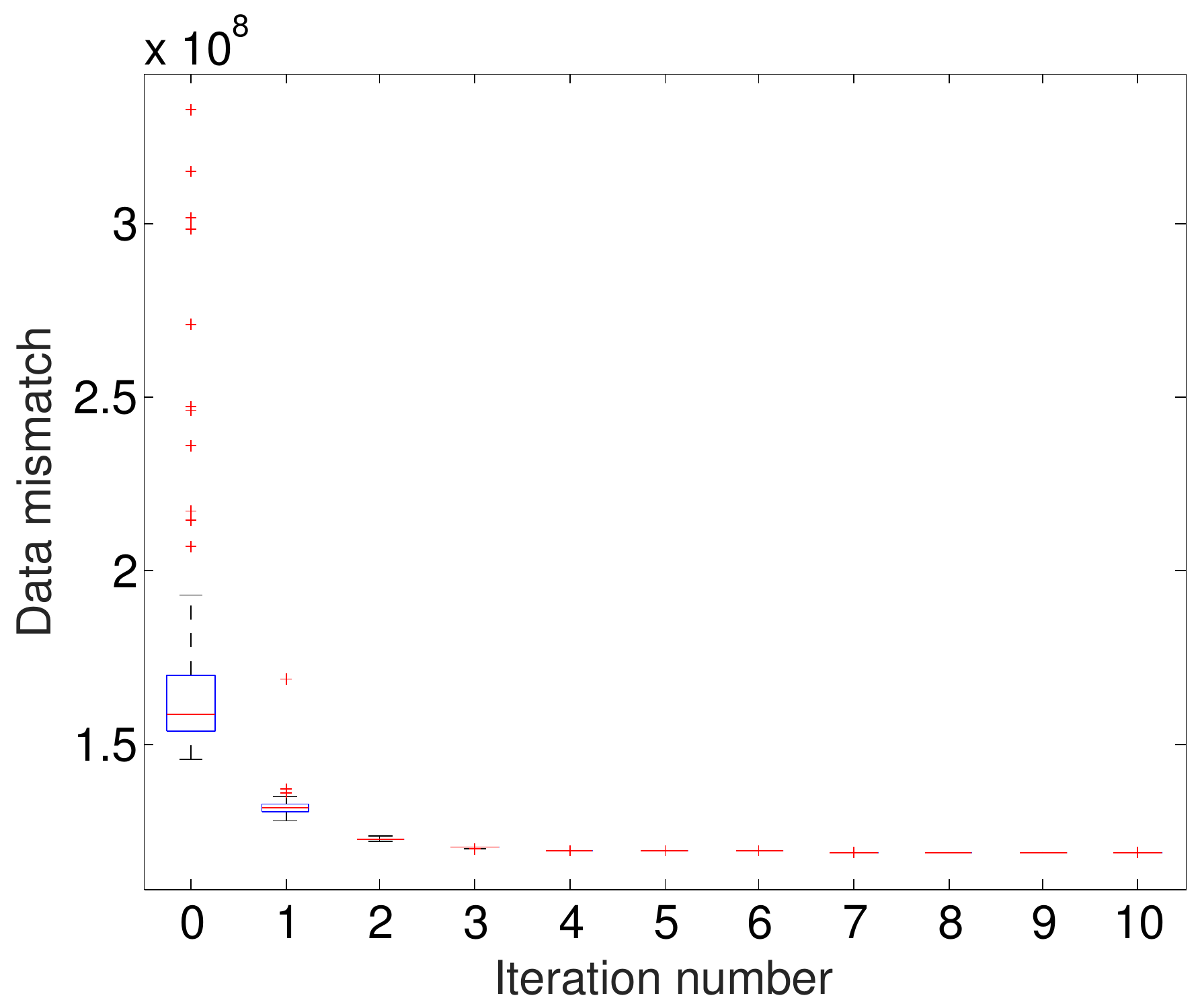}
				}
								
	\subfigure[Production data, sparse-data experiment]{ \label{subfig:Norne2D_objReal_prod_boxplot_objRealIter_S3_sparse}
					\includegraphics[scale=\nScale]{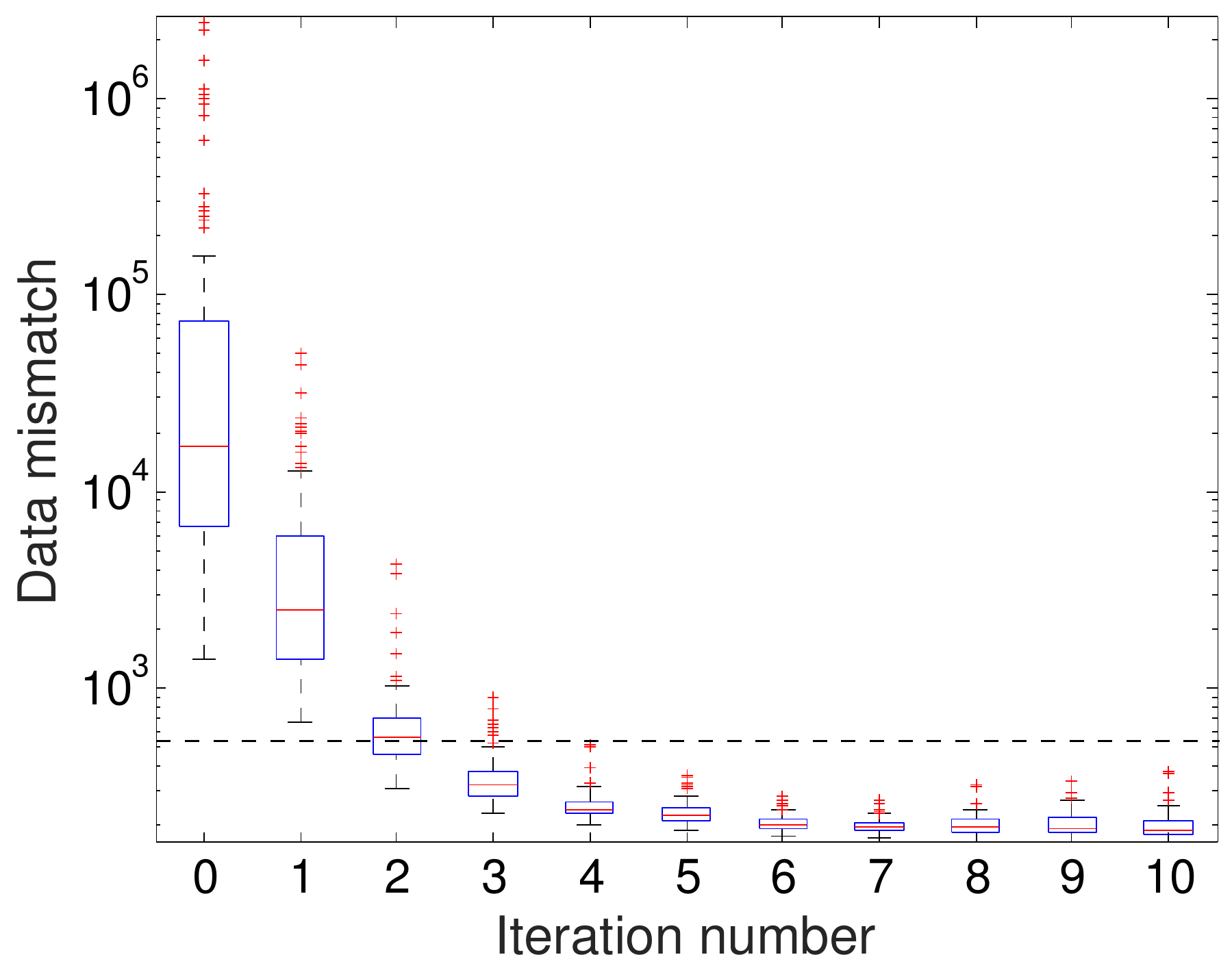}
				} 			
	\subfigure[Seismic data, sparse-data experiment]{ \label{subfig:Norne2D_objReal_seis_boxplot_objRealIter_S3_sparse}
					\includegraphics[scale=\nScale]{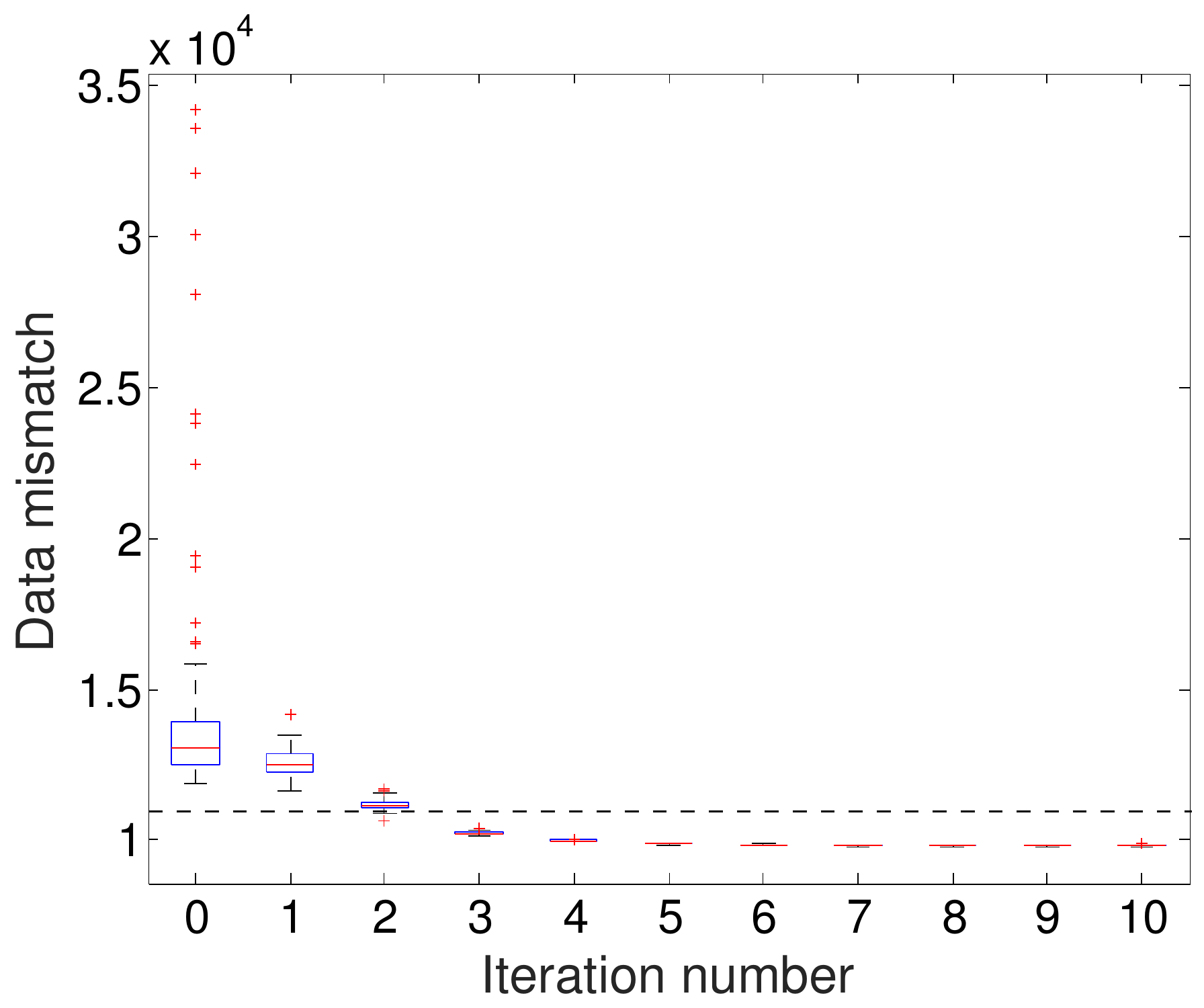}
				}				
	\caption{\label{fig:Norne2D_obj_S3} Boxplots of data mismatch as functions of iteration step (scenario S3) for (a) production data (full-data experiment), (b) seismic data (full-data experiment), (c) production data (sparse-data experiment) and (d) seismic data (sparse-data experiment). Except for (c), the horizontal dashed line in each sub-figure represents the threshold
	value of stopping criterion (\ref{eq:stopping_criterion_ndm}) for either production or seismic data, whereas data mismatch in (c) is too high to hit at a threshold value.}
\end{figure*} 

\renewcommand{\nScale}{0.4}
\begin{figure*} %
	\centering

    \subfigure[RMSEs of log PERMX (full-data)]{ \label{subfig:rmse_PERMX_boxplot_ensemble_S3_full}
					\includegraphics[scale=\nScale]{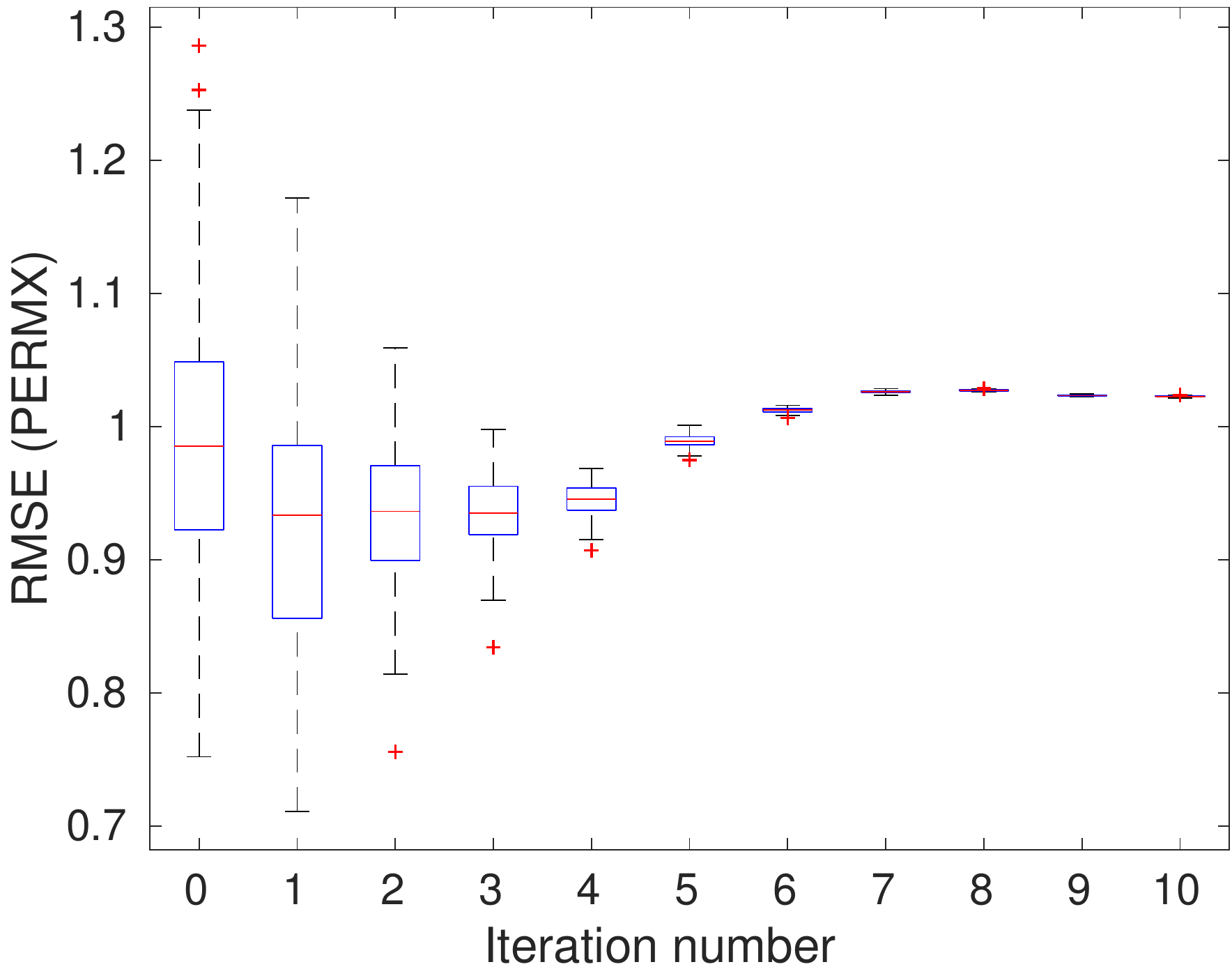}
				} 			
	\subfigure[RMSEs of PORO (full-data)]{ \label{subfig:rmse_PORO_boxplot_ensemble_S3_full}
					\includegraphics[scale=\nScale]{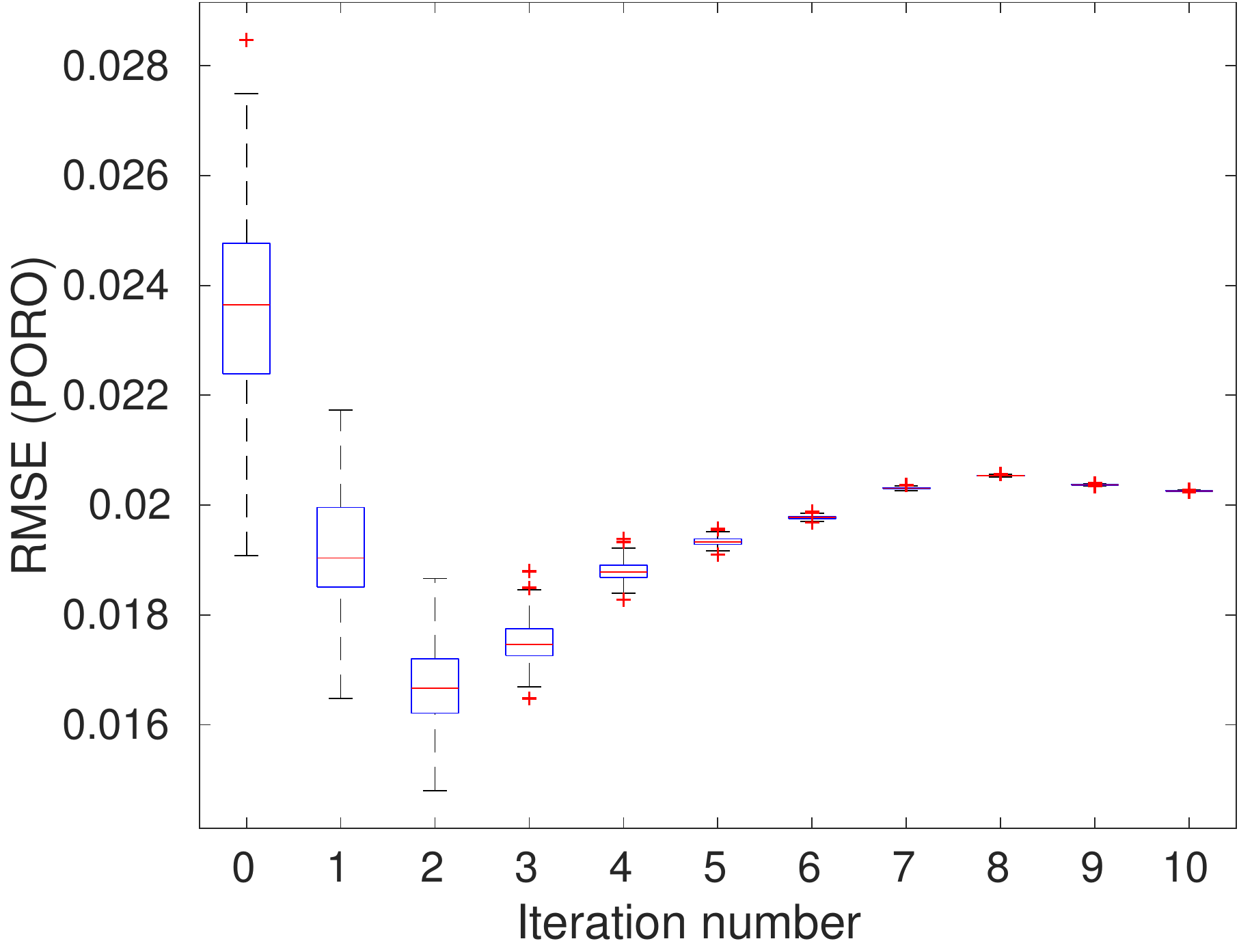}
				}	
				
	\subfigure[RMSEs of log PERMX (sparse-data)]{ \label{subfig:rmse_PERMX_boxplot_ensemble_S3_sparse}
					\includegraphics[scale=\nScale]{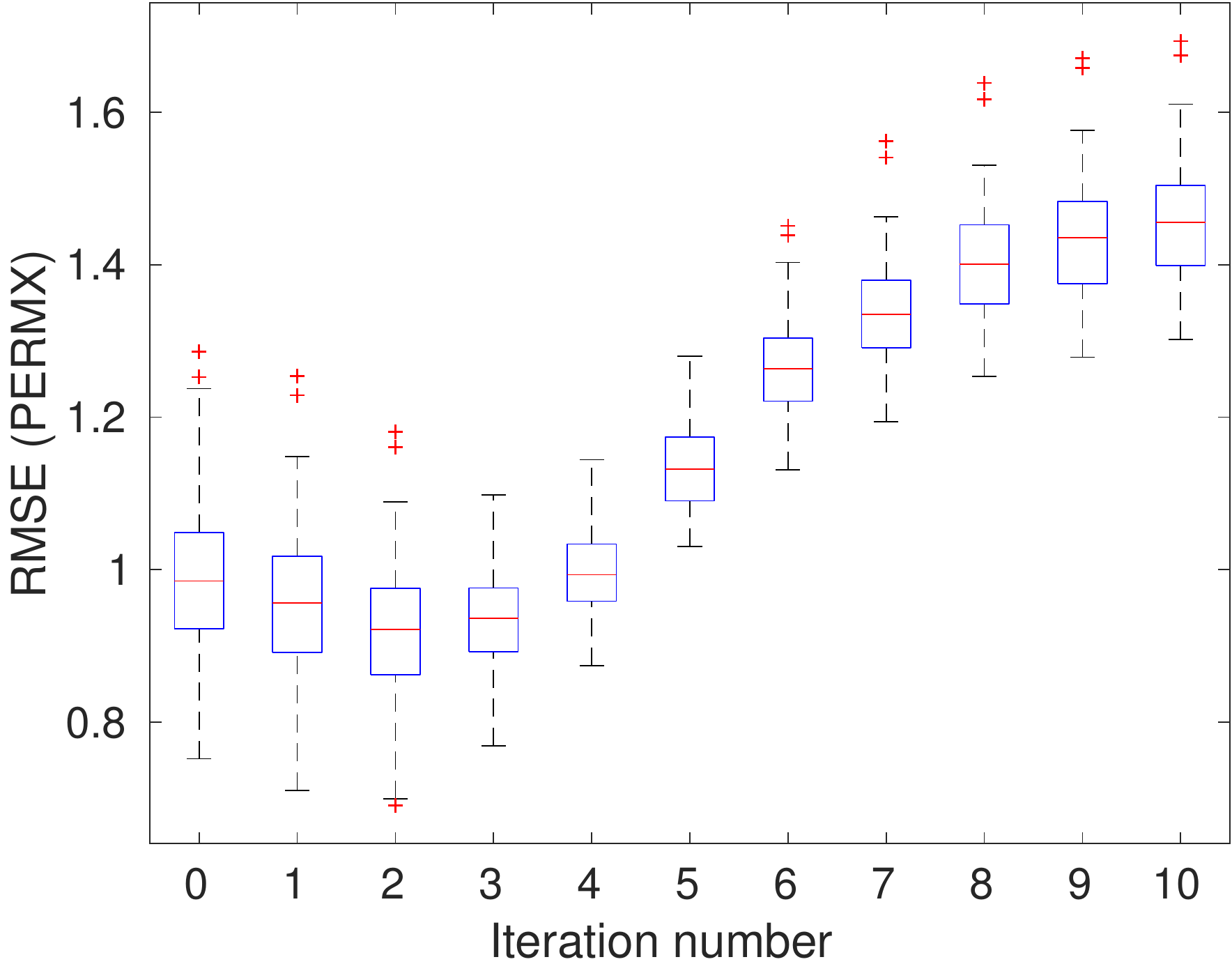}
				} 			
	\subfigure[RMSEs of PORO (sparse-data)]{ \label{subfig:rmse_PORO_boxplot_ensemble_S3_sparse}
					\includegraphics[scale=\nScale]{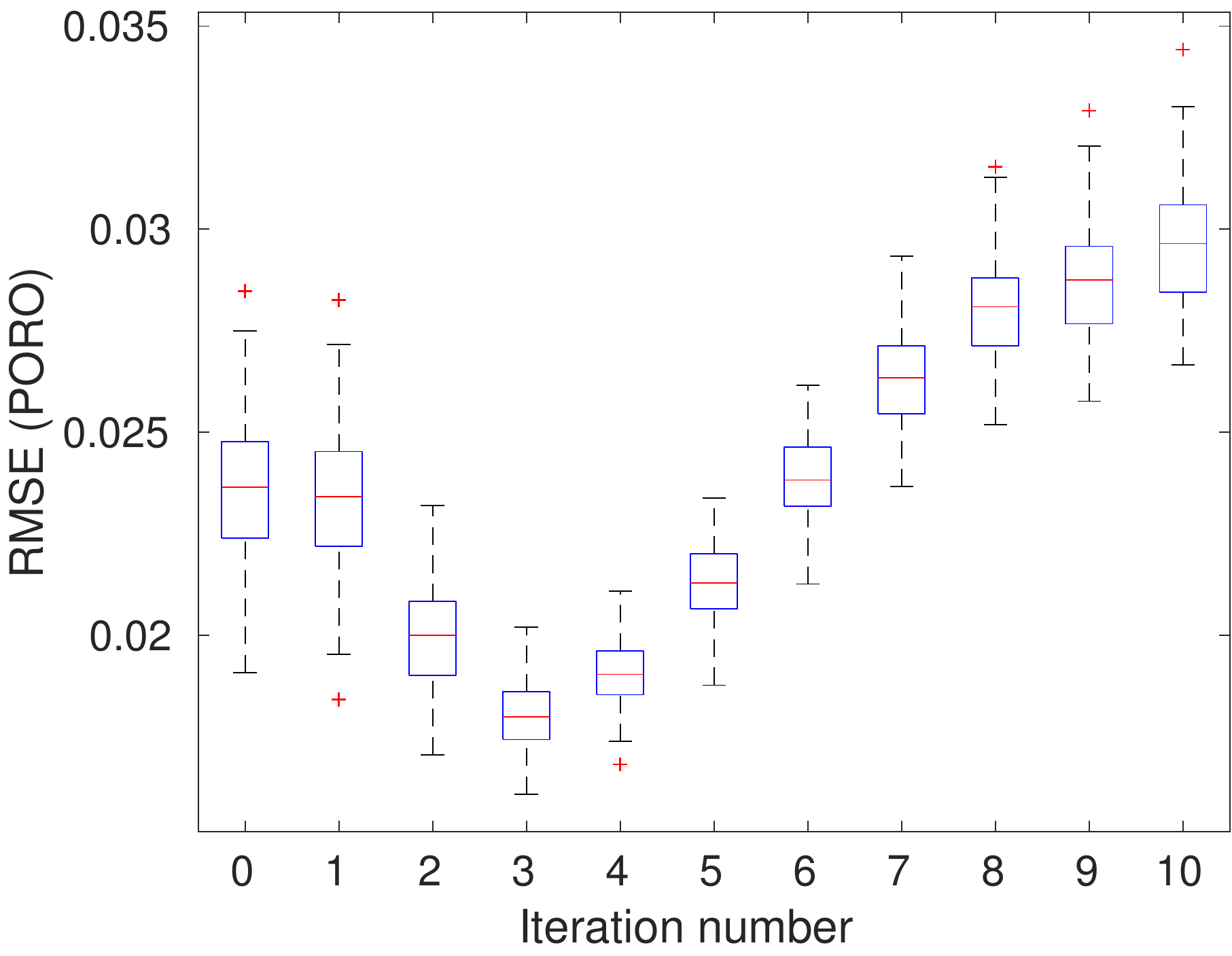}
				}		
											
	\caption{\label{fig:Norne2D_rmse_S3} Boxplots of RMSEs as functions of iteration step (scenario S3). Sub-figures (a) and (b) in the first row are for the results of full-data experiment, and (c) and (d) in the second row for the results of sparse-data experiment.}
\end{figure*} 

\renewcommand{\nScale}{0.3}
\begin{figure*} %
	\centering
	\subfigure[BHP (bar)]{ \label{subfig:WBHP_I1_final_forecasts_S3}
				\includegraphics[scale=0.3]{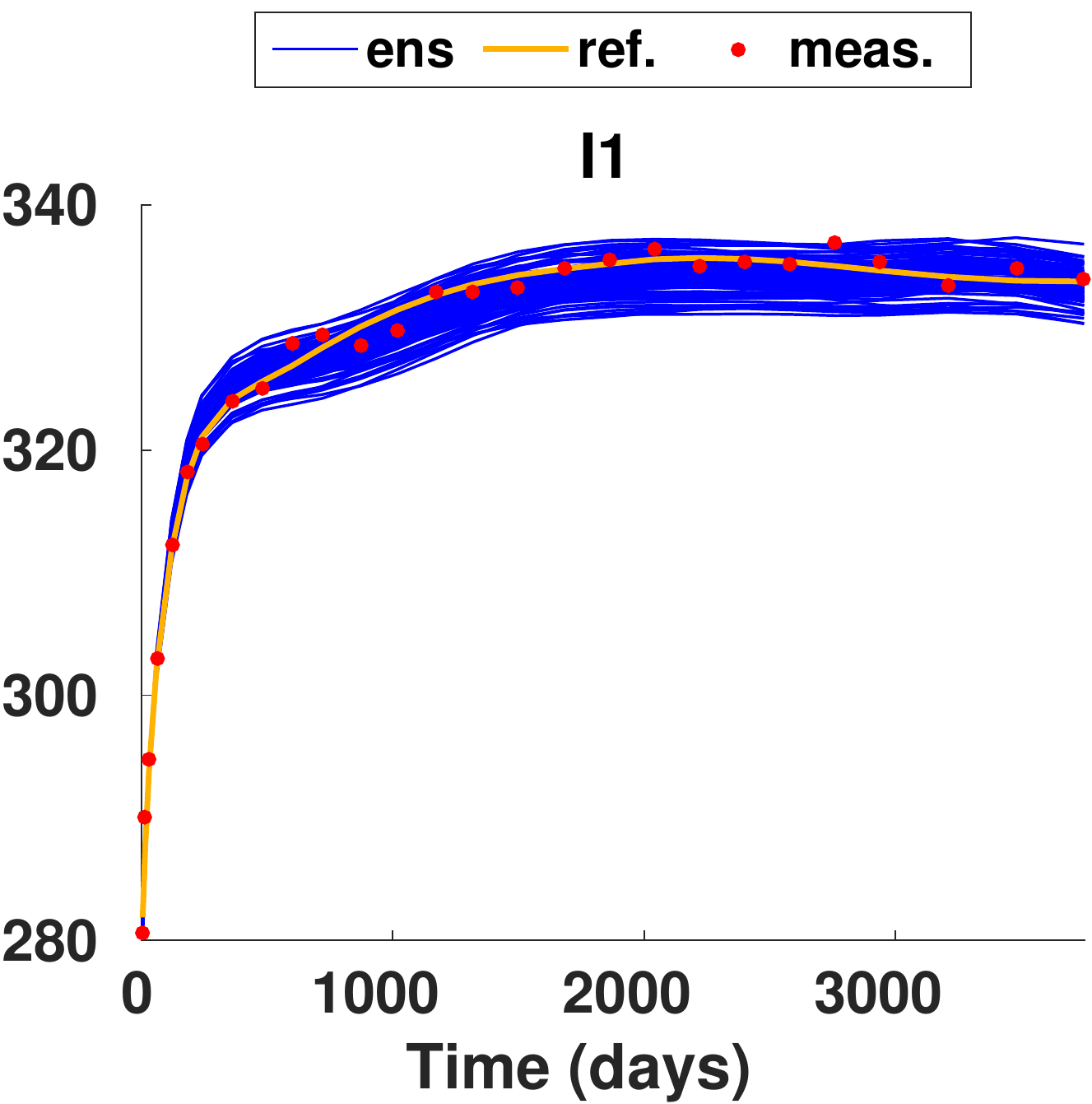}
			}%
	\subfigure[BHP (bar)]{ \label{subfig:WBHP_P1_final_forecasts_S3}
					\includegraphics[scale=0.3]{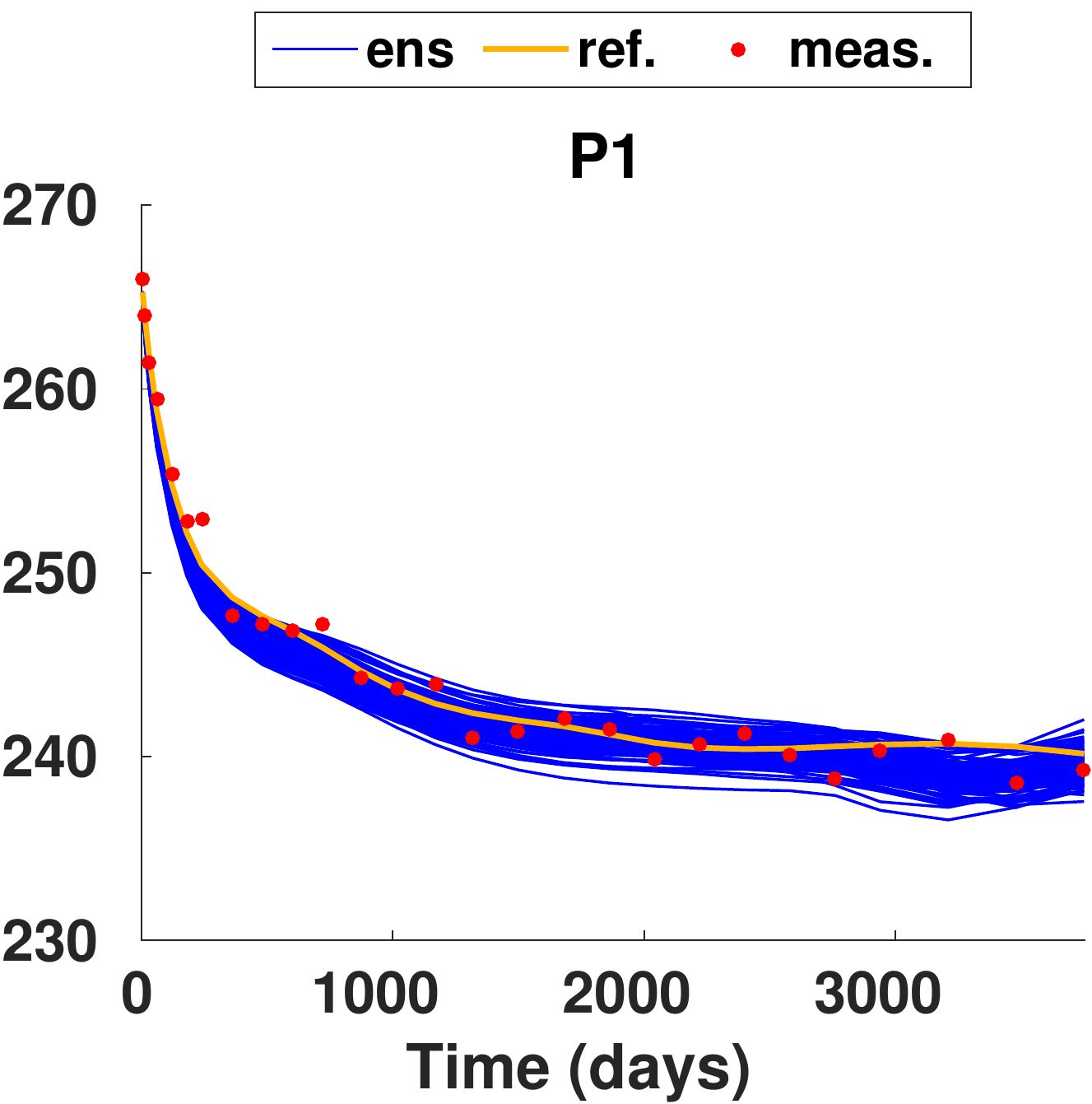}
				}
				
	\subfigure[GOR]{ \label{subfig:WGOR_P1_final_forecasts_S3}
					\includegraphics[scale=0.3]{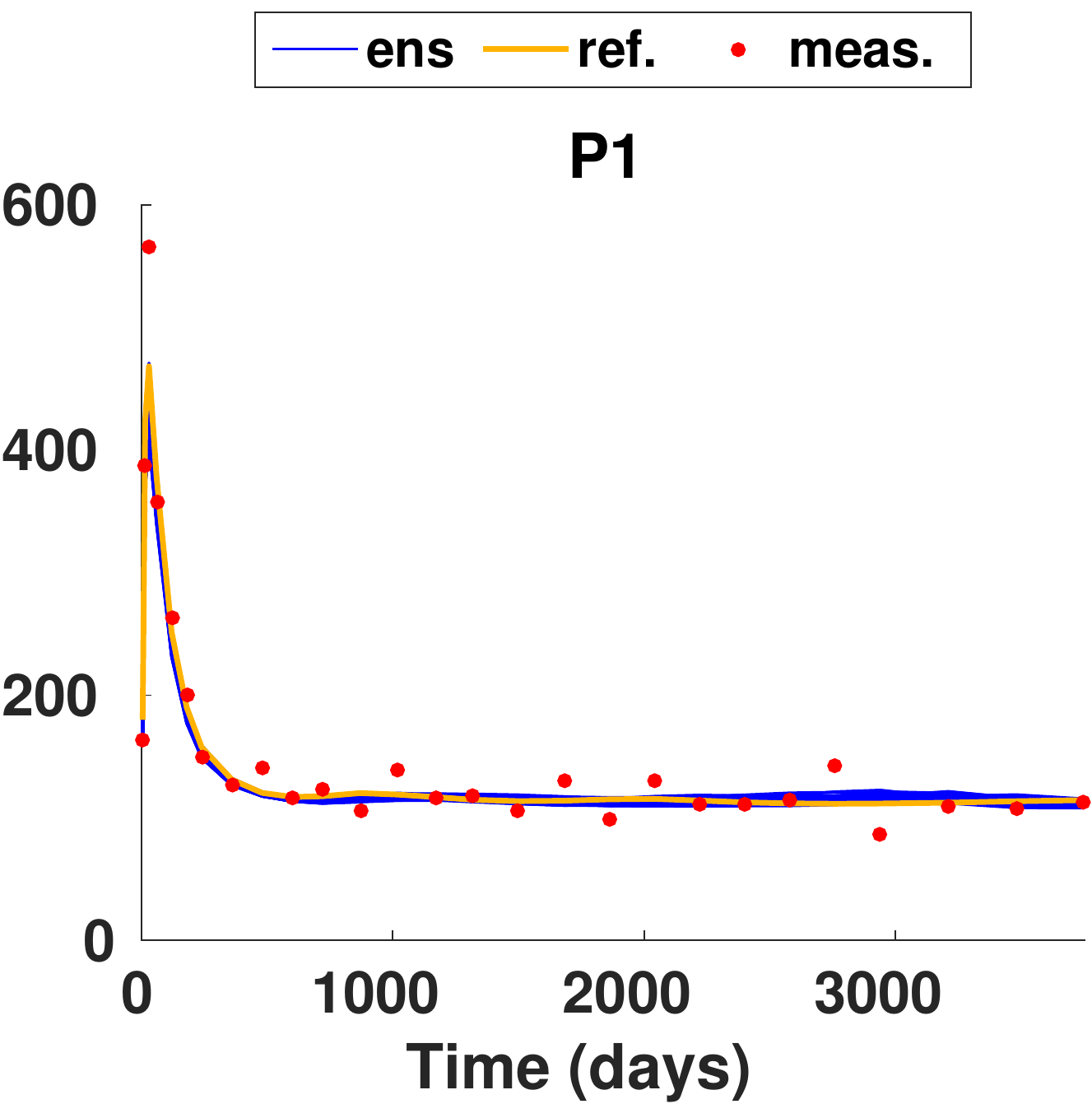}
				}%
	\subfigure[OPT (sm$^3$)]{ \label{subfig:WOPT_P1_final_forecasts_S3}
					\includegraphics[scale=0.3]{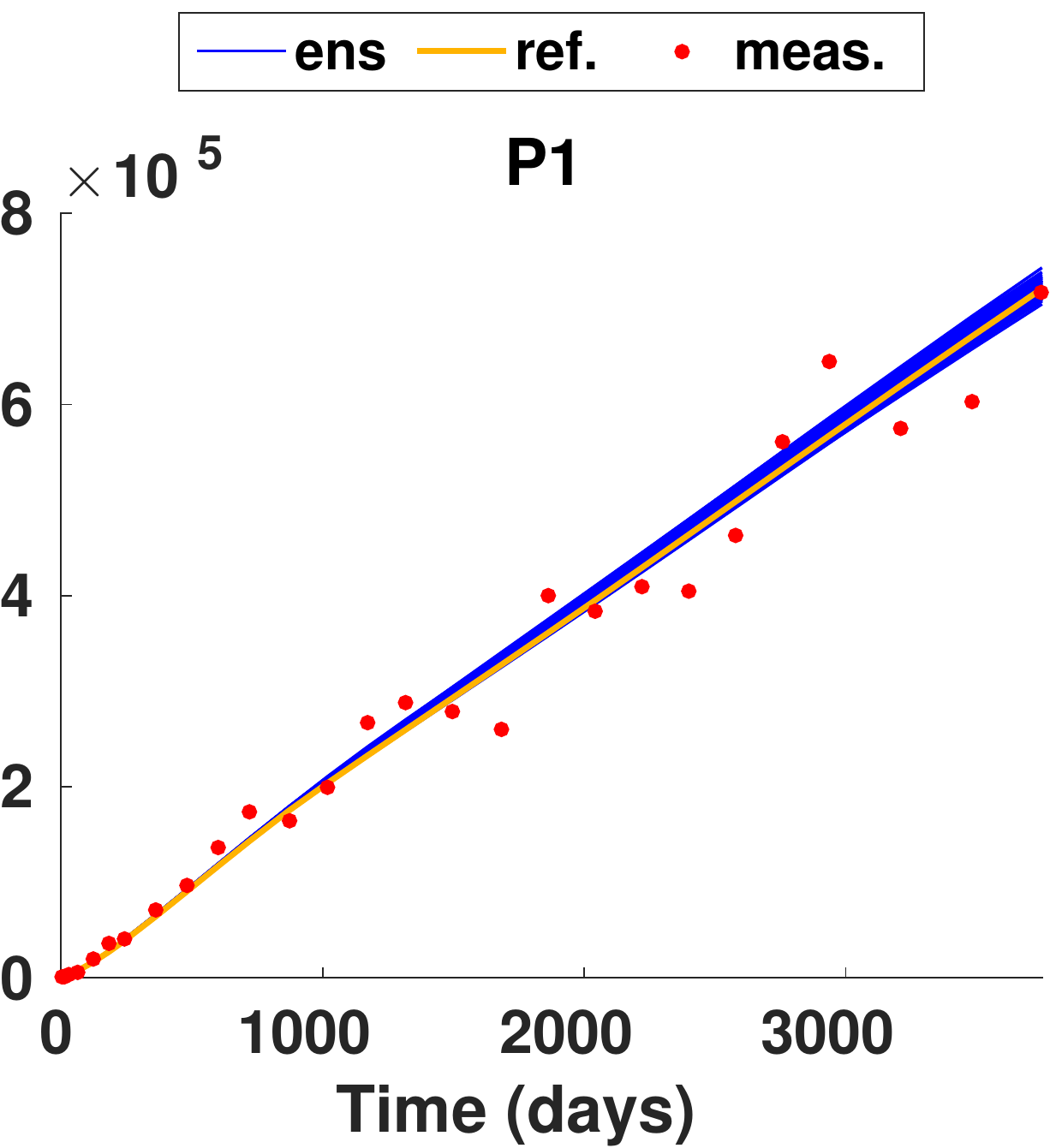}
				}%
	\subfigure[WCT]{ \label{subfig:WWCT_P1_final_forecasts_S3}
					\includegraphics[scale=0.3]{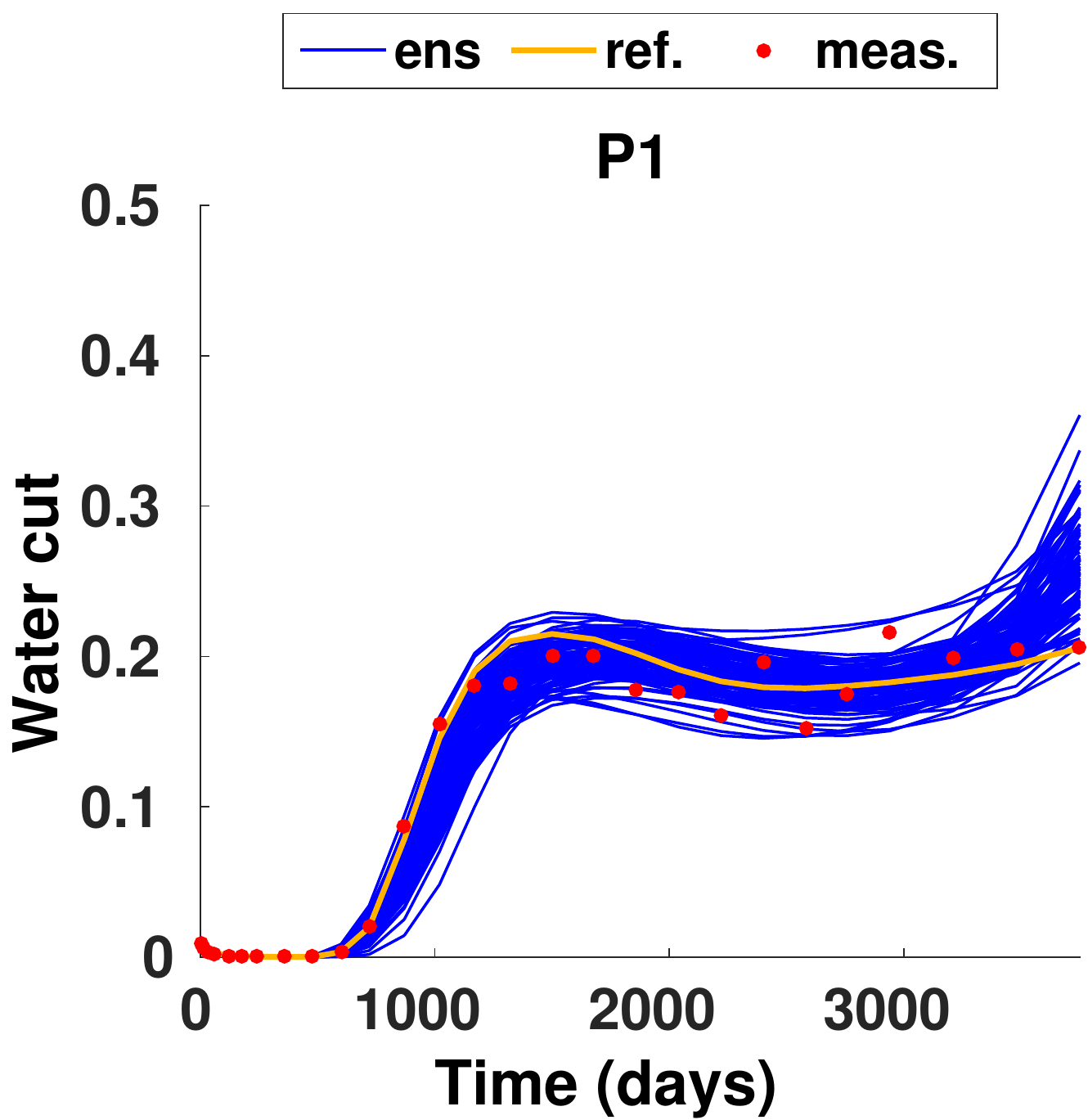}
				}											
	\caption{\label{fig:Norne2D_production_profile_final_S3} As in Figure \ref{fig:Norne2D_production_profile_final_S1}, but now blue curves correspond to production data forecasts with respect to the ensemble obtained at the 3rd iteration step, using both production and seismic data (scenario S3, sparse-data experiment).}
\end{figure*} 

\renewcommand{\nScale}{0.3}
\begin{figure*} %
	\centering
	
	\subfigure[Initial ensemble, base survey]{ \label{subfig:diff_initEnsMean_R0_traces_seisTimeStep1_S3}
					\includegraphics[scale=0.3]{./figures/diff_initEnsMean_R0_traces_seisTimeStep1_S2.eps}
				}%
	\subfigure[Initial ensemble, 1st monitor survey]{ \label{subfig:diff_initEnsMean_R0_traces_seisTimeStep2_S3}
					\includegraphics[scale=0.3]{./figures/diff_initEnsMean_R0_traces_seisTimeStep2_S2.eps}
				}%
	\subfigure[Initial ensemble, 2nd monitor survey]{ \label{subfig:diff_initEnsMean_R0_traces_seisTimeStep3_S3}
						\includegraphics[scale=0.3]{./figures/diff_initEnsMean_R0_traces_seisTimeStep3_S2.eps}
					} 		
				
	\subfigure[Final ensemble, base survey]{ \label{subfig:diff_finalEnsMean_R0_traces_seisTimeStep1_S3}
					\includegraphics[scale=0.3]{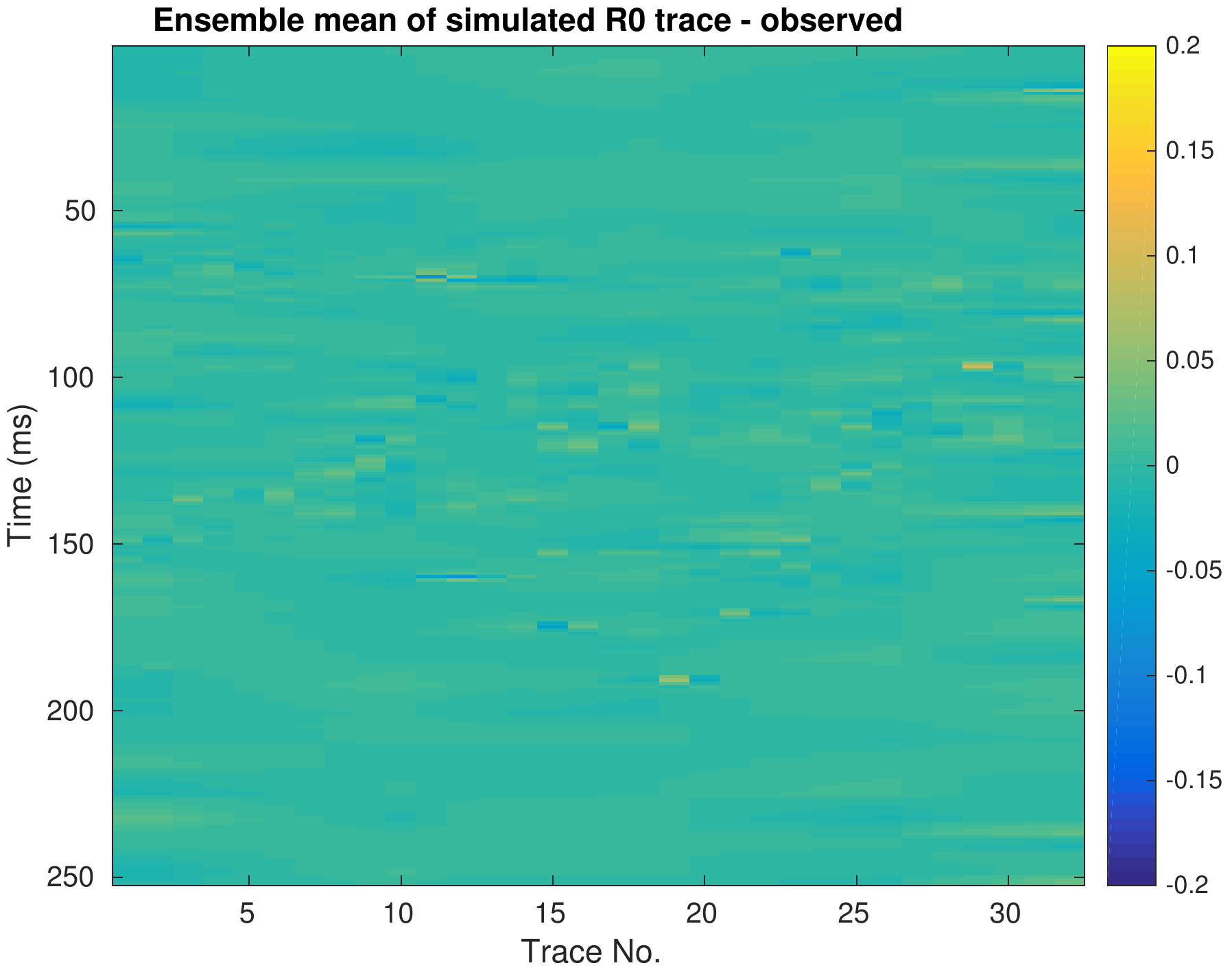}
				}%
	\subfigure[Final ensemble, 1st monitor survey]{ \label{subfig:diff_finalEnsMean_R0_traces_seisTimeStep2_S3}
					\includegraphics[scale=0.3]{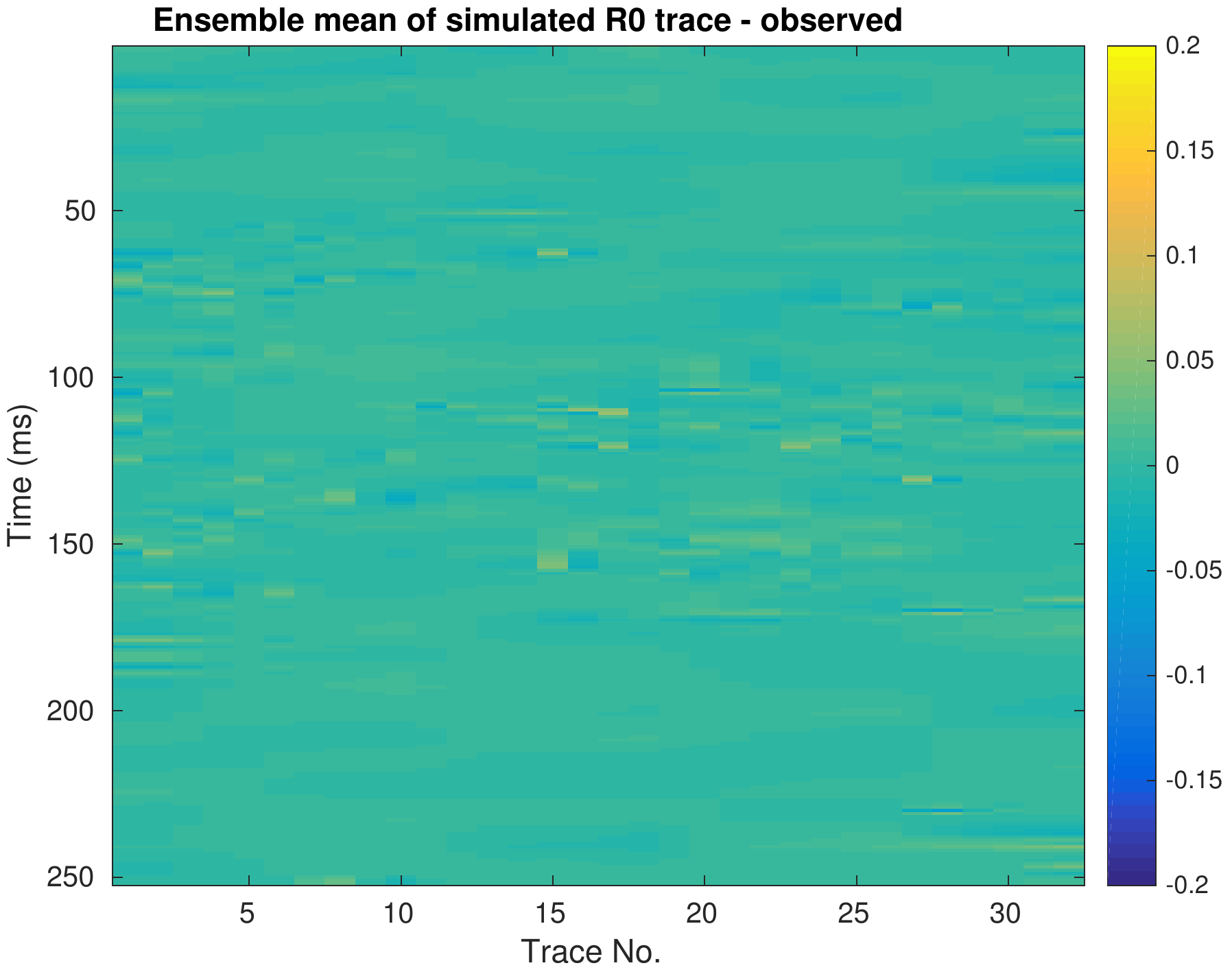}
				}%
	\subfigure[Final ensemble, 2nd monitor survey]{ \label{subfig:diff_finalEnsMean_R0_traces_seisTimeStep3_S3}
						\includegraphics[scale=0.3]{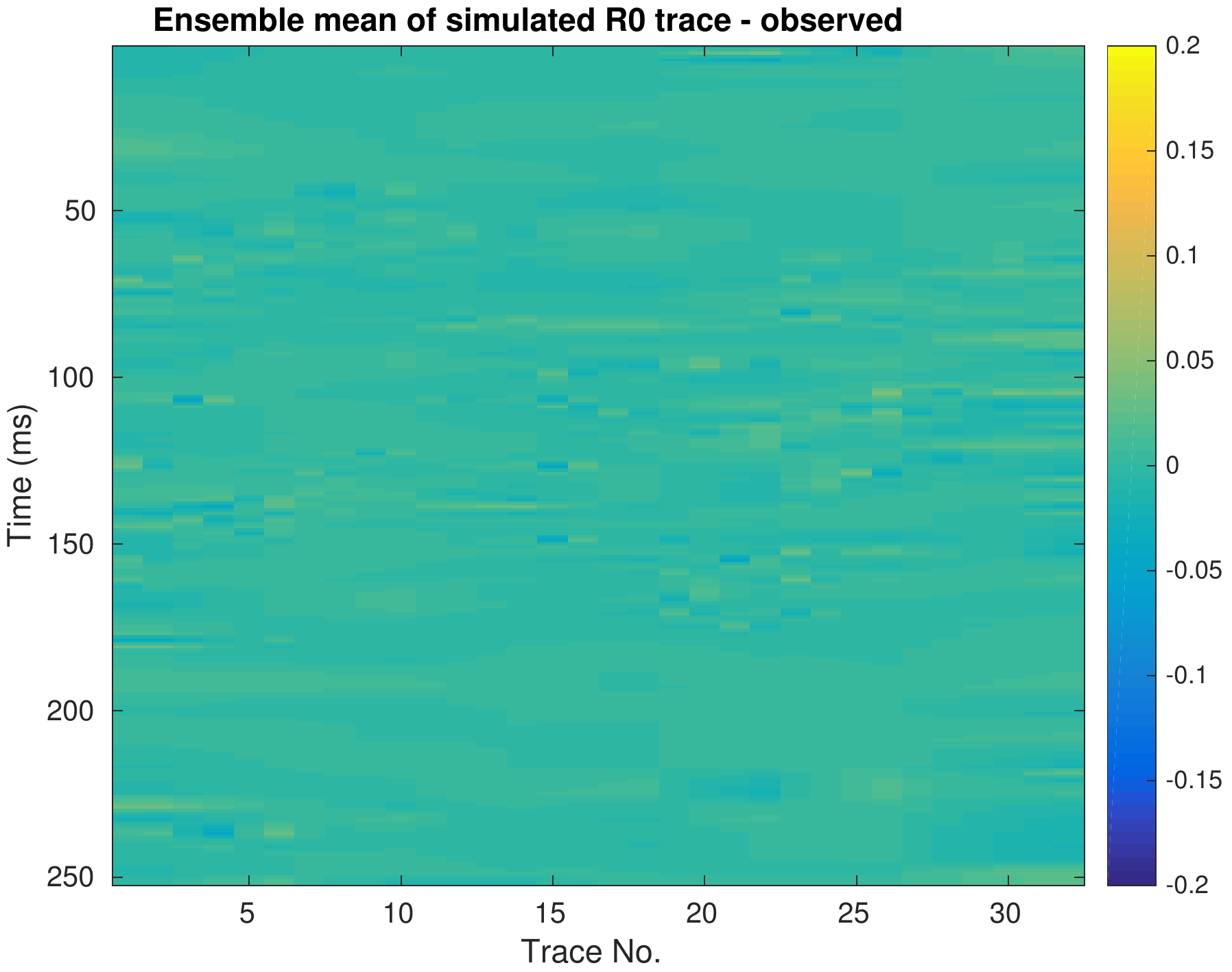}
					}						
	\caption{\label{fig:Norne2D_diff_R0_S3} Differences between ensemble means of reconstructed simulated intercepts and reconstructed observed intercepts at three survey time instances in scenario S3. Sub-figures in the first row (a--c) are for the results with respect to the initial ensemble, whereas those in the second row (d -- f) for the results with respect to the ensemble obtained at the 3rd iteration step in sparse-data experiment.}
\end{figure*} 

\renewcommand{\nScale}{0.3}
\begin{figure*} %
	\centering
	
	\subfigure[Initial ensemble, base survey]{ \label{subfig:diff_initEnsMean_G_traces_seisTimeStep1_S3}
					\includegraphics[scale=0.3]{./figures/diff_initEnsMean_G_traces_seisTimeStep1_S2.eps}
				}%
	\subfigure[Initial ensemble, 1st monitor survey]{ \label{subfig:diff_initEnsMean_G_traces_seisTimeStep2_S3}
					\includegraphics[scale=0.3]{./figures/diff_initEnsMean_G_traces_seisTimeStep2_S2.eps}
				}%
	\subfigure[Initial ensemble, 2nd monitor survey]{ \label{subfig:diff_initEnsMean_G_traces_seisTimeStep3_S3}
						\includegraphics[scale=0.3]{./figures/diff_initEnsMean_G_traces_seisTimeStep3_S2.eps}
					} 		
				
	\subfigure[Final ensemble, base survey]{ \label{subfig:diff_finalEnsMean_G_traces_seisTimeStep1_S3}
					\includegraphics[scale=0.3]{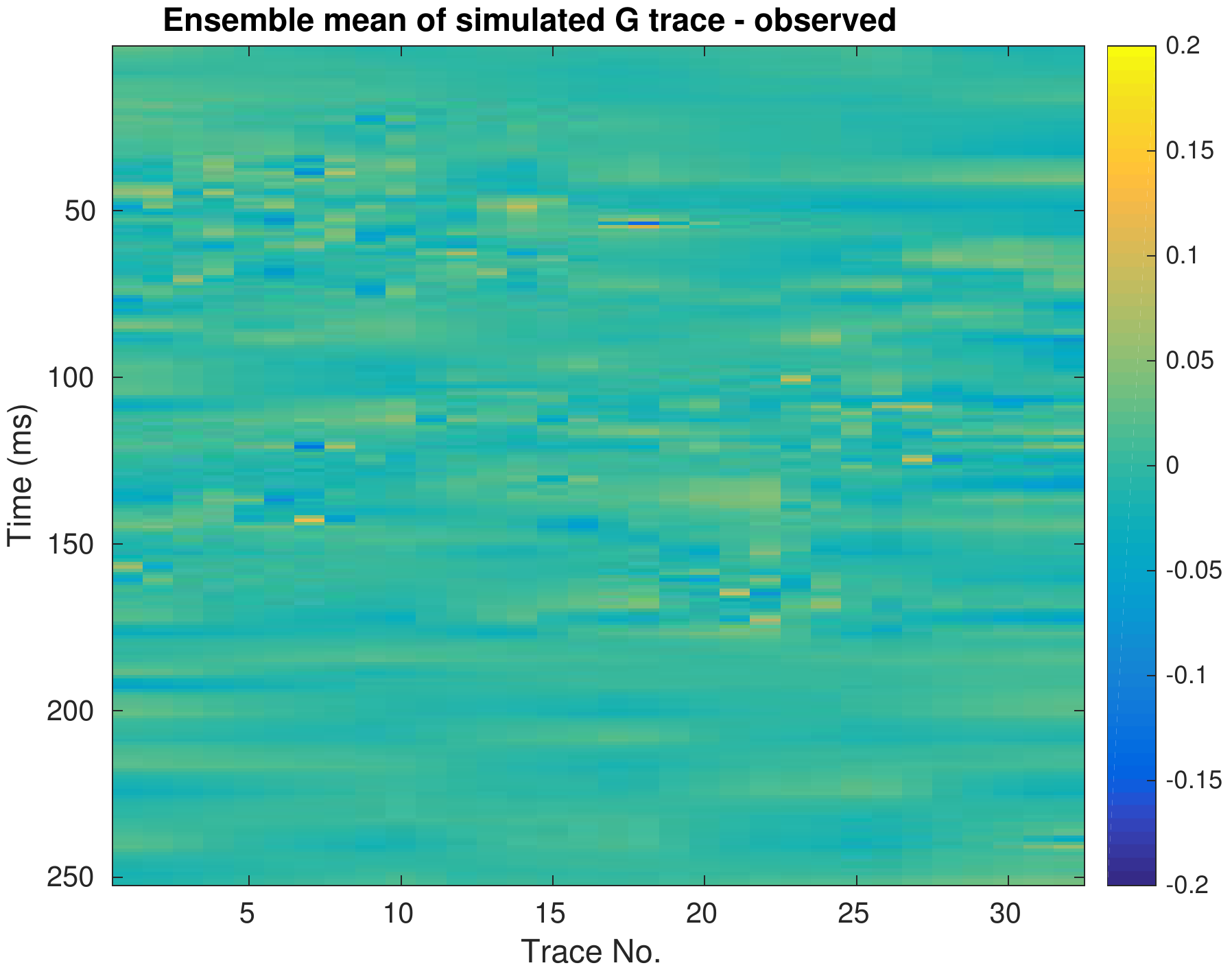}
				}%
	\subfigure[Final ensemble, 1st monitor survey]{ \label{subfig:diff_finalEnsMean_G_traces_seisTimeStep2_S3}
					\includegraphics[scale=0.3]{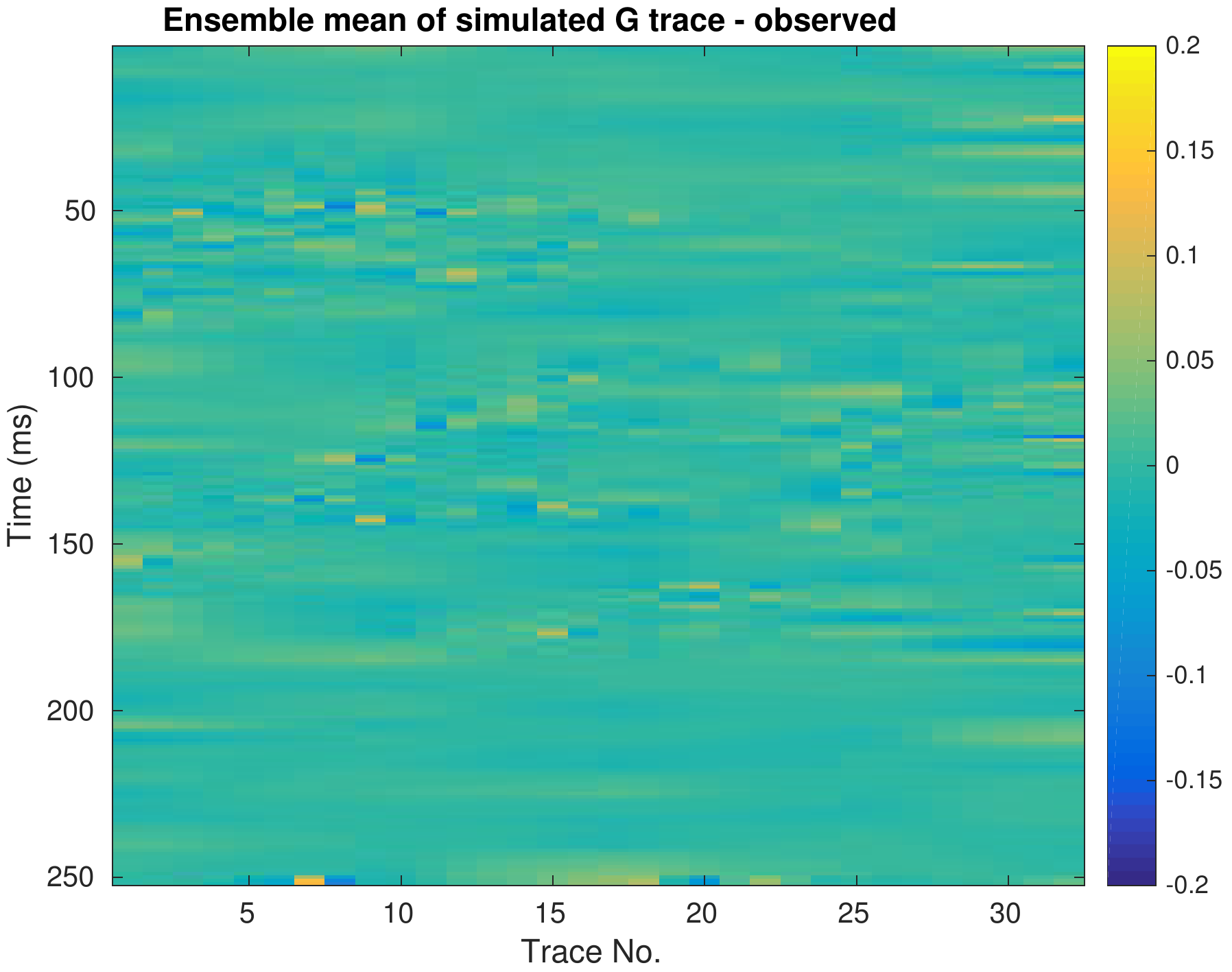}
				}%
	\subfigure[Final ensemble, 2nd monitor survey]{ \label{subfig:diff_finalEnsMean_G_traces_seisTimeStep3_S3}
						\includegraphics[scale=0.3]{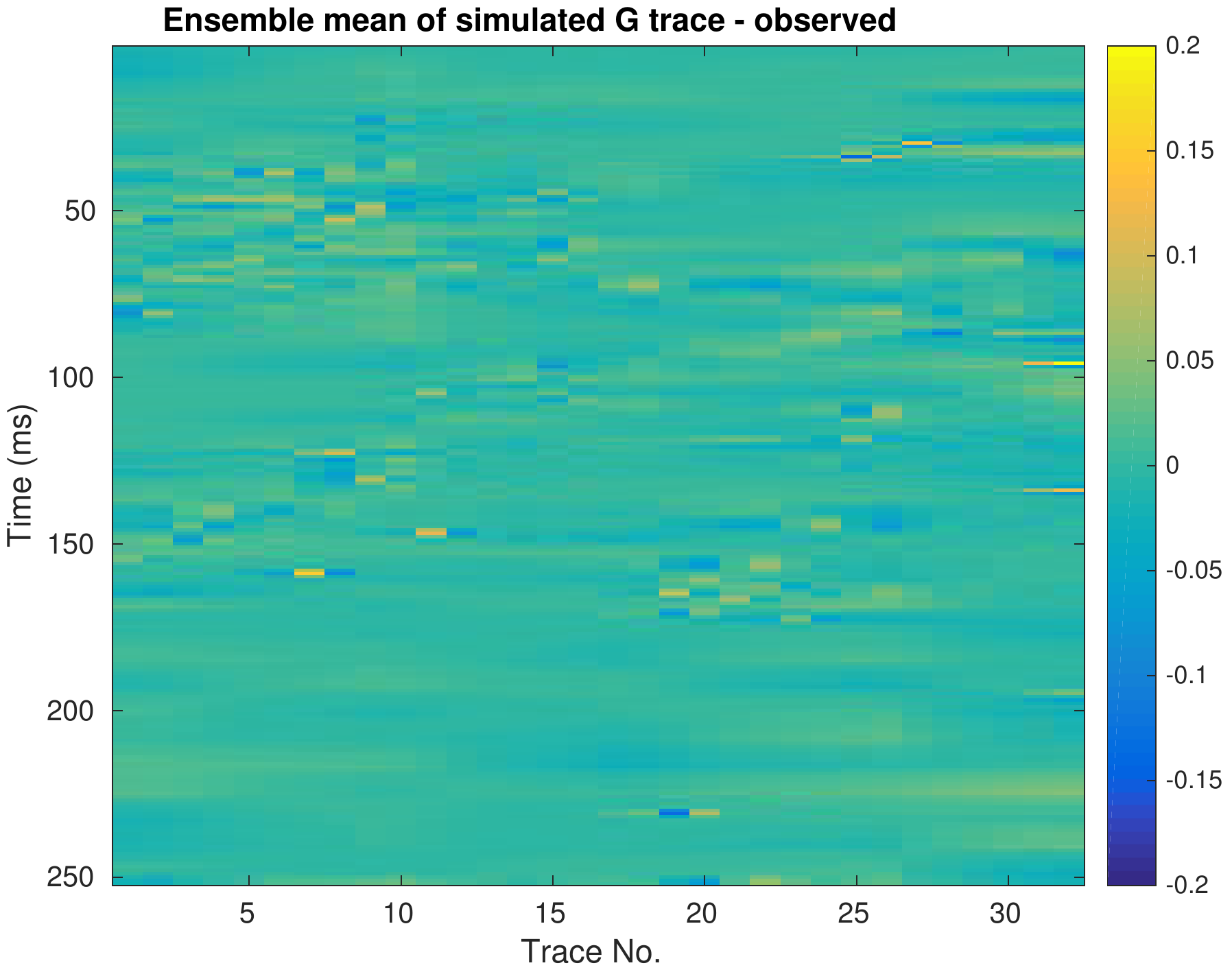}
					}						
	\caption{\label{fig:Norne2D_diff_G_S3} As in Figure \ref{fig:Norne2D_diff_R0_S3}, but now for gradient attributes in scenario S3.}
\end{figure*}  

\renewcommand{\nScale}{0.4}
\begin{figure*} 
	\centering
	\hspace{+1.5cm} \subfigure[Reference]{ \label{subfig:permx_ref_S3}
				\includegraphics[scale=0.32]{./figures/permx_ref_S2.eps}
			}  
				
	\subfigure[Initial mean]{ \label{subfig:field_PERMX_mean_init_ensemble_S3}
					\raisebox{2.5mm}{\includegraphics[scale=0.32]{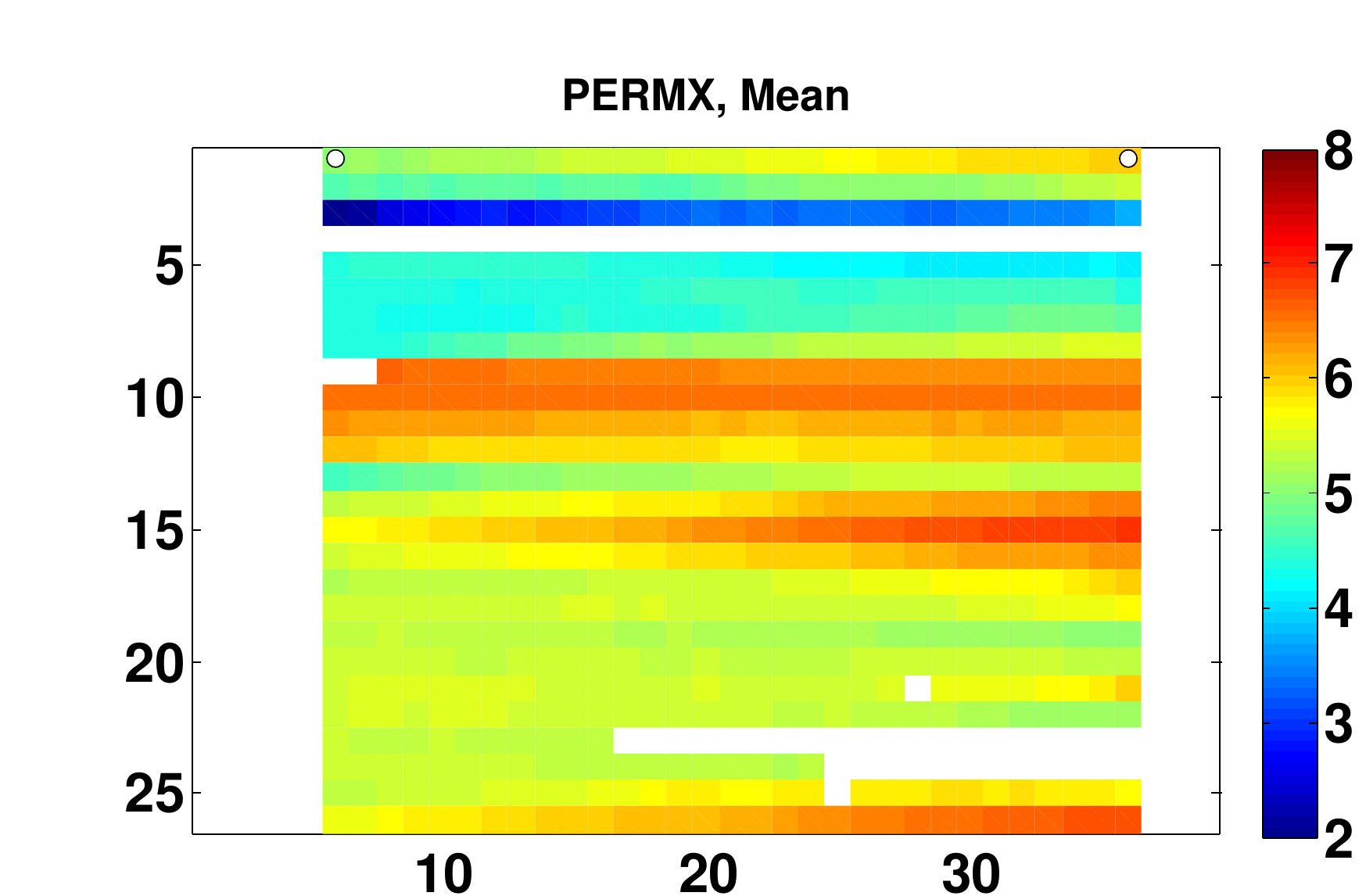}}
				}%
	\subfigure[Initial member 1]{ \label{subfig:field_PERMX_1_1_init_ensemble_S3}
					\includegraphics[scale=0.4]{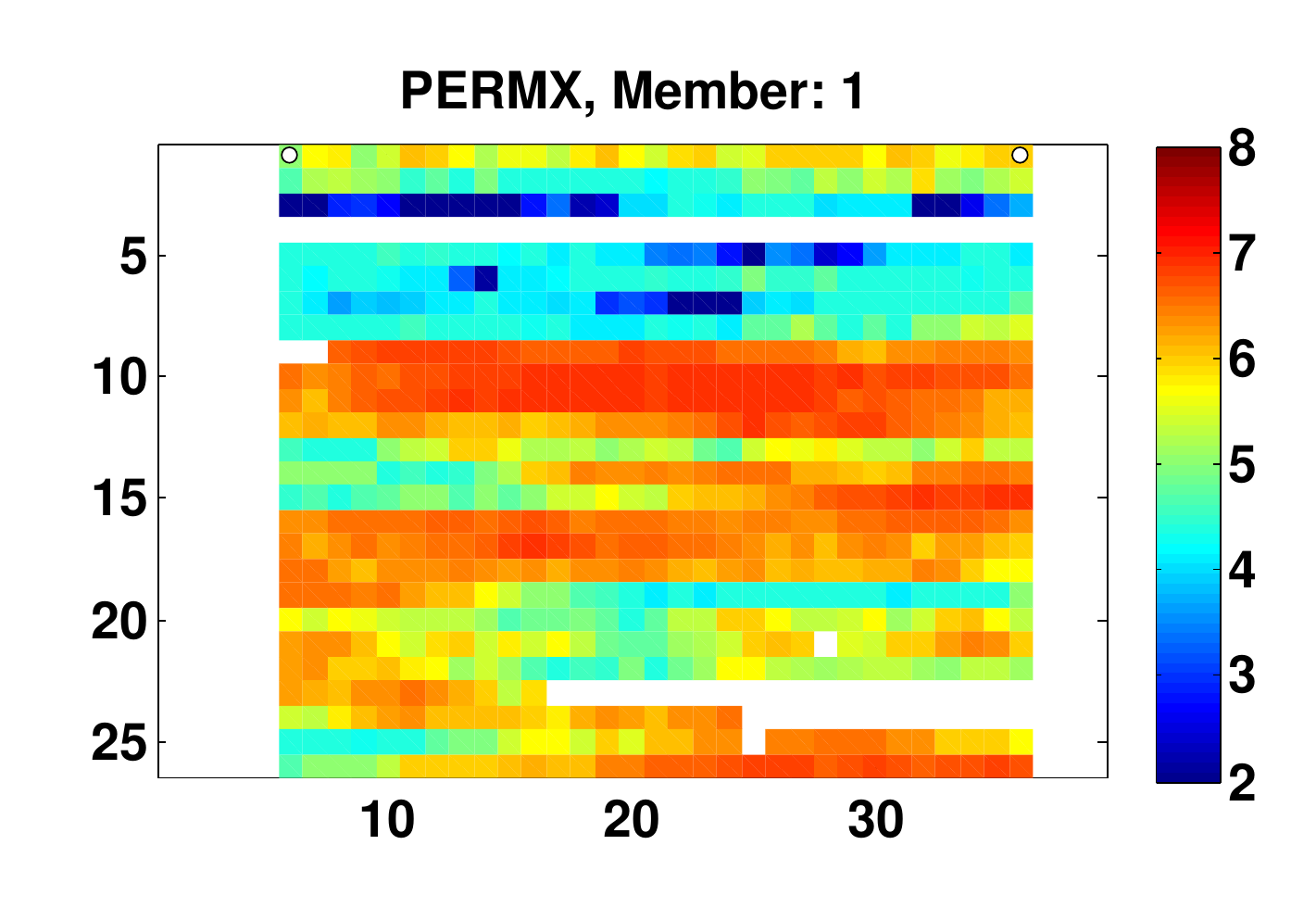}
				}%
	\subfigure[Initial member 2]{ \label{subfig:field_PERMX_1_2_init_ensemble_S3}
					\includegraphics[scale=0.4]{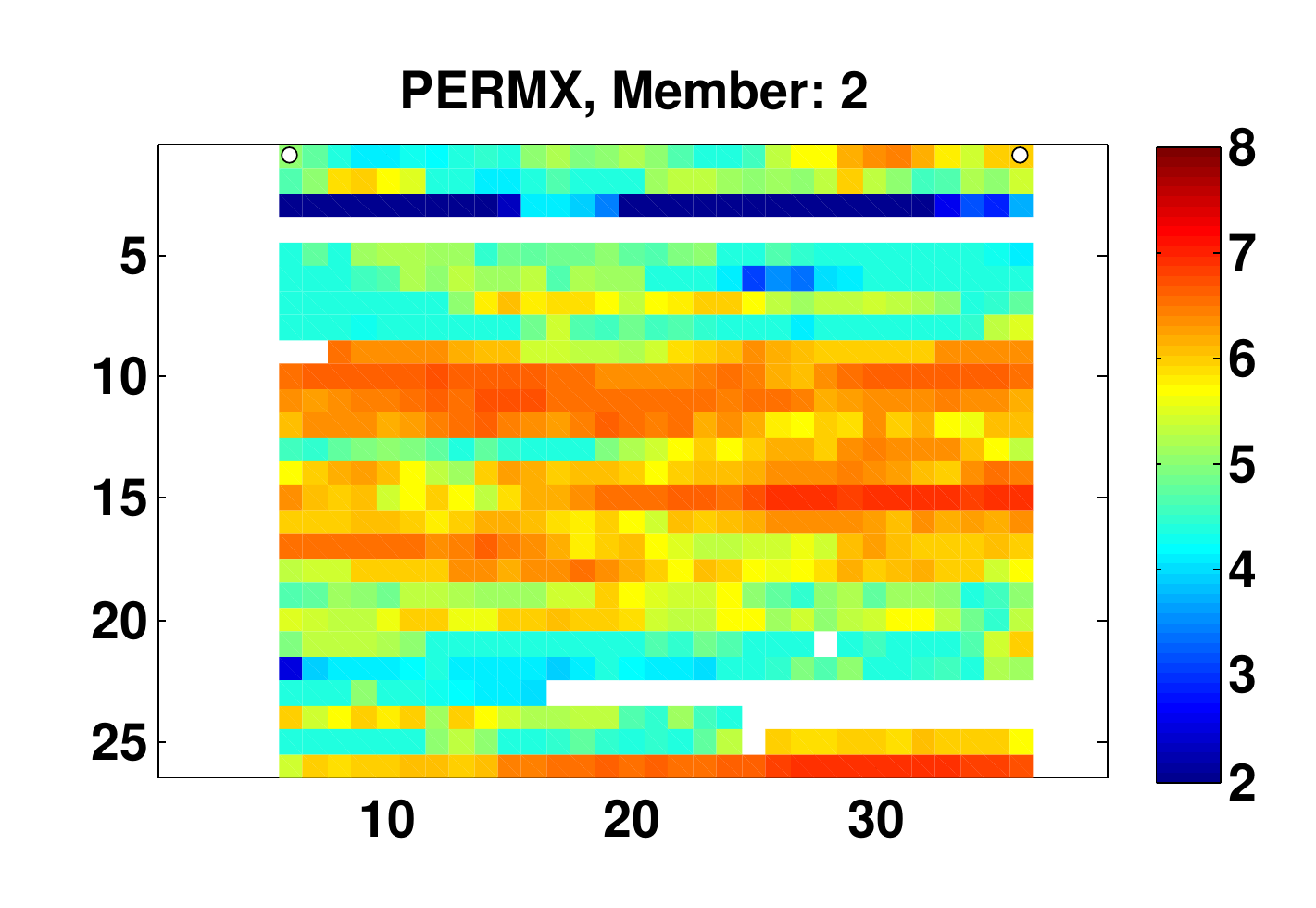}
				}		
				
	\subfigure[Final mean]{ \label{subfig:field_PERMX_mean_ensemble2_S3}
					\raisebox{2.5mm}{\includegraphics[scale=0.32]{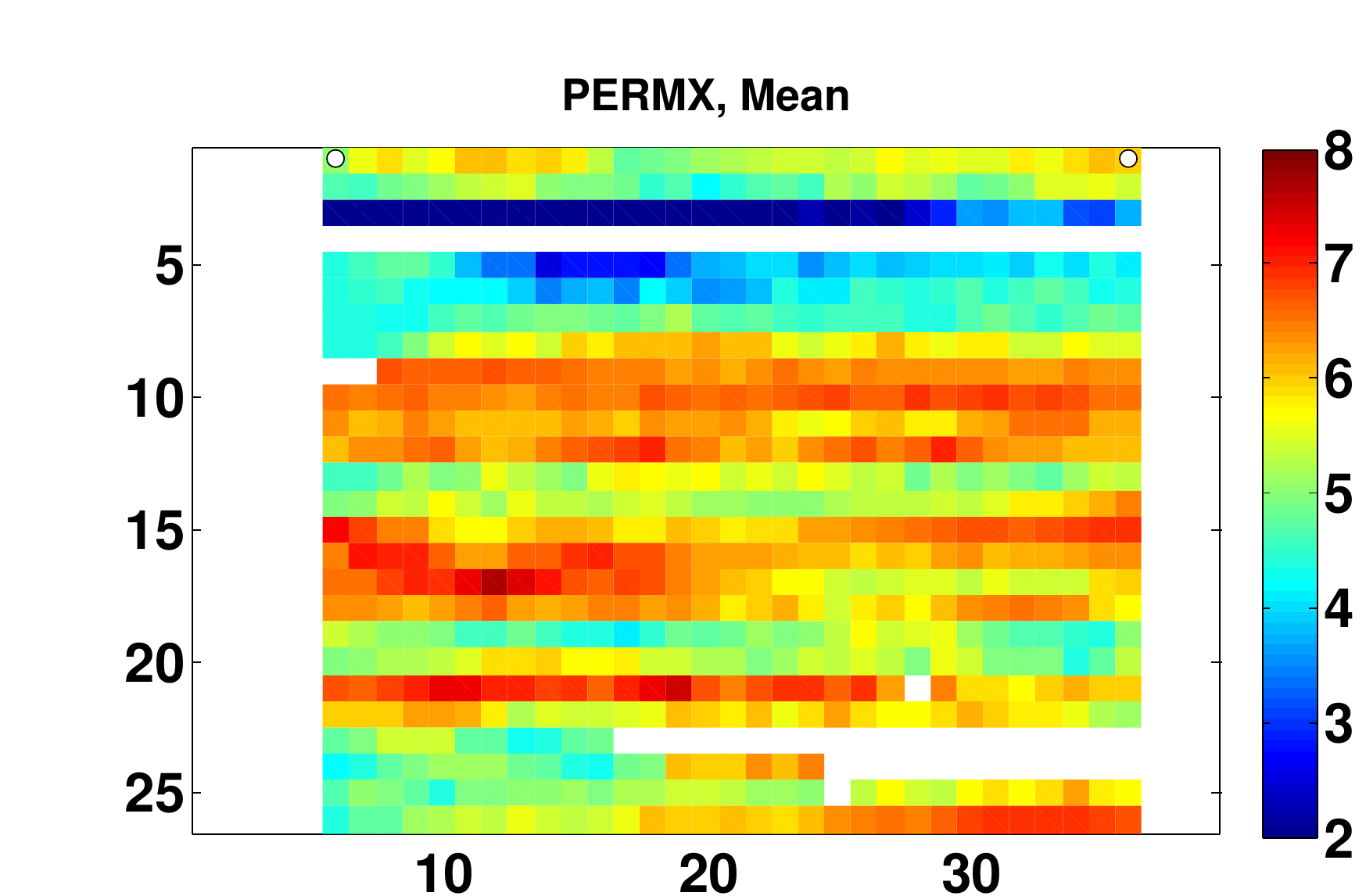}}
				}%
	\subfigure[Final member 1]{ \label{subfig:field_PERMX_1_1_ensemble2_S3}
					\includegraphics[scale=0.4]{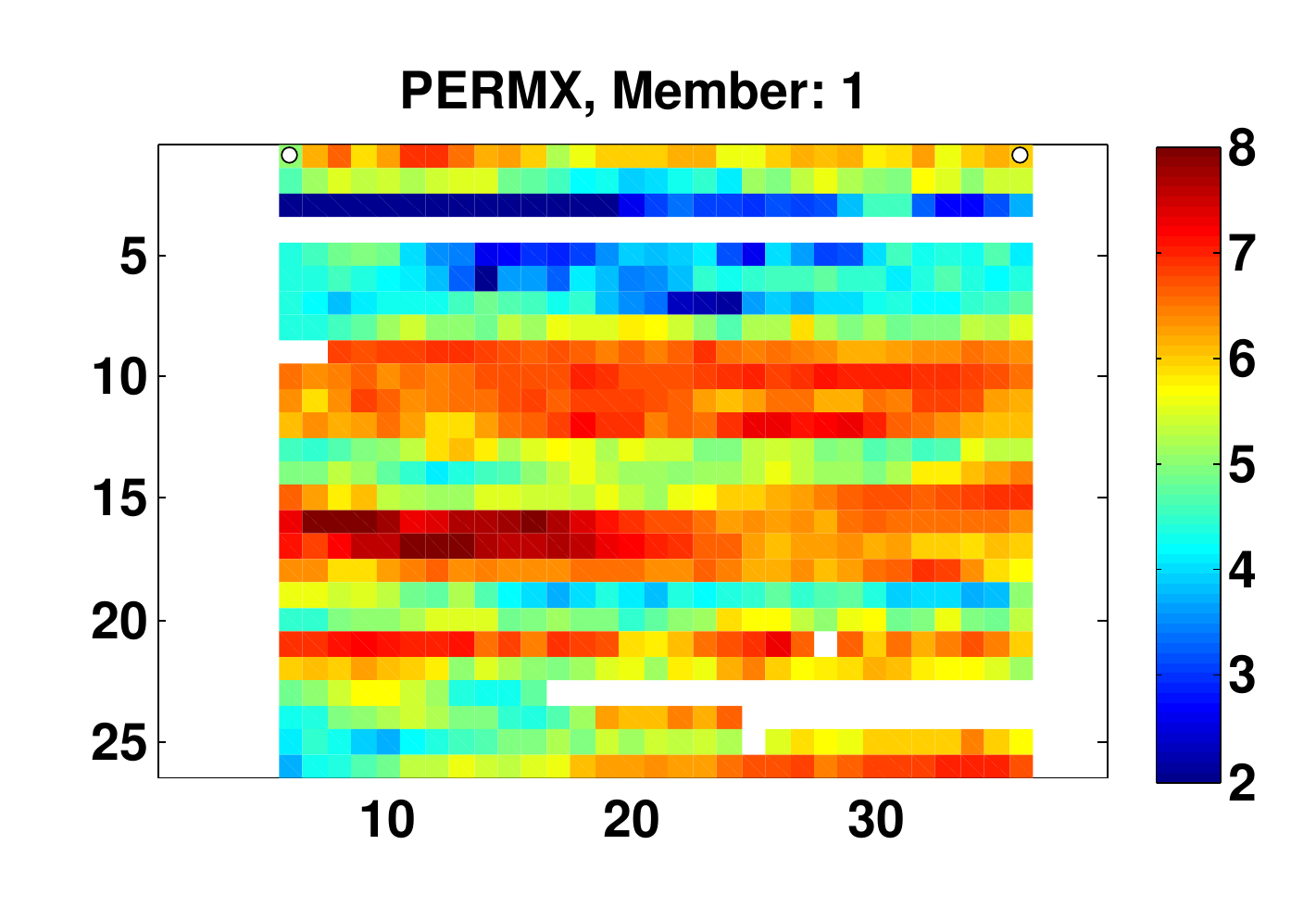}
				}%
	\subfigure[Final member 2]{ \label{subfig:field_PERMX_1_2_ensemble2_S3}
					\includegraphics[scale=0.4]{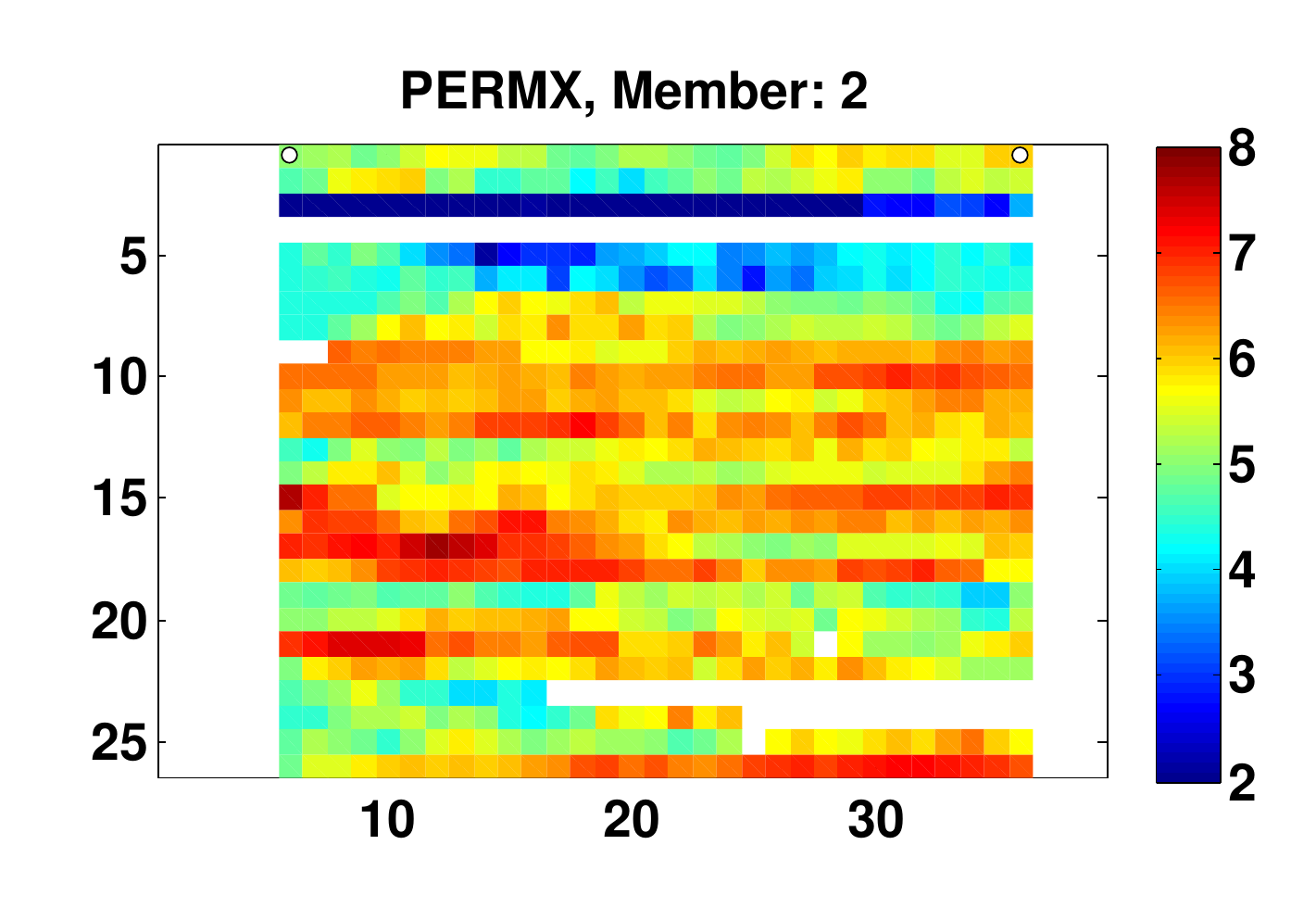}
				}															
	\caption{\label{fig:Norne2D_PERMX_S3} Distributions of log PERMX (scenario S3). (a) Reference model; (b) -- (d) Mean and 2 sample realizations of the initial ensemble of log PERMX; (e) -- (g) Corresponding mean and 2 sample realizations of the final ensemble.}
\end{figure*}   

\renewcommand{\nScale}{0.4}
\begin{figure*} %
	\centering
	\hspace{+2cm} \subfigure[Reference]{ \label{subfig:poro_ref_S3}
				\includegraphics[scale=0.32]{./figures/poro_ref_S2.eps}
			}
			
	\subfigure[Initial mean]{ \label{subfig:field_PORO_mean_init_ensemble_S3}
					\raisebox{2.5mm}{\includegraphics[scale=0.32]{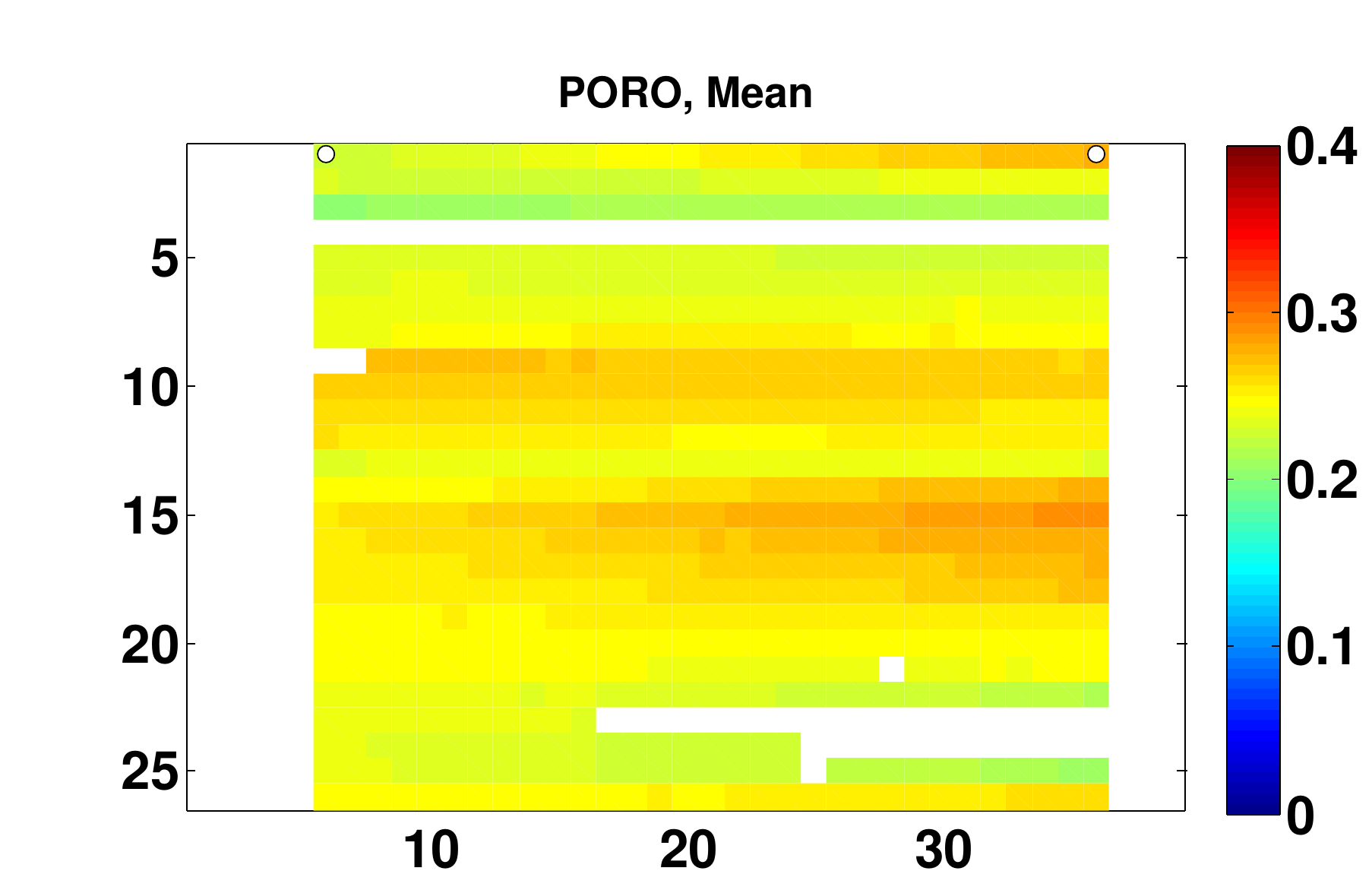}}
				}%
	\subfigure[Initial member 1]{ \label{subfig:field_PORO_1_1_init_ensemble_S3}
					\includegraphics[scale=0.4]{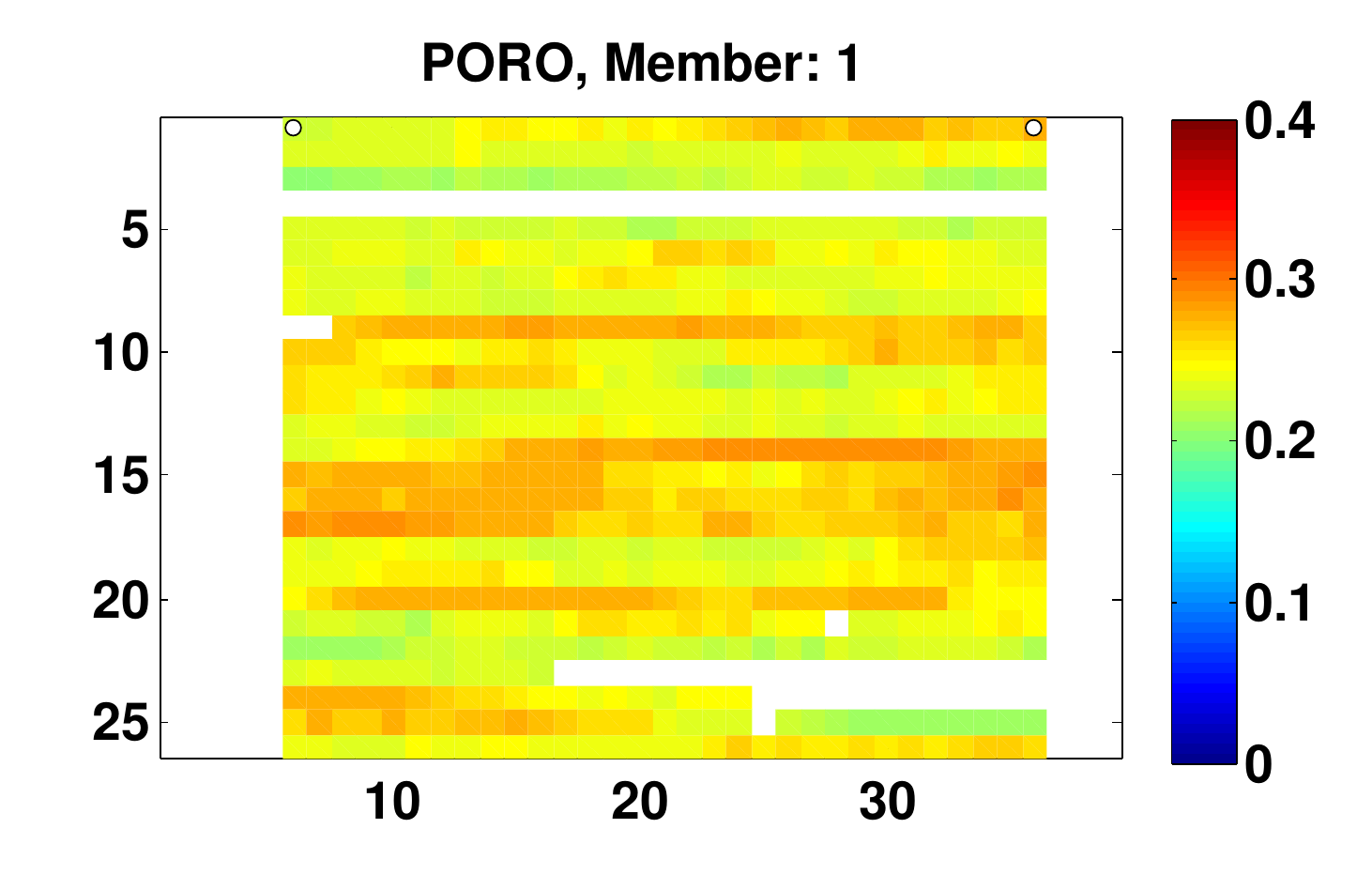}
				}%
	\subfigure[Initial member 2]{ \label{subfig:field_PORO_1_2_init_ensemble_S3}
					\includegraphics[scale=0.4]{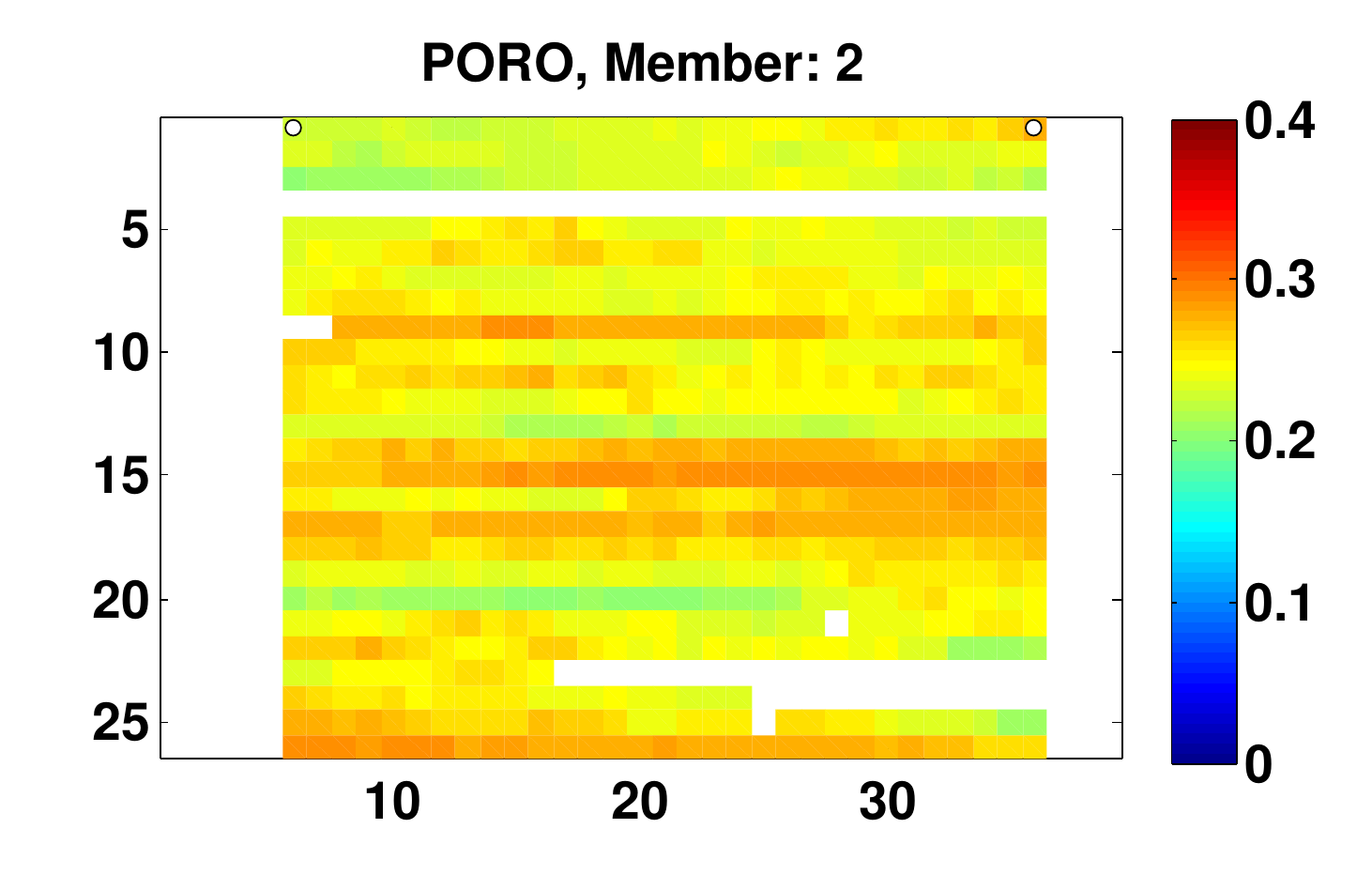}
				}		
				
	\subfigure[Final mean]{ \label{subfig:field_PORO_mean_ensemble2_S3}
					\raisebox{2.5mm}{\includegraphics[scale=0.32]{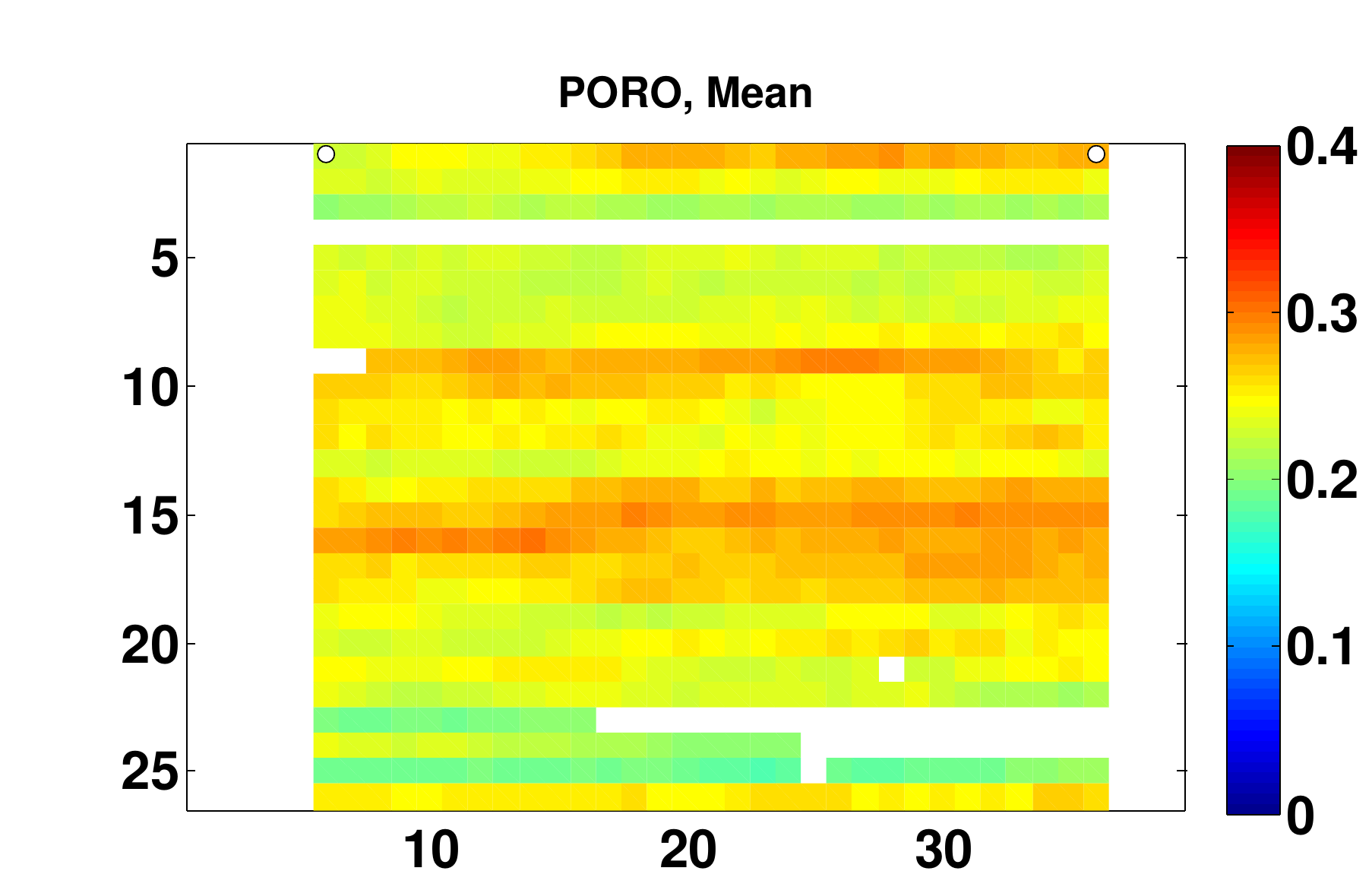}}
				}%
	\subfigure[Final member 1]{ \label{subfig:field_PORO_1_1_ensemble2_S3}
					\includegraphics[scale=0.4]{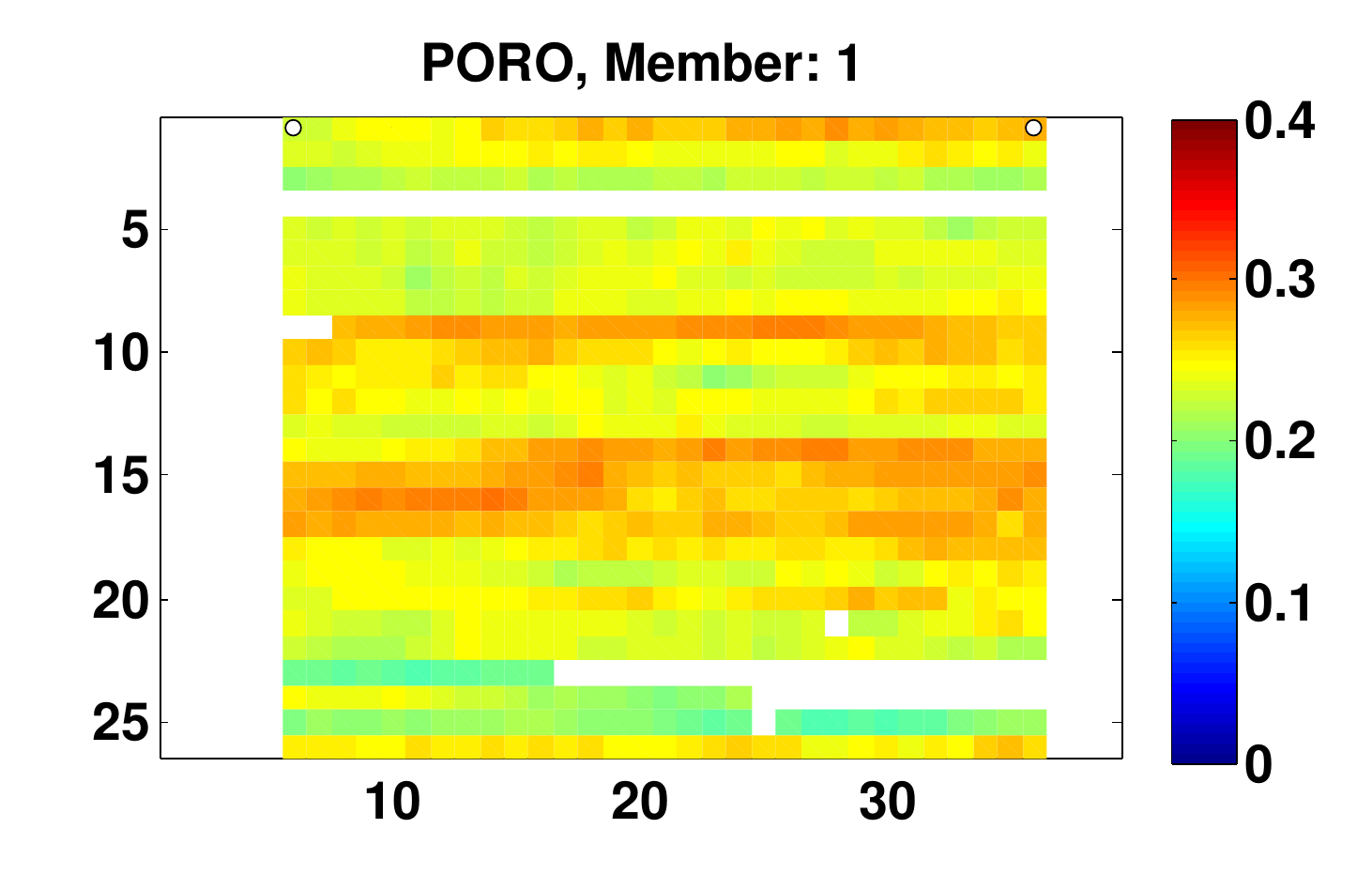}
				}%
	\subfigure[Final member 2]{ \label{subfig:field_PORO_1_2_ensemble2_S3}
					\includegraphics[scale=0.4]{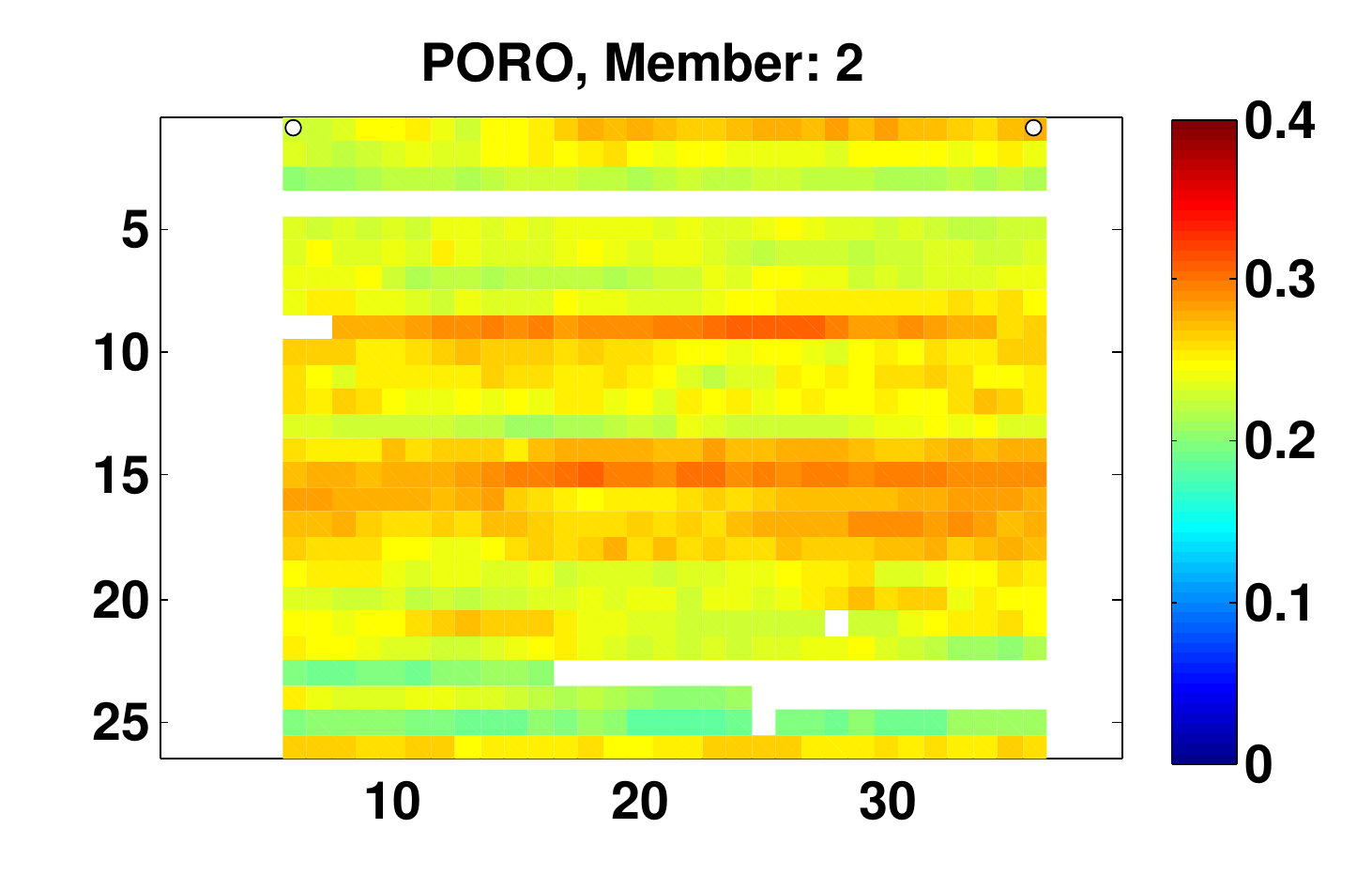}
				}															
	\caption{\label{fig:Norne2D_PORO_S3} As in Figure \ref{fig:Norne2D_PERMX_S3}, but now for the distributions of PORO in scenario S3.}
\end{figure*}            

Finally we consider the scenario in which both production and seismic data are used in history matching. In the context of iES, we build an augmented observation vector that concatenates all production and seismic data at different time steps, such that RLM-MAC can be applied in the same way as in the previous two scenarios. In this scenario, the stopping criterion (\ref{eq:stopping_criterion_ndm}) applies to individual data mismatch with respect to production and seismic data, respectively. That is to say, RLM-MAC stops if data mismatch of production and seismic data both satisfies (\ref{eq:stopping_criterion_ndm}) for the first time (if applicable). As discussed in the previous scenarios, the threshold value of (\ref{eq:stopping_criterion_ndm}) is $540$ for production data, and around $1.87 \times 10^5$ (full-data experiment) or $1.10 \times 10^4$ (sparse-data experiment) for seismic data.    

Figure \ref{fig:Norne2D_obj_S3} reports data mismatch as functions of iteration step. In full-data experiment, the average data mismatch of production data satisfies the stopping criterion (\ref{eq:stopping_criterion_ndm}) from the 4th iteration on, whereas the average data mismatch of seismic data never hits at the threshold value. As a result, RLM-MAC stops at the 10th iteration step using stopping conditions (C1) and (C2), which leads to final average data mismatch of $324.29$ for production data and around $1.19 \times 10^8$ for seismic data. Alternatively, since the average data mismatch of production data can be lower than the threshold value (Figure \ref{subfig:Norne2D_objReal_prod_boxplot_objRealIter_S3_full}), one may apply the stopping criterion (\ref{eq:stopping_criterion_ndm}) to production data only and let RLM-MAC stop at the 4th iteration step, and this leads to final average data mismatch of $380.13$ for production data and around $1.20 \times 10^8$ for seismic data. In sparse-data experiment, the average data mismatch of both production and seismic data can satisfy the stopping criterion (\ref{eq:stopping_criterion_ndm}). Figures \ref{subfig:Norne2D_objReal_prod_boxplot_objRealIter_S3_sparse} and \ref{subfig:Norne2D_objReal_seis_boxplot_objRealIter_S3_sparse} together suggest that RLM-MAC stop at the 3rd iteration, and this corresponds to final average data mismatch of $356.48$ for production data and around $1.02 \times 10^4$ for seismic data. In contrast, if using stopping conditions (C1) and (C2) only, RLM-MAC stop at the 10th iteration, which corresponds to final average data mismatch of $200.83$ for production data and around $9.82 \times 10^3$ for seismic data.             

Figure \ref{fig:Norne2D_rmse_S3} shows RMSEs of log PERMX and PORO as functions of iteration step. In full-data experiment (Figure \ref{subfig:rmse_PERMX_boxplot_ensemble_S3_full} and \ref{subfig:rmse_PORO_boxplot_ensemble_S3_full}), the RMSEs of log PERMX and PORO exhibit U-turn behaviours, such that the RMSEs at the last few iteration steps tend to be larger than those at the intermediate steps. In addition, there are also signs of ensemble collapses at the last few iteration steps. As a result, even though the stopping criterion (\ref{eq:stopping_criterion_ndm}) is valid for production data only, combining stopping conditions (C1) and (C2) and  (\ref{eq:stopping_criterion_ndm}) leads to improved estimation results in terms of RMSE, in comparison with the choice of using stopping conditions (C1) and (C2) only. In sparse-data experiment (Figure \ref{subfig:rmse_PERMX_boxplot_ensemble_S3_sparse} and \ref{subfig:rmse_PORO_boxplot_ensemble_S3_sparse}), we observe a similar effect of the stopping criterion (\ref{eq:stopping_criterion_ndm}) on the final estimation results, but in this case ensemble collapses are avoided.    

To see the effects of using production and/or seismic data on history matching in this particular example, in Table \ref{tab:average_RMSE} we compare the average RMSEs of log PERMX and PORO in all three scenarios (S1 -- S3), where the stopping criterion (\ref{eq:stopping_criterion_ndm}) is used as far as possible. When only using production data in S1 (Figure \ref{fig:Norne2D_RLM-MAC_RMSE_S1}), the average RMSEs of log PERMX and PORO (at the 3rd iteration) are 0.8941 and 0.0230, respectively, while those of the initial ensemble are 0.9952 and 0.0236, respectively. Therefore, the final estimations of both log PERMX and PORO are improved in comparison with the initial ensemble. The average RMSE of log PERMX in S1 is the lowest among all RMSEs of log PERMX in Table \ref{tab:average_RMSE}, and is about $10.2\%$ less than that of the initial ensemble. Nevertheless, the improvement for PORO appears less significant, and is only about $2.5\%$ less than that of the initial ensemble. When only using seismic data in S2 (Figure \ref{fig:Norne2D_rmse_S2}), in either full- or sparse-data experiment (with the final iteration steps being 10 and 2, respectively), the estimations of PORO are improved compared to the initial ensemble, whereas the estimations of log PERMX become worse. Finally, when using both production and seismic data in S3 (Figure \ref{fig:Norne2D_rmse_S3}), in either full- or sparse-data experiment (with the final iteration steps being 4 and 3, respectively), the estimations of log PERMX and PORO are both improved compared to the initial ensemble. The RMSEs of log PERMX and PORO in full-data experiment are $0.9454$ and $0.0188$, respectively, while the RMSEs in sparse-data experiment are $0.9389$ and $0.0180$, respectively, slightly lower than those in full-data experiment. The RMSE of PORO in sparse-data experiment of S3 is the lowest among all RMSEs of PORO in Table \ref{tab:average_RMSE}, and is about $23.7\%$ less than that of the initial ensemble. Meanwhile, The RMSE of log PERMX in sparse-data experiment of S3 is about $5.7\%$ less than that of the initial ensemble. Overall, in this particular example, it appears that production data are useful for the estimation of log PERMX, and that seismic data are important for the estimation of PORO. History matching both production and seismic data (in sparse-data experiment) leads to a better balance between the estimation accuracies of log PERMX and PORO.              


For succinctness, in what follows we show the results of sparse-data experiment in S3, where the final ensemble is the one obtained at the 3rd iteration step. The production data profiles with respect to the final ensemble are plotted in Figure \ref{fig:Norne2D_production_profile_final_S3}. Compared to the corresponding profiles of the final ensemble in S1 (Figure \ref{fig:Norne2D_production_profile_final_S1}), we see that they appear similar, but exhibit slight differences in some cases (e.g., Figure \ref{subfig:WWCT_P1_final_forecasts_S3} versus Figure \ref{subfig:WWCT_P1_final_forecasts_S1}). Figures \ref{fig:Norne2D_diff_R0_S3} and \ref{fig:Norne2D_diff_G_S3} show differences between ensemble means of reconstructed simulated attributes and reconstructed observed attributes at three survey time instances in sparse-data experiment of S3, and they are very similar to the results in S2 (Figures \ref{fig:Norne2D_diff_R0_S2} and \ref{fig:Norne2D_diff_G_S2}). Figures \ref{fig:Norne2D_PERMX_S3} and \ref{fig:Norne2D_PORO_S3} report some distributions of log PERMX and PORO. Comparing the results of initial and final ensembles (e.g., Figure \ref{subfig:field_PERMX_mean_init_ensemble_S3} versus Figure \ref{subfig:field_PERMX_mean_ensemble2_S3}), we also see substantial changes in both log PERMX and PORO after history matching, consistent with the results in Figure \ref{fig:Norne2D_rmse_S3}.

\section{Conclusion and future works}\label{sec:conclusion}
In this study we propose an ensemble 4D seismic history matching framework with wavelet multiresolution analysis. The seismic data are intercept and gradient attributes of amplitude versus angle (AVA) data. To reduce data size, we adopt sparse representation of seismic data in wavelet domain. In the history matching workflow, we apply wavelet transforms to seismic data, estimate the noise in wavelet domain, and use an iterative ensemble smoother to history-match leading wavelet coefficients of seismic data. 

As a proof-of-concept study, we apply the proposed framework to a 2D synthetic case. In the experiments, to prevent the iterative ensemble smoother from over-fitting the observations, we introduce an extra stopping criterion (\ref{eq:stopping_criterion_ndm}), apart from the stopping conditions (C1) and (C2). Moreover, to see the impacts of production and/or seismic data on the history matching performance, we investigate three scenarios (S1 -- S3) that involve (S1) production data only, (S2) 4D seismic data only, and (S3) both production and 4D seismic data. In addition, to examine the effects of sparse representation in scenarios S2 and S3, we also compare the history matching performance in two types of experiments: one is the full-data experiment where the original seismic attributes are used as the observations, and the other is the sparse-data experiment where leading wavelet coefficients are adopted as sparse presentation of seismic data. In this particular case study, our numerical results indicate that, 
\begin{enumerate}
	\item[(1)] combining the stopping criterion (\ref{eq:stopping_criterion_ndm}) and the stopping conditions (C1) and (C2) tends to make the iterative ensemble smoother stop earlier than only using stopping conditions (C1) and (C2), and this leads to better estimation results in terms of RMSEs;
	\item[(2)] using both production and seismic data in history matching results in more balanced improvements on average RMSEs of log PERMX and PORO than only using either production or seismic data;
	\item[(3)] provided that the stopping criterion (\ref{eq:stopping_criterion_ndm}) is adopted, using sparse representation substantially reduces the data size in history matching, while renders better estimation results (in terms of average RMSEs) than using the original seismic data. 
\end{enumerate}     

An extension of the proposed framework to 3D case studies will be carried out in the future. In addition, it is expected that the idea of sparse representation -- either based on wavelet or other basis functions -- may also be applied to different types of seismic data, for instance, acoustic impedance or saturation map. This is another topic that will be covered in our future investigations. 

\begin{table*}
	\centering
	\caption{\label{tab:average_RMSE} Average RMSEs of log PERMX and PORO of the initial ensemble and final ones obtained in different scenarios. The last row also lists the corresponding iteration steps at which these ensembles are obtained.}
	\begin{tabular}{|c|c|c|c|c|c|c|}
		\hline 
		& initial & S1  & S2 (full)  & S2 (sparse)  & S3 (full) & S3 (sparse) \\ 
		\hline  
		Average RMSE (log PERMX) & 0.9952  & 0.8941  & 1.0519 & 1.0088 & 0.9454  &  0.9389   \\ 
		\hline  
		Average RMSE (PORO) &  0.0236 & 0.0230 & 0.0205 & 0.0181 & 0.0188 & 0.0180  \\ 
		\hline
		Stopping step & 0  & 3 & 10  & 2  & 4 & 3 \\ 
		\hline 
	\end{tabular} 
\end{table*} 
  
\paragraph{{\bf Acknowledgement}} We would like to thank Dr. Mohsen Dadashpour for providing us the reservoir model of 2D Norne field, and thank Schlumberger for providing us academic software licenses to ECLIPSE and PETREL. XL acknowledges partial financial supports from the CIPR/IRIS cooperative research project ``4D Seismic History Matching'' which is funded by industry partners Eni,  Petrobras, and Total, as well as the Research Council of Norway (PETROMAKS). All authors acknowledge the Research Council of Norway and the industry partners -- ConocoPhillips Skandinavia AS, BP Norge AS, Det Norske Oljeselskap AS, Eni Norge AS, Maersk Oil Norway AS, DONG Energy A/S, Denmark, Statoil Petroleum AS, ENGIE E\&P NORGE AS, Lundin Norway AS, Halliburton AS, Schlumberger Norge AS, Wintershall Norge AS -- of The National IOR Centre of Norway for financial supports.

\bibliographystyle{TUPREP}
\bibliography{references}

\begin{thebibliography}{46}
\expandafter\ifx\csname natexlab\endcsname\relax\def\natexlab#1{#1}\fi

\bibitem[{Aanonsen et~al.(2009)Aanonsen, N{\ae}vdal, Oliver, Reynolds, and
  Vall\`{e}s}]{Aanonsen-ensemble-2009}
Aanonsen, S., G.~N{\ae}vdal, D.~Oliver, A.~Reynolds, and B.~Vall\`{e}s, The
  ensemble {K}alman filter in reservoir engineering: a review, \emph{SPE
  Journal}, \textbf{14}, 393--412, 2009, {SPE-117274-PA}.

\bibitem[{Abadpour et~al.(2013)Abadpour, Bergey, and Piasecki}]{abadpour20134d}
Abadpour, A., P.~Bergey, and R.~Piasecki, {4D} seismic history matching with
  ensemble {K}alman filter-assimilation on {H}ausdorff distance to saturation
  front, in \emph{SPE Reservoir Simulation Symposium}, Society of Petroleum
  Engineers, 2013, {SPE-163635-MS}.

\bibitem[{Awotunde(2014)}]{awotunde2014multiresolution}
Awotunde, A.~A., A multiresolution adjoint sensitivity analysis of time-lapse
  saturation maps, \emph{Computational Geosciences}, \textbf{18}, 677--696,
  2014.

\bibitem[{Buland and Omre(2003)}]{buland2003bayesian}
Buland, A. and H.~Omre, Bayesian linearized {AVO} inversion, \emph{Geophysics},
  \textbf{68}, 185--198, 2003.

\bibitem[{Chen et~al.(1974)Chen, Gavalas, Seinfeld, and
  Wasserman}]{chen1974new}
Chen, W.~H., G.~R. Gavalas, J.~H. Seinfeld, and M.~L. Wasserman, A new
  algorithm for automatic history matching, \emph{SPE Journal}, \textbf{14},
  593--608, 1974, {SPE-4545-PA}.

\bibitem[{Chen and Oliver(2013)}]{chen2013-levenberg}
Chen, Y. and D.~Oliver, {L}evenberg-{M}arquardt forms of the iterative ensemble
  smoother for efficient history matching and uncertainty quantification,
  \emph{Computational Geosciences}, \textbf{17}, 689--703, 2013.

\bibitem[{Chen and Oliver(2012)}]{chen2012multiscale}
Chen, Y. and D.~S. Oliver, Multiscale parameterization with adaptive
  regularization for improved assimilation of nonlocal observation, \emph{Water
  Resources Research}, \textbf{48}(4), {W04,503}, 2012.

\bibitem[{Dadashpour(2009)}]{dadashpour2009reservoir}
Dadashpour, M., \emph{Reservoir characterization using production data and
  time-lapse seismic data}, Ph.D. thesis, Norwegian University of Science and
  Technology, 2009.

\bibitem[{Dadashpour et~al.(2008)Dadashpour, Landr{\o}, and
  Kleppe}]{dadashpour2008nonlinear}
Dadashpour, M., M.~Landr{\o}, and J.~Kleppe, Nonlinear inversion for estimating
  reservoir parameters from time-lapse seismic data, \emph{Journal of
  Geophysics and Engineering}, \textbf{5}, 54, 2008.

\bibitem[{Domenico(1974)}]{Domenico:1974}
Domenico, S.~N., Effect of water saturation on seismic reflectivity of sand
  reservoirs encased in shale, \emph{Geophysics}, \textbf{39}, 759--769, 1974.

\bibitem[{Donoho and Johnstone(1995)}]{donoho1995adapting}
Donoho, D.~L. and I.~M. Johnstone, Adapting to unknown smoothness via wavelet
  shrinkage, \emph{Journal of the American Statistical Association},
  \textbf{90}, 1200--1224, 1995.

\bibitem[{Donoho and Johnstone(1994)}]{donoho1994ideal}
Donoho, D.~L. and J.~M. Johnstone, Ideal spatial adaptation by wavelet
  shrinkage, \emph{Biometrika}, \textbf{81}, 425--455, 1994.

\bibitem[{Emerick and Reynolds(2012{\natexlab{a}})}]{emerick2012ensemble}
Emerick, A.~A. and A.~C. Reynolds, Ensemble smoother with multiple data
  assimilation, \emph{Computers \& Geosciences}, \textbf{55}, 3--15,
  2012{\natexlab{a}}.

\bibitem[{Emerick and Reynolds(2012{\natexlab{b}})}]{emerick2012history}
Emerick, A.~A. and A.~C. Reynolds, History matching time-lapse seismic data
  using the ensemble {K}alman filter with multiple data assimilations,
  \emph{Computational Geosciences}, \textbf{16}, 639--659, 2012{\natexlab{b}}.

\bibitem[{Emerick et~al.(2013)Emerick, Reynolds et~al.}]{emerick2013history}
Emerick, A.~A., A.~C. Reynolds, et~al., History-matching production and seismic
  data in a real field case using the ensemble smoother with multiple data
  assimilation, in \emph{SPE Reservoir Simulation Symposium}, Society of
  Petroleum Engineers, 2013, {SPE-163675-MS}.

\bibitem[{Engl et~al.(2000)Engl, Hanke, and Neubauer}]{Engl2000-regularization}
Engl, H.~W., M.~Hanke, and A.~Neubauer, \emph{Regularization of Inverse
  Problems}, Springer, 2000.

\bibitem[{Evensen and van Leeuwen(2000)}]{Evensen2000}
Evensen, G. and P.~J. van Leeuwen, An ensemble {K}alman smoother for nonlinear
  dynamics, \emph{Mon. Wea. Rev.}, \textbf{128}, 1852--1867, 2000.

\bibitem[{Fahimuddin et~al.(2010)Fahimuddin, Aanonsen, and
  Skjervheim}]{fahimuddin2010ensemble}
Fahimuddin, A., S.~Aanonsen, and J.-A. Skjervheim, Ensemble based {4D} seismic
  history matching--integration of different levels and types of seismic data,
  in \emph{72nd EAGE Conference \& Exhibition}, 2010.

\bibitem[{Gassmann(1951)}]{Gassmann:1951}
Gassmann, F., {\"U}ber die {E}lastizit\"at por\"oser {M}edien,
  \emph{Vierteljahresschrift der Naturforschenden Gesellschaft}, \textbf{96},
  1--23, 1951.

\bibitem[{Gentilhomme et~al.(2015)Gentilhomme, Oliver, Mannseth, Caumon, Moyen,
  and Doyen}]{gentilhomme2015ensemble}
Gentilhomme, T., D.~S. Oliver, T.~Mannseth, G.~Caumon, R.~Moyen, and P.~Doyen,
  Ensemble-based multi-scale history-matching using second-generation wavelet
  transform, \emph{Computational Geosciences}, \textbf{19}, 999--1025, 2015.

\bibitem[{Gosselin et~al.(2003)Gosselin, Aanonsen, Aavatsmark, Cominelli,
  Gonard, Kolasinski, Ferdinandi, Kovacic, Neylon et~al.}]{gosselin2003history}
Gosselin, O., S.~Aanonsen, I.~Aavatsmark, A.~Cominelli, R.~Gonard,
  M.~Kolasinski, F.~Ferdinandi, L.~Kovacic, K.~Neylon, et~al., History matching
  using time-lapse seismic ({HUTS}), in \emph{SPE Annual Technical Conference
  and Exhibition}, Society of Petroleum Engineers, 2003, {SPE-84464-MS}.

\bibitem[{Gu and Oliver(2007)}]{Gu2007-iterative}
Gu, Y. and D.~Oliver, An iterative ensemble {K}alman filter for multiphase
  fluid flow data assimilation, \emph{SPE Journal}, \textbf{12}, 438--446,
  2007, {SPE-108438-PA}.

\bibitem[{Jack(2001)}]{jack2001coming}
Jack, I., The coming of age for {4D} seismic, \emph{First Break}, \textbf{19},
  24--28, 2001.

\bibitem[{Jansen(2012)}]{jansen2012noise}
Jansen, M., \emph{Noise reduction by wavelet thresholding}, vol. 161, Springer
  Science \& Business Media, 2012.

\bibitem[{Jin(2010)}]{jin2010regularized}
Jin, Q., On a regularized {L}evenberg--{M}arquardt method for solving nonlinear
  inverse problems, \emph{Numerische Mathematik}, \textbf{115}, 229--259, 2010.

\bibitem[{Johnstone and Silverman(1997)}]{johnstone1997wavelet}
Johnstone, I.~M. and B.~W. Silverman, Wavelet threshold estimators for data
  with correlated noise, \emph{Journal of the royal statistical society: series
  B (statistical methodology)}, \textbf{59}, 319--351, 1997.

\bibitem[{Katterbauer et~al.(2015)Katterbauer, Hoteit, and
  Sun}]{katterbauer2015history}
Katterbauer, K., I.~Hoteit, and S.~Sun, History matching of electromagnetically
  heated reservoirs incorporating full-wavefield seismic and electromagnetic
  imaging, \emph{SPE Journal}, \textbf{20}, 923 -- 941, 2015, { SPE-173896-PA}.

\bibitem[{Landr{\o}(2001)}]{landro2001discrimination}
Landr{\o}, M., Discrimination between pressure and fluid saturation changes
  from time-lapse seismic data, \emph{Geophysics}, \textbf{66}, 836--844, 2001.

\bibitem[{Leeuwenburgh and Arts(2014)}]{leeuwenburgh2014distance}
Leeuwenburgh, O. and R.~Arts, Distance parameterization for efficient seismic
  history matching with the ensemble {K}alman filter, \emph{Computational
  Geosciences}, \textbf{18}, 535--548, 2014.

\bibitem[{Li and Reynolds(2009)}]{Li2009-iterative}
Li, G. and A.~Reynolds, Iterative ensemble {K}alman filters for data
  assimilation, \emph{SPE Journal}, \textbf{14}, 496--505, 2009,
  {SPE-109808-PA}.

\bibitem[{Luo and Moroz(2009)}]{Luo-ensemble}
Luo, X. and I.~M. Moroz, Ensemble {K}alman filter with the unscented transform,
  \emph{Physica D}, \textbf{238}, 549--562, 2009.

\bibitem[{Luo et~al.(2015)Luo, Stordal, Lorentzen, and
  N{\ae}vdal}]{luo2015Iterative}
Luo, X., A.~Stordal, R.~Lorentzen, and G.~N{\ae}vdal, Iterative ensemble
  smoother as an approximate solution to a regularized minimum-average-cost
  problem: theory and applications, \emph{SPE Journal}, \textbf{20}, 962--982,
  2015, {SPE-176023-PA}.

\bibitem[{Mallat(2008)}]{mallat1999wavelet}
Mallat, S., \emph{A wavelet tour of signal processing}, Academic press, 2008.

\bibitem[{Mavko et~al.(2009)Mavko, Mukerji, and Dvorkin}]{mavko2009rock}
Mavko, G., T.~Mukerji, and J.~Dvorkin, \emph{The rock physics handbook: {T}ools
  for seismic analysis of porous media}, Cambridge University Press, 2009.

\bibitem[{Mindlin(1949)}]{Mindlin:1949}
Mindlin, R.~D., Compliance of elastic bodies in contact, \emph{Journal of
  Applied Mechanics}, \textbf{16}, 259--268, 1949.

\bibitem[{N{\ae}vdal et~al.(2005)N{\ae}vdal, Johnsen, Aanonsen, Vefring
  et~al.}]{naevdal2005reservoir}
N{\ae}vdal, G., L.~M. Johnsen, S.~I. Aanonsen, E.~H. Vefring, et~al., Reservoir
  monitoring and continuous model updating using ensemble {K}alman filter,
  \emph{SPE journal}, \textbf{10}, 66--74, 2005, {SPE-84372-PA}.

\bibitem[{Nur(1989)}]{Nur:1989}
Nur, A., Four-dimesional seismology and (true) direct detection of
  hydrocarbons: {T}he petrophysical basis, \emph{The Leading Edge}, \textbf{8},
  30--36, 1989.

\bibitem[{Oliver and Chen(2010)}]{Oliver-recent-2010}
Oliver, D. and Y.~Chen, Recent progress on reservoir history matching: a
  review, \emph{Computational Geosciences}, \textbf{15}, 185--221, 2010.

\bibitem[{Reuss(1929)}]{reuss1929berechnung}
Reuss, A., Berechnung der flie{\ss}grenze von mischkristallen auf grund der
  plastizit{\"a}tsbedingung f{\"u}r einkristalle., \emph{ZAMM-Journal of
  Applied Mathematics and Mechanics/Zeitschrift f{\"u}r Angewandte Mathematik
  und Mechanik}, \textbf{9}, 49--58, 1929.

\bibitem[{Sahni and Horne(2005)}]{sahni2005multiresolution}
Sahni, I. and R.~N. Horne, Multiresolution wavelet analysis for improved
  reservoir description, \emph{SPE Reservoir Evaluation \& Engineering},
  \textbf{8}, 53--69, 2005, {SPE-87820-PA}.

\bibitem[{Skjervheim and Evensen(2011)}]{skjervheim2011ensemble}
Skjervheim, J.-A. and G.~Evensen, An ensemble smoother for assisted history
  matching, in \emph{SPE Reservoir Simulation Symposium}, The Woodlands, Texas,
  USA, 2011, {SPE-141929-MS}.

\bibitem[{Skjervheim et~al.(2007)Skjervheim, Evensen, Aanonsen, Ruud, and
  Johansen}]{skjervheim2007incorporating}
Skjervheim, J.-A., G.~Evensen, S.~I. Aanonsen, B.~O. Ruud, and T.-A. Johansen,
  Incorporating {4D} seismic data in reservoir simulation models using ensemble
  {K}alman filter, \emph{SPE Journal}, \textbf{12}, 282 -- 292, 2007,
  {SPE-95789-PA}.

\bibitem[{Smith and Gidlow(1987)}]{Smith:1987}
Smith, G. and P.~Gidlow, Weighted stacking for rock property estimation and
  detection of gas, \emph{Geophysical Prospecting}, \textbf{35}, 993--1014,
  1987.

\bibitem[{Tarantola(2005)}]{Tarantola-inverse}
Tarantola, A., \emph{Inverse Problem Theory and Methods for Model Parameter
  Estimation}, SIAM., 2005.

\bibitem[{Trani et~al.(2012)Trani, Arts, and Leeuwenburgh}]{trani2012seismic}
Trani, M., R.~Arts, and O.~Leeuwenburgh, Seismic history matching of fluid
  fronts using the ensemble {K}alman filter, \emph{SPE Journal}, \textbf{18},
  159--171, 2012, {SPE-163043-PA}.

\bibitem[{Van~Leeuwen and Evensen(1996)}]{van1eeuwen996data}
Van~Leeuwen, P.~J. and G.~Evensen, Data assimilation and inverse methods in
  terms of a probabilistic formulation, \emph{Mon. Wea. Rev.}, \textbf{124},
  2898--2913, 1996.

\end{thebibliography}
\end{document}